\documentclass[3p]{elsarticle}

\usepackage{graphicx}
\usepackage{dcolumn}
\usepackage{bm}
\usepackage{amsmath}
\usepackage{amsfonts}
\usepackage{xcolor}
\usepackage{etoolbox}

\usepackage[T1]{fontenc}
\usepackage[utf8]{inputenc}
\usepackage{hyperref}
\usepackage{caption}
\myfooter[C]{Final version published as Phys. Rep. \textbf{1183}, 1--98, 2026.}
\graphicspath{{figures/}{.}}
\setkeys{Gin}{width=\linewidth,height=0.78\textheight,keepaspectratio}
\providecommand{\rmmu}{\mathrm{\mu}}

\providecommand{\tsup}[1]{\textsuperscript{#1}}

\providecommand{\tplus}{+}
\providecommand{\tmin}{-}
\providecommand{\degree}{\ensuremath{^\circ}}

\providecommand{\nf}{}
\providecommand{\svert}{\vert}
\providecommand{\grantsponsor}[1]{#1}
\providecommand{\grantnumber}[2]{#2}

\newcommand{\citeA}{\citet}

\begin{document}
\raggedbottom

\begin{frontmatter}
\title{Self-Organized Pattern Formation  in \\Geological Soft Matter}

\author[1,2]{Julyan H. E. Cartwright}
\address[1]{Instituto Andaluz de Ciencias de la Tierra, CSIC, 18100 Armilla, Granada, Spain}
\ead{julyan.cartwright@csic.es}
\address[2]{Instituto Carlos I de F\'isica Te\'orica y Computacional, Universidad de Granada, 18071 Granada, Spain}
\author[3]{Charles S. Cockell}
\address[3]{School of Physics and Astronomy, University of Edinburgh, Edinburgh, UK}
\author[4]{Lucas Goehring}
\address[4]{School of Science and Technology, Nottingham Trent University, Nottingham NG11 8NS, UK} 
\author[5]{Silvia Holler}
\address[5]{Cellular Computational and Biology Department, CIBIO, Laboratory for Artificial Biology, University of Trento, Via Sommarive 9, Povo, 38123, Italy}
\author[6]{Sean F. Jordan}
\address[6]{Life Sciences Institute, School of Chemical Sciences, Dublin City University, Glasnevin, Dublin 9, Ireland}
\author[3]{Pamela Knoll}
\author[7]{Electra Kotopoulou}
\address[7]{Ecologie Soci\'et\'e Evolution, CNRS, Universit\'e Paris-Saclay, AgroParisTech, Gif-sur-Yvette, France}
\author[8]{Corentin C. Loron}
\address[8]{School of Natural Sciences, Trinity College Dublin, Dublin, Ireland}
\author[3]{Sean McMahon}
\author[9]{Stephen W. Morris}
\address[9]{Department of Physics, University of Toronto, 60 St. George St., Toronto, ON, Canada, M5S 1A7}
\author[10]{Anna Neubeck}
\address[10]{Department of Earth Sciences, Uppsala University, Uppsala, Sweden}  
\author[11]{Carlos Pimentel}
\address[11]{Departamento de Mineralog\'ia y Petrolog\'ia, Facultad de Ciencias Geol\'ogicas, Universidad Complutense de Madrid, 28040 Madrid, Spain}
\ead{cpimentelguerra@geo.ucm.es}
\author[1]{C. Ignacio Sainz-D\'{\i}az}
\author[12]{Noushine Shahidzadeh}
\address[12]{University of Amsterdam, Institute of Physics-Van der Waals-Zeeman Institute,  Science Park 904, 1098 XH Amsterdam, The Netherlands}
\author[13]{Piotr Szymczak}
\address[13]{Institute of Theoretical Physics, Faculty of Physics, University of Warsaw, Poland}
\ead{piotrek@fuw.edu.pl}

\begin{abstract}
Geological materials are often seen as the antithesis of soft; rocks are hard. However, during the formation of minerals and rocks, all the systems we shall discuss, indeed geological materials in general, pass through a stage where they are soft. This occurs either because they form at a high temperature --- igneous or metamorphic rock --- or because they form at a lower temperature but in the presence of water --- sedimentary rock. For this reason it is useful to introduce soft-matter concepts into the geological domain. There is a universality in the diverse instances of geological patterns that may be appreciated by looking at the common aspect in their formation of having passed through a stage as soft matter.
\end{abstract}

\end{frontmatter}

\tableofcontents

\section{Introduction}

Pattern formation occurs at all scales in our universe, from the very largest of the large-scale structures of the cosmos, to the very smallest: how quarks organize themselves within atomic nuclei. In this review, we concentrate on an intermediate range spanning millimetres to metres. In this mesoscopic domain, the governing processes are no longer the chemistry of individual atoms, molecules, or crystals. Instead, patterns emerge from collective behaviour: interfaces, filaments, channels, and networks that organize under modest driving forces. 

This is the realm of soft matter. Here, the stresses and energies that act are often comparable to $k_B T$, so thermal fluctuations, weak elasticities, and gentle flows can reorganize material configurations. A defining feature is the appearance of structures produced by self-organization and self-assembly; features large enough to be directly imaged and manipulated, yet still far below the overall extent of the material. These mesoscopic entities mediate transport, reactivity, and mechanics, and they often display universal motifs --- branching, coarsening, fingering, rippling --- that recur across disparate systems.

Our focus is on rocks; materials that hardly seem ``soft''. Yet at certain stages of their formation they behaved as soft matter, when thermal or mechanical stresses were sufficient to reshape their mesoscopic structure. Thus, understanding geological soft-matter systems is foundational for explaining how past processes produced the shapes and patterns we see in rocks and minerals today.

Although there is a large literature in physics on self-organization and pattern formation,
it is rarely presented in relation to geology. There are a few notable exceptions. The first
is Peter Ortoleva's \emph{Geochemical Self-Organization}~\cite{Ortoleva1994}; important in
scope but  mathematically intensive so that few geoscientists make it through. A second
thread is \emph{Growth, Dissolution and Pattern Formation in Geosystems}~\cite{Jamtveit1999},
followed by the review \emph{Sculpting of Rocks by Reactive Fluids}~\cite{Jamtveit2012};
These works focus largely on dissolution--precipitation processes mediated by fluid flow.
While many geological patterns arise this way, a wide class of phenomena lies outside that
framework.
A closely related, soft-matter-oriented contribution is \emph{Soft Matter Physics of the
Ground Beneath Our Feet} by~\citeA{voigtlander2024}, which centres on soils and sediments. By
contrast, we turn to lithified systems: patterns  preserved in hard rock that bear the
imprint of earlier, transient soft regimes. Our aim is to show how signatures of that early
softness govern the genesis and present architecture of these rock structures.

The recent advances in soft matter physics have revealed numerous new processes that provide critical insights into long-standing puzzles in geological pattern formation. These developments offer a fresh perspective, potentially unlocking key mechanisms behind the dynamics of mesoscale patterns in Earth's systems by linking them to the stage when they were soft. This focus on soft matter introduces novel insights that go  beyond the scope of previous reviews, addressing complexities in geological phenomena that have long resisted explanation.

Our aim is that soft-matter physicists with an interest in geological applications will  find this review of use.
At the same time, we aspire to indicate to geologists the range of mechanisms in soft-matter physics that are involved in geological self organization. 

 It should be noted that some pattern-forming phenomena in rocks may involve biological contributions. This includes a broad grey zone  of  cases that may be biogenic or may instead reflect physical processes with microbial mediation when present. We discuss these in a separate
 review~\cite{geo_bio_review}.

\section{Soft matter}
\label{sec:soft}

\begin{figure}
\centering
 \includegraphics{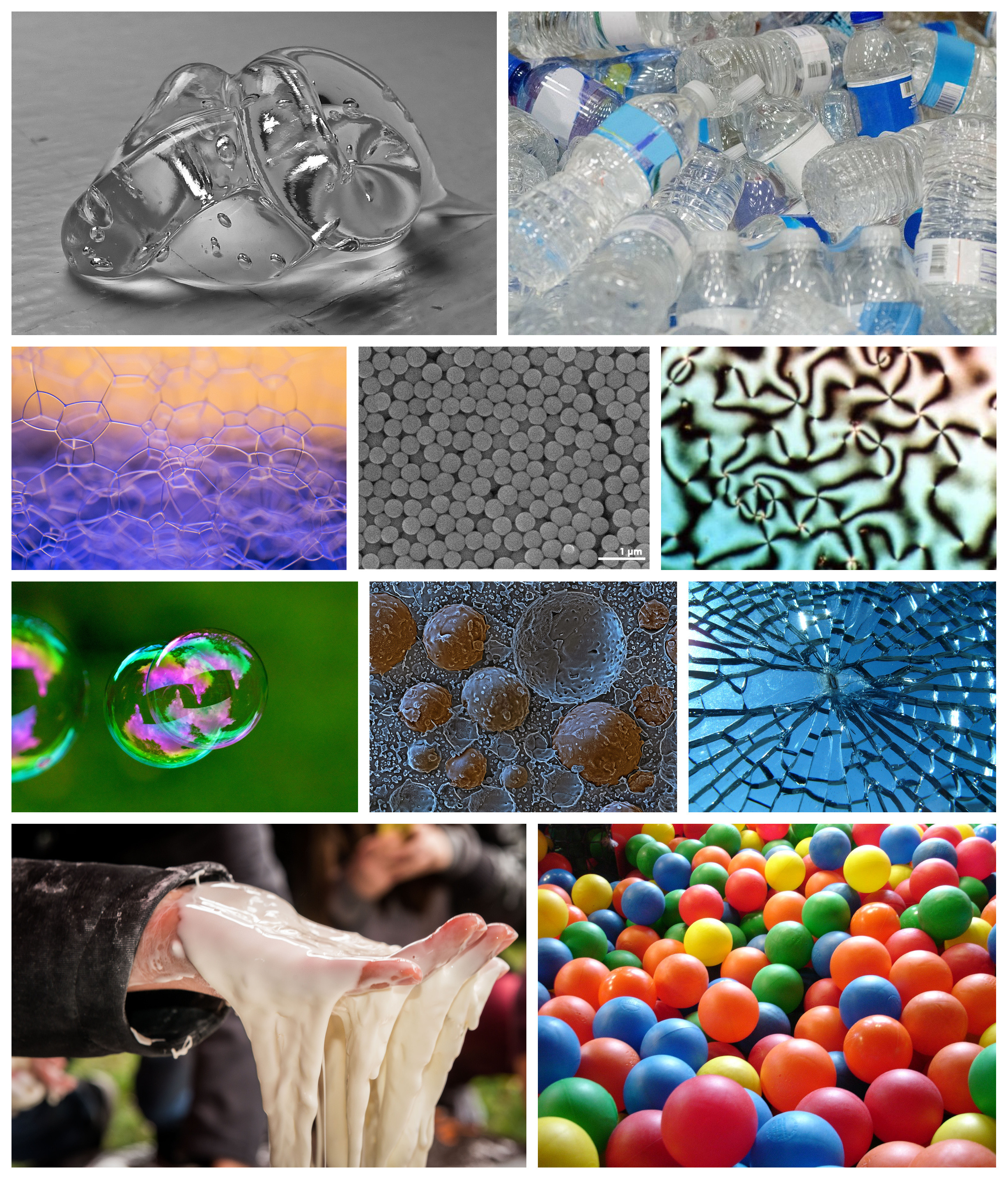}
\caption{\label{fig:soft_matter}Examples of soft matter. From top left to bottom right: gel, polymer, foam, colloid, liquid crystal, membrane, emulsion, glass, 
viscoelastic material,
granular medium.
(Images: From top left to bottom right: 
 Steve Johnson; CC-BY-2.0,
 public domain,
 AdaptaLux; CC-BY-2.0,
 Yasrena;  CC-BY-SA-4.0,
 Minutemen; CC-BY-SA-2.5,
 public domain,
 BASF; CC-BY-NC-ND-2.0,
 Jef Poskanzer; CC-BY-2.0,
 Victor Wong,; CC-BY-SA 2.0,
 Julie Kertesz; CC-BY-2.0.)
}
\end{figure}

Soft matter is a relatively young field of condensed matter physics~\cite{de1992soft}. But
what is soft matter? 
Common examples are complex fluids, gels, liquid crystals, polymers, foams, and colloids (Fig.~\ref{fig:soft_matter}).
 What these have in common is that the  material properties are not based directly on the molecular-scale structure, but on mesoscopic structures that self organize in the system.
 
A number of  definitions have been suggested to delimit the field: that soft matter is soft; that it is a complex liquid; that the building blocks are intermediate in size between atoms and the system size; that forces between the building blocks are of the magnitude of thermal fluctuations; and that soft matter shows a large response to weak forces.
 All of these definitions seek to characterize systems in which mesoscopic, intermediate-scale structures self organize and self assemble and can display a sensitive spatial dependence on the initial conditions that is similar to the  sensitive temporal dependence that nonlinear chaotic systems experience. 
 All of these definitions also work reasonably well to differentiate soft matter from traditional solid-state condensed matter like metals,
semiconductors, superconductors, etc. That is to say, all give a cut-off at the small scale, but what about a cut-off at larger length-scales?

At large scales, things are less clear. 
Patterns form in condensed matter up to the large-scale structure of the universe. And the building blocks of those patterns --- stars in a galaxy, galaxies in a cluster, etc --- are effectively intermediate in scale.
In this sense, it would suggest that there is no upper limit. Then, in the geological
sciences, the whole of geomorphology, being deformable and sensitive to collective effects,
might be viewed as soft matter. Indeed, this has been argued~\cite{jerolmack2019}.
For example, during erosion, sediment transport and accumulation, relatively weak forces,
acting over longish times, make dramatic changes to the
landscape~\cite{nash2007geochemical,budd2016introduction}.
It is clear, however, that this time-scale is different to that the pioneers of the field of soft matter had in mind. They were thinking of 
relatively short time-scales over which soft materials deform while geology deals with exceedingly long time-scales over which
weak forces can make substantial changes. 
In this sense, the mesoscale in geology is especially broad and context dependent---and, more generally, reminds us that in physics any form of condensed matter can exhibit ``soft'' behaviour when viewed at sufficiently large length- and time-scales.

We argue that it is reasonable to include  these large-scale phenomena as \textit{soft systems}, though not strictly as \textit{soft materials}. Nevertheless, doing so opens up the field so much that it is in danger of becoming a catch-all. To remain grounded, we adopt an operational definition: soft matter is what is studied within soft-matter groups and discussed at soft-matter conferences (Fig.~\ref{fig:soft_matter}).

Admittedly, softness may seem a peculiar notion in the case of rocks, which are hard and not easily deformed by thermal fluctuations. Yet they were soft when they were formed, and they became hard after. In other words, many geological patterns originate through deformation, diffusion or flow in a liquid or gel medium that later solidifies. 

In the following, we consider some common, and overlapping, classes of soft matter systems
and examine how geological phenomena fit into these categories as forms of geological soft
matter. As~\citeA{evans2019} put it, ``predicting the structure and dynamics of such complex
phases of matter from the constituent building blocks and their interactions defines
soft-matter science''.

\section{Physical types of geological  soft  matter}
\label{sec:types}

Let us review some common types of soft matter and note instances in which they are found in geology. Note that as these types overlap,  many examples can fall into several of these categories.

\subsection{Gels}

\begin{figure}
\centering
\includegraphics{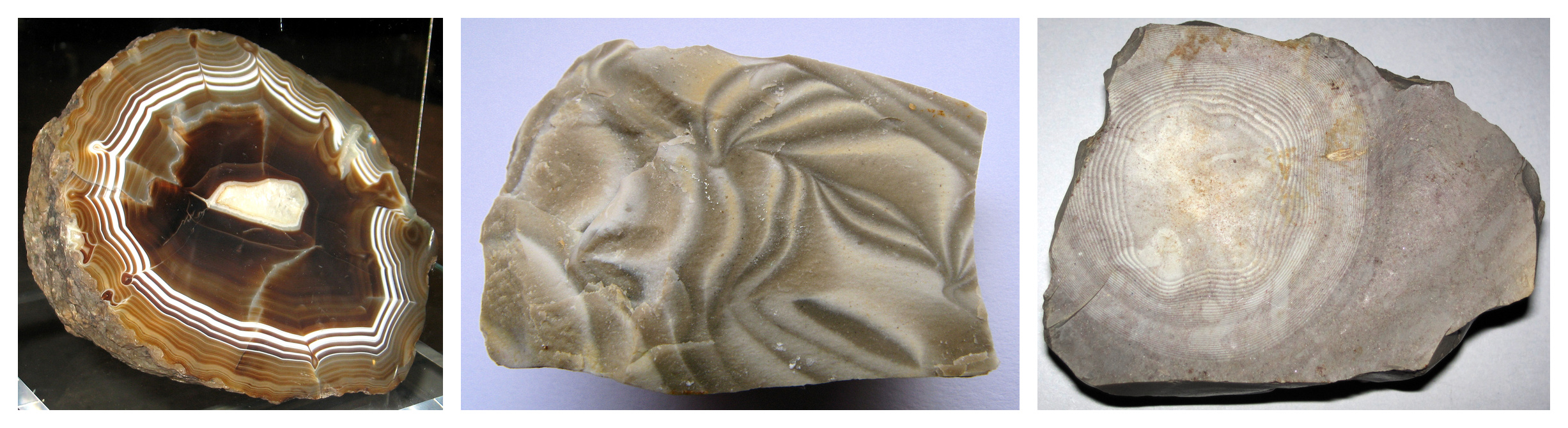}
\caption{\label{fig:gel}
Geological gels: these geological materials most probably passed through a gel stage in their formation. 
(left) Agate-filled geode, Brazil. 
(centre) Striped Flint, Poland.
(right) Grey chert, Knox County, Ohio, USA.
(Images: Left: James St. John; CC-BY-2.0;
 Centre: Ra'ike;
CC BY-SA 3.0,
Right: James St. John; CC-BY-2.0.)
}
\end{figure}

A gel is a network of particles suspended in a fluid medium.
Several geological materials passed through a gel stage during their formation, including
agates (Sec.~\ref{sec:agates}), flints  and cherts
(Sec.~\ref{sec:flint})~\cite{oehler1976,howard2018} (Fig.~\ref{fig:gel}). Dendritic patterns
originating from viscous fingering (Sec.~\ref{sec:fingering}) likewise reflect gel
dynamics~\cite{langer1989}.
Basalt columns (Sec.~\ref{sec:basalt}) form when molten lava cools and
cracks~\cite{Mallet1875,Spry1962} near the glass transition
temperature~\cite{Ryan1981,Goehring2008}, making them  both soft and fluidic during
formation. Gold dendrites (Sec.~\ref{sec:nuggets}) in rocks have also been shown to form within
silica gel~\cite{Monecke2023}. Silica gels may even play a role in earthquakes through
weakening faults~\cite{kirkpatrick2013,borhara2020}.

\begin{figure}
\centering
\includegraphics{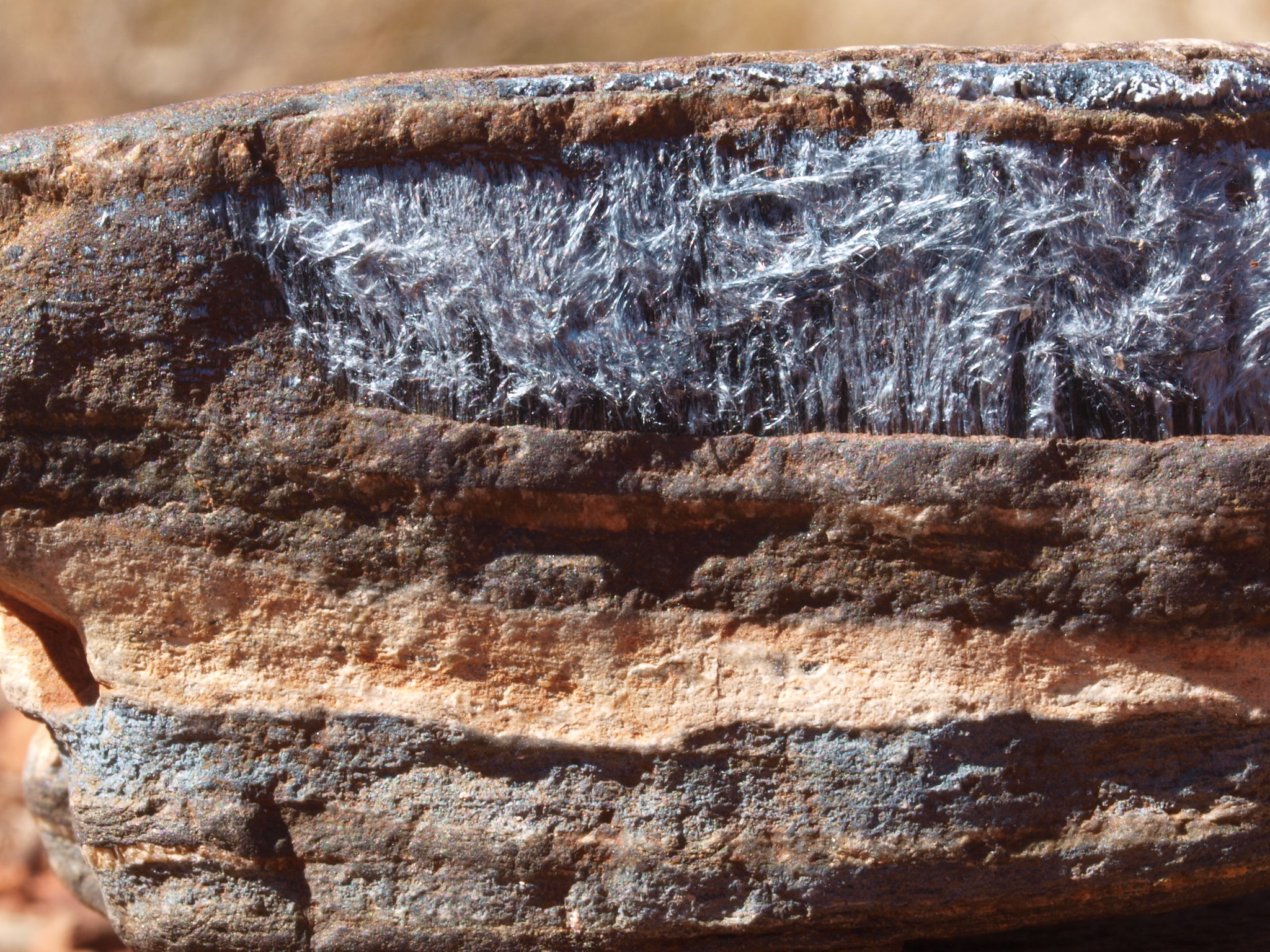}
\caption{\label{fig:polymer}
Geological polymers:
Crocidolite, blue asbestos, an amphibole mineral. 
(Image: Edgar Vonk; CC-BY-NC-SA-2.0.)
}
\end{figure}

\subsection{Polymers}

A polymer is a macromolecule composed of repeating structural subunits.
In the Earth's crust, many alumino-silicate minerals, including clays, quartz, feldspar, and
amphiboles, have polymeric atomic structures and thus can be regarded as geological polymers
(Fig.~\ref{fig:polymer}). Agates (Sec.~\ref{sec:agates}) and opals (Sec.~\ref{sec:opals}) are examples.
The term \emph{geopolymer} is also used in materials research, though here it usually refers
to synthetic substances, typically alumino-silicate in composition~\cite{davidovits1994},
with petroleum-based variants also described~\cite{kim2006}.

\subsection{Foams}

\begin{figure}
\centering
\includegraphics{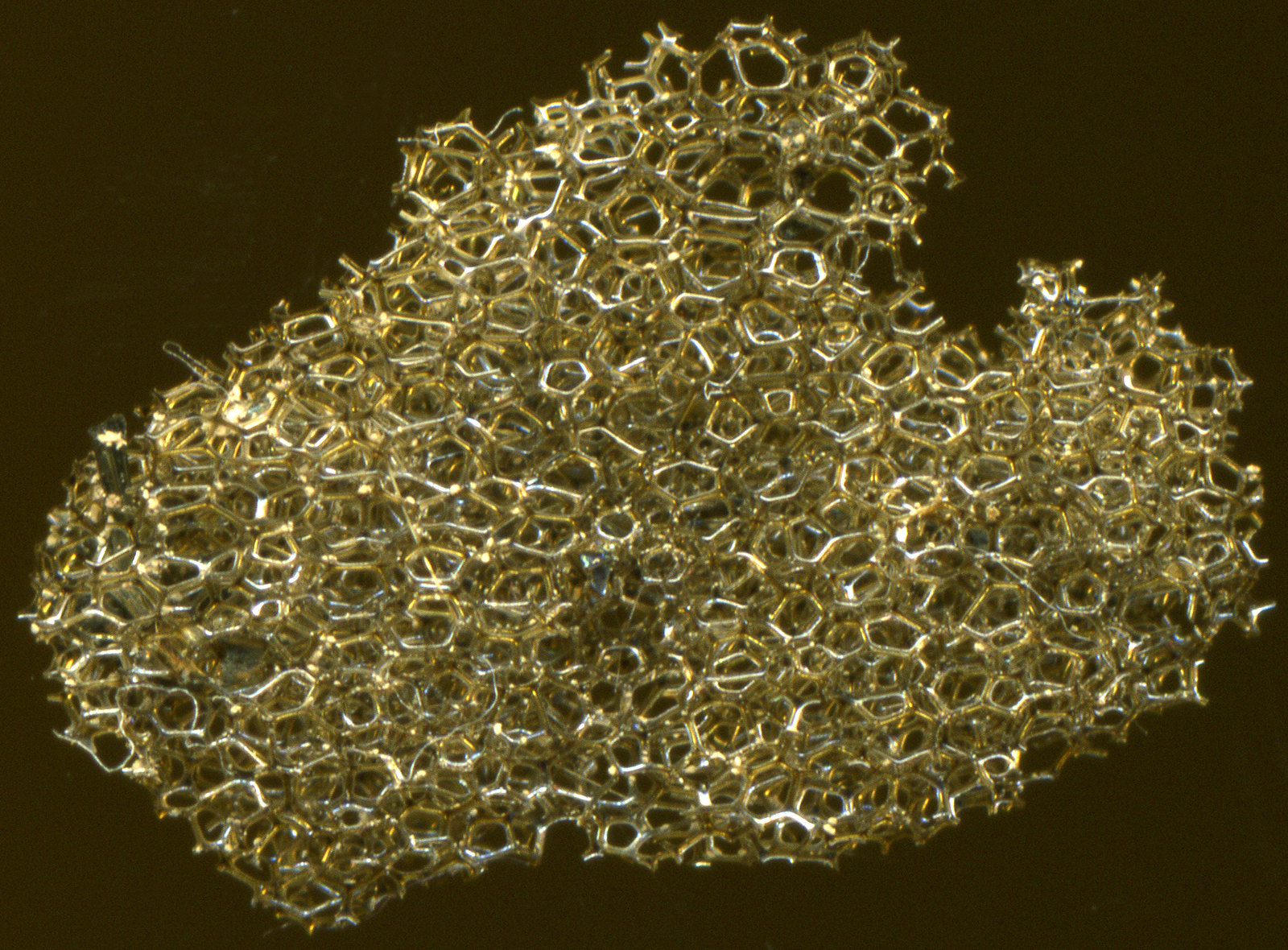}
\caption{\label{fig:foam}
Geological foam:
Reticulite from the Holocene of Hawaii, 1.8 centimetres across.
(Image: James St. John; CC-BY-2.0.)
}
\end{figure}

A foam consists of gas bubbles dispersed in a liquid or solid.
Geological foams (Fig.~\ref{fig:foam}) include bubbly volcanic rocks such as pumice and
scoria~\cite{jaupart1989,pal2003,vasseur2020}. 
Pumice is light and highly porous, formed when molten lava cools rapidly, trapping gas bubbles  and producing a foam-like structure. 
Amygdaloidal basalts form when such bubbles later become infilled with secondary minerals
precipitated from circulating fluids~\cite{morris1930amygdules}, as with agates
(Sec.~\ref{sec:agates}).

\subsection{Colloids}

\begin{figure}
\centering
\includegraphics{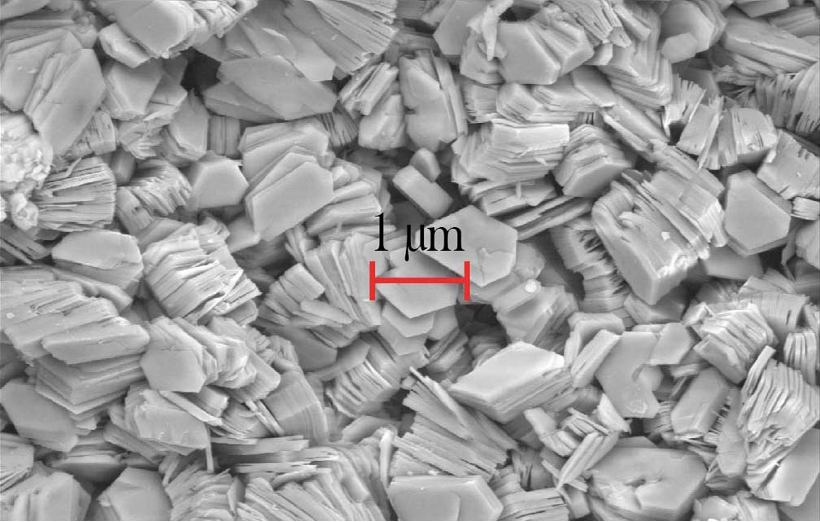}
\caption{\label{fig:colloid}
Geological colloids: phyllosilicates.
(Image: Paedona; CC-BY-SA-4.0.)
}
\end{figure}

A colloid is a two-phase system in which particles of one substance are dispersed within a continuous medium of another.
Silica (amorphous $\text{SiO}_2$) and many clay minerals commonly occur as colloidal dispersions
in water, with particle sizes on the order of ${\sim}$1--1000~nm~\cite{yariv1979}
(Fig.~\ref{fig:colloid}).

Clays are aggregates of fine-grained minerals, typically less than $2~\mu$m in size,  so they often behave as colloids in water. Their small size and plate-like shapes give them very high specific surface area and, through structural and edge charges, substantial surface charge. These features govern their interactions with water and dissolved ions (e.g.,~hydration, double-layer forces, ion exchange), and can help keep particles suspended (alongside turbulence; Brownian motion dominates mainly for the finest fractions).

The colloidal nature of clays underlies several key properties. First, plasticity: with added water, clays deform readily and can be shaped, which is central to ceramics and pottery. Second, swelling: smectitic clays (e.g.,~montmorillonite) intercalate water and expand; behaviour important in drilling muds and landfill liners. Third, adsorption and ion exchange: charged surfaces and interlayers bind ions and organic molecules, enabling contaminant removal and catalytic uses.

\begin{figure}
\centering
\includegraphics{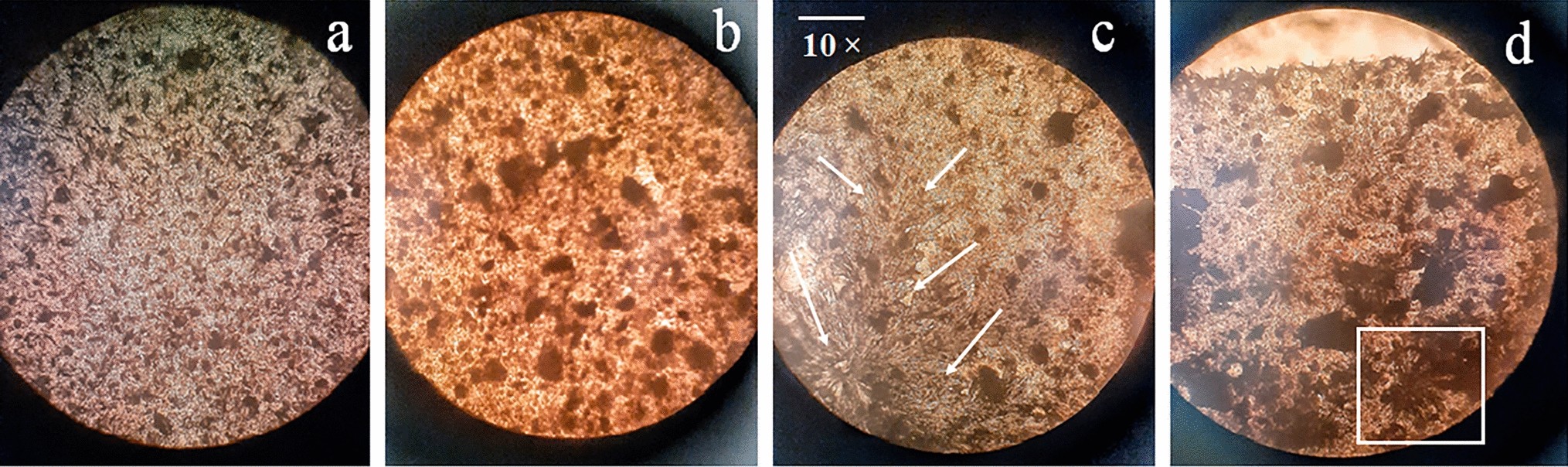}
\caption{\label{fig:liquid_crystal}
Geological liquid crystals.
POM images observed for 0.055~g of clay in different Na$_2$SO$_4$ solvent concentrations. 
A: In 0.05 M Na$_2$SO$_4$ 
B: In 0.1 M Na$_2$SO$_4$ 
C:
In 0.2 M Na$_2$SO$_4$, showing lyotropic aggregates with nematic phase; 
D: In 0.2 M Na$_2$SO$_4$, showing lyotropic nematic micelle formation in the lower
right corner.
(Source: \citeA{neelamma2022bentonite}.)
}
\end{figure}

Clay minerals such as montmorillonite or kaolinite are important colloidal materials in soil and sedimentary rocks. They have a layered structure and can form colloidal suspensions due to their small particle size and surface charge.
Colloidal silica, also known as silica gel,  can be found in various geological settings, including hydrothermal systems, sedimentary environments, and volcanic ash deposits.
Opals (Sec.~\ref{sec:opals}) are colloidal crystals of silica nanospheres~\cite{jones1964}.
Colloidal iron and manganese oxides often occur in soils, sediments, and weathering profiles. They contribute to the colouration of rocks such as agates (Sec.~\ref{sec:agates}) and play a role in the transport and sequestration of trace elements.
Colloidal gold nanoparticles are thought to play a significant role in ore formation
(Sec.~\ref{sec:nuggets})~\cite{Saunders2022}.

\subsection{Liquid crystals}

A liquid crystal is a phase of matter that exhibits fluidity together with long-range
orientational, and sometimes positional, order. Suspensions of anisotropic mineral particles
can form lyotropic liquid-crystal phases; this was noted a century ago~\cite{zocher1925} and
is now an active area in chemistry and materials science~\cite{davidson2005}. Clays provide a
key geological example (Fig.~\ref{fig:liquid_crystal}). When dispersed in water above certain
concentrations, plate-like clay particles align due to their anisotropic shape and
interparticle interactions, producing ordered phases. In nematic phases, particles share a
common director without positional order; at higher concentrations they can form smectic
phases, with parallel layers exhibiting both orientational and one-dimensional positional
order. Such liquid-crystal textures have been observed in laboratory studies of aqueous clay
gels~\cite{gabriel1996,michot2006}.

The liquid-crystalline behaviour of clays is significant in both natural and industrial contexts. In sedimentary basins, the alignment of clay particles can affect the porosity and permeability of the rock, influencing fluid flow and the extraction of resources. In industrial processes, understanding the liquid-crystal phases of clays can help in designing materials with specific mechanical or optical properties.

It has been proposed that zebra rock patterns (Sec.~\ref{sec:Zrock}) may have originated as
liquid crystals~\cite{mattievich2003}.

\subsection{Membranes}

\begin{figure}
\centering
\includegraphics{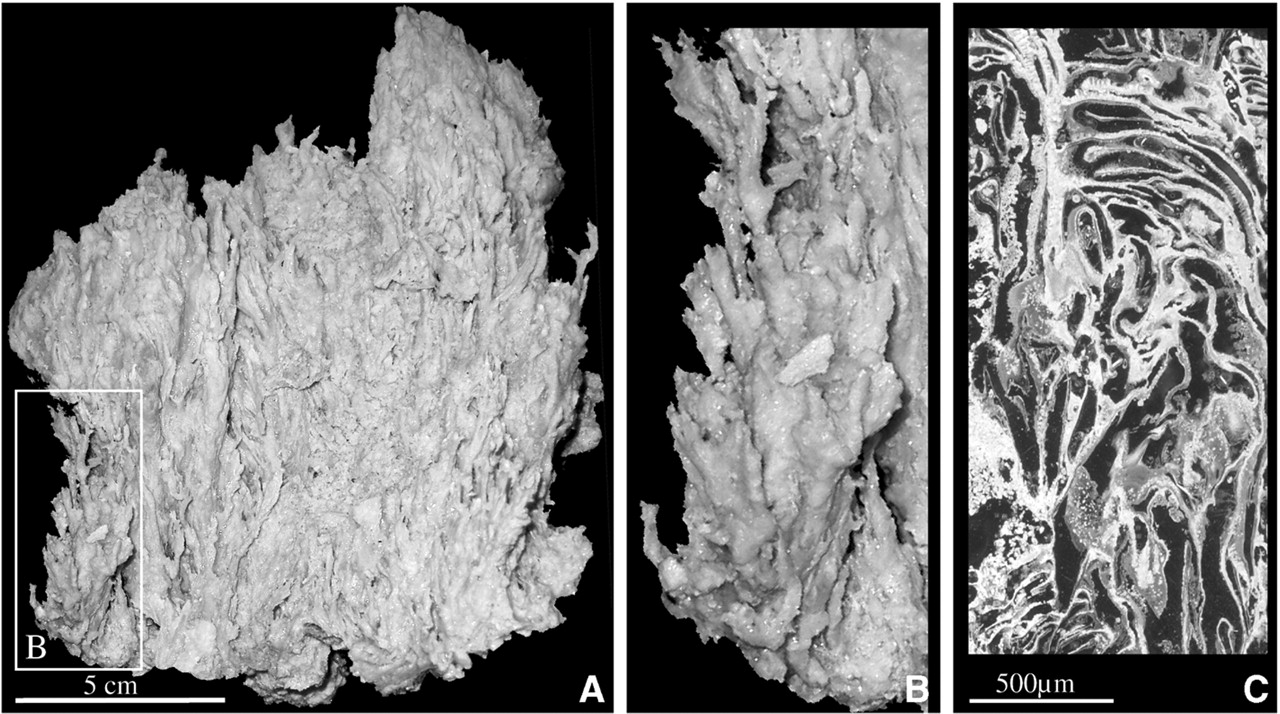}
\caption{\label{fig:membrane}
Geological membranes. A--C progressively zoom in from the centimetre to the micrometre scale on
mineral membranes at Lost City hydrothermal field.
(Source: \citeA{kelley2005serpentinite}.)
}
\end{figure}

A membrane is a two-dimensional interface that acts as a selective barrier, allowing some molecules and ions to pass through while inhibiting others.
Geological mineral membranes occur in clays~\cite{kharaka1973,fritz1986,neuzil2000osmotic},
shales, siltstones, and zeolites, and in hydrothermal vents  and vent fields
(Sec.~\ref{sec:hydrothermalvents})~\cite{russell1994,barge2015chemical} (Fig.~\ref{fig:membrane}).
These mineral membranes are often semipermeable, allowing only the solvent and not the solute
species to pass, generating osmotic pressures in geological settings where fluids and
semipermeable media interact. Such osmotic effects (Sec.~\ref{sec:osmotic}) can influence fluid
flow, chemical transport, and hence the formation of minerals and ore
deposits~\cite{barge2015chemical}.
Mechanisms involving osmotic membranes have been proposed for the formation of agates (Sec.~\ref{sec:agates}) and menilites (Sec.~\ref{sec:menilites}).

\begin{figure}
\centering
\includegraphics{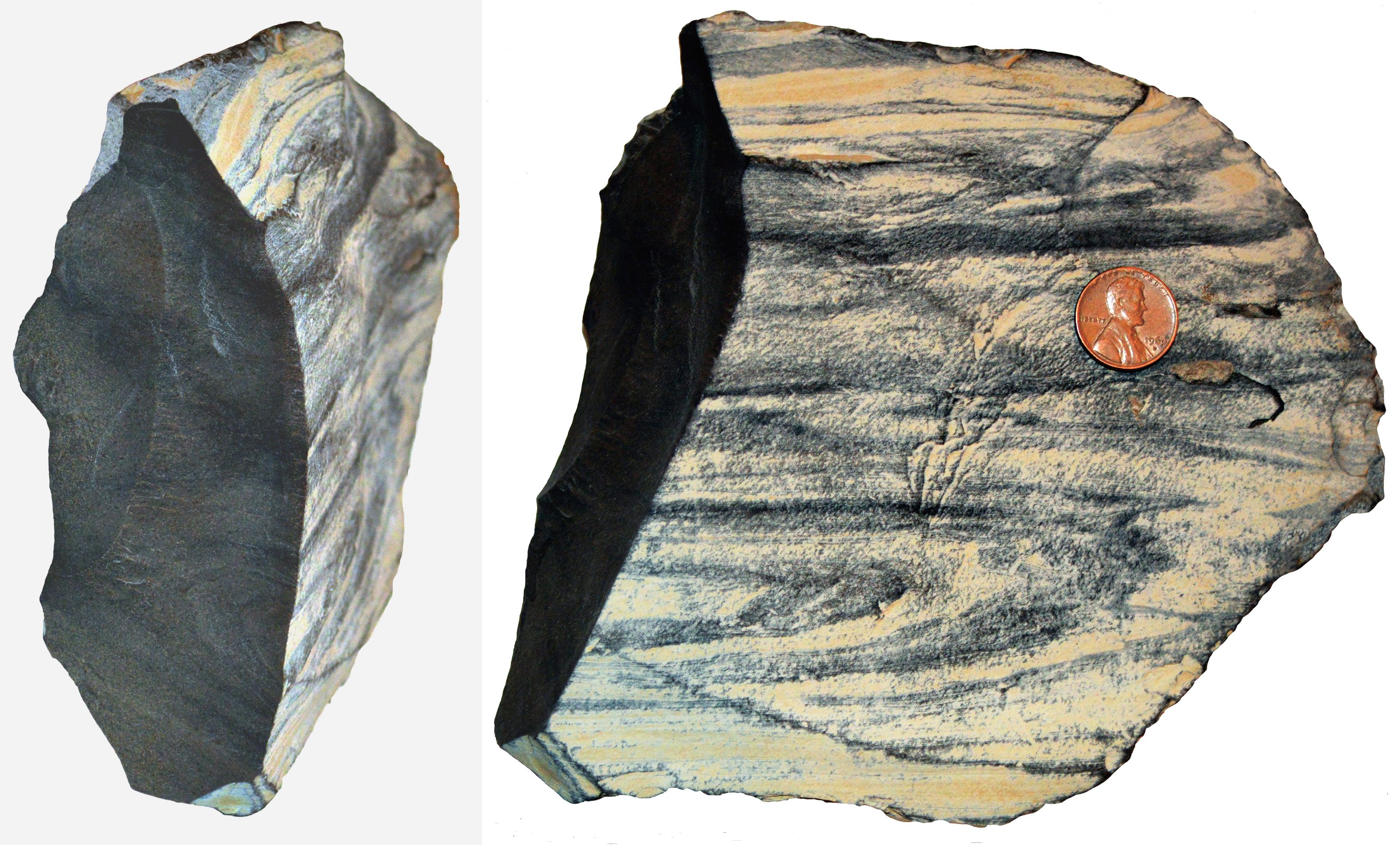}
\caption{\label{fig:emulsion}
Geological emulsion:
oil shale from the Mahogany Zone of the Green River Formation, Colorado, USA.
(Images: Georgialh; CC-BY-SA-3.0.)
}
\end{figure}

\subsection{Emulsions}

An emulsion is a colloid of liquid droplets dispersed in an immiscible liquid. 
Like foams, emulsions occur in magmas, and may be preserved in events such as meteorite impacts. 
It has been proposed that the Sudbury Igneous Complex in Canada was formed through differentiation
of a superheated impact melt sheet into a  viscous emulsion~\cite{zieg2005}.
Magma mingling has been proposed to occur when two magmas of different compositions come into contact
but do not fully mix into a single homogeneous melt, leading to emulsion-like textures in
igneous rocks; an example that has been proposed is amphibole--biotite veins~\cite{gogoi2019}.

In sedimentary settings, water--oil--clay emulsions are common (e.g.,~in heavy-oil reservoirs
and oil sands), and early accounts describe organic--aqueous emulsions associated with oil
shales~\cite{cunningham1916iv}. 
Upon burial and diagenesis, however, the organic phase polymerizes into kerogen dispersed within a fine-grained mineral matrix --- yielding oil shale --- so the end product is not a preserved emulsion (Fig.~\ref{fig:emulsion}).

\subsection{Glasses}

\begin{figure}
\centering
\includegraphics{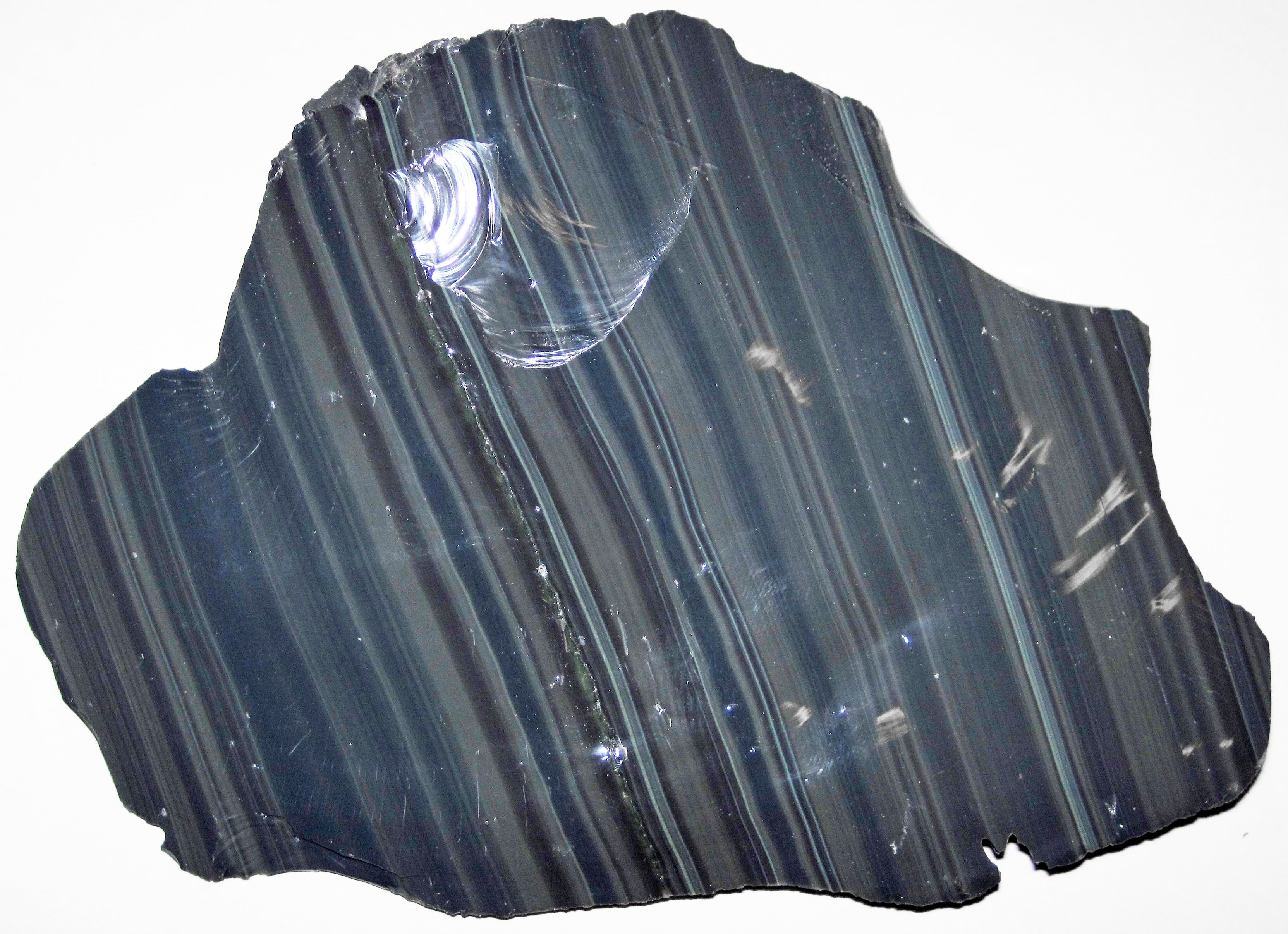}
\caption{\label{fig:glass}
Geological glass:
gray obsidian.
(Image: James St. John;
CC-BY-2.0.)
}
\end{figure}

A glass is an amorphous (non-crystalline) solid. Natural glasses predate human
manufacture~\cite{cicconi2019}, and geological glasses occur in igneous, metamorphic, and
sedimentary contexts~\cite{heide2011}. Obsidian is a volcanic glass that forms when
silica-rich (felsic) lava cools rapidly, suppressing crystallization; it is typically dark,
vitreous, and may contain vesicles or mineral inclusions, with occasional
patterning~\cite{ma2001micro} (Fig.~\ref{fig:glass}).
Tektites are impact glasses formed when meteorite strikes melt target rocks and soils; the ejecta cools in flight to glassy bodies  (e.g.,~splashform, teardrop, or aerodynamically sculpted shapes). 
Fulgurites (Sec.~\ref{sec:Fulgurites}) form when lightning melts sandy soil or sandy near-surface
sediments, producing  hollow glassy tubes and branching
structures~\cite{pasek_fulgurite_2012}.
Agates (Sec.~\ref{sec:agates}) and basalt columns (Sec.~\ref{sec:basalt}) may involve geological glasses in their formation.

\begin{figure}
\centering
\includegraphics{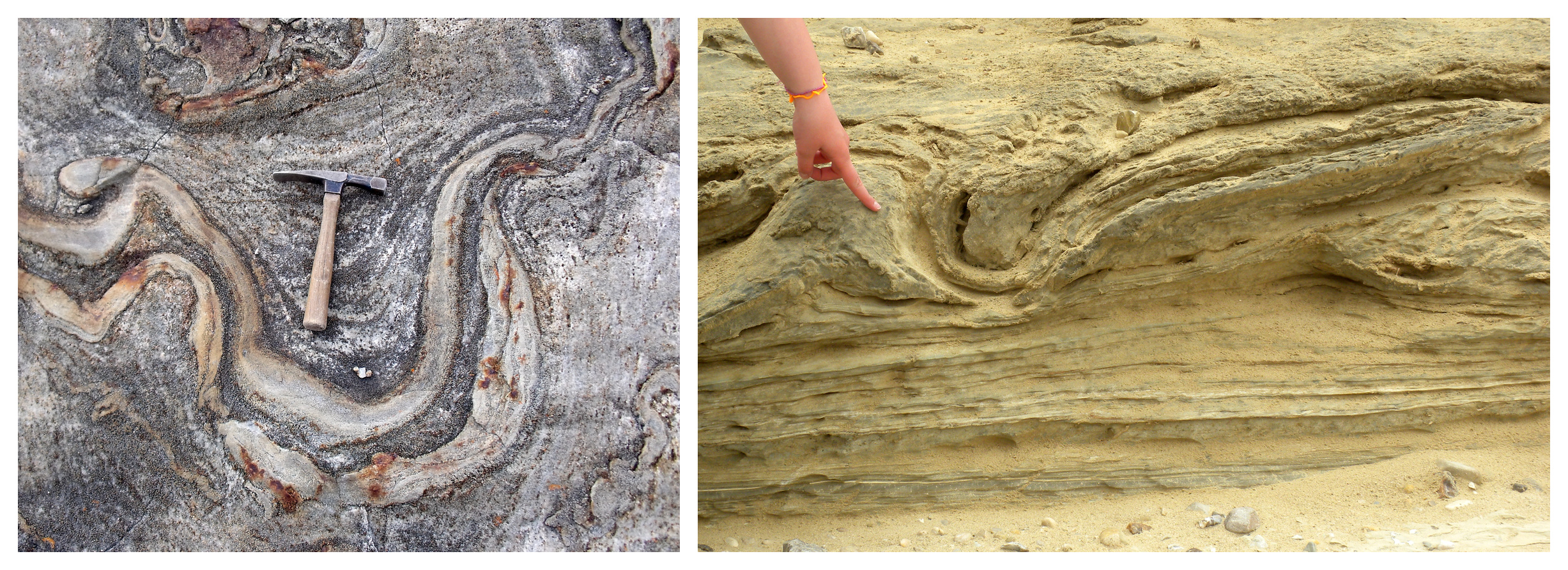}
\caption{\label{fig:visco}
Geological viscoelasticity  and viscoplasticity.
Left: Aphebian-aged marble showing plastic deformation (image:
Mike Beauregard.
Right: Soft sediment deformation in exposed Dead Sea sediment, Israel, gives rise to a Kelvin--Helmholtz shear instability (Sec.~\ref{sec:KH}).
(Images: Left: Mike Beauregard; CC-BY-2.0; Right: Mark A. Wilson; CC-BY-SA-3.0.)
}
\end{figure}

\subsection{Viscoelastic and viscoplastic materials}

The nonlinear rheology of geological materials has been studied for
centuries~\cite{boswell1951} and underpins theories of tectonics and
orogenesis~\cite{biot1961}.  Rocks commonly exhibit both viscoelastic and viscoplastic
behaviour (Fig.~\ref{fig:visco}). 
Viscoelasticity combines a viscous response (flow under sustained stress) with an elastic response (instantaneous, recoverable strain). 
This behaviour is time-dependent, meaning that the material response varies depending on the rate and duration of applied stress.
Quicksands are a well-known example of complex viscoelastic behaviour associated with shear
thickening~\cite{matthes1953quicksand,khaldoun2005liquefaction,kadau2009living}.
In rocks, viscoelastic behaviour is observed when the applied stress lies within a certain range and the loading period is relatively short. Under these conditions, deformation is partly elastic and partly viscous: the elastic component allows partial recovery of the original shape after unloading, whereas the viscous component leads to permanent deformation (creep).

Viscoplasticity refers to deformation that involves both viscous flow (fluid-like, lacking a fixed shape) and plasticity  (inelastic, non-recoverable strain) once a yield stress is exceeded. Unlike viscoelasticity, viscoplasticity is governed
primarily by the magnitude of the applied stress relative to the yield threshold (often with additional rate and temperature dependence), rather than by the loading duration itself. 
In rocks, viscoplastic behaviour is observed when the stress level exceeds a material-specific threshold: at high stresses, rocks deform permanently without elastic recovery. Such plastic deformation reflects the breaking, slip, and rearrangement of grains or crystal lattices, resulting in irreversible strain. Viscoplastic behaviour is prominent in geological processes such as folding, faulting, and the flow of rock masses under high-pressure conditions.
Flowing lava frequently exhibits this type of behaviour and can be modelled as a yield-stress
(viscoplastic) fluid~\cite{griffiths2000dynamics}.

Viscoelastic constitutive models have been applied to problems such as sinkhole
formation~\cite{shalev2012} and the propagation and emplacement of dikes and diapirs within
viscoelastic host rock~\cite{rubin1993}.
Viscoplastic soils, such as clays, display striking pattern formation, for example
shrinkage-crack networks (Sec.~\ref{sec:crackpatterns}) that emerge as inelastic strains localize
during drying or unloading~\cite{veveakis2021}. More broadly, both viscoelastic and
viscoplastic responses are recognized as important for the development of compositional
banding and foliation in rocks, where time-dependent relaxation, yield, and irreversible flow
interact to reorganize fabric and mineral distributions~\cite{burnley2013}. Gneiss is one
example (Sec.~\ref{sec:gneiss}).

\subsection{Granular media}

A granular medium is composed of solid grains  interacting primarily through contact forces
and friction, so that assemblies can behave like solids or like fluids, depending on loading
and confinement~\cite{andreotti2013granular}.
Many granular systems are treated within soft-matter physics~\cite{gravish2016entangled};
geological examples  (sand, soils, tephra) display solid--fluid duality and phenomena with no
direct molecular analogue, such as jamming, shear banding, and
avalanches~\cite{zheng2021mechanics}.

\begin{figure}
\centering
\includegraphics{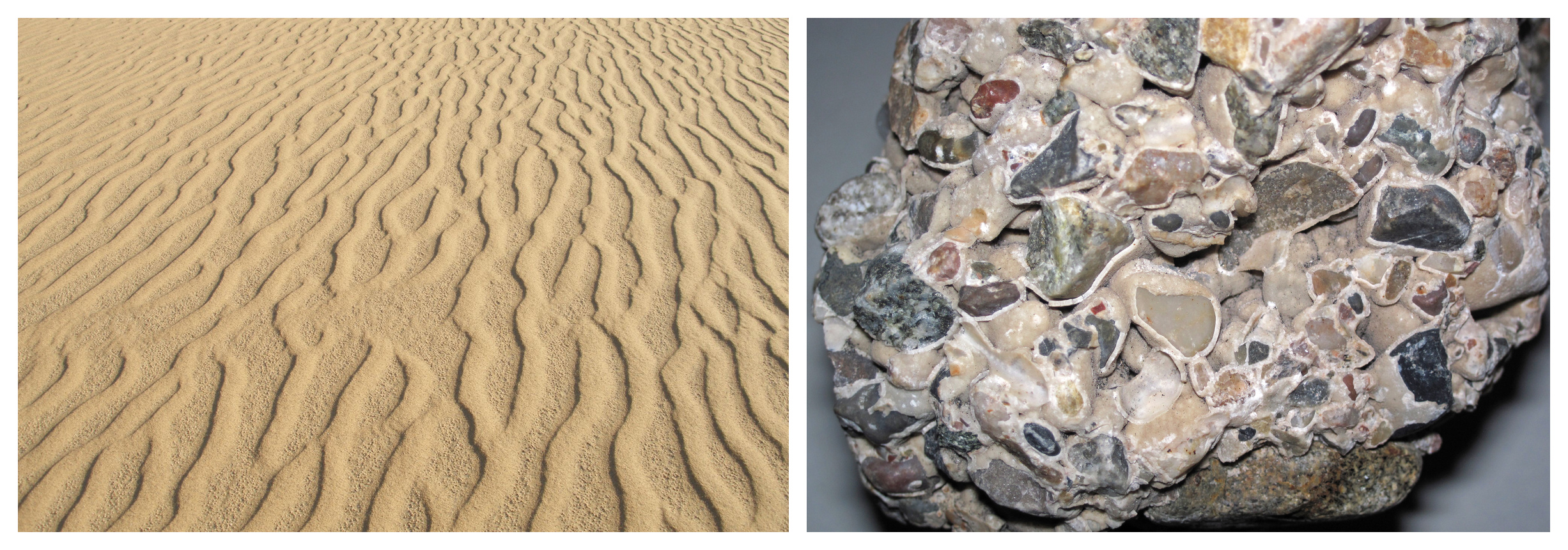}
\caption{\label{fig:granular}
Geological granular media.
Left: 
sand dunes near Dakhla Oasis, Western Desert, Egypt;
Right: travertine-cemented conglomerate.
(Images: Left: Vyacheslav Argenberg;  CC-BY-2.0; Right: James St. John; CC-BY-2.0.)
}
\end{figure}

Granular materials pervade Earth's surface environments (Fig.~\ref{fig:granular}). All clastic sediments are granular: they form sedimentary rocks (sandstone, conglomerate, breccia) by accumulation and lithification; build river beds, bars, and deltas through erosion--deposition cycles; shape aeolian landforms (dunes, sand sheets, desert pavements); create glacial deposits (moraines, eskers, kames, drumlins); constitute volcanic fall and flow products (ash, lapilli, pyroclasts); and dominate coastal features (beaches, barrier islands, sandbars, shoreface deposits). In the marine realm they blanket continental shelves and abyssal plains and are funnelled through canyons as turbidites.

Many of the examples of geological patterns we describe involve granular media at some stage in their formation, including concretions and nodules (Sec.~\ref{sec:concretions}), efflorescence patterns (Sec.~\ref{sec:roses}), fulgurites (Sec.~\ref{sec:Fulgurites}), opals (Sec.~\ref{sec:opals}), porphyroblasts (Sec.~\ref{sec:porphyroblasts}), sedimentary crack patterns (Sec.~\ref{sec:crackpatterns}), and stylolites (Sec.~\ref{sec:stylolites}).

\section{Physical mechanisms underlying mesoscale geological self-organization}
\label{sec:physical}

We hope that the geological examples presented above have convinced the reader that soft matter provides a helpful physical classification applicable to a wide variety of geological systems.  
What, then, are the common underlying physical mechanisms responsible for the emergence of patterns in these systems? 
In what follows we ``round up the usual suspects'' to indicate the breadth of mechanisms involved in geological self-organization; for fuller treatments, we refer the reader to the physics literature.

\subsection{Reaction--diffusion systems}
\label{sec:RD}

Reaction--diffusion systems describe processes where substances spread out in space (diffuse) and interact with each other through chemical reactions. These have the general mathematical form
\begin{equation}
\partial_t u =D \nabla^2 u + R(u),
\end{equation} 
where $D$ is the diffusion matrix and $R$ represents the reaction component.
Often, we associate diffusion with the smoothing of concentration gradients over time, leading to a uniform homogenized soup, as substances move from regions of high concentration to low concentration.
However, introducing chemical reactions into these systems may disrupt uniformity and drive the spontaneous formation of complex spatial structures.  Such systems occur in various physical, chemical, and biological settings.

\begin{figure}
\centering
\includegraphics{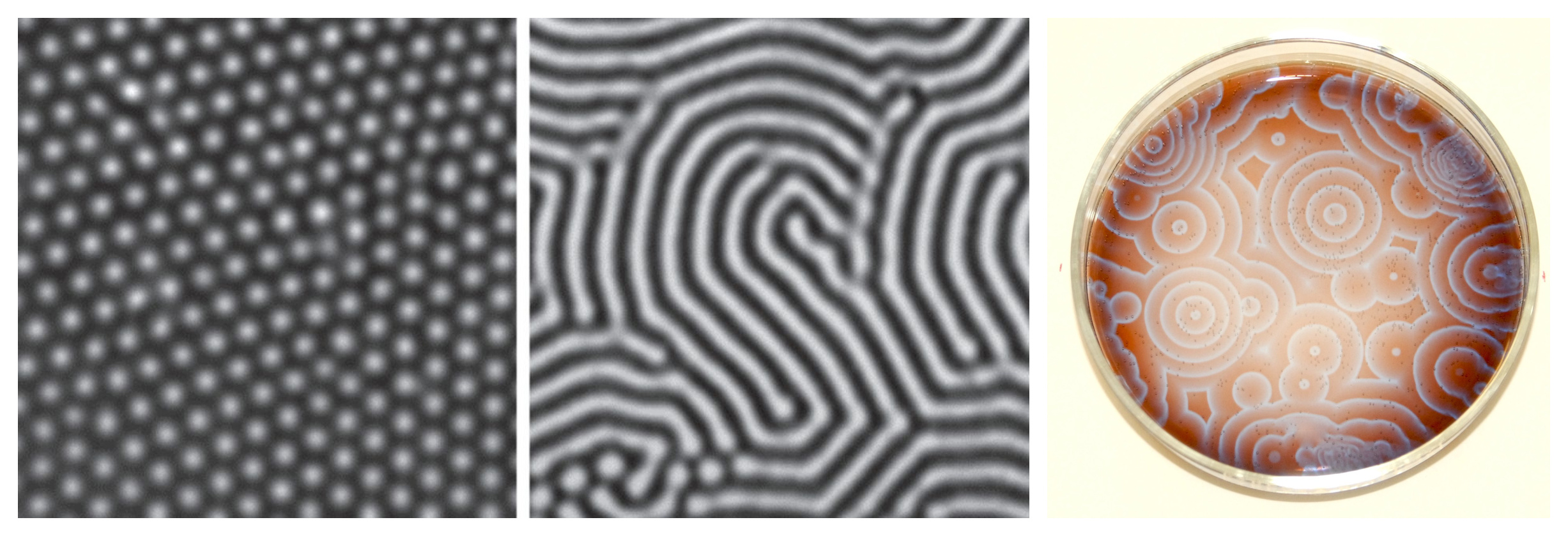}
\caption{\label{fig:reaction-diffusion}Reaction--diffusion systems.
(left and centre) Turing patterns.
(right) Belousov--Zhabotinsky target and spiral patterns.  
(Images: Left, centre: Courtesy of Jacques Boissonade and Patrick De Kepper, right: Stephen Morris.)
}
\end{figure}

\subsubsection{Turing patterns}

At the core of pattern formation is the concept of Turing instability. In 1952, Alan
Turing~\cite{turing1952chemical} demonstrated that diffusion, when combined  with nonlinear
chemical reactions involving two interacting species, can spontaneously break the symmetry of
an initially uniform mixture of chemical compounds. 
In the simplest case of a Turing model, one species acts as an activator, promoting its own production, while the other functions as an inhibitor, suppressing the activator's growth. Both diffuse, but the inhibitor has a larger diffusion coefficient than the activator: $D_A > D_I$.
Through the coupling of diffusion and reactions, small perturbations from uniformity can grow, leading to the spontaneous appearance of spatial patterns.
Depending on the parameters, a Turing system can generate spots, stripes, or even labyrinthine patterns, reflecting the self-organizing capacity of reaction--diffusion dynamics (Fig.~\ref{fig:reaction-diffusion}A,B).

The mechanism behind pattern formation in such systems was creatively illustrated by James
Murray in his book \emph{Mathematical Biology}~\cite{Murray2002} with an analogy of
``sweating grasshoppers'': Imagine a dry grassy field with grasshoppers scattered throughout.
If a fire starts in one part of the field, it will naturally spread, burning everything in
its path. However, grasshoppers, sensing the approaching flames, react by moving ahead of the
fire and sweating profusely, which dampens the grass and prevents it from burning. In this
analogy, the fire is the ``activator'', spreading destruction, while the grasshoppers are the
``inhibitors'', stopping the fire from spreading by moistening the grass. The key to the
pattern's emergence is that the grasshoppers react faster than the fire spreads, creating a
non-uniform, repeating pattern of charred and unburnt areas, similar to the spotted patterns
on animal coats.  Two mechanisms are key to the creation of a pattern: self-activation (a
short-range positive feedback loop), where small perturbations in the uniform state tend to
grow over time, similar to fire spreading, and screening (long-range inhibition), where the
formation of a structure at one point reduces the likelihood of similar structures appearing
nearby. In the grasshopper example, this corresponds to grasshoppers surrounding the fire and
stopping its spread.

\subsubsection{Short-range activation and longer-range inhibition}

What we have just described is akin to how many natural patterns emerge from generic mechanisms of self-organization: the interplay between local positive feedback and long-range suppression. 
Geological examples include patterns in sand (Fig.~\ref{fig:granular}): ``As a ridge gets
bigger, it enhances its own growth by capturing more sand
from the moving air. But in doing so it acts as a sink, removing
sand from the wind and suppressing the formation of other ripples nearby. The balance between these two processes
establishes a roughly constant mean distance between ripples''~\cite{ball2015forging}.
Honeycomb weathering 
(Fig.~\ref{fig:tafoni})~\cite{mcbride2004origin} and travertine patterning~\cite{hammer2008calcite} are further geological examples of 
 patterns emerging from the combination of short-range activation and longer-range inhibition.

\subsubsection{Excitable media}

Another related type of reaction--diffusion system is termed an excitable
medium~\cite{meron1992pattern}.  In excitable systems a small perturbation leads only to a
small transient effect; however a perturbation above a certain threshold excites the system,
which then remains in its excited state for some time before returning to quiescence, and
upon this return has a refractory period during which it cannot be excited a second time. A
classic example is the Belousov--Zhabotinsky (BZ) reaction~\cite{zhabotinsky1991history}.
Geologically relevant instances of excitable media include the example of crystal
growth~\cite{cartwright2012crystal}. Earthquake dynamics, as modelled by the
Burridge--Knopoff model~\cite{burridge1967model} describing a chain of blocks connected by
springs and pulled across a surface, 
may be viewed as a type of  
excitable medium with elastic rather than the usual diffusive
coupling~\cite{cartwright1997burridge}.
Also in the temporal domain, it has been suggested that the oceanic carbon cycle may be
excitable, and that mass extinction events may correspond to large perturbations that excite
the system~\cite{rothman2019characteristic}.
As for spatio-temporal geological patterns, an excitable model for the coupled evolution of
ice-surface temperature and elevation produces spiral waves  similar to those observed in the
Martian north polar ice cap~\cite{pelletier2004spiral}.

\begin{figure}
\centering
\includegraphics{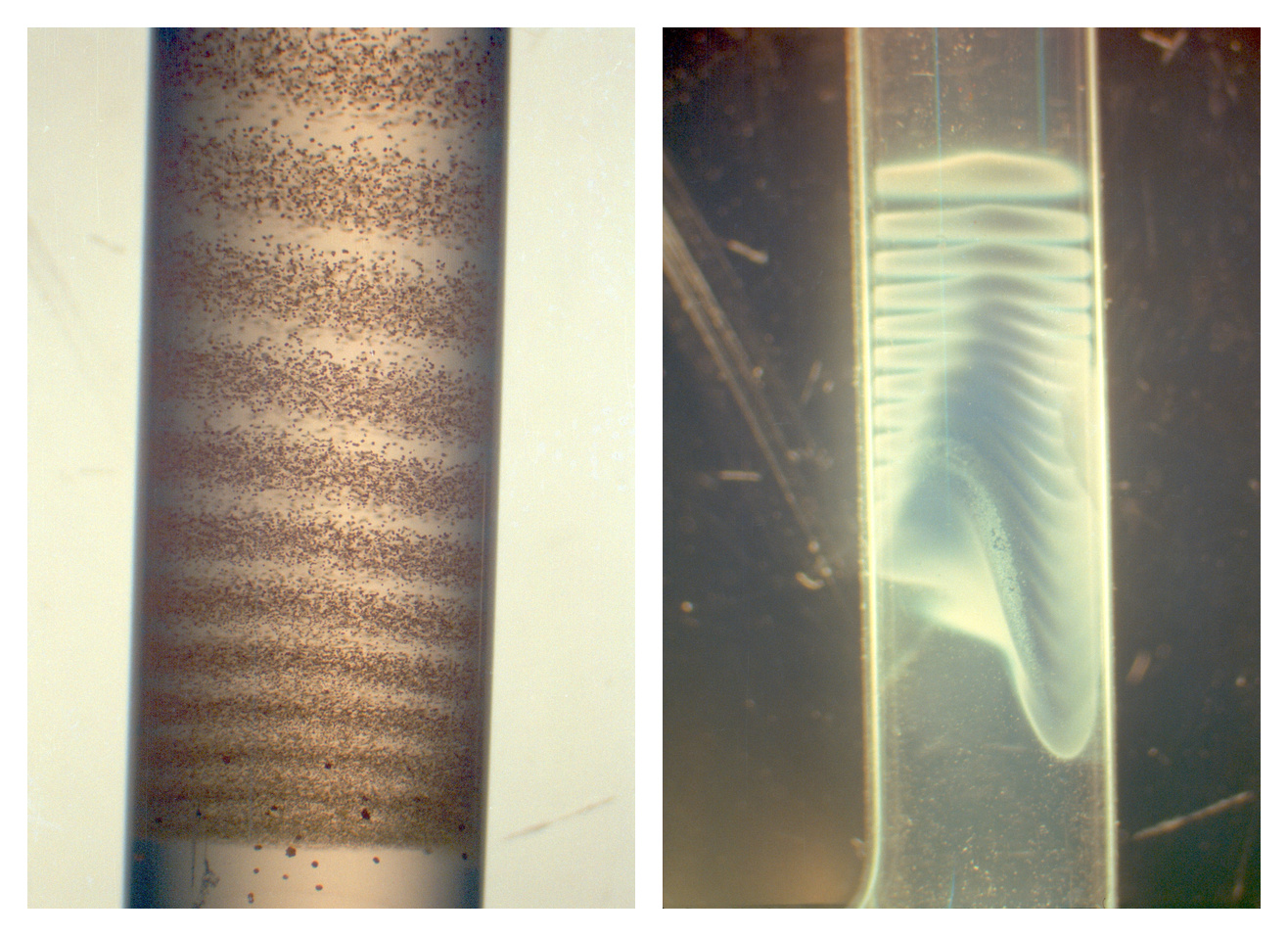}
\caption{\label{fig:liesegang}Liesegang bands in laboratory experiments.
(Images: Julyan Cartwright.)}
\end{figure}

\subsubsection{Geological reaction--diffusion systems}

Reaction--diffusion systems remain an active area of research in both biological and geological pattern formation. While the complexity of biological systems often makes it difficult to identify the specific morphogens responsible for a given pattern, geological systems are usually  simpler, involving just a handful of chemicals. Examples such as zebra rock (Fig.~\ref{fig:zebra}) or target patterns in sandstone rocks (Fig.~\ref{fig:target}) strongly suggest a reaction--diffusion origin, although other banded patterns, like striped flints (Fig.~\ref{fig:Flint}) or sphalerites (Fig.~\ref{fig:Blende}) elude such a simple description.
However, there are two important factors to consider when thinking about geological systems.
First, in order for a reaction--diffusion pattern to survive the millennia, it needs to be
literally turned to stone, sculpted into the rock~\cite{Jamtveit2012}. This means that not
only bulk reactions need to be involved but also surface reactions, such as precipitation.
The second key factor is the importance of fluid flow. For large enough system size
$L$, the advective time-scale, $\tau_a = L/v$, is always shorter than the diffusive
time-scale, $\tau_d = L^2/D$. Thus, it is invariably advection --- usually through fractures,
joints, and bedding planes --- that brings reactants to rock strata, leading to its chemical
transformation. Diffusion can be a controlling factor, particularly on smaller spatial
scales, such as in the cracks observed in  Fig.~\ref{fig:target}A. We  discuss
fluid-flow-mediated instabilities in  Sec.~\ref{sec:instabil}.

\subsubsection{Liesegang patterns}
\label{sec:liesegang}

One particular reaction--diffusion system is Liesegang banding (Fig.~\ref{fig:liesegang}).
Liesegang band formation is a rhythmic precipitation phenomenon that occurs when a solute
diffuses through a gel and reacts with another solute, resulting in distinct bands or rings.
The spacing of these bands increases geometrically outward from the point of
diffusion~\cite{Jablczynski1926,matalon1955liesegang},
\begin{equation}
x_n \approx x_0(1+p)^n,\label{eq:Jablczynski}
\end{equation}
where $x_n$ is the position of the $n$th band and $p$  a spacing coefficient. 
Although the phenomenon was initially reported by~\citeA{Runge1855} and~\citeA{Ord1879}, it
became more widely known after the systematic experiments conducted
by~\citeA{liesegang1896uber}. In his experiments, Liesegang observed the diffusion of silver
nitrate in a gelatin layer containing potassium dichromate, leading to visually striking
rhythmic precipitates oriented parallel to the diffusion front. Rings are formed in a radial
geometry, while bands are found in linear geometries. Often, these structures are referred to
as periodic precipitation structures~\cite{henisch2014periodic}, which is somewhat
misleading; as we have noted, the primary defining characteristic of Liesegang structures is
that they are generally not simple periodic patterns but instead have aperiodic spacings that
follow a geometric series. To complicate matters further, the term \emph{Liesegang} is
frequently found in geological publications to describe any banding observed in rocks,
without regard to its provenance. 

Despite the phenomenon being known for well over a century, the mechanism of  Liesegang band
formation is still a matter of discussion. Two main groups of theories have been proposed,
termed pre- and post-nucleation models, differentiated by the timing of precipitate
nucleation in the system~\cite{LHeureux1999,nabika2019pattern}.

In the pre-nucleation models, following~\citeA{Ostwald1897} and~\citeA{Prager1956}, bands are
proposed to result from a feedback loop between nucleation and diffusion. Precipitation
starts at a specific value of supersaturation, the nucleation threshold, and subsequently
depletes the solute around it, preventing it from reaching the nucleation threshold in
adjacent regions. This process is followed by renewed nucleation farther from the initial
band, resulting in a rhythmic pattern that regularly obeys quantitative spacing laws.

In contrast, post-nucleation models consider the interaction between growth, diffusion, and
surface tension effects. The pattern is generated through Lifshitz--Slyozov
instability~\cite{Lifshitz1961}, in which   homogeneous systems with uniformly distributed
precipitate particles may become unstable due to Ostwald ripening (Sec.~\ref{sec:ostwald}).
Smaller particles tend to dissolve because their surface energy is higher due to their
greater curvature. The dissolved material then diffuses through the surrounding matrix and
redeposits onto larger particles, which have a lower surface energy. This process results in
the growth of larger particles over time at the expense of smaller ones. In an initially
non-uniform system, this can generate bands that evolve through coarsening after the
nucleation phase is over~\cite{Boudreau1995}. In contrast to the pre-nucleation models,
post-nucleation models can generate patterns even in the absence of an initial concentration
gradient.

It is possible that both these mechanisms may be  found in nature. 
Experiments and simulations show examples of various different morphologies beyond the classical bands and rings, including
spirals, rods and spots~\cite{krug_morphological_1999,dayeh_transition_2014,papp_fine_2020}
as well as bands of two crystalline phases
\cite{cartwright1999pattern}.

One of our aspirations with this review is to discourage  the indiscriminate use of the word \emph{Liesegang}  in geology for any spatial oscillation; as we show here, the Liesegang pattern formation mechanism is just one of a variety of physical mechanisms that can produce banding, both periodic and aperiodic.

\begin{figure}
\centering
\includegraphics{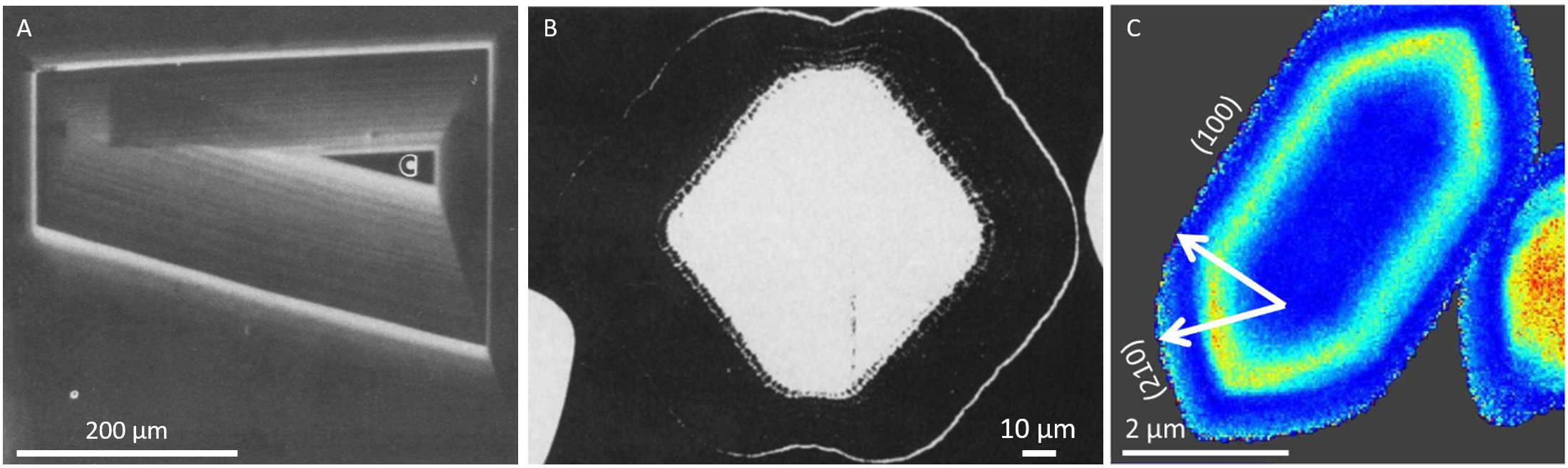}
\caption{\label{fig:crystals}Examples of synthetic crystals showing oscillatory zoning. 
A: Calcite crystal with white areas enriched in Mn. B:
(Ca,Cd)CO$_3$ crystal, white areas are enriched in Cd. C:
Ba(SO$_4$,HAsO$_4$) crystal, blueish areas are depleted in As, while
yellowish--greenish areas are enriched.
(Sources: 
A:~\citeA{reeder_oscillatory_1990},
B:~\citeA{prieto_nucleation_1997},
C: ~\citeA{ling_nanospectroscopy_2018}.)
}
\end{figure}

\subsubsection{Oscillatory zoning}
\label{sec:zoning}

Classical crystallization is itself a form of self-organization in condensed matter. Growth from a melt or a solution may also be regarded as a soft-matter process, as it gives rise to mesoscopic structures such as growth steps, target patterns, and spirals. The more complex growth modes that produce mesoscale compositional patterns, such as oscillatory zoning, are even more clearly manifestations of soft-matter behaviour.

Oscillatory zoning (Fig.~\ref{fig:crystals}) is another reaction--diffusion pattern-formation mechanism observed in geological systems.
Depending on the formation conditions, three different zoning patterns have been described in
crystals:  normal, reverse and oscillatory zoning~\cite{ginibre_crystal_2007}. Normal and
reverse zoning involve monotonic changes: in normal zoning, the crystal core is
compositionally less evolved than the rim (e.g.,~reflecting cooling of a magma chamber),
whereas in reverse zoning the trend is opposite (e.g.,~indicating heating).  Oscillatory
zoning, by contrast, consists of repetitive, concentric compositional bands. 

For natural crystals showing oscillatory zoning, both extrinsic and intrinsic mechanisms have
been proposed~\cite{shore_oscillatory_1996}. Extrinsic mechanisms reflect system-wide
physical and chemical changes (e.g.,~fluid mixing, pressure or temperature shifts), whereas
intrinsic mechanisms arise from crystal growth dynamics and local
phenomena~\cite{shore_oscillatory_1996}. Although oscillatory zoning is common, identifying
the dominant controls is difficult. 
Laboratory syntheses, where all parameters can be controlled, are particularly valuable for studying cases in which oscillatory zoning develops through intrinsic mechanisms, in consequence, offering clear insight into self-organization during crystal growth.

Among the internal mechanisms proposed to explain oscillatory zoning, two are particularly well established.
One invokes a feedback between growth rate and the partition coefficient of the substituting
ion~\cite{reeder_oscillatory_1990}. For example, in manganese-rich calcite, incorporation of
Mn slows the calcite growth rate; the slower growth increases the Mn partition coefficient,
enhancing Mn incorporation into the lattice. This depletes Mn in solution, which in turn
accelerates calcite growth and lowers the Mn partition coefficient. The cycle then repeats,
producing oscillatory zoning (Fig.~\ref{fig:crystals}A).

Another mechanism relies on the difference in solubilities of the solid-solution
end-members~\cite{prieto_nucleation_1997}. If this difference is large, the less soluble
phase precipitates first until it becomes undersaturated owing to depletion of one ion;
growth then switches to the other phase until depletion of its controlling ion allows the
first phase to reappear. This behaviour has been documented for (Cd,Ca)CO$_3$, where
otavite forms first; as [Cd] falls, calcite grows, and when [Ca] later drops, a new otavite
layer nucleates (Fig.~\ref{fig:crystals}B). An analogous mechanism operates in anionic solid
solutions, e.g.,~Ba(SO$_4$,HAsO$_4$) (Fig.~\ref{fig:crystals}C).

\subsection{Fluid-flow instabilities and convective processes}
\label{sec:instabil}

Because many geological processes operate on long time-scales and at low velocities, they can
often be modelled as viscous fluid flows
(e.g.,~\cite{wollkind1982kelvin,fletcher1977folding}). 
A common pathway to geological self-organization in such cases is a physical instability arising from fluid interactions. Such instabilities may be categorized by the nature of the driving mechanism.

\begin{figure}
\centering
\includegraphics{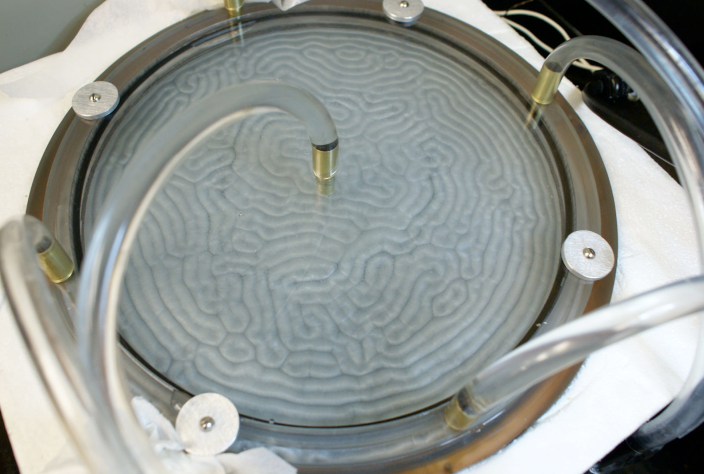}
\caption{\label{fig:RB}Rayleigh--B\'enard convection seen in an experimental set-up with a rigid boundary condition at the transparent top wall ($Ra = 2490$).
(Image: \citeA{Matsson2008}.)
}
\end{figure}

\subsubsection{Rayleigh--B\'{e}nard convection}
\label{sec:RB}

Rayleigh--B\'enard instability, or  buoyancy- (density)-driven convection, 
 occurs when a fluid layer is heated from below and cooled from above, leading to the
formation of convection cells (Fig.~\ref{fig:RB}) as the fluid moves (convects) under this
forcing~\cite{getling1998rayleigh,bodenschatz2000recent}. 
 This occurs above a critical Rayleigh number, the ratio of the time-scale for diffusive heat transport to the time-scale for convective heat transport in the fluid,
\begin{equation}
Ra= \frac{g \beta \Delta T L^3} {\nu \alpha}.
\end{equation}
Here  $g$ is the acceleration due to gravity, $\beta$ is thermal expansivity, $\Delta T$ is temperature difference, $L$ is a characteristic length-scale, $\nu$ is the kinematic viscosity, and $\alpha$ is the thermal diffusivity.

On a planetary scale, Rayleigh--B\'enard instability is a key process in the Earth's thermal
evolution and in the dynamics of fluid movement within geological
settings~\cite{busse1989fundamentals}. It provides a framework for understanding how heat is
transported in the Earth's interior, influencing everything from plate
tectonics~\cite{anderson2002plate} to the cooling of magmatic bodies and the circulation of
fluids in hydrothermal systems.

The RB instability plays a central role in the formation of layered geological structures, particularly within igneous bodies such as plutons and layered intrusions. These sheet-like bodies of igneous rock  display distinct mineralogical layering produced by convection-driven differentiation during magma cooling. In a magma chamber, buoyant, hot magma at the base rises while cooler, denser magma near the top sinks, establishing convective currents.  As the melt cools, minerals crystallize at characteristic temperatures: early-forming, high-temperature  minerals like olivine and pyroxene settle towards the base, while lower-temperature phases such as plagioclase accumulate above.  The resulting
convective motion produces stratified structures that reflect successive stages of crystallization. 
The resulting layers may be sharply bounded or more diffuse, depending on the vigour of convection and the cooling rate, and they often exhibit variations in mineral proportion, texture, and chemical composition

In some intrusions, cyclic layering occurs, with sequences of mineral bands repeated at
different levels due to periodic fluctuations in temperature or episodic magma recharge. The
Bushveld Complex, containing extensive layers of chromite, magnetite, and other minerals, is
a classic example in which convection driven by RB instability contributed to magmatic
differentiation~\cite{Wager1968,Cawthorn1994}. Similarly, the Skaergaard Intrusion in
Greenland exhibits well-defined layers of gabbro formed through this process~\cite{Thy2023}.

\begin{figure}
\centering
\includegraphics{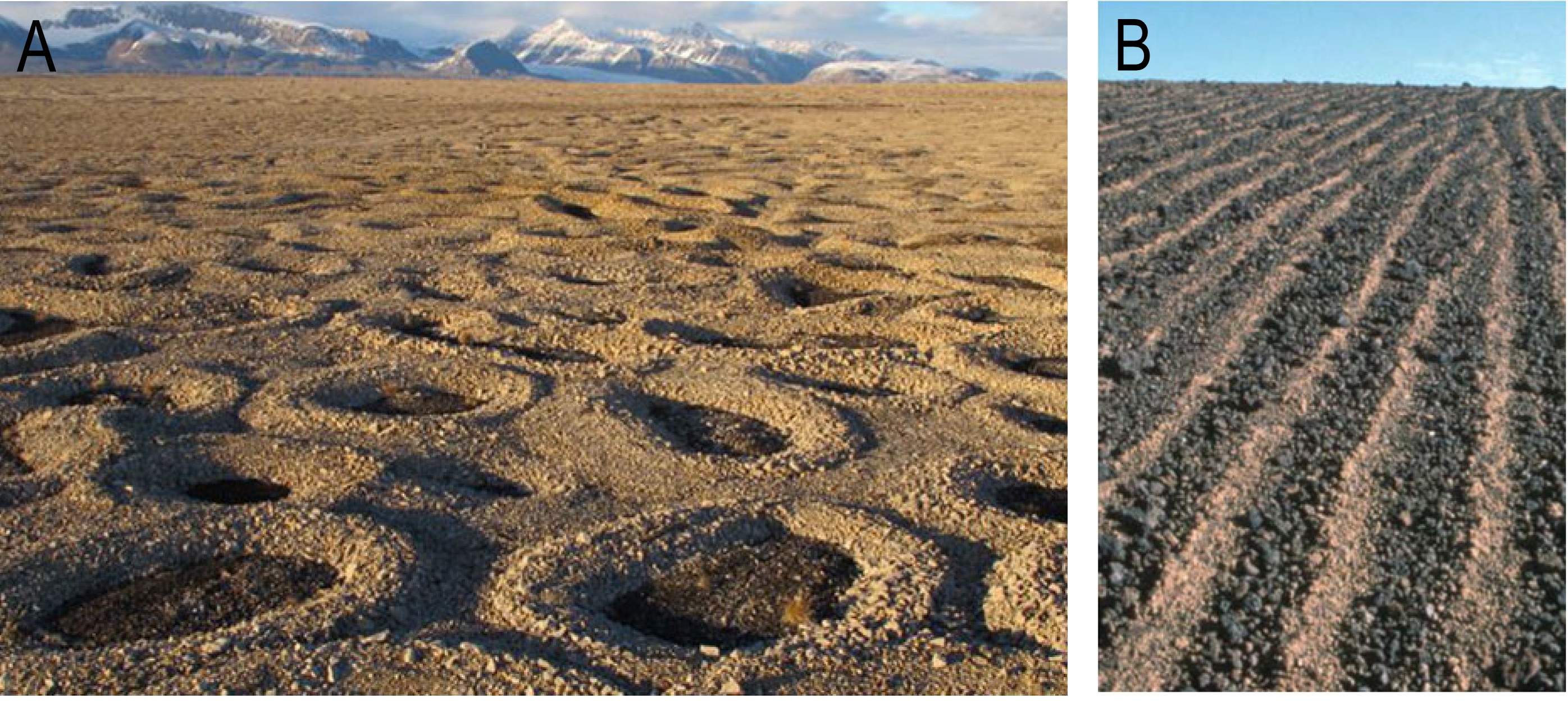}
\caption{\label{fig:rings}Patterned ground: (A) Circular domains of fine-grained soil approximately 2~m in
diameter ringed by gravel approximately 0.25~m high, Broggerhalvoya, NW Spitsbergen. (B) Narrow sorted stripes approximately 0.15~m wide, near summit of Mauna Kea, HI, USA.
(Source:\citeA{li2021ice}.)
}
\end{figure}

 RB instability also operates at smaller scales and lower temperatures. During early
diagenesis, the process by which sediments are lithified into rock, RB instability can drive
pore-fluid motion in fine-grained sediments heated from below, e.g.~from underlying
geothermal gradients.  Upward migration of warm, less dense fluids and sinking of cooler,
denser ones may lead to the formation of soft-sediment deformation features, such as load
casts or convolute bedding. In permafrost regions, RB instability in the seasonally thawed
active layer can  promote soil mixing, leading to patterned ground forms such as stone
circles and stripes~\cite{Kessler2003,Hallet2013,li2021ice} (Fig.~\ref{fig:rings}). More generally,
fluid-venting structures  such as hydrothermal vents (Sec.~\ref{sec:hydrothermalvents}), mud
volcanoes and cold seeps (Sec.~\ref{sec:mudvolcanoes}), are also ultimately sustained by
buoyancy-driven convection of this type.

\begin{figure}
\centering
\includegraphics{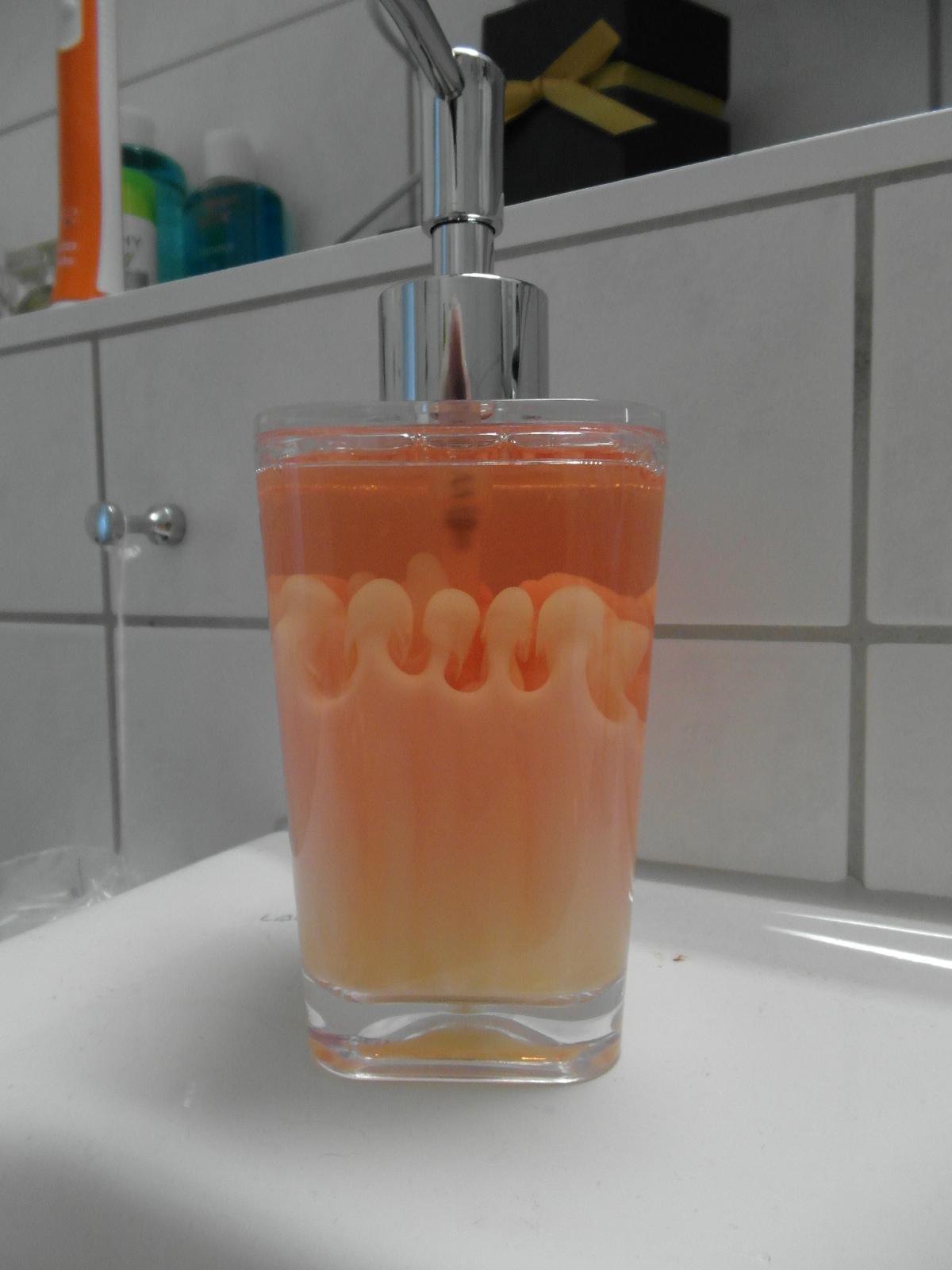}
\caption{\label{fig:RT}Rayleigh--Taylor instability, seen here between two different viscous soap solutions of slightly different densities, the denser overlying the lighter.
(Image: Courtesy of Thermoskanne.)}
\end{figure}

\begin{figure}
\centering
\includegraphics{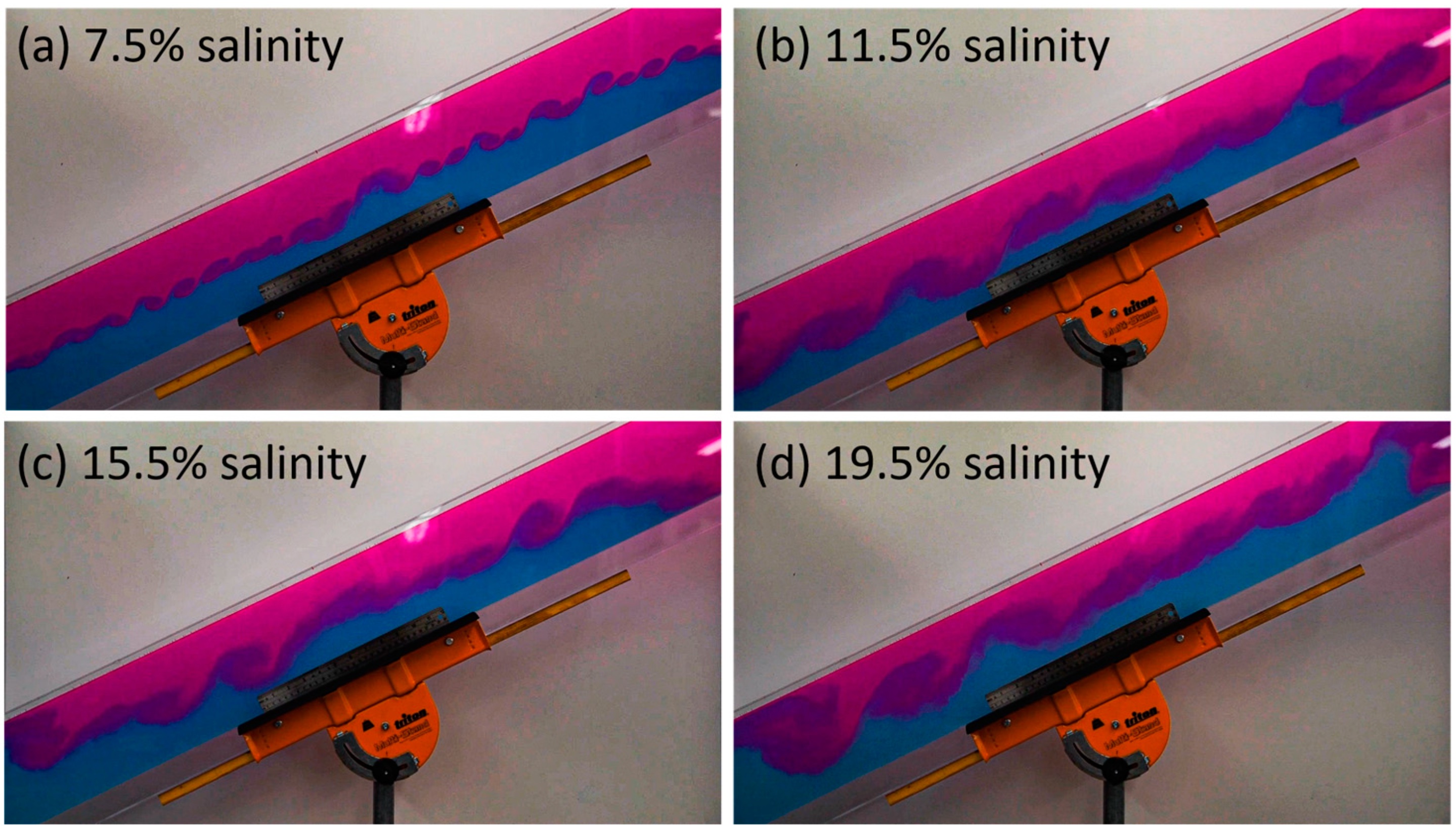}
\caption{\label{fig:KH}Kelvin--Helmholtz instabilities in
 a laboratory experiment in which the interface between the two fluids is sheared by tilting the container.
 (Source: \citeA{gibbons2023demonstrating}.)}
\end{figure}

\subsubsection{Rayleigh--Taylor instability}
\label{sec:RT}

The Rayleigh--Taylor (RT) instability occurs when a denser fluid is accelerated into a
lighter one, producing characteristic interpenetrating structures as the fluids
mix~\cite{kull1991theory};   Fig.~\ref{fig:RT}. 
In the absence of surface tension at the interface, all wave-lengths are unstable; surface tension suppresses short wave-lengths, introducing a cut-off to the instability.

In geology, RT-like overturning arises whenever density contrasts act under gravity. 
In seismites (Sec.~\ref{sec:seismites}),  seismic shaking transiently accelerates denser slurries into lighter layers, triggering RT deformation.  
The RT instability is also relevant to mantle dynamics: buoyant, hot mantle material rising
through cooler, denser mantle rock can form mushroom-shaped plumes, a hallmark of RT
behaviour~\cite{whitehead1975dynamics}. 
In evaporite basins, less dense salt buried beneath denser sedimentary rocks can  pierce the
overburden to form salt diapirs (domes)~\cite{trusheim1957halokinese},  that can trap oil and
gas, making them important in petroleum geology~\cite{Hudec2007,Davison2000}.

\subsubsection{Kelvin--Helmholtz instability}
\label{sec:KH}

Kelvin--Helmholtz (KH) instability,  Fig.~\ref{fig:KH}, a lateral shear instability often seen in
the atmosphere and ocean~\cite{de1996evolution}, is not uncommon in large-scale geological
settings, where earthquake-induced, layer-parallel displacement of soft sediments creates
strong interfacial shear (e.g.,~\cite{heifetz2005soft}). In seismites (Fig.~\ref{seismite}) KH
deformation commonly occurs hand in hand with Rayleigh--Taylor overturning: RT generates load
casts, flames, and pillows under normal acceleration, while synchronous shear along bedding
planes superimposes KH-type wavy or rolled interfaces.

\begin{figure}
\centering
\includegraphics[width=\linewidth,height=0.78\textheight,keepaspectratio]{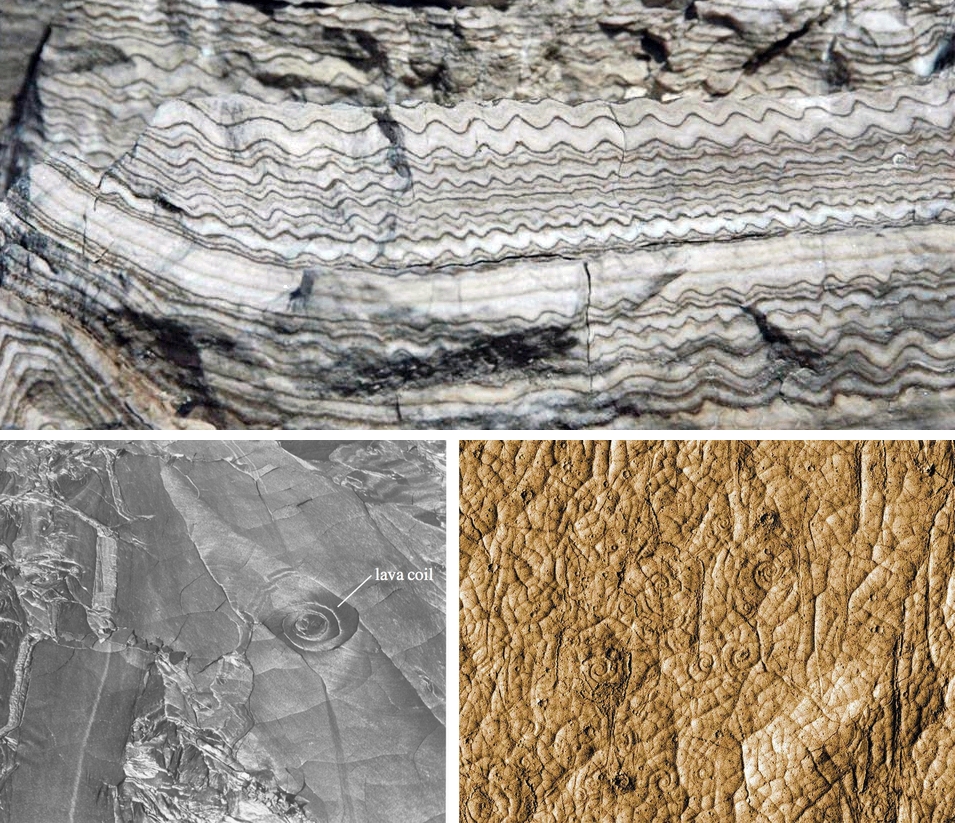}
\caption{\label{fig:KH2}Geological Kelvin--Helmholtz instabilities. 
(Above) Castile Formation gypsum--calcite laminites (white gypsum; dark calcite, Upper Permian), from New Mexico--Texas border, USA. Finely laminated gypsum--calcite couplets with pronounced crinkling and small-scale folding, interpreted as a Kelvin--Helmholtz shear instability in soft sediments.
(Below left) Lava coil approximately 10~m in diameter in pahoehoe lava
north east of Pu'u Koa'e, Hawaii.
(Below right) lava coils in Cerberus Palus, a volcanic region on Mars. Image taken from HiRISE camera on the Mars Reconnaissance Orbiter; field of view approximately 500~m.
(Images:
Above: James St.\ John CC-BY-2.0;
Below left: USGS photo by Elliot Endo; public domain;
Below right: NASA; public domain.)
}
\end{figure}

On smaller scales, a related shear-driven mechanism has been proposed as a driver for mesoscale
rock folding (Fig.~\ref{fig:visco}B).~\citeA{wollkind1982kelvin}  incorporated surface tension
into earlier  laboratory folding models, deriving instability  onset conditions consistent
with capillary-modified KH instability and successfully accounting for the characteristic
undulatory patterns in gypsum--limestone couplets of the Castile Formation in New Mexico
(Fig.~\ref{fig:KH2}).
KH dynamics also appear in volcanic contexts. Lava coils --- helical surface textures formed
where relatively low-viscosity lava solidifies within slow, shear-dominated flow --- are well
documented on Earth and Mars~\cite{ryan2012coils} (Fig.~\ref{fig:KH2}). At the interface of
geological and biological  soft matter, KH instability has been invoked to explain the
formation of the (pseudo)fossil Kinneyia~\cite{thomas2013formation,herminghaus2016kinneyia}:
passive deformation of a viscoelastic microbial mat under shallow, flowing water produces
ripple-like corrugations that can later be preserved diagenetically~\cite{fabbri2017fluid}.

\subsubsection{B\'enard--Marangoni convection}

\begin{figure}
\centering
\includegraphics{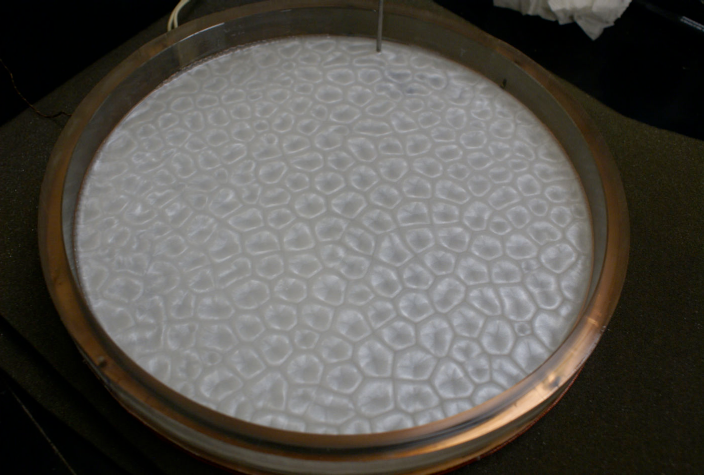}
\caption{\label{fig:BM}B\'enard--Marangoni convection seen in the same experimental set-up as seen in  Fig.~\ref{fig:RB}, but with a free liquid surface.
(Image: \citeA{Matsson2008}.}
\end{figure}

B\'enard--Marangoni convection, or  Marangoni flow,  occurs when surface tension gradients
drive fluid motion~\cite{schatz2001,craster2009dynamics} (Fig.~\ref{fig:BM}). Temperature or
concentration gradients along a free surface create variations in surface tension, inducing
flow from regions of lower surface tension towards regions of higher surface tension.

B\'enard--Marangoni instability also appears in geological contexts. 
It has been proposed to influence segregation of  basaltic lithologies in the lower
mantle~\cite{baron2022melting}. At Earth's surface, clay rafts floating on shallow lake water
can experience Marangoni-driven circulation near shorelines, producing  complex,
Suminagashi-like depositional patterns upon precipitation 
\cite{rouwet2017sedimentation}.

\subsubsection{Double-diffusive convection and mushy layers}
\label{sec:double-diffusive}

\begin{figure}
\centering
\includegraphics{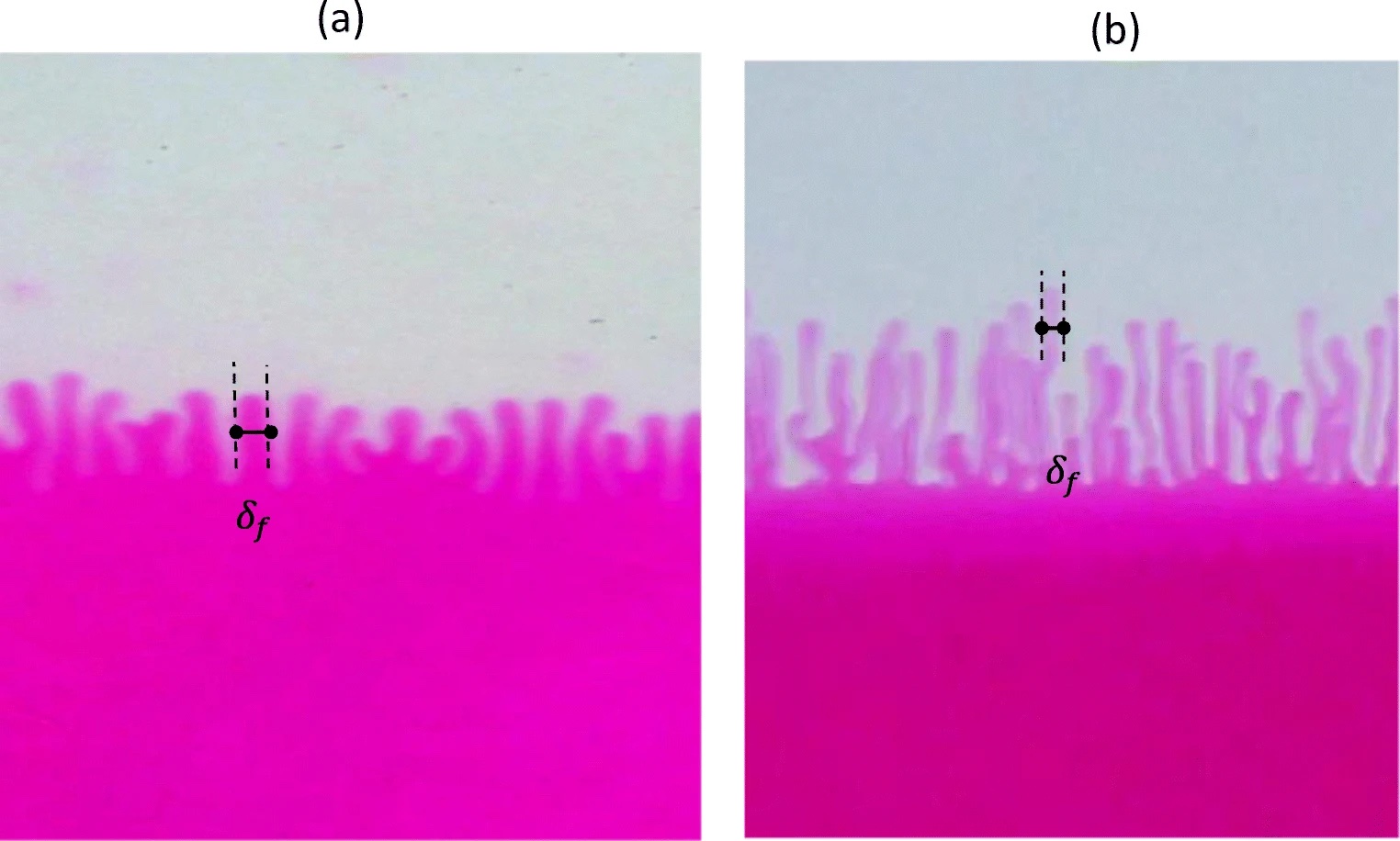}
\caption{\label{fig:double_diffusive}
Double-diffusive convection; salt fingers in an experimental study illustrating how
finger structures and time-scales vary with Rayleigh number.
A: low Rayleigh number $Ra=4.3\times10^3$, evolution time 1~h 4 min, 
 wavelength $\delta_f = 2~{\rm mm}$; 
B: High Rayleigh number $Ra=4.3\times10^5$, evolution time 2~min 33 s, wavelength
$\delta_f= 1.25~{\rm mm}$.
(Source: \citeA{rehman2018role}.)}
\end{figure}

\begin{figure}
\centering
\includegraphics[width=\linewidth,height=0.78\textheight,keepaspectratio]{gr22}
\caption{\label{fig:igneous_layers}
Igneous layering. 
A: Layer width $\Delta x_n$ (cm) versus numbers indicating the order from the reference layer
$n$ at Ogi picrite sill, Sado Island, Japan. The trend approximately follows $\log \Delta x_n = \log \Delta x_0 + 0.003n$. Photographs showing cyclic layering: 
B: Picrite sill at Ogi, Sado Island, Japan. Layer spacing is approximately
30~cm. 
C: Dolerite sill at Duntulm Castle, Isle of Skye, Scotland. A geologist in a  blue circle provides scale. 
D: Dolerite
sill at Konpiraiwa, Atsumi, Japan. Layer spacing is approximately 20~cm. 
E: Tertiary basaltic andesitic dyke at Shibushi,
Hirado Island, Japan. The country rocks are dolerite. The scale bar is 37~cm long.
(Source: \citeA{toramaru2012numerical}.)}
\end{figure}

Double-diffusive convection  is driven by  buoyancy forces arising from two components with
different diffusivities~\cite{Huppert1981,Schmitt1995,radko2013double}. Unlike convection
where density is determined only by temperature, double-diffusive convection occurs when, for
example in the ocean, thermal gradients and compositional gradients (like salt or dissolved
minerals) meet each other. Because heat diffuses much faster than molecular components, this
process can trigger instability even in otherwise stably stratified fluids, often resulting
in layered, staircase structures. 
Double-diffusive convection thus arises 
from a difference in diffusion rates, and there are two possibilities. The finger regime (salt fingers;  Fig.~\ref{fig:double_diffusive}) occurs when hotter, salty water overlies colder, fresh water. This provides an unstable salinity gradient alongside a stable thermal gradient; in other words the instability is driven by the slower diffuser. The other, 
diffusive regime, occurs when cold, fresh water overlies hot, salty water. This situation couples a stable salinity gradient to an unstable thermal gradient; the instability here is driven by the faster diffuser. The instability looks very different in the two instances. 

Double-diffusive convection appears 
in geological settings where multiple components affect fluid density, such as in magma chambers
\cite{huppert1984double}, lavas~\cite{turner1986komatiites},
 the planetary core--mantle boundary~\cite{hansen1990nonlinear}, and hydrothermal
systems~\cite{bischoff1989salinity}. However, 
 while double-diffusive processes may certainly play a role in the dynamics of molten and otherwise fluidized rock, there is less evidence of rock \emph{patterning} linked to double-diffusive convection.

There have been suggestions 
that large-scale layering in igneous rocks~\cite{charlier2015layered}
(Fig.~\ref{fig:igneous_layers}~\cite{toramaru2012numerical})
might arise, like thermohaline staircases in the ocean, from double-diffusive
convection~\cite{kerr1982layered}. This is now viewed as being an
oversimplification~\cite{naslund1996mechanisms}, 
and mechanisms involving 
coupled solidification and desorption processes
 producing a 
geometric pattern of layer spacing have been derived
\cite{rogerson2000patterns}.
When considering patterning in rocks one should keep in mind the differences between bulk
fluids and porous media~\cite{nield2017convection}.
However, it is clear that spatio-temporal concentration variations in the liquid phase should affect a forming solid phase, and this links double-diffusive convection to oscillatory zoning,  Sec.~\ref{sec:zoning}.
Some soil patterns of the finger type may  have an origin in double-diffusive convection in
porous media~\cite{imhoff1988experimental}, but it should be noted that  there are 
other reactive-infiltration instabilities, as we discuss below in  Sec.~\ref{sec:fingering} and Sec.~\ref{sec:pipes}.

During the solidification of multicomponent solutions, such as in magma chambers, lava lakes, or sea ice,  a solidifying matrix of crystals --- a \emph{mushy  layer} --- forms alongside interstitial liquid. 
Convection can cause local melting or increased solidification, depending on whether the rising fluid is hotter or colder than the surrounding, affecting the permeability and structure of the layer. Convection in mushy layers 
 often leads to the formation of tubular \emph{chimneys}, vertical channels within the mushy layer, in which fluid plumes rise, enhancing transport and modifying solidification rates. 
In magma chambers, convection in mushy zones can drive the formation of \emph{freckles}, chains of mafic minerals and other localized chemical heterogeneities,  impacting the differentiation of igneous rocks.  
During sea ice formation, when seawater freezes, it forms a mushy layer of ice crystals and saline brine. The brine, being denser, can convect, affecting the rate of ice growth and the transport of salt.
As mushy layers are characterized by the interaction between compositional (solutal) buoyancy --- due to the rejection of lighter or denser elements during freezing --- and thermal buoyancy, creating a density difference, one can consider the possibility of double-diffusive convection 
 because heat and solute diffuse at different rates.
However, the question is complicated by the consideration of whether temperature and concentration fields are linked through 
 local thermodynamic equilibrium. It turns out that double-diffusive effects may appear in a
ternary case~\cite{anderson2020convective}.
  Although geological systems might well be good places to look for three coupled elements,  this idea has as yet been little investigated in a geological context.

\subsubsection{Capillary flow and coffee-ring instabilities}
\label{sec:capillary}

Salt dissolution in water enables long-distance transport of ions, with subsequent crystal precipitation occurring upon evaporation or cooling. This process manifests itself in various scenarios, such as the capillary rise of groundwater to soil or rock surfaces and evaporation near salt lakes (Sec.~\ref{sec:saltpan}). Crystal precipitation generally occurs in regions where evaporation is strongest and ions are advected by the liquid flow. The increase in ion concentration in these areas raises the probability of crystal nucleation.

\begin{figure}
\centering
\includegraphics{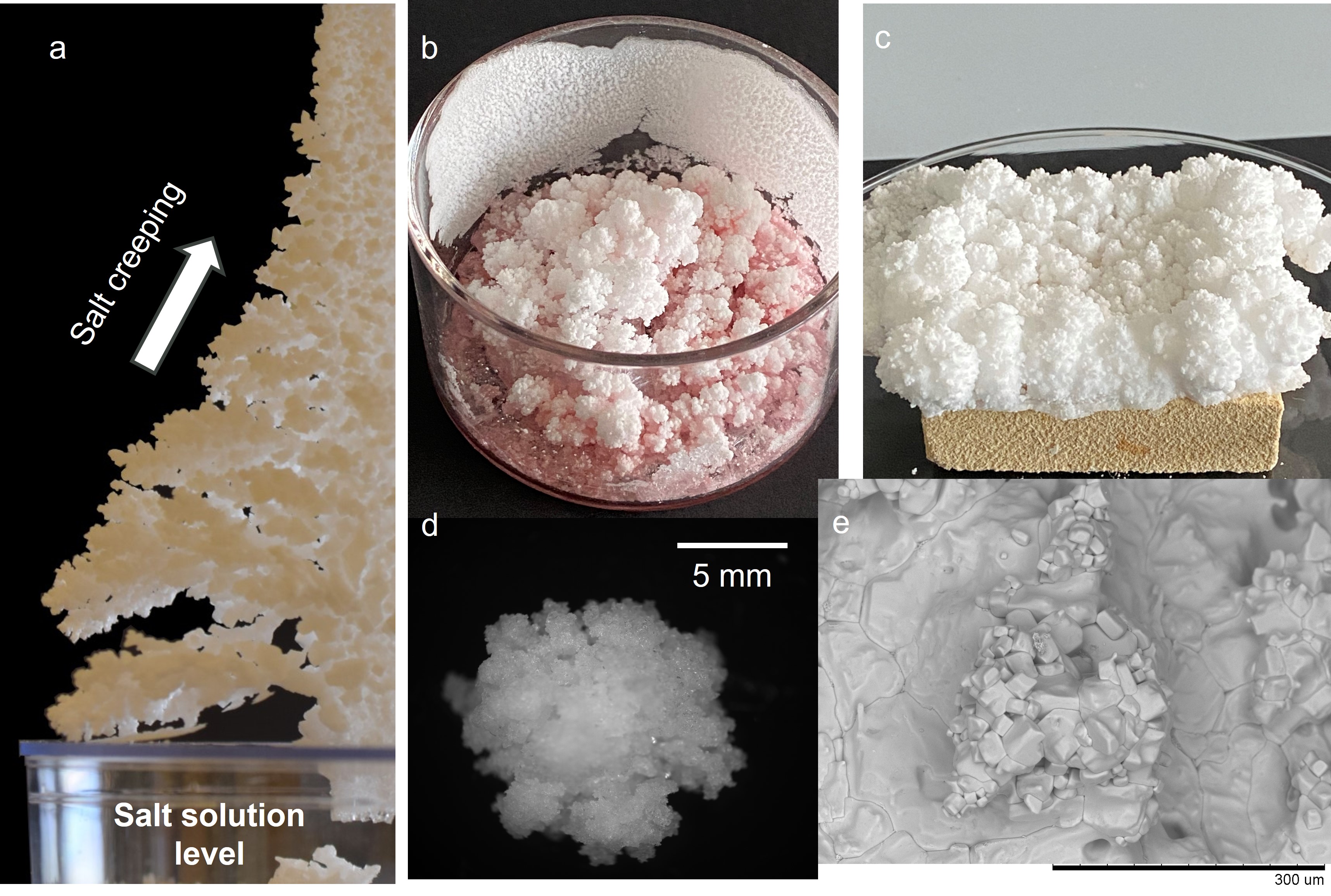}
\caption{\label{fig:creeping1}
Salt crystallization patterns at different scales. 
A: laboratory experiment showing salt creeping on a cylindrical glass rod, above the flat surface and far from the evaporating salt solution in the Petri dish. 
B:salt crystallization as efflorescence on large NaCl crystals, and salt creeping along the glass wall of an evaporating NaCl solution in a beaker, in a laboratory-scale analogue of the drying of a salty lake. C: Salt efflorescence on a limestone fragment 
D,E: close-up of the morphology of salt efflorescence; SEM images of its porous microstructure, showing an assembly of microcrystals.
(Images: Noushine Shahidzadeh.)}
\end{figure}

This is  similar to the coffee-ring effect, the pattern left behind when a particle-laden
liquid droplet dries, characterized by a ring-like deposit along its
perimeter~\cite{craster2009dynamics,Shahidzadeh2015Salt}. It arises because evaporation is
faster at the droplet's edges, and liquid from the centre flows outward to replenish the
edge. This outward flow carries suspended particles to the droplet's perimeter and, as
evaporation finishes, particles accumulate at the edge, forming a ring. 

In the case of salt solutions, the accumulation of ions induces multiple nucleation events at the
evaporation front. This generates salt creeping, a self-amplifying process that forms
three-dimensional porous crystalline networks at considerable distances from the original
solution source~\cite{qazi2019salt}. While salt creeping  occurs on flat surfaces, a similar
process known as salt efflorescence develops on porous surfaces such as soils or stones,
driven primarily by capillary flow of the salt solution towards the evaporative region at the
top of the porous medium. Studies show that the ion concentration near a receding meniscus in
pores is controlled by evaporation and can be quantified by the P\'eclet number. This
dimensionless number, the ratio of advective to diffusive transport of ions, is central in
continuum transport~\cite{Wijnhorst2024}. When $Pe>1$, advection dominates diffusion and
promotes crystal precipitation. Early large crystals precipitating at the evaporation front
create macropores that enable subsequent nucleation of smaller crystals, which in turn
generate progressively smaller pores. In addition, crystal-mediated spreading of the salt
solution increases the invaded area by capillarity. Marangoni flows are negligible in salt
creeping dynamics because the concentration gradient between the bulk salt solution and the
liquid film in contact with precipitated crystals at the evaporative points is small,
preventing high supersaturation and leading to immediate secondary nucleation of crystals. 

 Fig.~\ref{fig:creeping1} illustrates the dynamic nature of salt creeping and efflorescence
formation and their ability to propagate well beyond the initial boundaries. Scanning electron
microscopy (SEM) images of sodium chloride (NaCl) creeping efflorescence reveal porous
polycrystalline structures and show that this phenomenon spans many orders of magnitude in
length-scale, so that it appears fractal~\cite{Wijnhorst2024}. The crystalline porous
structure, with a pore-size distribution set  by the range of crystal sizes, allows saline
water to be drawn in by capillary pumping (wicking). Consequently, crystallization continues
further upward at the top of the efflorescence. The rate of evaporation plays a critical role
in the development of salt creeping, efflorescence, and salt pillar
formation~\cite{desarnaud2015drying}. For creeping to occur, the evaporation rate should be
high enough to induce multiple nucleation processes at the evaporation front ($Pe>1$).
This can be achieved with  air flow (wind), low relative humidity, or elevated temperature
during evaporation. By contrast, a slow evaporation rate induces the precipitation of fewer
nuclei which grow larger and evolve towards their equilibrium shape; this can suppress
creeping. Very low relative humidity may also suppress the salt creeping process, as
precipitation kinetics can change at very high evaporation rates, promoting crust
formation~\cite{desarnaud2015drying}. Another important factor is the type of porous medium or surface on which salt precipitation occurs, which can strongly affect the
final deposition morphology~\cite{qazi2019salt}.

These mechanisms  can  result in spectacular formations such as desert roses and other efflorescence structures in natural environments
in arid regions 
(Sec.~\ref{sec:roses}).

\subsubsection{Saffman--Taylor instability, viscous fingering, and reactive-infiltration instability}
\label{VFRI}\label{sec:fingering}

\begin{figure}
\centering
\includegraphics{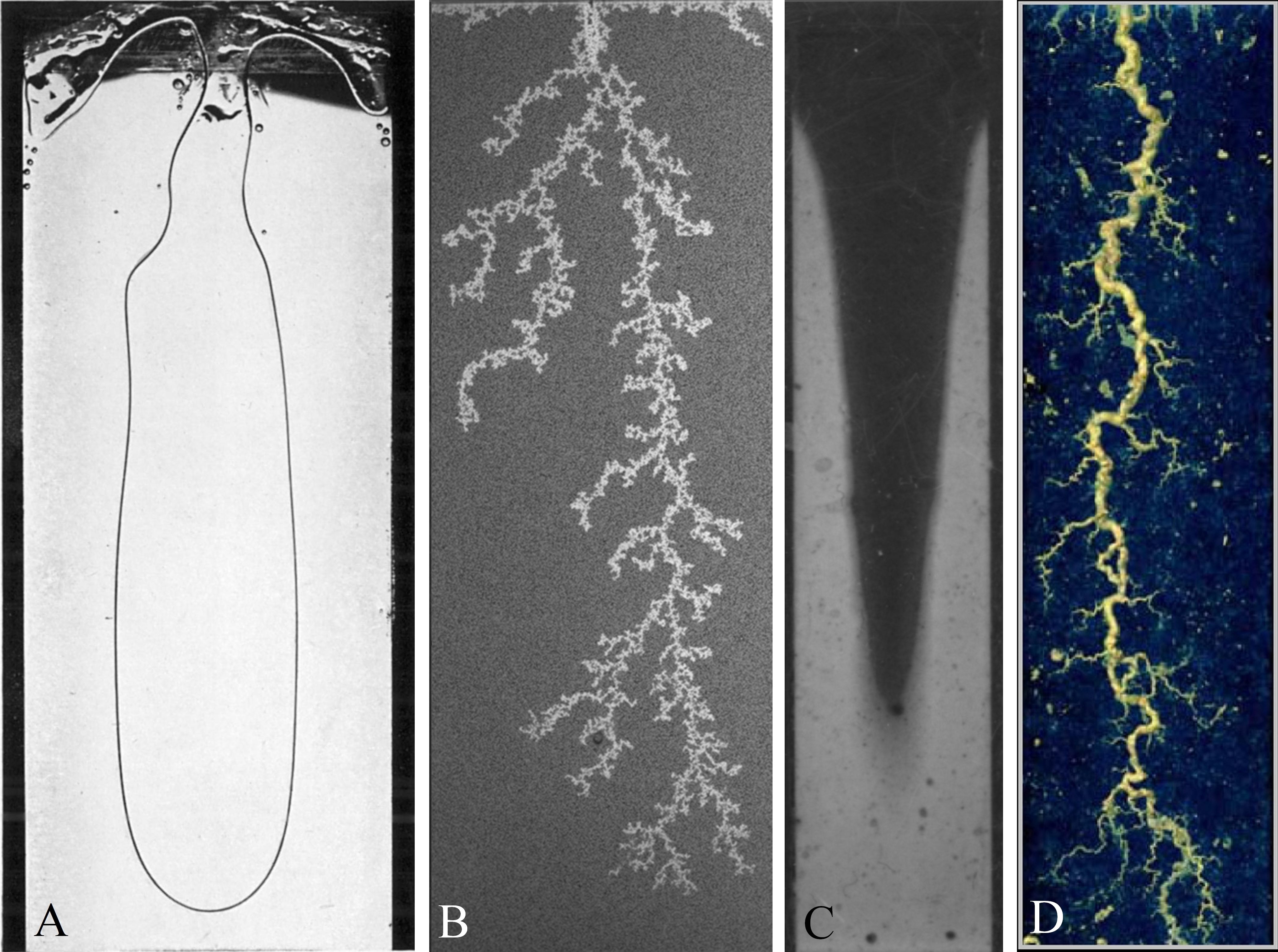}
\caption{\label{fig:single}Fingering. A:  Saffman--Taylor finger that appears in the
long-time limit in the viscous fingering experiment. B: The irregular,
fractal finger formed in viscous fingering experiment in porous medium, where air was
displacing glycerine in a Hele-Shaw cell filled with glass beads. C: A
smooth dissolution finger arising in a Hele-Shaw cell with a soluble
bottom. D: A fractal dissolution finger (wormhole) formed during
acidization of limestone. 
}
\par\smallskip{\footnotesize\noindent\raggedright\emph{Sources:} A:~\citeA{Saffman1958}, B:~\citeA{Lovoll2004}, C:~\citeA{Zukowski2025},   D:~\citeA{mcduff2010}.
\par}
\end{figure}

Whenever a more mobile fluid is rapidly injected into a less mobile one, the interface between them becomes unstable. This occurs because the fluid tends to flow along the path of least resistance. Therefore, if there is a protrusion on the advancing interface, it will focus the flow, leading to a faster growth of the protrusion and transforming it into a finger of the more mobile phase propagating within the less mobile one.

Examining Darcy's law, which relates the flow rate $\mathbf{u}$ to the pressure gradient $\nabla p$ in a porous medium
\begin{equation}
\mathbf{u} = -M \nabla p = -\frac{K(\phi)}{\eta} \nabla p,
\end{equation}
we see that mobility $M$ can be controlled in two ways: by altering the viscosity of
the fluid $\eta$ or by changing the permeability of the medium $K$, which
depends on the porosity $\phi$. The first mechanism gives rise to viscous fingering~\cite{Hill1952,Saffman1958,Chuoke1959}, in which a less viscous fluid displaces a more viscous one, and the initially flat interface develops growing protrusions.

The second mechanism arises when a chemically active fluid dissolves the porous medium as it reacts with the pore surfaces.
This is so-called reactive-infiltration
instability~\cite{Ortoleva1987b,Hinch1990,Szymczak2012,Szymczak2014}.
A classic example here are karst processes, initiated by the dissolution of limestone by
$\mathrm{CO}_2$ saturated water. Flow focusing due to the reactive-infiltration instability
impacts  cave system formation~\cite{Groves1994a,Hanna1998,Szymczak2011} and surface karst
features such as solution pipes (Sec.~\ref{sec:pipes})~\cite{lipar2021}. In industrial
applications, much stronger acids, such as HCl or HF, are used for stimulation of petroleum
reservoirs~\cite{Hoefner1988,Fredd1998,Economides2000} or sustaining fluid circulation in
geothermal systems~\cite{Charalambous2021,Sutra2017}. 

It is interesting to note what happens to the initial protrusions on the unstable front
during the course of the evolution of the instability. As they grow larger, they transform
into fingers that begin to interact with each other. Two processes take place here: one is
the competition of the fingers for the flow, causing the longer ones to advance ahead of the
shorter ones. The other is the merging of  fingers, reducing their total number. As a result,
the pattern coarsens, and eventually, a single finger emerges. In the theory of viscous
fingering such a final, stable finger-like structure  is known as a Saffman--Taylor finger
(Fig.~\ref{fig:single}A). If the Saffman--Taylor experiment is conducted in a Hele-Shaw cell
filled with glass beads, mimicking a porous medium, the finger appearing in the system has a
highly ramified, fractal structure, not unlike diffusion-limited aggregates~\cite{Witten1981}
(Fig.~\ref{fig:single}B). This is due to the fact that the local interface curvature controlling
the capillary pressure drop depends on the local pore geometry and is thus sensitive to the
frozen disorder associated with the structure of this system~\cite{Lovoll2004}.

A similar situation occurs in the case of reactive-infiltration instability. If we conduct
experiments in a homogeneous system, such as a Hele-Shaw cell with a soluble bottom, we end
up with relatively regular fingers, which merge and transform into a single dissolution
finger in the long-time limit~\cite{Zukowski2025} (Fig.~\ref{fig:single}C); compare the
geological  solution pipes of  Fig.~\ref{fig:Pipes}. On the other hand, in the dissolution of
porous rocks, twisted and ramified wormholes form (Fig.~\ref{fig:single}D;   Sec.~\ref{sec:pipes}).

So far, the discussion has concerned fingering driven by viscosity contrast or by reaction-induced changes in permeability under imposed through-flow. In carbon sequestration, however, fingering arises in a different way: reaction couples to buoyancy-driven convection of dissolved CO$_2$ in porous media. As CO$_2$ dissolves into brine, the fluid becomes denser, which may destabilize the diffusive boundary layer and trigger downward-propagating convective fingers. Chemical reactions between the dissolved CO$_2$ and the host rock may then alter the hydrodynamic behaviour of this layer and either promote or suppress fingering.
 The relevant control parameter is the Cardoso number, defined as the ratio of the Damk\"ohler number to the square of the solutal Rayleigh--Darcy number,
\begin{equation}
Ca=Da/Ra^2=k_raD\varphi/ (k\Delta\rho_0 g/\mu)^2,
\end{equation}
where $\Delta\rho_0$ is the maximum density contrast between pure and solute-saturated fluid, $D$ 
is the diffusivity of the solute in the aqueous layer, $\varphi$ is the porosity of the
rock matrix and $g$ is the acceleration due to gravity. The permeability of the
porous medium $k$ and the fluid viscosity $\mu$ are constant. The Cardoso
number is a measure of the relative magnitude between the timescale for onset of convection
and the timescale for reaction; a small Cardoso number reflects a slow reaction with little
effect on the development of convection, while a large Cardoso number implies a fast reaction
with consequent impact on the intensity of the instability~\cite{cardoso2014geochemistry}.

\begin{figure}
\centering
\includegraphics[width=\linewidth,height=0.78\textheight,keepaspectratio]{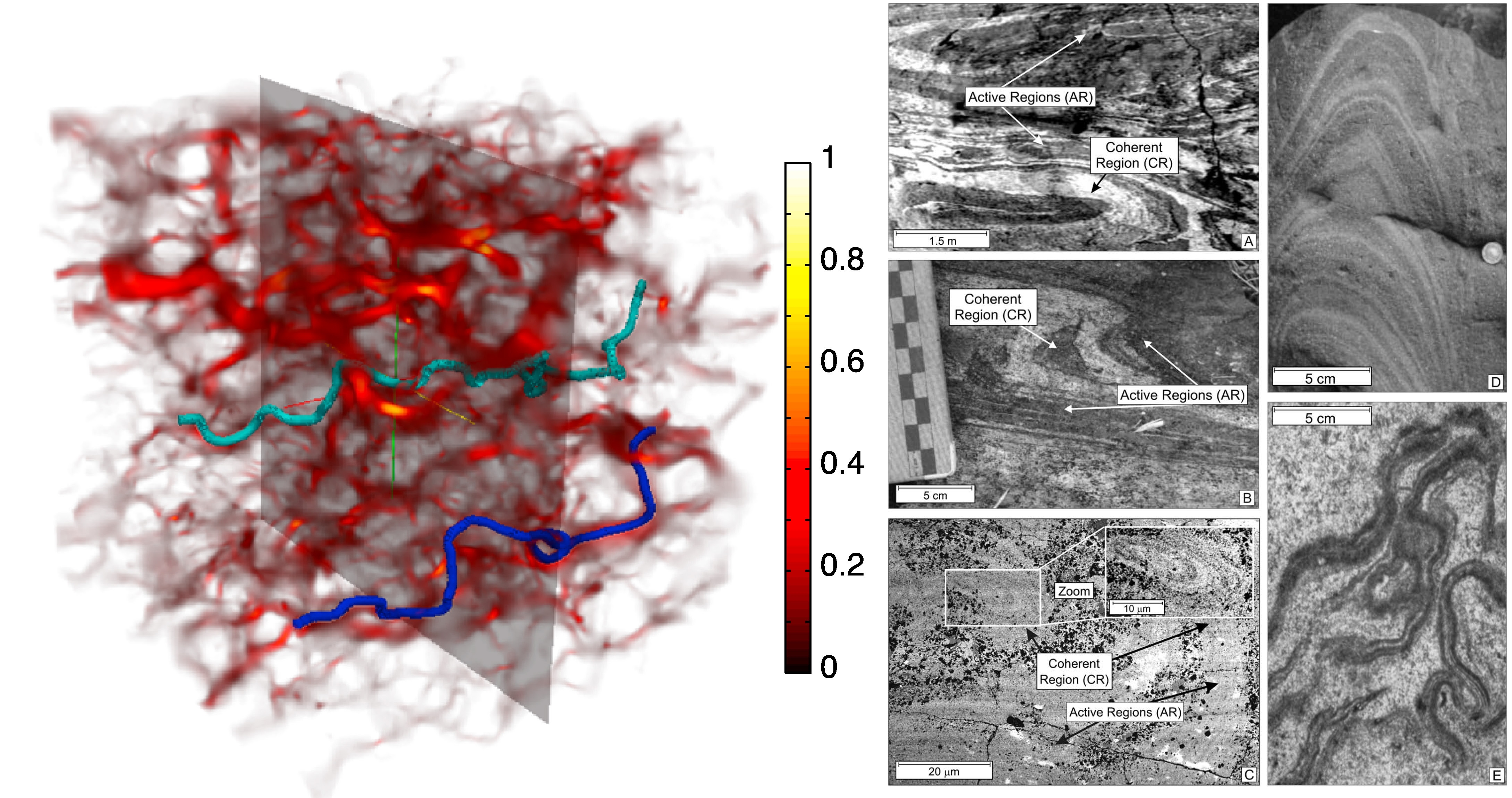}
\caption{\label{fig:advection}Chaotic advection.
Left:
Digital reconstruction of the 3D flow field through a Berea sandstone block (1.6~mm cube).
The normalized velocity magnitude is rendered according to the colour bar; yellow zones
indicate high velocity bursts. Two flow paths (blue and cyan) show evidence of branch and
merge behaviour. 
Right: Examples of magma mixing structures in lava flows from the islands of Lesbos (A--C),
Vulcano (D) and Salina (E). The dark flow structures consist of B magmas dispersed through
light-coloured A magmas. In all lava flows mixing structures have the same patterns at many
scales of magnification.}
\par\smallskip{\footnotesize\noindent\raggedright\emph{Sources:} Left:~\citeA{kang2014pore};
    Right:~\citeA{perugini2003chaotic}.  
\par}
\end{figure}

\subsubsection{Chaotic advection}
\label{sec:chaotic}

Chaotic advection, a phenomenon recognized in fluid mechanics since the
1980s~\cite{aref1984stirring}, shows  that deterministic nonlinear dynamics can render
fluid trajectories chaotic~\cite{cartwright1999introduction}. Stretching and folding generate
intricate, often fractal structures that greatly enhance mixing, with consequent enhancements
in heat transfer and reaction rates. 
Chaotic advection is distinct from turbulence: it arises in slow  (low-Reynolds-number)
flows, whereas turbulence characterizes fast, inertially dominated
flows~\cite{aref2017frontiers}. 

Since a considerable portion of geophysical flows are slow, chaotic advection has important consequences in geological systems. 
One key application is pore-scale fluid flow in rocks (Fig.~\ref{fig:advection}). A series of
studies has addressed this question and shown its importance for macroscopic heat and solute
transport~\cite{metcalfe2010partially,kang2014pore,lester2016chaotic,lester2016chaotic2}.
\citeA{lester2012mechanics} show how
fluid mixing and chemical reactions  significantly affect the mechanics of hydrothermal systems. 
\citeA{trefry2019temporal} discuss how temporal fluctuations, together with poroelasticity,
generate chaotic advection in natural groundwater systems.
\citeA{schoofs1999chaotic} show that,  for typical geological parameters,  thermohaline
circulation in the Earth's crust is intrinsically chaotic. 
 This dynamics impacts heat transport at mid-ocean ridges,  ore genesis,  metasomatism and metamorphism, and the diagenetic history of sediments in subsiding basins.
\citeA{oberst2018detection} propose that feedbacks between thermochemical and deformation
processes are important for the deposition of vein-filling minerals such as pyrite and gold.
Further work has examined chaotic advection in magma flow (Fig.~\ref{fig:advection}),  its role
in mixing, and its effects
on the time-scales of cooling and crystallization of
magma~\cite{yuen1992strongly,perugini2003chaotic,perugini2008virtual,de2015chaotic,petrelli2016effects}.

\subsection{Osmotic processes}
\label{sec:osmotic}

\begin{figure}
\centering
\includegraphics{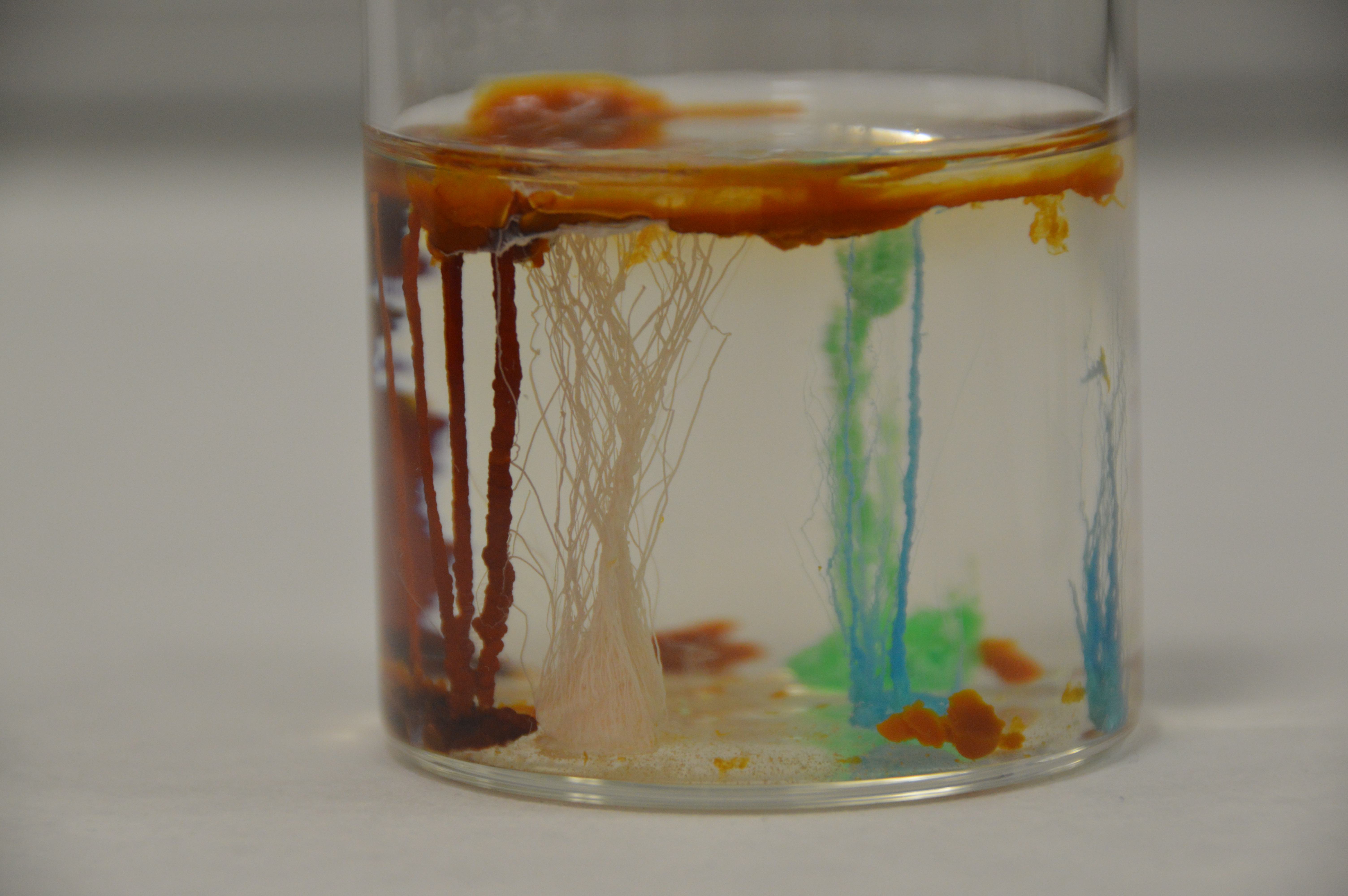}
\caption{\label{fig:osmosis}Osmotic processes.
The chemical laboratory demonstration known as a  chemical garden depends on osmotic forces for its growth.}
\par\smallskip{\footnotesize\noindent\raggedright\emph{Image:} Julyan Cartwright.\par}
\end{figure}

Osmosis is a phenomenon that arises from the differential mobility of molecular species in a fluid. In the simplest case, if the solvent (water) can pass through a membrane, whereas a solute (e.g.,~metal ion) cannot, then exclusion of the solute generates an osmotic pressure across the membrane. In this case the membrane is termed semipermeable. Such a barrier to movement can  result from pore size (sieving), but also  chemical affinities and electrical potentials.

Osmosis and osmotic pressure are ubiquitous in biology, since cells possess membranes, and
are both influenced by and exploit these forces. However, osmosis also occurs in geological
systems. Clays and shales are typical instances of geological semipermeable porous
media~\cite{neuzil2000osmotic}. Wherever water interacts with a porous medium that acts as a
semipermeable substrate, osmotic pressure can develop~\cite{cardoso2014dynamics}.  
Osmosis, alongside buoyancy forces, may drive the convective processes of submarine mud
volcanoes and seeps~\cite{cardoso2016increased,rocha2021formation}.
 Modelling studies indicate that osmosis may be at the root of permafrost explosions
producing large craters in the high arctic of Siberia~\cite{morgado2024osmosis}.

Osmotic processes can also drive striking self-organized precipitation phenomena, most notably in so-called chemical gardens.
Chemical gardens are biomimetic structures that form tubular solids which twist into multiple
forms resembling plants. This phenomenon has fascinated researchers for centuries, from
Glauber in 1646~\cite{glauber_furni_1646} to the present~\cite{Pimentel2023Chemobrionics}.
Their formation begins by adding a solid seed crystal of a metal salt to a solution, often
silicate, but also carbonate, phosphate, and others. The surrounding liquid partially
dissolves the seed surface, forming a semipermeable membrane. External solvent enters the
interior of the membrane by osmosis. This influx increases the internal volume, stressing the
membrane until it ruptures. Then, the internal solution is expelled under osmotic and
buoyancy forces. When the internal solution contacts the external solution, a precipitation
reaction occurs, forming the tube walls. These osmotic and buoyancy forces, together with
precipitation, produce structures  with continuously evolving  direction and morphology that
resemble biological gardens. This phenomenon is a self-organizing, non-equilibrium,
reaction--advection process~\cite{barge2015chemical}.

Many studies on this phenomenon have been performed at the laboratory scale
(Fig.~\ref{fig:osmosis}). However, chemical gardens can also be found in nature as geological
structures. Although the length- and time-scales are clearly different, the same physical
drivers recur in natural settings: semipermeable membranes, osmotic pressure, concentration
and pH gradients, buoyancy, fluid dynamics, and precipitation. Some examples of geological
chemical gardens are hydrothermal vents on the seafloor
(Sec.~\ref{sec:hydrothermalvents})~\cite{cardoso2017differing,sainz2018growth,Cartwright2019}. In
the cryosphere, related structures include  brinicles in seawater~\cite{brinicle,teston2024experimental} and analogous formations on
other celestial bodies~\cite{vance2019self}.
Furthermore, such structures, particularly  alkaline submarine hydrothermal
vents, are now regarded among the prime candidates for the environments where life may have
originated on Earth~\cite{russell1994,russell1997,Cartwright2019,cardoso2020}, within 
inorganic pores and vesicles~\cite{ding2019intrinsic,ding2024dynamics}, via processes that
embed patterns and information at interfaces between mineral
phases~\cite{cartwright2012beyond,cartwright2016dna,cartwright2024information}.

\subsection{Diffusion-limited aggregation}
\label{sec:DLA}

\begin{figure}
\centering
\includegraphics{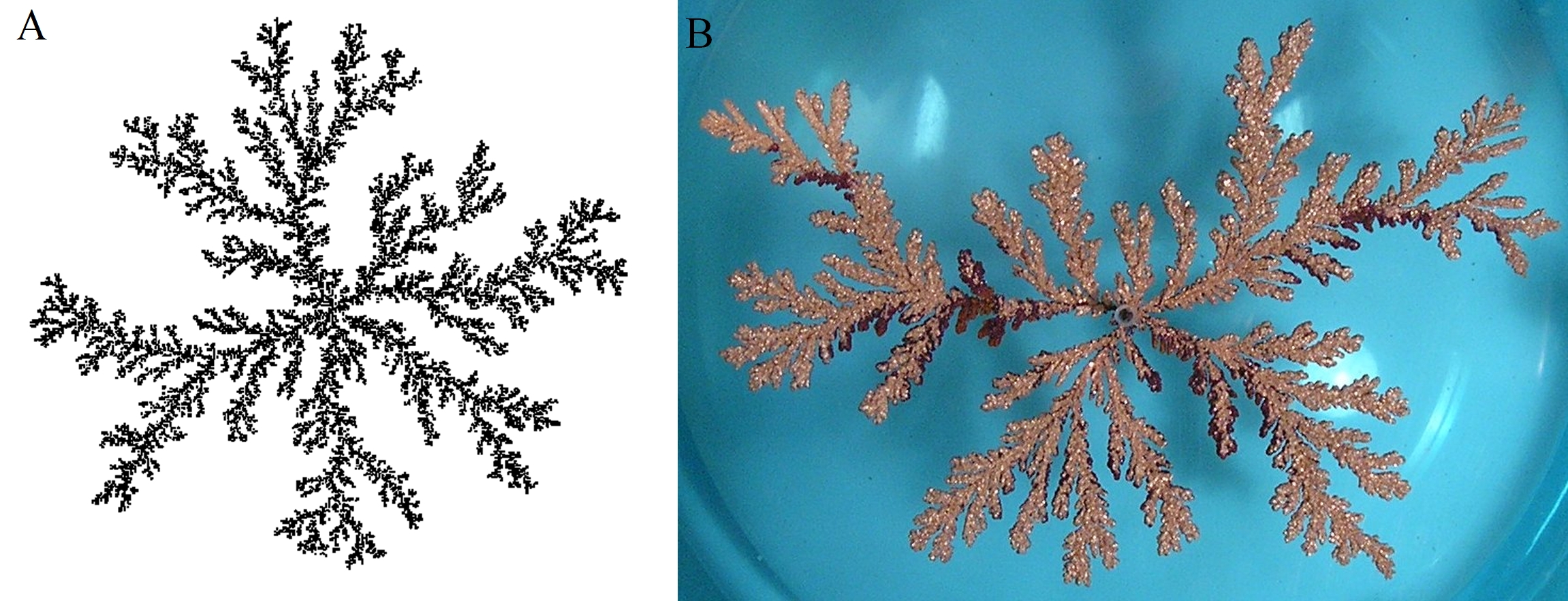}
\caption{\label{fig:DLA}Left: A cluster created by the diffusion-limited aggregation (DLA) model. Right: A dendritic copper aggregate, produced through electrodeposition from a copper sulphate solution, which can be described by the DLA model.}
\par\smallskip{\footnotesize\noindent\raggedright\emph{Images:} Left: Piotr Szymczak, Right: Kevin R Johnson; CC-BY-2.5.\par}
\end{figure}

 Diffusion-limited aggregation, DLA~\cite{Witten1981,witten1983diffusion},  Fig.~\ref{fig:DLA},
is a numerical model that describes how complex, disordered structures, often fractal-like in
nature, form through the random movement and attachment of particles to a growing cluster.
The simulation starts with a seed or nucleation site, and particles are added one 
at a time via random-walk paths starting outside the region occupied by the cluster.
If a particle comes close to the cluster, it attaches to the perimeter, contributing to cluster growth. This process is repeated, and over time, the cluster develops a complex, ramified structure.

DLA can be a good model for aggregation from solution, provided that certain conditions are
met. First, it assumes that diffusion of the particles is the slowest process shaping the
cluster, while the attachment process is faster. Second, since only one random walker is
released at a time, the model corresponds to the infinite dilution limit of the particles. In
such a limit, the cluster growth time-scale is much longer than the relaxation time of the
particle concentration to the stationary profile. Finally,  aggregation  is considered to be
irreversible, and the cluster does not interact with the incoming particles. Many
generalizations of the DLA model exist, which relax one or more of these
assumptions~\cite{Meakin1983,Garik1985,Erlebacher1993,Meakin1998}, thereby extending the
applicability of these models to a broader range of phenomena.

The origin of the complexity of  DLA aggregates lies in the competition between different regions of the growing cluster. A branch extending from the centre is able to capture more particles since random walkers tend to stick to the first part of the cluster they encounter. Between the growing branches, there are regions, known as fjords, that remain shielded from incoming particles,  which become more pronounced as the branches extend.

The limited penetration of the fjords results from a strong screening
effect~\cite{Meakin1998}. Particles following random-walk paths tend to become trapped along
the edges of the fjords, making it highly unlikely for them to reach the interior of the
cluster. This phenomenon is analogous to the screening of a Laplacian field, with absorbing
boundary conditions applied to the surface of the DLA fractal, resulting in an exponentially
small probability of penetration as a function of the penetration depth.

Similar screening mechanisms apply to other fields, such as pressure fields or electric potentials, leading to a resemblance between diffusion-limited aggregates and other fractal growth patterns, such as viscous fingers or electrodeposition structures.

Although DLA is widely used to model far-from-equilibrium growth in physical and biological
systems, it remains a theoretical challenge to fully describe its structure using scaling
models. Most approaches assume a self-similar fractal pattern, but this does not fully
capture the complexity of DLA~\cite{Meakin1998,Halsey2000}. The mathematical structure
underlying DLA, as well as more general Laplacian growth models, is quite involved, with the
link to the theory of integrable systems and even quantum gravity
models~\cite{Mineev-Weinstein2000,Mineev-Weinstein2008}, providing further evidence that a
simple model can indeed generate complex structures.
In geological systems, DLA provides a useful idealized description of the growth of some mineral dendrites (Sec.~\ref{sec:dendrites}).

\subsection{Piezoelectricity}
\label{sec:piezoelectricity}

\begin{figure}
\centering
\includegraphics{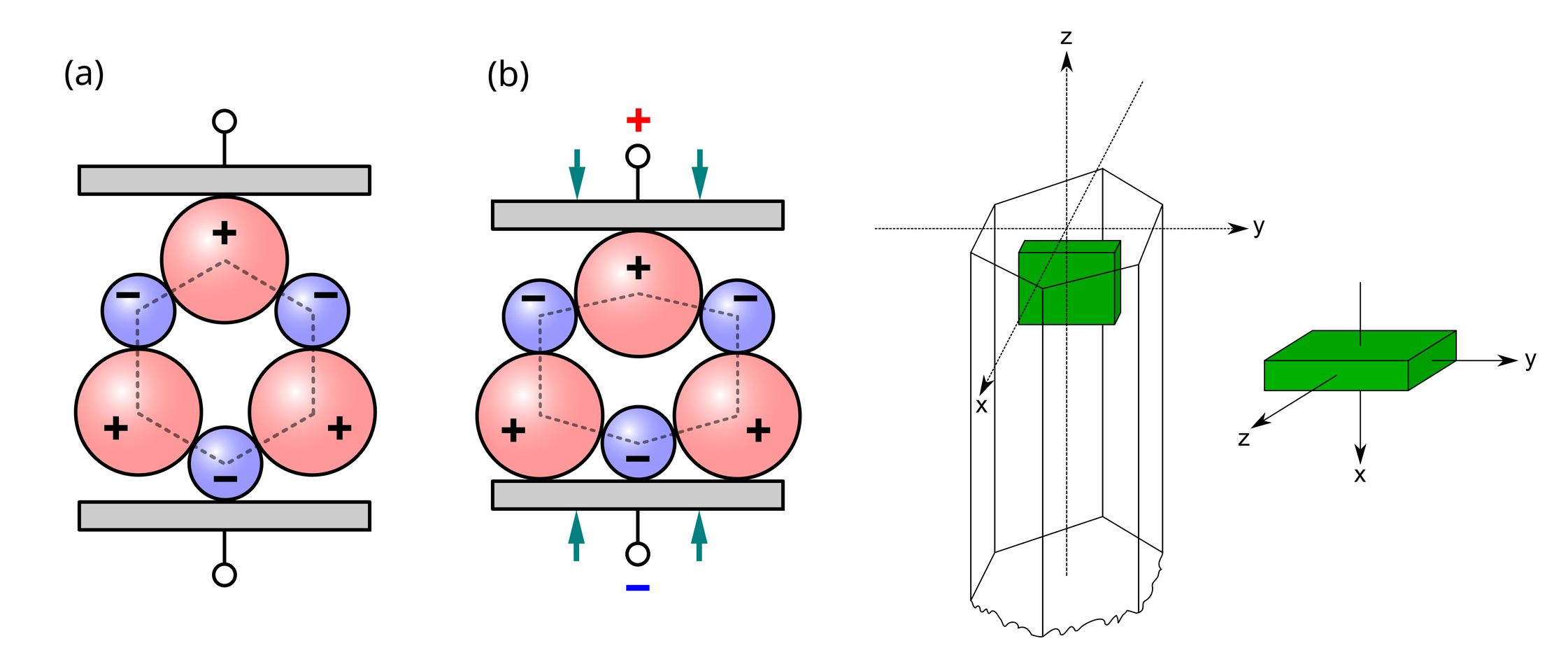}
\caption{\label{fig:piezoelectricity}Piezoelectricity:
Left: (a) simplified model of a quartz  (SiO$_2$) crystal between two electrodes. 
 Mechanical pressure (b) shifts the positive and negative centres of charge. This creates an electrical dipole, inducing a voltage at the electrodes.
Right:
Part of a quartz crystal with marked crystallographic axes and a piezoelectric plate (Curie cut) cut from it.
(Images: 
Left: MikeRun;  CC-BY-SA-4.0;
    Right:  CLI; CC-Zero.
)
}
\end{figure}

Piezoelectricity, the generation of electric charge in response to mechanical stress, is a less commonly discussed  mechanism of pattern formation. The piezoelectric effect arises from broken inversion symmetry and is observed in non-centrosymmetric crystals (lacking a centre of symmetry). Under stress, such crystals polarize, producing measurable charge.  Quartz is a particularly common geological example (Fig.~\ref{fig:piezoelectricity}).

Piezoelectricity is used in engineering applications, for example to  form patterned thin
films via piezoelectric printing~\cite{khan2016piezoelectric} and in functional soft-matter
systems~\cite{pishvar2020foundations}. As described in  Sec.~\ref{sec:nuggets}, it may be
involved in concentrating gold nanoparticles into nuggets~\cite{voisey2024gold}.

\subsection{Ostwald ripening}
 \label{sec:ostwald}

\begin{figure}
\centering
\includegraphics{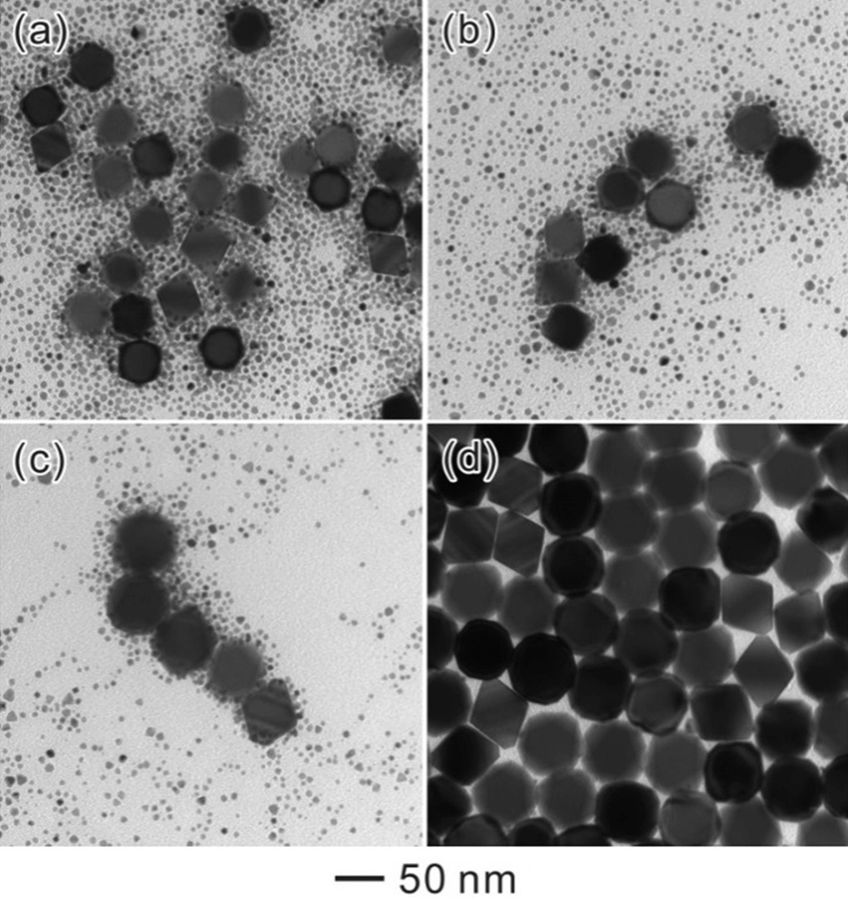}
\caption{\label{fig:ripening}Ostwald ripening of palladium nanoparticles in formaldehyde. In (A), two particle populations with different sizes can be seen. During the reaction, as shown in (B) and (C), the smaller particles dissolve while the larger ones grow. By the end of the process (D), only the population of larger particles remains, having increased in size.}
\par\smallskip{\footnotesize\noindent\raggedright\emph{Source:} \citeA{zhang_redox_2015}.\par}
\end{figure}

The Ostwald ripening process was first described in the late 19th century by Wilhelm
Ostwald~\cite{Ostwald1897}, from whom it takes its name. Under identical conditions,
differences in particle size alone drive mass transfer: subcritical (smaller) crystals
dissolve while larger ones grow at their expense (Fig.~\ref{fig:ripening}). Ostwald ripening
constitutes the late stage of a first-order phase transition, during which domains coarsen by
material redistribution. Its governing physics is classically captured by
Lifshitz--Slyozov--Wagner (LSW) theory, which predicts self-similar coarsening with growth
rates set by either diffusion or interfacial
kinetics~\cite{Lifshitz1961,wagner_theorie_1961}. 
In LSW theory, the particle-size distribution approaches a self-similar form when rescaled by
a characteristic length (e.g.,~the mean radius). That characteristic size grows as
$t^{1/3}$ for diffusion-controlled ripening~\cite{lifshitz_kinetics_1961} and as
$t^{1/2}$ for interface-kinetics-controlled ripening~\cite{wagner_theorie_1961}. These
predictions have been observed qualitatively in experiments, but quantitative discrepancies
remain, so Ostwald ripening continues to be an active subject of research
today~\cite{van_westen_effect_2018}

\begin{figure}
\centering
\includegraphics{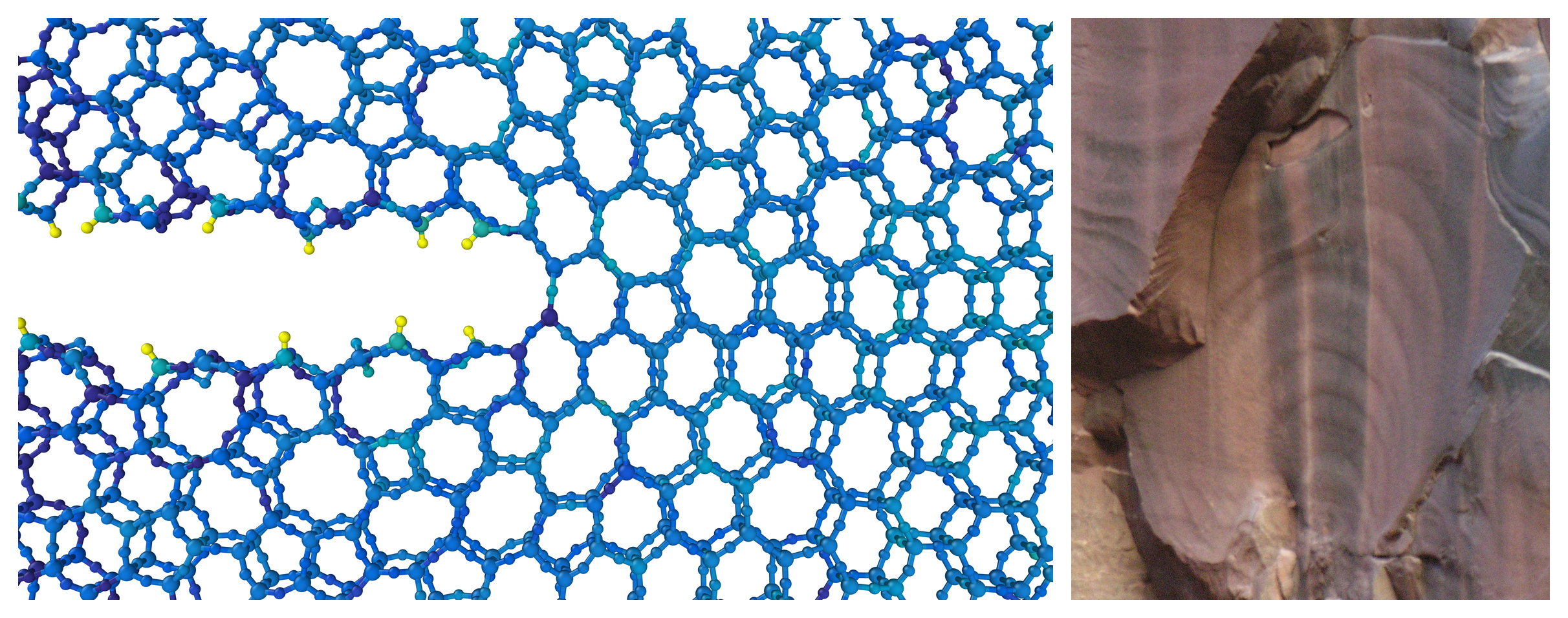}
\caption{\label{fig:cracking}
Fracture processes.
Left: SiO$_2$ fracture.
Crack propagation in a two-dimensional bilayer of amorphous silicon dioxide modelled with a
polarizable interatomic potential~\cite{caccin2015framework}. Atoms are coloured by their
local potential energy on a scale normalized from dark blue, lowest energy, to yellow,
highest energy.
 Right:  Plumose fracture patterns in sandstone, North Canyon, Arizona, USA.}
\par\smallskip{\footnotesize\noindent\raggedright\emph{Images:} Left: Marco Caccin and Alessandro De Vita, King's College London; James Kermode, University of Warwick;  CC-BY-NC-SA-2.0;
Right: Awickert, CC-BY-SA-3.0.\par}
\end{figure}

Geologically-relevant instances of Ostwald ripening include processes occurring in clays minerals and metamorphic rocks
\cite{eberl1990ostwald},
mineral paragenesis in sediments~\cite{morse1988ostwald},
olivine and plagioclase in magmas
\cite{cabane2005experimental},
coarsening of gold nanoparticles
\cite{hastie2021transport}
and 
advection-mediated chiral autocatalysis in which complete chiral
symmetry breaking is attained from an initially unbiased mixture of seed crystals
\cite{cartwright2007ostwald}.
Banded sphalerite,  Sec.~\ref{sec:sphalerite}, may be another geological example.

\subsection{Fracture processes}

When a material is subjected to stress, the internal forces may exceed its cohesive strength,
causing fracture~\cite{anderson2017fracture},  Fig.~\ref{fig:cracking}. The way a material
fractures is governed by the material's properties and the applied loading conditions,
leading to a variety of patterns.

Fracture typically begins with the nucleation of a crack~\cite{broberg1999cracks}, which then
propagates through the material. The path of this crack is controlled by the stress field
within the material. As a crack advances, it relieves stress in its immediate vicinity yet
can concentrate stress at its tip, which can cause the crack to branch or change direction,
depending on the material heterogeneity and the external loading conditions. For example, in
brittle materials like glass, cracks tend to propagate approximately straight under uniform
stress, whereas in more ductile materials, they may follow more tortuous paths, creating
complex patterns.

Several mechanisms contribute to  fracture-pattern formation in materials. One such mechanism
is the interaction between multiple cracks~\cite{meyer2000crack}. When multiple cracks
propagate simultaneously, their paths can intersect or mutually influence each other, leading
to branching~\cite{sun2021state} and the formation of intricate networks of cracks. Another
important factor is material anisotropy~\cite{hakim2005crack}, which can cause cracks to
preferentially propagate along certain directions, resulting in directional patterns.

Fracture patterns often exhibit
self-similarity~\cite{shcherbakov2003damage,tarasovs2014self}, meaning they appear similar at
different scales. This self-similarity is governed by scaling laws that describe how the
fracture process operates similarly across length-scales~\cite{bavzant1993scaling}. For example, both the branching patterns in a shattered car windscreen and the crack network in dried mud exhibit fractal characteristics, with smaller branches or cracks echoing the structure of the pattern as a whole. These scaling laws highlight the universal nature of fracture processes, showing how
similar patterns can emerge in different materials under varying conditions.

Chemical processes significantly influence the development of fracture patterns in geological materials, a factor that has often been underestimated. There exists an intricate relationship between mechanical and geochemical processes, and  progress in understanding geological fracture patterns requires  integration of these two domains 
\cite{laubach2019role}.
One illustrative setting where reactions reorganize fracture patterns is mineral replacement. Mineral replacement processes are common in various geological settings and involve the transformation of one mineral into another, often under conditions that change solid volume.
Such reaction-driven volume changes generate internal stresses within the mineral framework.
These stresses can fracture the parent mineral, thereby influencing the development of
replacement minerals~\cite{jamtveit2009reaction}. 

In the following, we discuss two geological manifestations of fracture pattern formation: sedimentary crack patterns, (Sec.~\ref{sec:crackpatterns}) and basalt columns (Sec.~\ref{sec:basalt}).

\section{Mesoscale self-organized pattern-forming systems in geology}
\label{sec:patterns}

\begin{figure}
\centering
\includegraphics{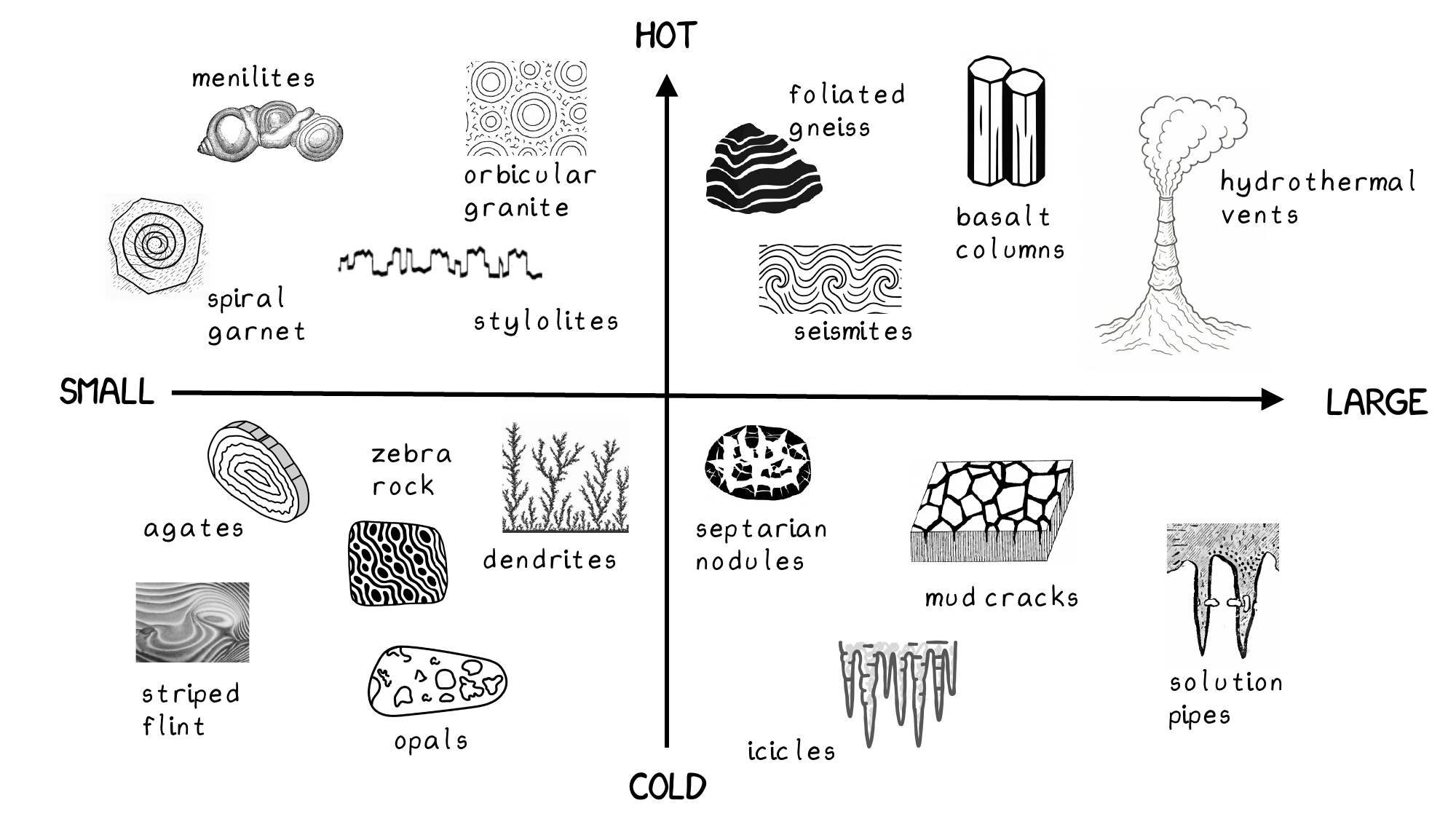}
\caption{\label{fig:patterns}
 Energy--size morphospace for self-organized geological patterns.
The horizontal axis records the characteristic spacing or wave-length of each structure and is broken at ${\sim}$10~cm to distinguish small- from large-scale patterns; the vertical axis records the dominant energy density that drives formation, rising from low-temperature, diffusion-controlled systems at the base to high-temperature or high-stress regimes at the top. 
}
\par\smallskip{\footnotesize\noindent\raggedright\emph{Image:} Text in this figure was rendered using the xkcd font from the xkcd-font project, licensed under CC-BY-NC-3.0.\par}
\end{figure}

Geological pattern formation refuses to confine itself to a single yardstick: rhythmic structures appear at wave-lengths that range from microns to metres, and they are powered by energy sources that vary just as widely, from the modest chemical free energy of supersaturated groundwater to the latent heat released when magma solidifies or the mechanical work done under tens of megapascals of burial stress. To keep that broad vista in focus we introduce a two-axis framework in which the abscissa records the characteristic length-scale of a pattern --- which might be its band spacing, crack pitch or convection-cell diameter --- while the ordinate records the governing energy-scale, defined as the dominant energy density required to create or sustain the structure. Plotting the examples on this length- and energy-scale map,  Fig.~\ref{fig:patterns}, makes it clear which physical ingredients a successful model must capture and how apparently disparate phenomena relate to one another.

In the lower-left quadrant, low-energy and small-scale coincide. Here we find  the sub-millimetre banding of agates (Sec.~\ref{sec:agates}); 
branching dendrites on bedding planes (Sec.~\ref{sec:dendrites});
and
the rhythmic micro-bands of 
striped flint (Sec.~\ref{sec:flint})
and
zebra rock (Sec.~\ref{sec:Zrock}). 
Although none of these structures requires elevated temperatures, each is controlled by reaction--diffusion or dissolution--precipitation feedbacks that act over millimetres to centimetres.

Sliding rightwards along the low-energy tier brings us to patterns that remain cool but express themselves on larger spatial scales. Icicles (Sec.~\ref{sec:icles}) that grow centimetre-wide flutings while releasing latent heat to sub-zero air;
vertical solution pipes that punch metre-wide shafts through limestone (Sec.~\ref{sec:pipes});
the familiar polygonal mud-cracks of drying lake beds (Sec.~\ref{sec:crackpatterns});
and 
concentric shrinkage cracks that fill septarian nodules (Sec.~\ref{sec:concretions}), 
all share this quadrant: the energy budget is modest, yet the geometric wave-length is measured in tens of centimetres or more.

Ascending into the higher-energy half of the figure we meet structures whose formation budgets are dominated by latent heat or by high mechanical work. At small length-scales this quadrant hosts the 
serrated teeth of stylolites (Sec.~\ref{sec:stylolites}), which record tens of megapascals of differential stress even though their amplitude is only a few millimetres; 
spiral
garnets (Sec.~\ref{sec:porphyroblasts}), the formation of which taps both metamorphic heat and considerable differential stress; 
the concentric shells of orbiculate granitoids (Sec.~\ref{sec:orbicule}); 
and 
the glassy menilite nodules thought to be silica bodies sculpted by earthquake-triggered fluidization (Sec.~\ref{sec:menilites}).

Finally, the high-energy, large-scale quadrant hosts patterns whose spacing climbs to decimetres or metres while their formation taps substantial heat or stress reservoirs. Among these are 
the columnar joints that fracture cooling basalt into metre-wide prisms (Sec.~\ref{sec:basalt});
hydrothermal-vent chimneys that build porous spires centimetres across yet tower metres above the seafloor (Sec.~\ref{sec:hydrothermalvents});
 the broad compositional layering of foliated gneiss (Sec.~\ref{sec:gneiss}), generated under the combined heat and stress of deep crustal metamorphism; as well as  seismites: soft-sediment convolutions shaken into metre-scale relief by strong earthquakes (Sec.~\ref{sec:seismites}).

Placing the examples on this length- and energy-scale map signals which physical ingredients a successful model must capture: diffusion lengths and reaction kinetics govern the lower band, whereas latent heat, fracture mechanics, and differential stress dominate the upper; meanwhile, the horizontal position reflects the geometric scale that any theory must reproduce.

In the pages that follow we tour the four quadrants of the energy--size morphospace. For each quadrant we highlight one or two patterns in depth to expose the underlying physics in measurable detail, then move more briskly through a suite of additional examples; sometimes because their mechanics are now well established, but just as often because they remain open puzzles awaiting decisive experiments or field evidence. 
We hope that this tour will let the reader see both the secure foundations and the outstanding questions that make pattern formation in geology such a lively research frontier.

\subsection{Small and cold}

Chemical supersaturation and diffusion feedbacks dominate at sub-centimetre wave-lengths. Agate and opal bands, mineral dendrites, striped flint and the rhythmic micro-laminae of zebra rock and zebra textures all form in cool diagenetic environments. The same low-energy regime fosters microporous gold-nugget overgrowths and  cavity-filling geodes
 controlled by millimetre-scale reaction fronts.

\subsubsection{Agates}
\label{sec:agates}

\begin{figure}
\centering
\includegraphics[width=\linewidth,height=0.78\textheight,keepaspectratio]{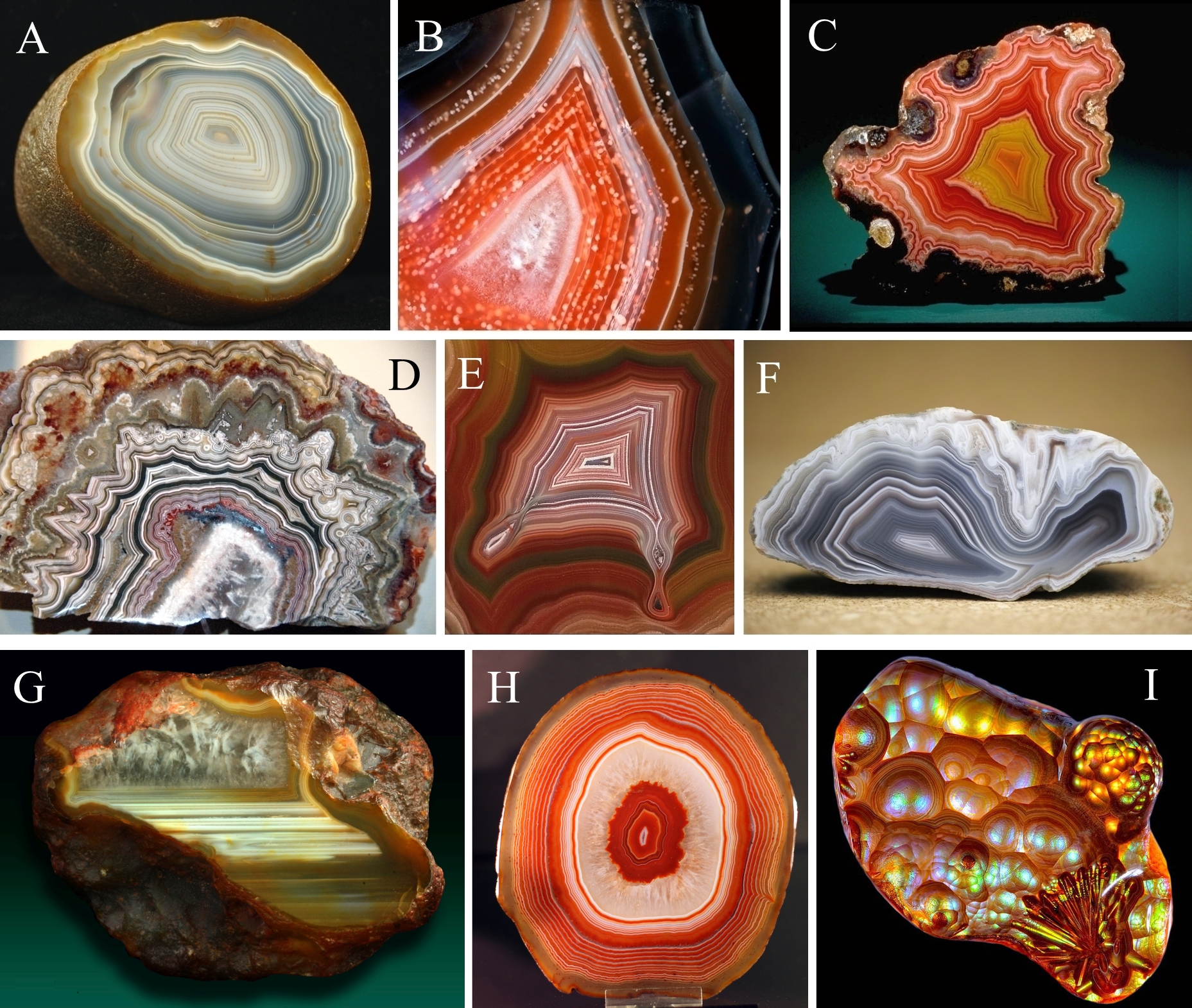}
\caption{\label{fig:Agates}Agates and their banded patterns.
A: Black River agate,
B: Lake Superior agate with a smoky quartz exterior and carnelian centre with mineral spherulites,
C: Laguna agate, Mexico,
D: Crazy Lace agate, Mexico, 
E: Laguna agate (Conejeros Claim), from Chihuahua, Mexico, 
F: Botswana agate from near Bobanong, Botswana
G: Agate with Uruguay-type banding and a quartz crystal zone (Agate Creek, Australia)
H: Agate from Idar-Oberstein, Germany 
I: Iridescent fire agate with botryoidal structure and a radiating spray of needle-like inclusions
\label{agates1}}
\par\smallskip{\footnotesize\noindent\raggedright\emph{Images:} 
    A: Courtesy of Matthew Wood,
    B: Courtesy of Karen Brzys, Gitche Gumee Museum/Agatelady Rock Shop, Grand Marais, MI, USA, www.agateladyrockshop.com,
    C: Chip Clark, Smithsonian Museum, public domain,
    D: James St. John, distributed under CC-BY 2.0 licence,
    E: specimen from Hannes  Holzmann collection, photographed by Albert Russ,
    F: Takehiro Toge, distributed under CC-BY-SA 3.0 licence,
    G: Lech Darski, distributed under CC-BY-SA 4.0 licence,
    H: Sailko, distributed under CC-BY 3.0 licence,
    I: Thomas Shearer, distributed under CC-BY 2.0 licence.
\par}
\end{figure}

\begin{figure}
\centering
\includegraphics{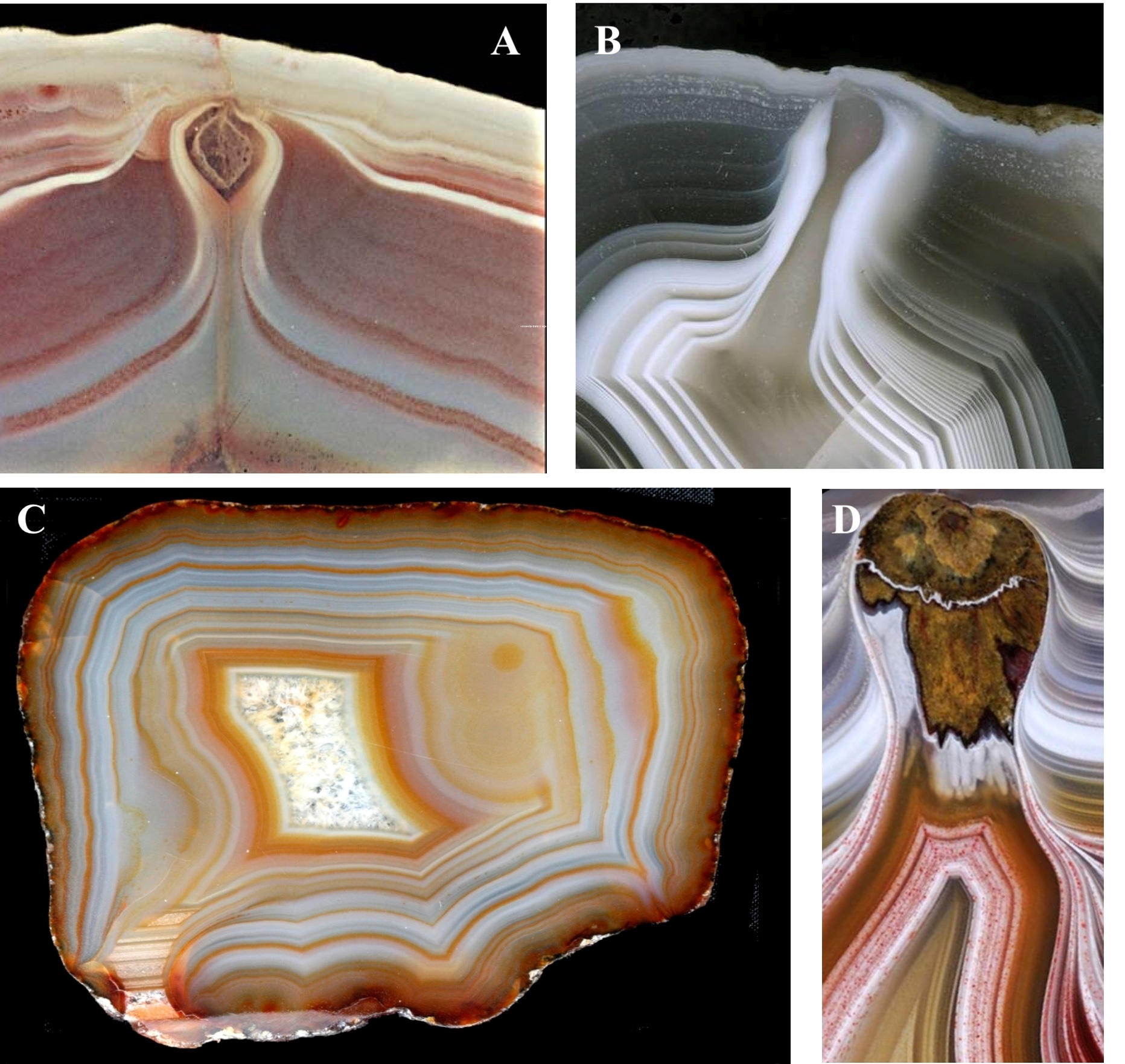}
\caption{\label{fig:Infiltration}Close-up views of infiltration channels in agates.
A: Infiltration channel in an agate from Neuwald near Flonheim, Germany; 
B: Infiltration channel in an agate from Scotland, UK; 
C: Infiltration channel in Brazilian agate (Rio Grande do Sul); 
D: Close-up of an infiltration channel in Laguna Agate (Mexico).
\label{infilt}}
\par\smallskip{\footnotesize\noindent\raggedright\emph{Sources and images:} A:~\citeA{Walger2009}, 
B: courtesy of Brian Jackson~\cite{Jackson2005}, 
C:~\citeA{Moxon2020}), 
D: 
Courtesy of Karen Brzys, Gitche Gumee Museum/Agatelady Rock Shop, Grand Marais, MI, USA, www.agateladyrockshop.com.\par}
\end{figure}

Agates,  Fig.~\ref{agates1}, are a banded type of chalcedony, a cryptocrystalline form of silica.
They are renowned for their stunning appearance, characterized by unique, translucent layers
of multiple colours. The term `agate' originates from the Greek word \textit{akh\'ates},
referring to a river in Sicily where agates were commonly found. Agates form in cavities
created by gaseous bubbles trapped in erupted lavas, which eventually become filled with
silica. While the characteristic layered structures lining the walls are perhaps the most
well-known features of agates, other interesting patterns also require explanation, such as
the horizontal, parallel layers known as Uruguay or gravitational banding (Fig.~\ref{agates1}G)
and infiltration channels (Fig.~\ref{infilt}): bulbous structures that develop near the boundary
of the agate-containing amygdale (a vesicle in an igneous rock containing secondary
minerals). The origin of colours in agate is linked to the presence of pigments on one hand
and variations in crystallite size and microstructure, including water content, on the other~\citeA{Goetze2020}. The most common colours in agate are white, grey, and blue.
White agate bands are associated with the presence of thick, plate-like crystallites, which
can totally reflect incident light, while the surrounding clear areas contain globular
crystallites that allow a portion of white light to be transmitted. The blue colour is caused
by the dispersion of light by the microparticles; the Tyndall effect. On the other hand, the
colour pigments are mostly iron and manganese oxides and hydroxides, responsible for red and
yellow colourations. This results in a wide range of colours and patterns, making each agate
distinct and visually appealing.

\begin{figure}
\centering
\includegraphics{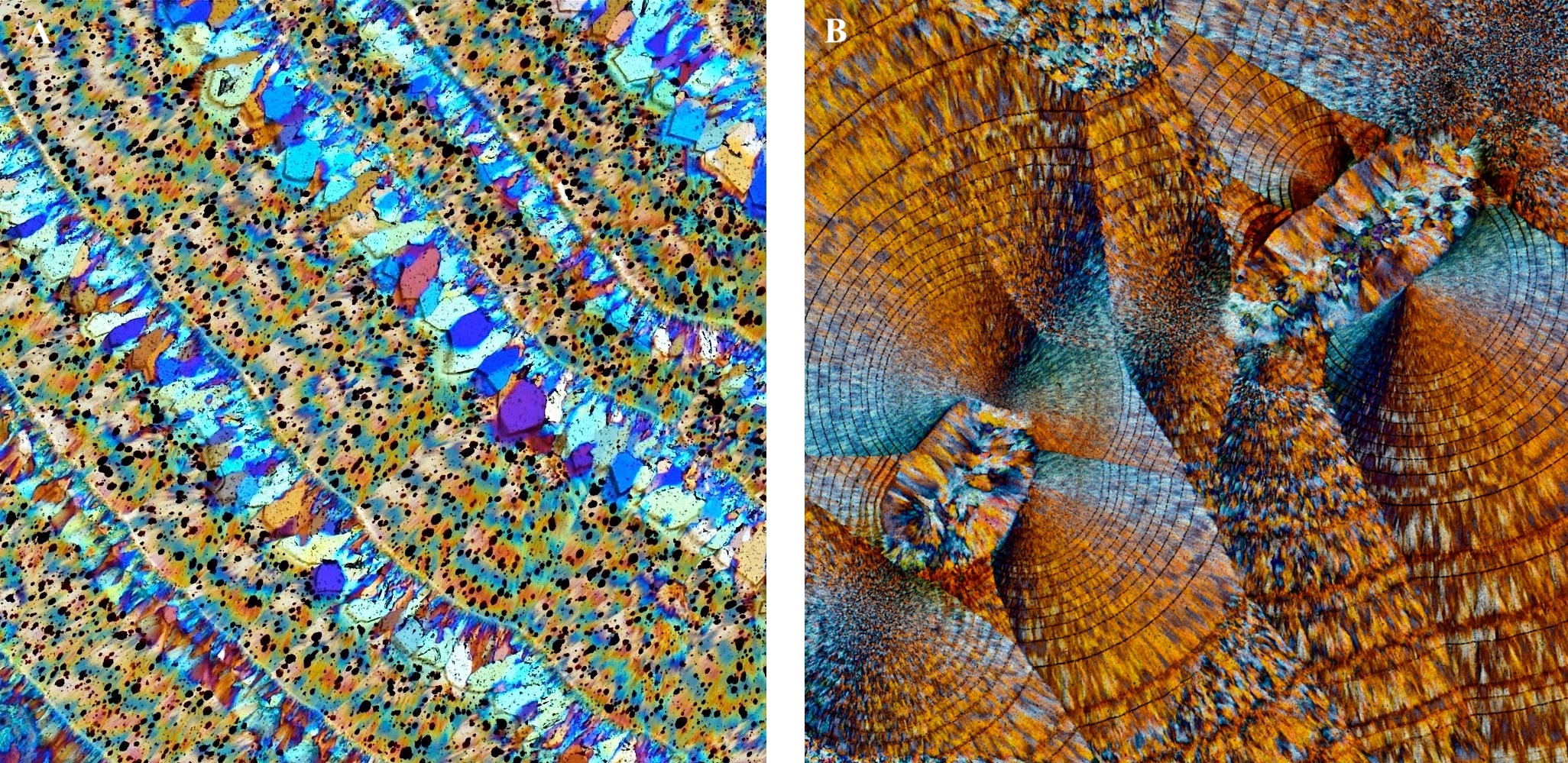}
\caption{\label{aga2}A: Agate micro-banding with fibrous chalcedony interspersed with macrocrystalline quartz. B: Spherulitic pattern in an agate with chalcedony fibres radiating from the nucleation points on the rim of an agate.}
\par\smallskip{\footnotesize\noindent\raggedright\emph{Images:} Courtesy of Bernardo Cesare, University of Padova, Italy\par}
\end{figure}

\begin{figure}
\centering
\includegraphics{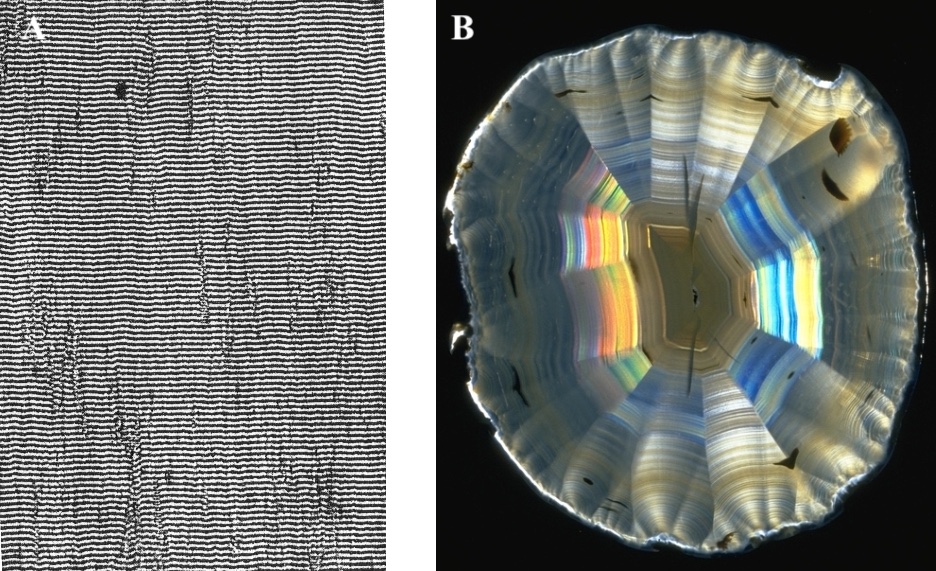}
\caption{\label{aga3}A: Micron-scale banding in iris agate. The distance between the
bands is ${\sim} 2~\rmmu\mathrm{m}$~\cite{Jones1952}. B: Iris agate produces a rainbow of colours in
transmitted light.}
\par\smallskip{\footnotesize\noindent\raggedright\emph{Sources and images:} A:~\citeA{Jones1952},
    B: Chip Clark, Smithsonian Institution, public
domain.
\par}
\end{figure}

However, there is more to an agate pattern than is visible to the naked eye; microscopic
studies reveal a much finer banded structure with alternating layers of defect-rich
chalcedony and macrocrystalline quartz~\cite{Heaney1995,Cady1998,Frondel1985,French2013}, as
illustrated in  Fig.~\ref{aga2}. The banding in  Fig.~\ref{aga2} has a wave-length in the range of
approximately $100$~$\rmmu$m, but similar oscillatory zonation in defect
concentration has been reported at the length-scale of a
0.5--5$~\rmmu$m~\cite{Jones1952,Frondel1978,Heaney1995}, clearly visible in cross-polarized
light (Fig.~\ref{aga3}A). The change of refractive index along these  bands produces a
rainbow-like display of colours in iris agates (Fig.~\ref{aga3}B). Zones with iris banding exist
in nearly every agate, although they are often concealed by pigmentation.

\begin{figure}
\centering
\includegraphics{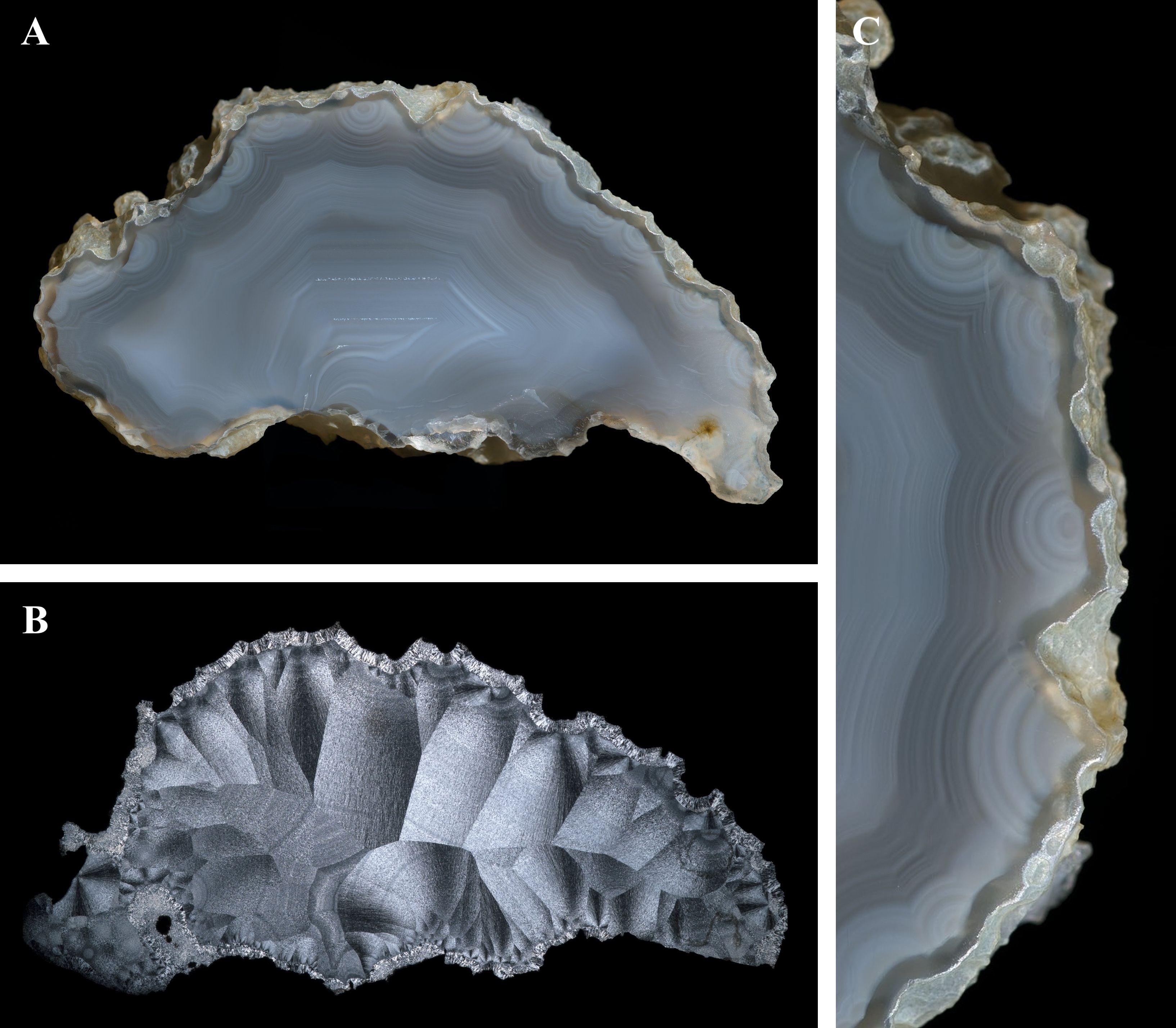}
\caption{\label{aga5}Agate from Ashland, Oregon in plane light (A) and cross-polarized light (B). The enlargement on the right, C, shows the growth of spherulitic nuclei on the boundary of the agate. }
\par\smallskip{\footnotesize\noindent\raggedright\emph{Images:} Courtesy of A. C. Akhavan (quartzpage.de).\par}
\end{figure}

 Microscopic images  suggest that microcrystalline chalcedony fibres nucleate on the cavity
walls and grow inward towards the centre of the cavity. Initially, there is spherulitic
growth, with the fibres radiating outward from separate nucleation points at the edges of the
system. Soon, a common banding is formed, running equidistant to the cavity
walls~\cite{Landmesser1984,Goetze2020}. The underlying spherulitic structure is particularly
striking when observed in cross-polarized light (Fig.~\ref{aga2}B and Fig.~\ref{aga5}C), which shows
the chains of quartz crystals of similar orientation growing from different nucleation
points.

The debate surrounding the genesis of agate is both challenging and contentious, primarily
due to the lack of a genuine laboratory synthesis that replicates natural agate formations.
As a result, numerous theories and approaches have emerged, many of which contradict each
other. Consequently, debates about the origin of agate remain as heated today as they were in
the 19th century, when most of the agate-genesis models were first formulated. An unsurpassed
review of these debates can be found in the paper by~\citeA{Landmesser1984}; more recent
reviews can be found in~\citeA{Goetze2020}, \citet{Moxon2020}.

\begin{figure}
\centering
\includegraphics{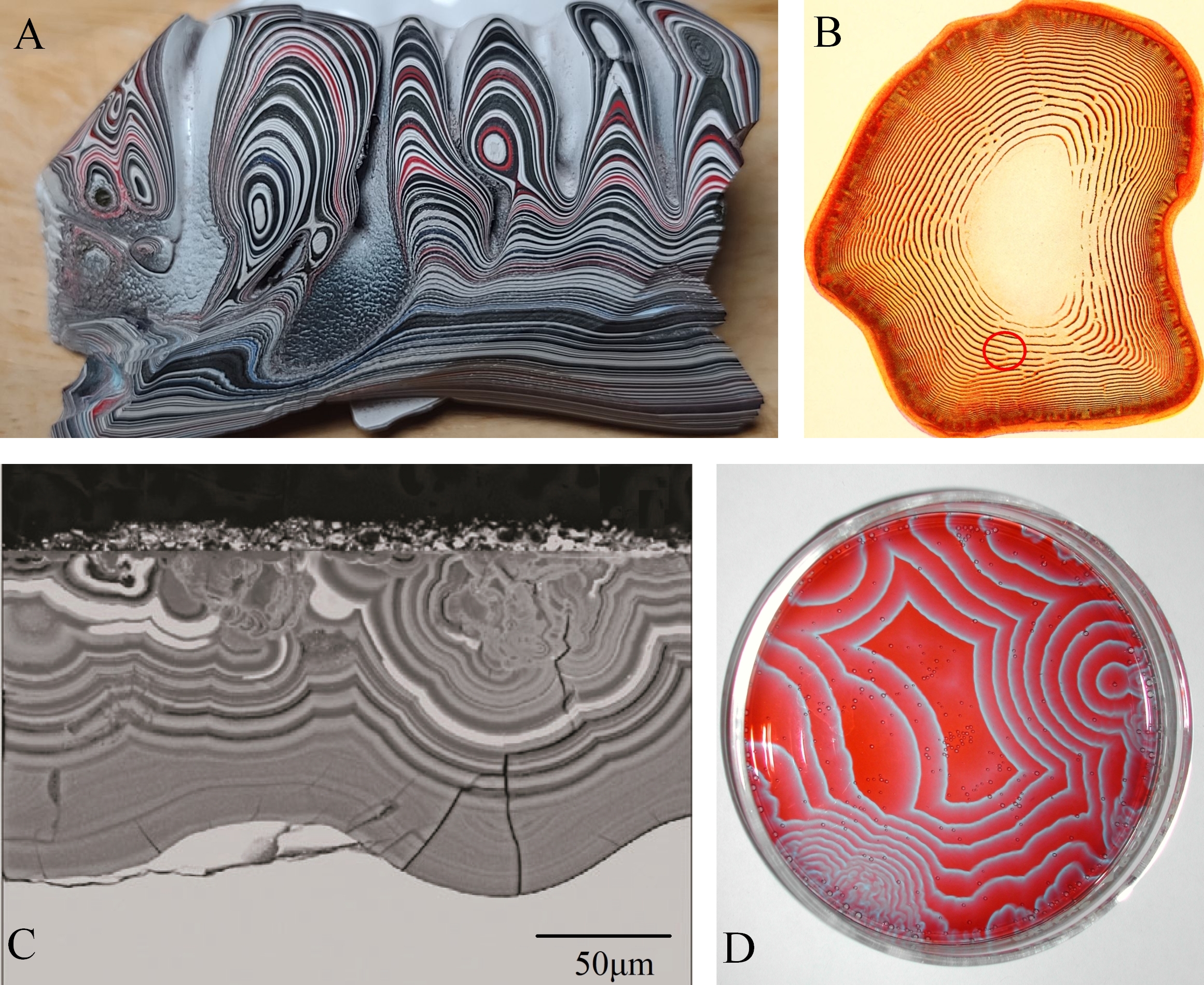}
\caption{\label{belo1}Four banded patterns resembling agates. 
A: Fordite, hardened enamel paint that has accumulated in hundreds of layers in vehicle factories during the lacquering process. 
B: Liesegang rings; rhythmic precipitation zones from supersaturated solutions in colloidal
systems. The circle marks a defect in the structure. 
C: A banded structure in a leached layer of an archaeological glass fragment from a
15th-century stained glass window. 
D: Patterns arising in the Belousov--Zhabotinsky reaction.}
\par\smallskip{\footnotesize\noindent\raggedright\emph{Sources and images:} A:  Jeremy Cook
(Willett Creek Agate Co.) distributed under CC BY-NC-SA 4.0 licence, 
B:~\citeA{Farrington1927}, 
C:~\citeA{Schalm2011}, 
D: Stephen Morris.\par}
\end{figure}

In general, theories about the genesis of agate patterns can be divided into two broad
classes based on what they consider to be the cause of the banding~\cite{Liesegang1910}.
External rhythm models hypothesize that agate is an accumulation structure, its pattern
reflecting alternating or periodic changes in environmental conditions. Agate, in this view,
could be seen as a geological counterpart to the pseudo-mineral fordite (Fig.~\ref{belo1}), an
accumulation of enamel paint deposited in hundreds of layers in automobile factories during
lacquering~\cite{hsu2016fordite,nova2021bestiary}. However, questions remain about how this
geological `lacquering process' is carried out, including the temperature and form in which
silica enters the cavity---here, numerous theories and ideas abound. Internal rhythm
theories, on the other hand, link the formation of the agate structure to self-organization
processes within the developing agate itself, treating it as a closed system. They usually
assume that the agate was filled first by  gelatinous silica, and then a reaction--diffusion
process of some kind took place that resulted in the formation of the banded pattern.

The first external rhythm theories were formulated in the 19th century.~\citeA{Haidinger1848}
hypothesized that agate patterns form where aqueous silica solutions seep into rock pores,
precipitating on the cavity walls. This led to a debate about how the silica solution could
enter the cavity after the layers had solidified.~\citeA{Noeggerath1849} proposed
infiltration channels as possible pathways for the solution, facilitating the continuous
development of precipitate layers. On the other hand,~\citeA{Kenngott1851} argued that the
silica deposits in a gel state and only gradually hardens due to contraction, suggesting that
the deposition, while not yet solidified, remains permeable to seeping solutions.

The above models involve ambient temperatures and pressures, but there are also several
theories involving elevated temperatures.~\citeA{Reusch1864} proposed that the cavities were
repeatedly filled and then emptied due to intermittent hydrothermal activity in the rock.
Solutions fill the cavities when driven upward by advancing steam; once the steam reaches the
cavity, it empties again. Each infiltration produces a single layer of silica, again in
close resemblance to the fordite lacquering process. More recently,~\citeA{Harris1989}
concluded, based on an oxygen isotope study of Jurassic agates,  that the agate formation
process involved periodic boiling of the hydrothermal solutions, with quartzine layers in
agates forming from water vapour and the chalcedony layers crystallizing from liquid water.
With each filling process, a fine layer of agate forms from the solution adhering to the
cavity edge. 

\begin{figure}
\centering
\includegraphics{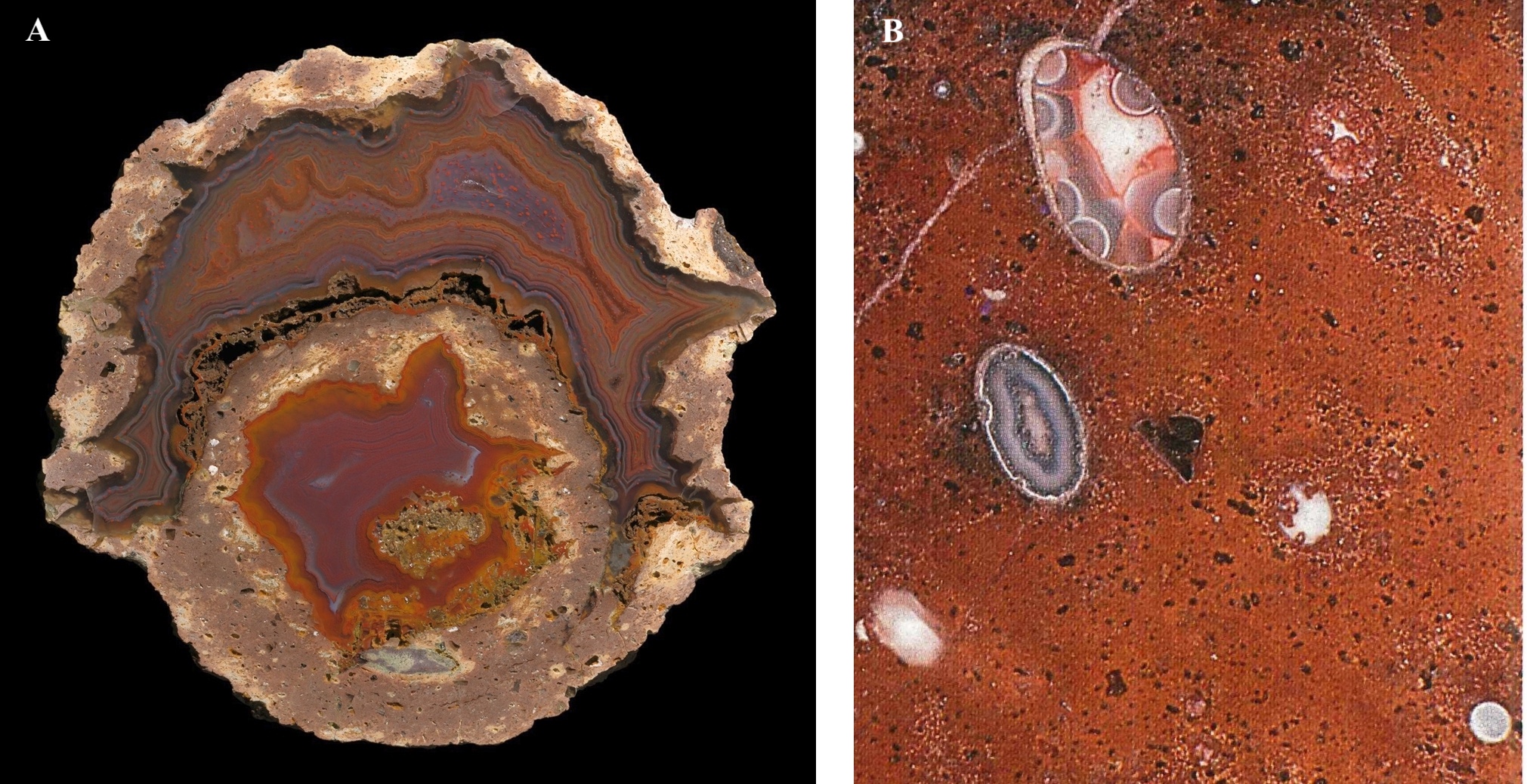}
\caption{\label{aga6}Two nearby agates do not exhibit the same banding structure.}
\par\smallskip{\footnotesize\noindent\raggedright\emph{Sources and images:} A:  Courtesy of A. C. Akhavan (quartzpage.de), B:~\citeA{Landmesser1984}.\par}
\end{figure}

In external rhythm theories, agate, as an accumulation structure, should reflect the changing hydrogeological conditions. Returning to the comparison with fordite, one can reconstruct the colours of cars lacquered at a particular workstation by analysing the successive layers. It is also expected that two fordites, created nearby, will have the same succession of stripes. However, this is not the case for agates; the banded structures of nearby agates are not correlated (Fig.~\ref{aga6}), an argument frequently put forward against the external rhythm concepts.

Among the internal rhythm theories, arguably the most famous is that  proposed
by~\citeA{Liesegang1914}, which links agate patterns to the banded structures
(Sec.~\ref{sec:liesegang}) that form when two components diffuse in a gel and react to form an
insoluble precipitate (Fig.~\ref{belo1}B). The similarity of patterns that can be created in this
way to the agate banding is at first glance striking. However, a closer inspection reveals a
number of important differences between these two systems, some of which were noted by
Liesegang himself~\cite{Liesegang1914}. Firstly,  Liesegang patterns lack the sharp corners
that are  characteristic of agate structures. Next, many structural defects are present in
Liesegang rings, such as one ring splitting into two or being displaced in a direction
perpendicular to the banding; these defects are absent in agates. Additionally, the distances
between successive Liesegang rings obey a well-defined functional relationship,
Jab{\l}czy\'nski's Law~\cite{Jablczynski1926}; Eq.~\eqref{eq:Jablczynski}, due to the
diffusive nature of the underlying process, which is not seen in
agates~\cite{Landmesser1984}. Moreover, for the rings to form, reacting species must diffuse
in the gel, and it is difficult to definitively identify them in the case of agates, and even
more challenging to associate their potential presence with the crystallite microstructure.
Finally, as with other internal rhythm models, these models encounter a volume problem: the
density of the gel is lower than that of the final chalcedony. As the gel dries, it would
then shrink, always leaving an empty space in the centre of the cavity. In real agates, this
is sometimes the case, but we equally often find fully filled specimens. Additionally, in all
actual experiments with Liesegang rings, the integrity of the banding is lost during the
desiccation process as the gel dehydrates and becomes cracked. Similar criticisms can be
levelled against the model of~\citeA{Pabian1994}  linking agate patterns with the
Belousov--Zhabotinsky reaction (Sec.~\ref{sec:RD})  (although the reader will notice that this
time the corners between the bands in  Fig.~\ref{belo1}D are sharp, resembling the agate
pattern). 

A more compelling internal rhythm theory was proposed by~\citeA{Wang1990}. The advantage of
this model is that this time the self-organization process is directly linked to
crystallization.  According to~\citeA{Wang1990}, silica gel has the potential to crystallize
in a patterned, repetitive manner. This is due to the acceleration of quartz crystallization
by Al\tsup{3{\tplus}} cations, which accumulate at the growth front. The model suggests that
fluctuations in the concentrations of trace elements could account for the alternating
formation of chalcedony and quartz bands, resulting in patterns similar to those depicted in
Fig.~\ref{aga2} and Fig.~\ref{aga3}.

The preceding models assumed that the precursor of an agate is a silica-gel-filled cavity. However, there are also models that assume agates to be transformed glass droplets.
The first model of this kind was proposed by~\citeA{Nacken1948}, inspired by his experimental
observation that silica glass crystallizes into chalcedony and quartz while preserving its
outer form when exposed to weak alkaline solutions at a temperature of approximately
400{\degree}C.  In this process, both the formation of spherulites at the edges of the
system and the spontaneous development of a banded structure are
observed~\cite{Nacken1948,White1961}. Based on this, Nacken proposed that already during the
magmatic stage, isolated molten silica droplets separated from the magma and rose upwards,
acquiring a shape resembling ascending gas bubbles.  With the influence of superheated water
vapour, agates then develop from these glass droplets. In fact, the transformation process
itself does not need to involve high temperatures, since a banding structure is known to
appear in  leached archaeological glass
\cite{Schalm2011};   Fig.~\ref{belo1}D. 

However attractive the high-temperature models with a glassy precursor may sound, as pointed
out by~\citeA{Landmesser1984}, such separation into pure glass does not occur in
multicomponent magma, even with small amounts of alkalis. Additionally, there have been cases
identified where agates formed in carbonate cements, thus being separated from the magmatic
rock. Furthermore, oxygen and hydrogen isotope analysis on Scottish agates
by~\citeA{Fallick1985} demonstrated that their formation temperature was in the range of
50\,\degree C. In general, there is a growing consensus in the literature that agates form
at temperatures below 100\,\degree C~\cite{Landmesser1984,Moxon2020}.

One of the most significant issues with the internal rhythm models is that of volume
preservation. Regardless of the assumptions made about the original filling of the cavity,
whether it be a solution, gel, or glass, its density is not sufficiently high to generate
enough silica to completely fill the entire cavity~\cite{Landmesser1984}. In the case of a
glass precursor, assuming the density of silica glass is 2.2~g/cm$^3$ and the density
of chalcedony is 2.6~g/cm$^3$, approximately 15\% of the mass is missing. On the other
extreme, if we suppose that silica precipitates from the solution under normal conditions
with a solubility of about 100~mg/L, then we would require about 1700~L of  solution to fill
a cavity with a 5~cm diameter.
 
It would thus seem that a successful agate formation theory should combine the
self-organization characteristic of internal rhythm theories with a continuous supply of
silica, typical for external rhythm theories. One such model was proposed
by~\citeA{Heaney1995}, who hypothesized that the observed alternating crystallization of
quartz and chalcedony in agate micro-banding is linked to varying degrees of silica
saturation in the solution. Cavity fluids that are partially polymerized rapidly precipitate
spherulitic chalcedony. If the depletion of silica near the fibre tips outpaces diffusion
towards these tips, then the reduced activity of silica in the solution facilitates the
progressively slower growth of larger crystals with fewer defects.~\citeA{Heaney1995} assume
that the crystallization proceeds directly from solution; other researchers suggest that the
sol--gel transition takes place first~\cite{Landmesser1984,Walger2009,Goetze2020}.

As for the mechanism that sustains a continual flow of silica into the cavity, one obvious
factor is the difference in solubility between the small pores and the large cavity space.
Owing to interfacial energy effects, the saturation concentration in small pores is
considerably higher~\cite{Emmanuel2007}. This not only can explain why silica precipitates in
the cavities while keeping the small pores free, but also provides a chemical potential
difference driving silica diffusion towards the cavity. However, this diffusion becomes
hindered as the first chalcedony layers are formed. An interesting mechanism, which can
significantly speed up the process, is the `reverse osmotic pump' concept proposed
by~\citeA{Walger2009}. Walger identifies the infiltration channels (Fig.~\ref{fig:Infiltration})
as potential entrance points for the silica solution, noting that they often correlate with
fractures in the first chalcedony layer. The solution would then enter the cavity and
precipitate on the walls, either as a gel or as a direct crystal layer. In any case, the
pores in the wall-lining layers are too small for the partially polymerized silica to pass
through, while still allowing water transport. This water then exits the system through the
walls. In this way, solvent circulation in the system is sustained, with the solution
entering through the infiltration channels and the solvent leaving through the walls. This is
similar to the flow pattern in osmotic pumps~\cite{Theeuwes1975}, but in a reversed
direction. In osmotic pumps, the solvent is drawn through the porous walls and leaves the
system, along with drugs, through a small orifice. As Walger points out, the 
presence of the flow in the infiltration channels would hinder the precipitation there, leading to
thinning out of the wall-lining layers and the formation of characteristic onion-like
shapes of the channels as observed in  Fig.~\ref{fig:Infiltration}. It is important to note that
there is no agreement in the literature about the role of these channels: some claim that
they are the solution entry points~\cite{Noeggerath1849,Kenngott1851,Reusch1864,Walger2009},
pressure release valves~\cite{Goodchild1899,Schlossmacher1960}  or gel deformation
structures~\cite{Landmesser1984,Goetze2020}. 

This necessarily brief overview of concepts regarding the genesis of agate shows the astonishing diversity of these concepts. Many of them explain certain aspects of agates, but usually not all, and unfortunately, none have resulted in the replication of agate-like pattern generation in the laboratory. 
However, it is evident that despite its appearance as a hard rock, soft-matter phenomena are underlying the agate formation process. Initially, silica enters the amygdales either as a suspension or a gel. Following this, its deposition on the walls involves crystallization competing with diffusion, potentially influenced by fluid flow. Some theories argue that the deformation and stretching of the gel shape the infiltration channels, while others point to osmosis as central to agate formation. Although the formation of agates still contains many unknowns, it is clear that the answer must ultimately be grounded in the physics of soft matter.

\begin{figure}
\centering
\includegraphics{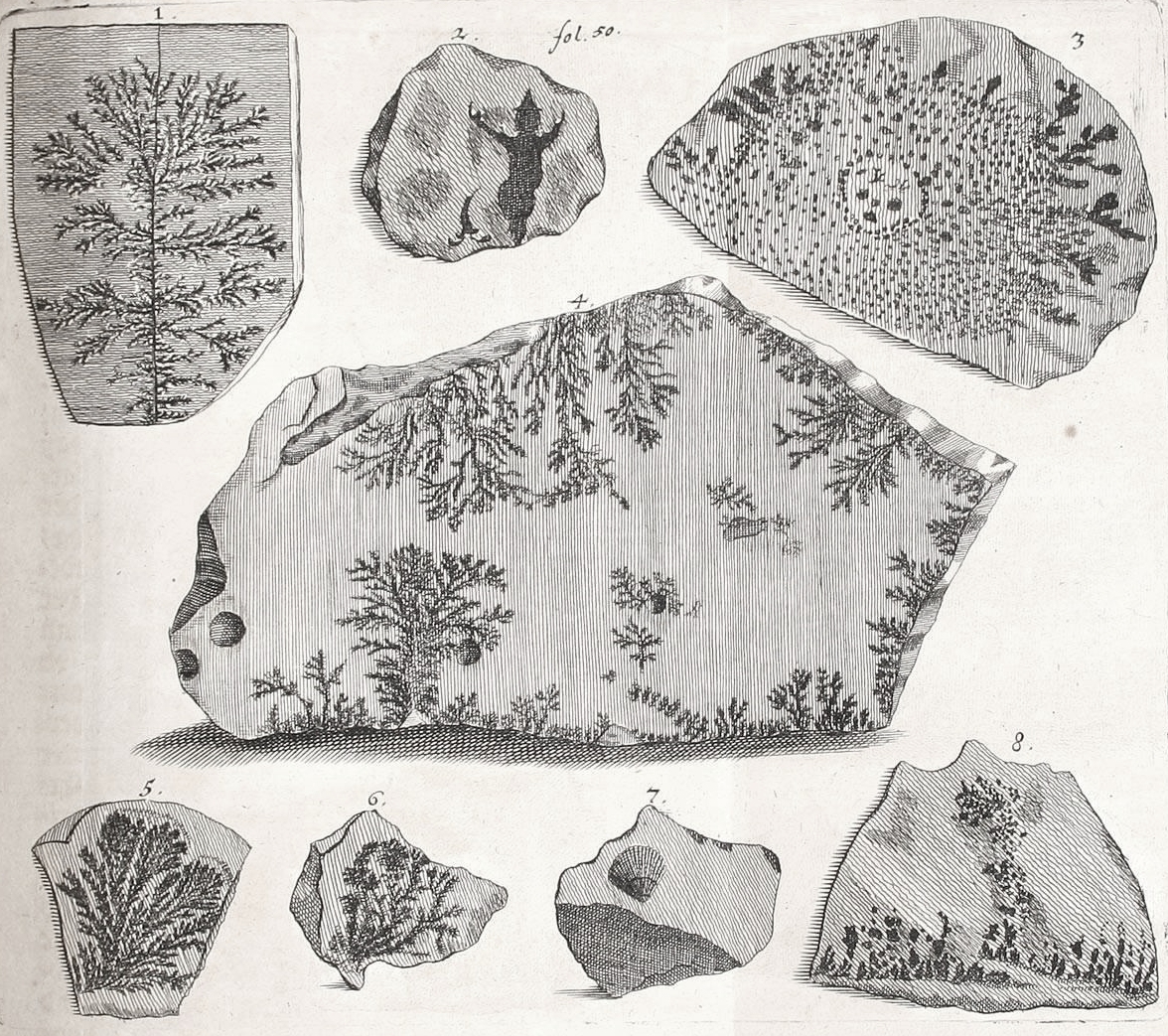}
\caption{Dendrite drawings in \emph{Memorabilium Saxoni\ae \ Subterrane\ae} by Gotfried Friedrich Mylius (1709)\label{fig:DendriteOld}}
\par\smallskip{\footnotesize\noindent\raggedright\emph{Source:} \citeA{Mylius1709}.\par}
\end{figure}

\subsubsection{Mineral dendrites}
\label{sec:dendrites}

Two-dimensional mineral dendrites --- the name is from the Greek \emph{dendrit\=es}
`treelike' --- in the form of black or red-brown deposits on the surfaces of limestones,
sandstones, opals or agates, have intrigued people for centuries. We can find  mention of
them as early as Pliny the Elder's \emph{Natural History} of 77 CE~\cite{Pliny77}, but more
detailed descriptions and classifications appeared with the rise of naturalism and mineral
collecting in the early 18th century~\cite{Scheuchzer1700,Mylius1709};
Fig.~\ref{fig:DendriteOld} shows a reproduction of a plate illustrating dendrites in Mylius'
\emph{Memorabilium Saxoni\ae \ Subterrane\ae} of 1709. The organic-like form of  dendrites
led some naturalists to the conviction that they are fossilized plants. However, by mid-18th
century, they were  pinpointed as patterns emerging as a result of physical processes. Some
early authors  even attempted to recreate this process; for example, Emanuel Mendes da Costa
in his 1757 \emph{Natural History of Fossils}~\cite{daCosta1757}  notes that similar patterns
can be created by inserting oil in between two sandstone plates and then pulling apart the
plates, making  this perhaps the first known experiment in viscous fingering
(Sec.~\ref{sec:fingering}). For a modern, detailed review of the various geochemical types and
shapes of mineral dendrites the reader is referred to the excellent paper
by~\citeA{Straaten1978}.

\begin{figure}
\centering
\includegraphics{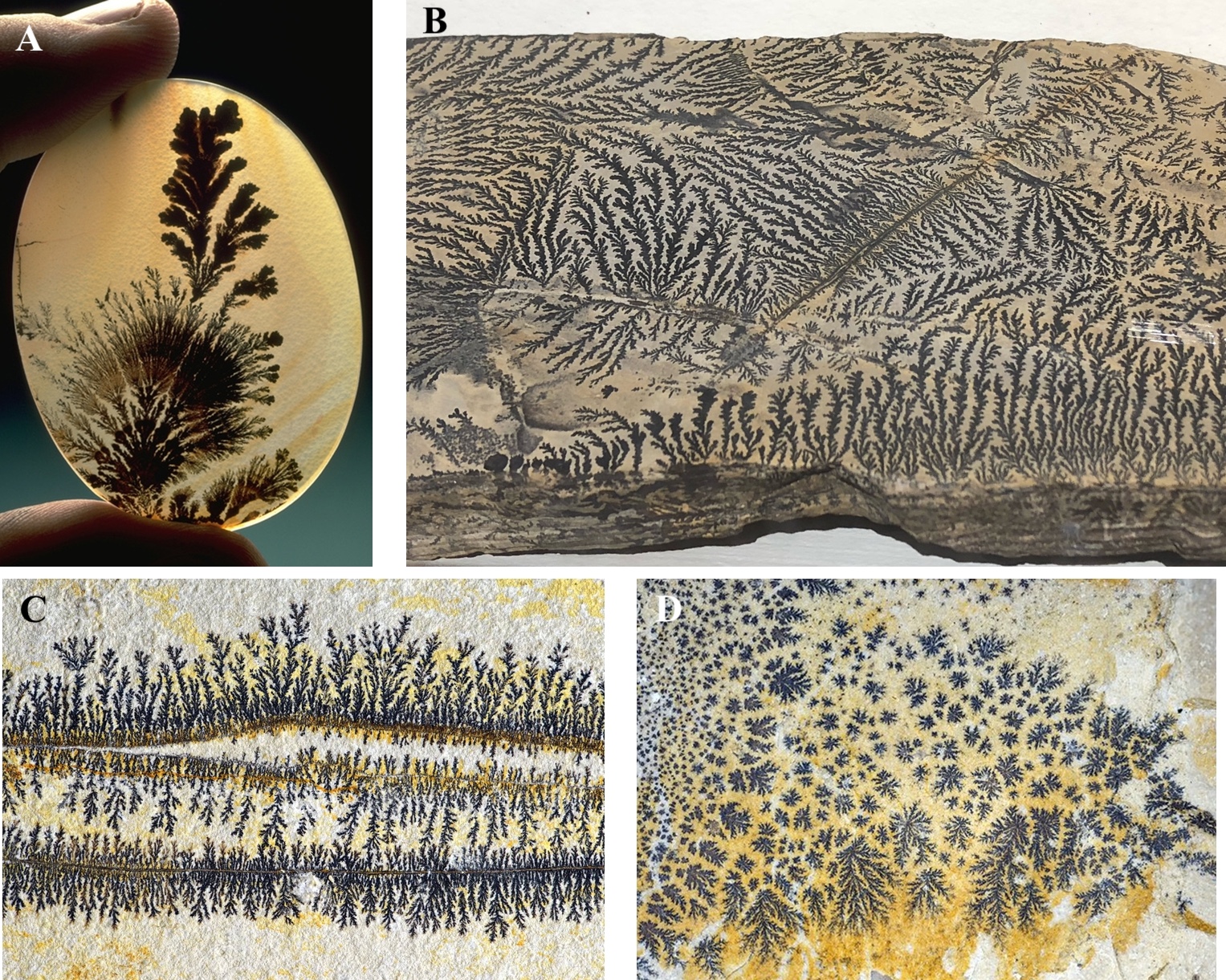}
\caption{\label{fig:Dendrite2d}Quasi-2D dendrites in agate (A), and limestone (B--D). In B and C they originate from fractures, whereas in D there are multiple nucleation points.}
\par\smallskip{\footnotesize\noindent\raggedright\emph{Images:} A: Chip Clark, Smithsonian Institution, public domain; 
 B: Courtesy of Jessica Rosenkrantz; 
 C: Adobe stock photography; 
 D: Courtesy of J\"urg Alean.\par}
\end{figure}

\begin{figure}
\centering
\includegraphics{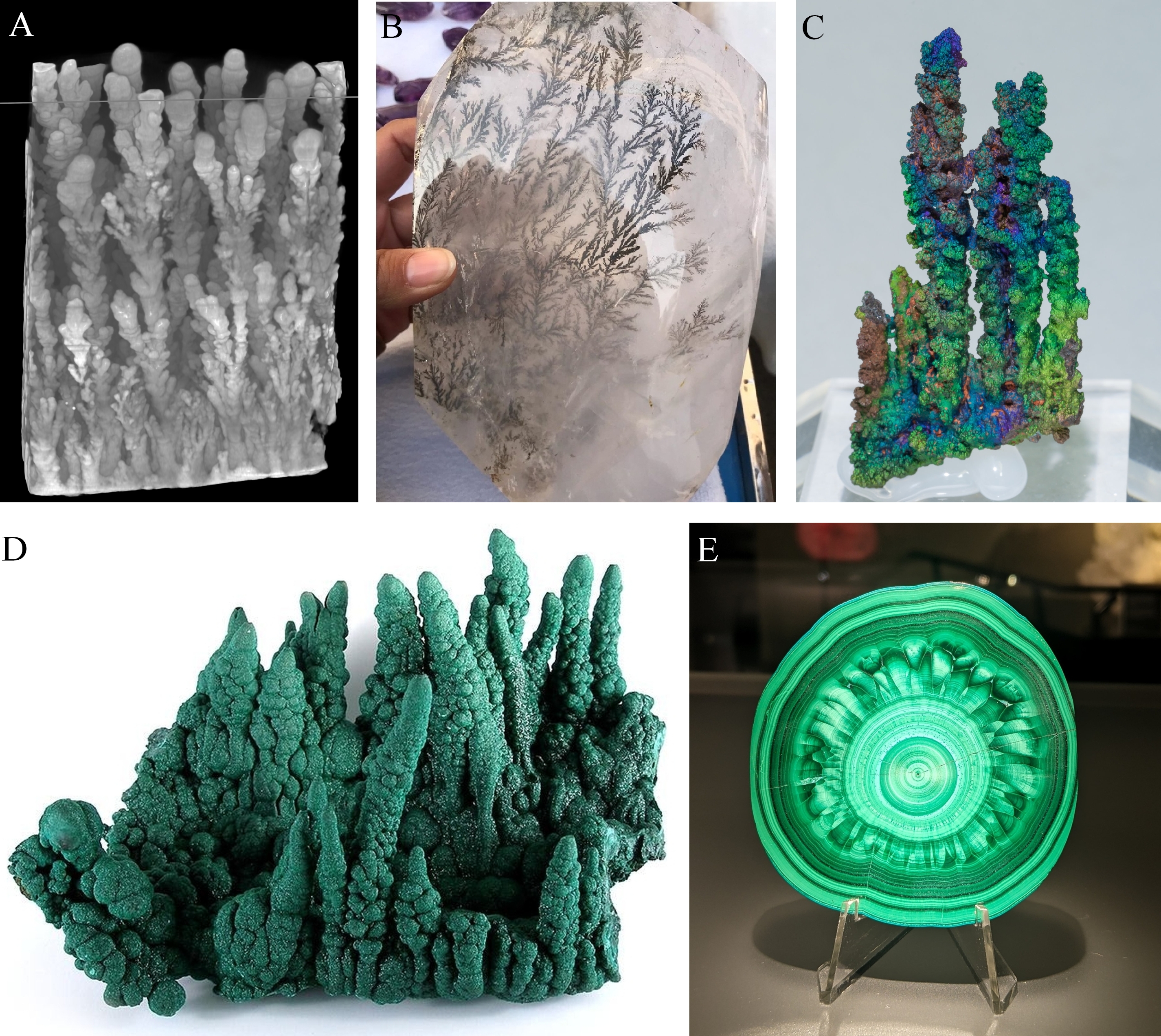}
\caption{\label{fig:Dendrite3d}Three-dimensional dendritic structures and morphologically related free-standing columnar forms. A: three-dimensional manganese oxide dendrites penetrating a zeolite matrix \cite{Hou2023}. B: manganese oxide dendrites in quartz. C: stalagmite-like columnar goethite. D: stalagmite-like columnar malachite from Kasompi Mine, Katanga Province, Democratic Republic of the Congo (size: 21.6?16.0?11.9 cm). E: transverse section of a malachite column showing concentric banding.
}
\par\smallskip{\footnotesize\noindent\raggedright\emph{Images:} 
A:~\citeA{Hou2023},
B: courtesy of Yuping Crystal Wholesale,
C: courtesy of  MVM (Minerals Virtual Museum),
D: Robert M. Lavinsky, distributed under CC BY-SA 3.0 licence, 
E: courtesy of  the  Alfie Norville Gem \& Mineral Museum of the University of Arizona.\par}
\end{figure}

Dendrites can be composed of different minerals, including manganese and iron oxides, copper, silver, and other metallic minerals. The most common are planar manganese and iron oxide dendrites formed along the bedding planes in limestones and sandstones, such as the ones shown in  Fig.~\ref{fig:Dendrite2d}B--D.  The bases of the dendrites are usually associated with the intersections of the bedding planes with joints or other fractures, from which they spread outward along the bedding plane (Fig.~\ref{fig:Dendrite2d}B--C).  However, other forms also occur, which consist of a multitude of small, radial clusters, not connected to the base (Fig.~\ref{fig:Dendrite2d}D).

Mineral dendrites are low-crystallinity solids, with the absence of even short-range three-dimensional periodic structure 
\cite{Garcia1994}, which suggests that they were formed by the precipitation of colloidal
particles. This distinguishes them from the crystalline dendrites growing from the supercooled melt, often encountered in
metallurgy~\cite{Trivedi1994}. 
The latter are usually needle-like, with a significant control of the crystalline
anisotropy and with thermal effects playing a key role during their formation. In the
petrological context, such dendritic crystals often form as a result of  magma
crystallization in rapidly cooled rocks~\cite{Fowler1989,Welsch2013,Barbey2019}.  Mineral
dendrites, on the other hand, have a branched,  fractal structure strikingly similar to
diffusion-limited aggregation, DLA~\cite{Witten1981};  Sec.~\ref{sec:DLA}. This observation
led~\citeA{Chopard1991} to propose a model in which two species, A and B, diffuse, forming
species C that then aggregates, although the geochemical details of this process and
the nature of the particles remain unspecified. This model has  been extended
by~\citeA{Hou2023}. In their model,  fluid pushes an oxygenated matrix pore fluid away from
the joints and mixes with it, generating an oversaturated manganese oxide solution.  
Laboratory studies on Mn oxide formation~\cite{Li2014,Huang2015} have confirmed that such
conditions promote the
initial growth of Mn-oxide nanoparticles. Such particles, once formed, diffuse through the
rock matrix, and become attached to the Mn-oxide-coated joint, initiating dendrite growth.
Dendrite formation would then proceed through  a non-classical crystallization pathway, via particle
attachment, which is increasingly recognized as an important and widespread type of crystal
growth~\cite{Coelfen2008,deYoreo2015}.~\citeA{Hou2023} emphasize that the shape of the
dendrite will also be a function of its interfacial energy as well as the concentration of
the manganese oxide particles around it. 

\citeA{Garcia1994} alternatively proposed that dendrites are a viscous fingering pattern
(Sec.~\ref{sec:fingering}) created when manganese-rich fluid infiltrated a fracture space filled
with presumably much more viscous, oxygenated fluid, possibly a colloidal suspension.  In
essence, this mechanism aligns with da Costa's 1757 conjecture~\cite{daCosta1757}.
However --- and frustratingly --- as they are in the same universality
class~\cite{Mathiesen2006}, both models, DLA and viscous fingering, give rise to patterns of
the same fractal dimension, around 1.70; thus geometric attributes of these patterns are
insufficient for discerning the correct physical growth process. Other arguments invoked in
this discussion are the lack of manganese outside the dendrites, which is easier to explain
in terms of a viscous fingering model, unless one assumes that the DLA mechanism is highly
effective in clearing the entire bedding plane of MnO particles formed in the space outside
the dendrites~\cite{Garcia1994}. On the other hand, viscous fingering cannot easily explain
the formation of isolated dendrites, not connected to the main joint, as in
Fig.~\ref{fig:Dendrite2d}D, or banding structures, like the ones in  Fig.~\ref{fig:Dendrite2d}A.
Additionally, one can raise a mass conservation issue. Manganese concentrations in  dendrites
are much higher than those in Mn-bearing groundwater solutions, thus a large volume of
solution, significantly larger than the volume of the visible pattern, is needed to create a
dendrite. 

In contrast to the quasi-2D planar dendrites growing along  bedding planes,  internal
dendrites are fully three-dimensional structures (Fig.~\ref{fig:Dendrite3d}), which start from
fissures and penetrate into the porous rock~\cite{Straaten1978,Potter1979,Hou2023}. Some of
these dendrites are composed of manganese and iron oxides, just like their 2D counterparts,
but there is also an interesting class of precious metal dendrites --- of gold, silver and
their alloy electrum --- which occur, in particular, in ultra-high-grade gold ``bonanza''
veins~\cite{Schoenly1993,Saunders2017,Saunders2022,Monecke2023} and are thus economically
important. Two models of their formation have been proposed. The first is the particle
attachment model, in which dendrites form by accretion of colloidal gold or silver
particles~\cite{Saunders1994,Saunders2020}. This is similar in spirit to the particle
attachment growth model of the manganese oxide dendrites described above, with the only
difference being that gold/silver particles are not produced locally, but are formed deep in
the hydrothermal system and then carried up by the fluid to the epithermal setting.  An
alternative mechanism for the growth of gold dendrites, following a classical crystallization
pathway based on ion addition, was proposed by~\citeA{Monecke2023} (see also
Sec.~\ref{sec:nuggets}).

As is often the case in earth sciences, a similar appearance of the structure does not
necessarily imply a similar genesis. The three-dimensional dendrites growing in a zeolite
matrix, as shown in  Fig.~\ref{fig:Dendrite3d}A, are similar in appearance to some of the
free-standing stalagmite-like forms of minerals, such as goethite (Fig.~\ref{fig:Dendrite3d}C) or
malachite (Fig.~\ref{fig:Dendrite3d}D), which form in rock cavities from slowly dripping
solutions. This might suggest a growth mechanism similar to that of stalactites, as discussed
in  Sec.~\ref{sec:icles}, although in the latter, CO$_2$ degassing is important, which
does not have a clear parallel in malachite, goethite, and other columnar forms. Instead,
they are most probably formed by an evaporative growth mechanism
(Sec.~\ref{sec:capillary})~\cite{Keller1990}, although other mechanisms, involving
reaction--diffusion self-organization (Sec.~\ref{sec:RD}), have also been
invoked~\cite{Papineau2020}. It is worth noting that, in contrast to agates
(Sec.~\ref{sec:agates}), malachite can be formed in the laboratory by evaporative growth, with
patterns virtually indistinguishable from  natural ones~\cite{Balitsky1987,Petrov2013}.

\begin{figure}
\centering
\includegraphics{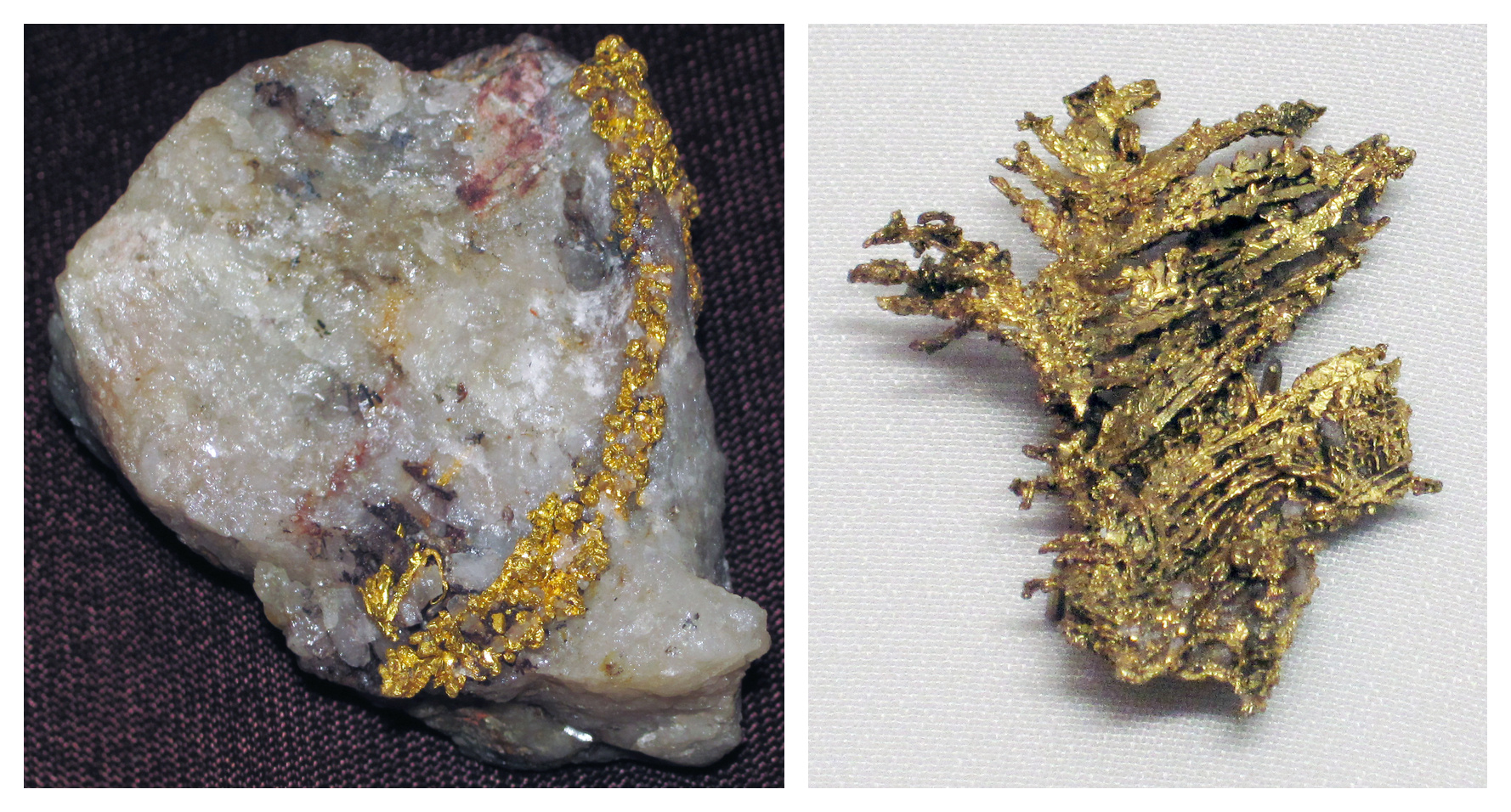}
\caption{\label{fig:nugget}Nuggets.
Left:
Gold-quartz hydrothermal vein from Colorado, USA; 
Right: 
dendritic gold nugget, California, USA.
}
\par\smallskip{\footnotesize\noindent\raggedright\emph{Images:} James St. John; CC-BY-2.0.\par}
\end{figure}

\subsubsection{Nuggets}
\label{sec:nuggets}

A nugget is a lump of precious metal, most commonly gold, found in geological settings; nuggets may also be composed of other metals such as copper, silver, and platinum.
Gold is a rare element, so how does it become concentrated into gold nuggets that occur
within quartz veins  (Fig.~\ref{fig:nugget}, left)~\cite{craw2016gold,butt2020gold}? During
earthquakes, seismic pumping, driven by pressure variations in fault zones, drives the
circulation of hydrothermal fluids, in which gold can be present under high-temperature,
high-pressure conditions. As faults experience repeated cycles of opening and closing during seismic and interseismic periods, cracks open and are gradually sealed via a fault-valve mechanism. During these processes, gold from the circulating fluids can precipitate and accumulate within faults. This mechanism leads to the episodic growth of gold veins or nuggets over time.
Quartz-rich rocks within fault zones can generate electric charges due to piezoelectric
effects, particularly during seismic deformation. These charges may influence the redox
conditions in the fluid, potentially facilitating the reduction of dissolved gold to its
elemental state, leading to the precipitation of gold particles~\cite{voisey2024gold}.

Gold nanoparticles can aggregate to form larger structures in colloidal gold suspensions. This process can be enhanced by the presence of carbon-rich fluids, which can stabilize the colloidal particles and promote the formation of high-grade gold deposits.
Diffusion-limited aggregation (Sec.~\ref{sec:DLA}) is one mechanism by which gold particles grow in size, resulting in the dendritic, branched structures observed in natural gold nuggets (Fig.~\ref{fig:nugget}, right).

\subsubsection{Opals}
\label{sec:opals}

\begin{figure}
\centering
\includegraphics{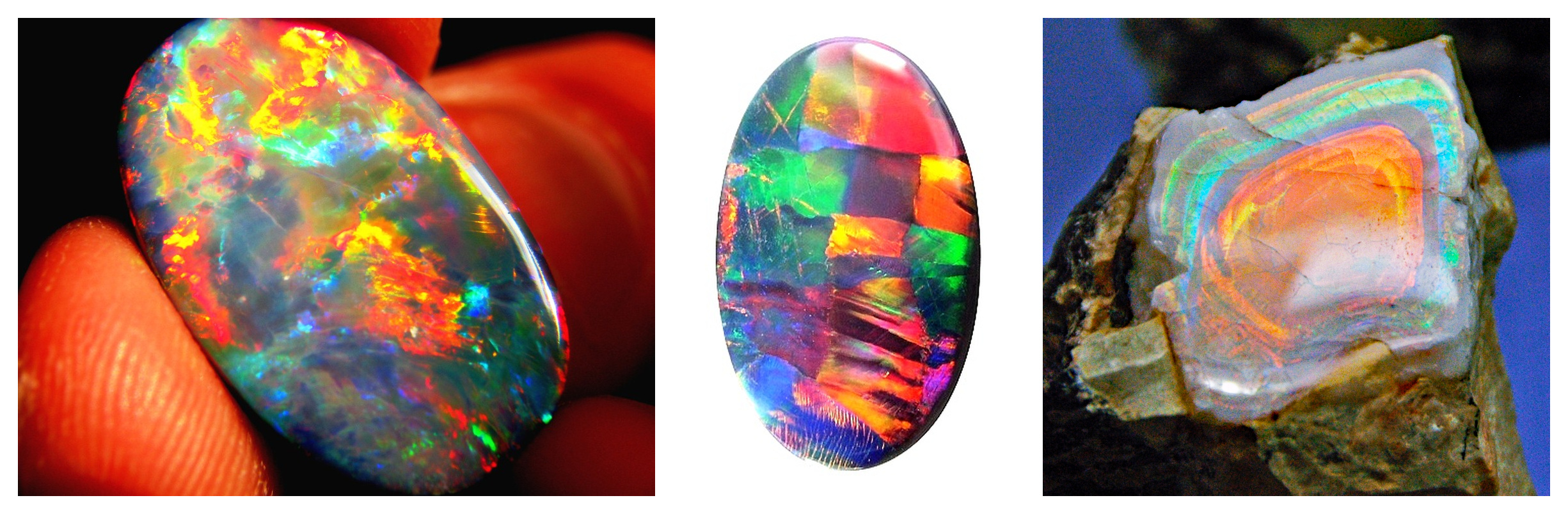}
\caption{\label{fig:opal}Opal.
Left: Cut and polished black opal from Lightning Ridge, Australia.
Centre: Harlequin opal.
Right: Zoned  opal from the Pliocene of Idaho, USA.
}
\par\smallskip{\footnotesize\noindent\raggedright\emph{Images:} Left:  Daniel Mekis; CC-BY-SA-3.0;
Centre:  Aisha Brown; CC-BY-SA-2.0;
Right:  James St. John; CC-BY-2.0.\par}
\end{figure}

Opal,  Fig.~\ref{fig:opal}, although considered a mineral for historical reasons by the
International Mineralogical Association (IMA), is in fact a mineraloid due to its amorphous (or
amorphous with paracrystalline components) nature. Even though opal is often referred to in
general terms, there are four different types of
opal~\cite{curtis_review_2019,gaillou_common_2008}: opal-A (amorphous), which is subdivided
into opal-AN (amorphous network) and opal-AG (amorphous gel); opal-CT
(cristobalite--tridymite); and opal-C (cristobalite). 

Opal-C is a rare phase in nature~\cite{curtis_review_2019}, and thus it has been little
studied. 
Opal-AN  does not exhibit micro- or nano-structural
order~\cite{gaillou_common_2008,curtis_silicon-oxygen_2021}. Opal-CT, which, in addition to
spherulitic morphologies, presents different forms of aggregation,  can form as disoriented
nanograins, fibrous nanograins, platy nanograins, and lepispheres (i.e.,~spheres composed of
silica plates, with a morphology similar to desert
roses)~\cite{gaillou_common_2008,curtis_silicon-oxygen_2021}.
In the following, we  focus on opal-AG, because it is characteristically formed by the
self-organized aggregation of silica
nanospheres~\cite{gaillou_common_2008,curtis_silicon-oxygen_2021}. Opal-A is often used as a
synonym for opal-AG~\cite{gaillou_common_2008}.

The formation of opals from colloidal silica involves the self-assembly of tiny silica
spheres, formed by polymerization or condensation of silica monomers, into a highly ordered,
three-dimensional structure; a form of colloidal (liquid) crystal. The process begins with
the formation of colloidal silica particles in a supersaturated silica solution. These
particles are typically in the nanometre size range and remain suspended in  solution owing
to their small size and the repulsive forces between them. Over time, these silica spheres
begin to organize themselves into a regular, close-packed lattice structure driven by the
minimization of free energy in the system. The formation of a densely packed colloidal
crystal has not yet been fully elucidated; two main mechanisms have been proposed: the older
and classical model is sedimentation due to
gravity~\cite{ILER1965Formation,Norris2004Opaline,gaillou_common_2008}, while a more recent
view holds that changes in solution conditions drive aggregation via electrostatic
forces~\cite{Stewart2010Self}. The spaces between the packed silica spheres are gradually
filled with silica gel which helps stabilize the structure and cements the spheres together,
leading to the formation of a rigid solid material~\cite{gaillou_common_2008}.

Opal formation can take place over long geological time-scales or be accelerated under
laboratory conditions. The specific conditions, such as temperature, pH, and solution
composition, control the final size of the silica spheres and the quality of the opal
produced~\cite{stober_controlled_1968,bogush_preparation_1988,Liesegang2014Australian,gao_facile_2016}.
The periodic arrangement of the silica spheres within the opal causes diffraction of light,
which is responsible for the characteristic play-of-colour observed in precious
opals~\cite{gaillou_common_2008}; opals are natural photonic crystals.

\subsubsection{Striped flint}
\label{sec:flint}

Layered, banded, striped, or wave-like patterns are common in rocks and are related to the processes that form them. 
Such patterns are particularly common in sedimentary rocks, where layers form due to the gradual deposition of sediments over time. Each layer represents a different period of deposition, possibly reflecting changes in the environment, such as water flow, climate, or sediment sources.

\begin{figure}
\centering
\includegraphics{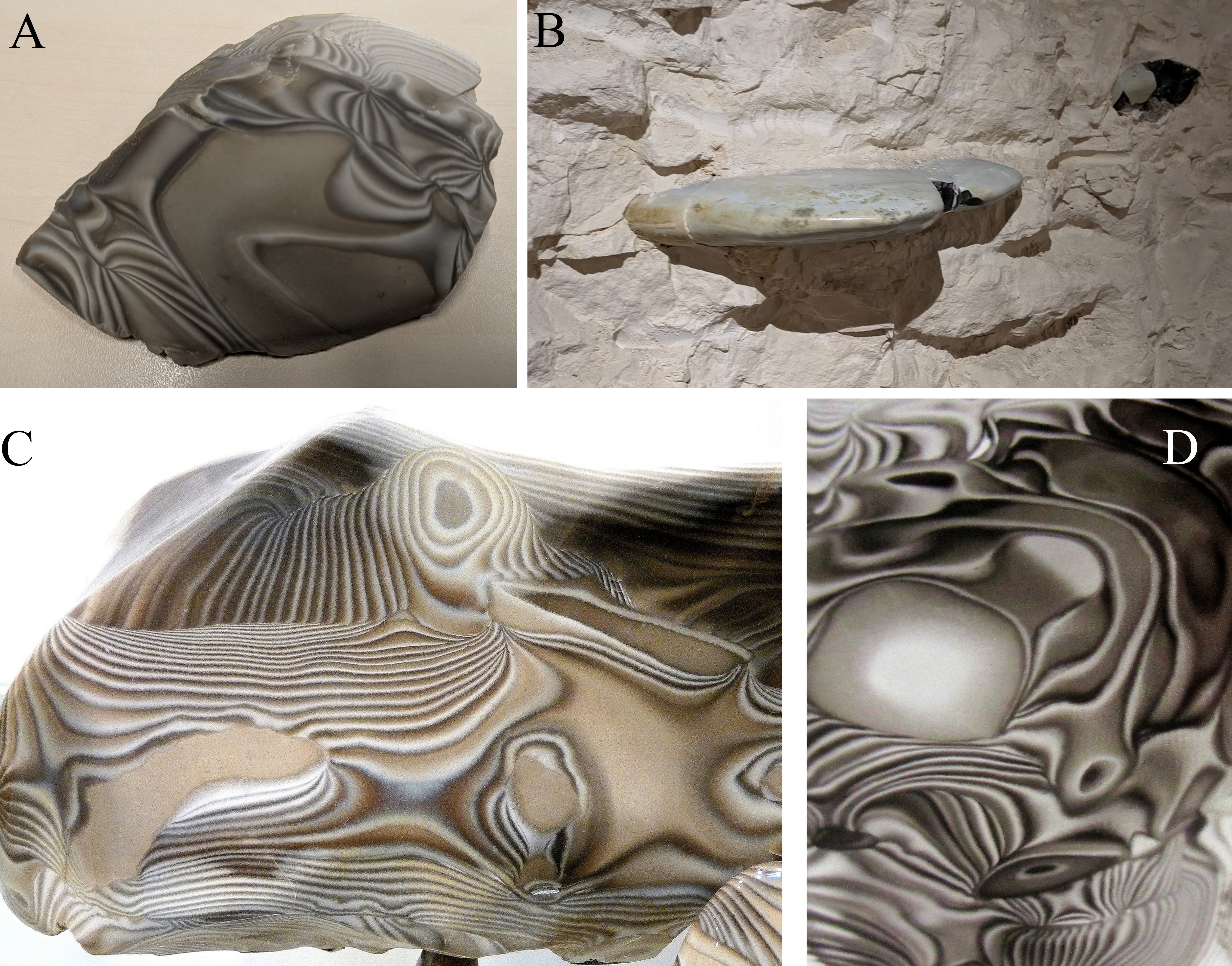}
\caption{\label{fig:Flint}Striped flint.
A,C,D: specimens from \'Swi\k etokrzyskie Mountains, Poland. 
B: Striped flint nodule at the limestone wall at the prehistoric flint mine in Krzemionki Opatowskie, Poland.}
\par\smallskip{\footnotesize\noindent\raggedright\emph{Images:} A, B:  Piotr Szymczak, 
C, D: courtesy of Ewa Siemo\'nska (Museum of Minerals and Fossils in \'Swi\k eta Katarzyna, Poland).\par}
\end{figure}

Nevertheless, the genesis of many patterns remains elusive. One of the most intriguing
examples is the highly anisotropic, interference-like pattern of striped flint
(Fig.~\ref{fig:Flint}), which resembles neither sedimentary banding, nor target patterns
(Sec.~\ref{sec:targetpatterns}) or agate banded patterns  (Sec.~\ref{sec:agates}). Surprisingly,
petrographic, crystallographic and chemical analyses show little difference between the bands
except for differences in pore distribution, with fewer and smaller pores in the darker
bands~\cite{Migaszewski2006}.

Striped or banded flint (Fig.~\ref{fig:Flint}) was first described in the 19th
century~\cite{sowerby1804british,woodward1864nature}. It is found in  localities around the
world, but the best-known occurence is in Poland~\cite{migaszewski2022geochemistry}, where it
is hosted in Middle Oxfordian to Lower Kimmeridgian carbonate sediments at the north-eastern
edge of the \'Swi\k etokrzyskie Mountains. 
It  belongs to a class of nodular cherts, occurring in nodules between a few centimetres and 2
metres in diameter (Fig.~\ref{fig:Flint}B), commonly more or less flattened parallel to their
horizontal axes and to the bedding of the host rock. The origin of silica forming nodular
cherts (not only striped)  is still under debate with some authors pointing to a hydrothermal
origin~\cite{Migaszewski2006}, while others invoke a diagenetic alteration product of
biogenic oozes~\cite{Knauth1994}. The origin of nodular structures in cherts  is also
unknown, with "self-organization processes of enigmatic character''~\cite{Mcbride1999}
invoked, and no theoretical framework predicts the sizes of the nodules and characteristic
distances between them.  
As to the formation of the nodule body itself, most often coupled dissolution--precipitation
processes are considered~\cite{Knauth1994} with replacement of original limestone with silica
on a volume-for-volume basis, analogous to petrified wood. Other theories assume that silica
crystallized from a gel~\cite{Oehler1971} or precipitated from
solution~\cite{Migaszewski2006}.

\subsubsection{Zebra rock}
\label{sec:Zrock}

Before addressing zebra rocks and zebra textures (Sec.~\ref{sec:Ztextures}), it is necessary to mention that the term \emph{zebra rocks} is often found used colloquially for any banded rocks. However, in this review we use the term \emph{zebra} to refer to two specific structures: the zebra rocks of Australia, the subject of this section, and the zebra textures characteristic of carbonate rocks,   Sec.~\ref{sec:Ztextures}.

\begin{figure}
\centering
\includegraphics{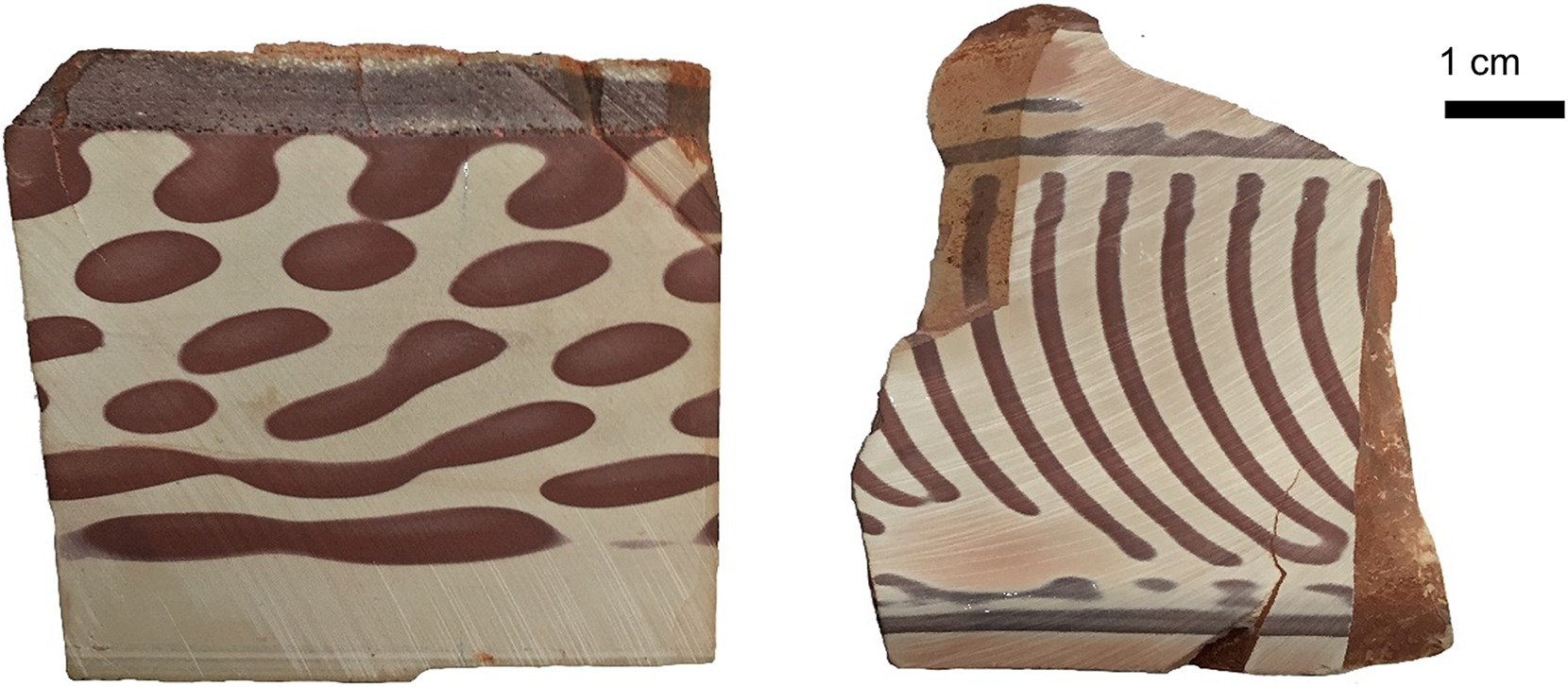}
\caption{\label{fig:zebra}
Zebra rock,  
from Kimberley, Western Australia. 
}
\par\smallskip{\footnotesize\noindent\raggedright\emph{Source:} \citeA{coward2023}.\par}
\end{figure}

Zebra rocks, characterized by their alternating bands of light and dark hues,
Fig.~\ref{fig:zebra}, are found only in the Kimberly region of Western
Australia~\cite{larcombe_rocks_1927,kawahara_hydrothermal_2022,coward2023}. Although these
bands give the rock its name, other patterns, such as spots, pillars, or rods, can also be
observed. Zebra rocks are composed of a fine-grained, clay-rich siltstone; the lighter bands
are mainly quartz and clay minerals, whereas the darker bands are enriched in iron oxides
(hematite)~\cite{retallack2021,kawahara_hydrothermal_2022}. What makes these rocks
remarkable, apart from the fact that they appear in only one very specific place, is the
degree of symmetry, regularity, and morphological pattern observed, which is unparallelled
elsewhere~\cite{coward2023}.

Furthermore, the geological processes that led to the formation of these rocks remain
uncertain, although several mechanisms have been proposed since their first description in
1925~\cite{larcombe_rock_1925}. In this review, we will mention those that seem to have
become obsolete and briefly describe the formation processes that have been worked on in
recent years. However, all these hypotheses remain controversial.

The first hypothesis of the origin of these rocks was proposed by~\citeA{larcombe_rock_1925,larcombe_rocks_1927}, who described how zebra rocks would have to have formed in deep, calm
waters, far from land. Soon after,~\citeA{trainer_zebra_1931} suggested four possible origins
for these zebra rocks: (1) original sedimentary banding followed by deformation, (2)
crystallization of a fine-grained flow rock, (3) infiltration of iron-bearing solutions into
a white rock, and (4) leaching of hematite from a red rock. Currently, three main hypotheses
are considered to explain the origin of these rocks: (1) hydrothermal
alteration~\cite{kawahara_hydrothermal_2022}, (2) acid sulphate soil
weathering~\cite{retallack2021}, and (3) liquid crystals~\cite{mattievich2003}.

The first proposed process, i.e.,~formation by hydrothermal alteration, can be summarized in
the following steps~\cite{kawahara_hydrothermal_2022,coward_how_2022}: first,
Fe\tsup{2{\tplus}}-rich acidic hydrothermal fluids infiltrated into micaceous siltstones and
shales. This fluid interacted with carbonate minerals, raising the pH and causing iron
precipitation as Fe-oxyhydroxide at the reaction front. Continued diffusion of the
Fe\tsup{2{\tplus}}-bearing fluid led to successive neutralization reactions, creating a
rhythmic banding pattern. This hydrothermal fluid also altered other primary silicates in the
bedrock, leading to the formation of secondary clay minerals. Over time, Fe-oxyhydroxides
were converted to hematite as oxidation progressed.

The second proposed mechanism relates the origin of  zebra rocks to the presence of Ediacaran
palaeosols with iron oxides~\cite{retallack2021,coward_how_2022}. These palaeosols were
subsequently weathered by sulphate-bearing acid solutions, which leached iron from some of
the soil horizons, changing their colour from red to white.

The third proposed process invokes liquid-crystal phases in the rock's mineral
structure~\cite{mattievich2003}. This theory stems from the observation that the banding
patterns in zebra rocks resemble those seen in certain liquid-crystal materials, which
exhibit regular, repeating structures when undergoing phase changes. In the case of zebra
rocks, it is hypothesized that liquid-crystal behaviour could have occurred in the minerals
present during the rock's diagenesis or early lithification stages. During these stages,
mineral-rich fluids could have flowed through  sedimentary siltstone, and under certain
physical and chemical conditions, these minerals might have organized themselves into
liquid-crystal phases. The theory posits that as the minerals cooled and solidified, they
could have preserved these liquid-crystal patterns as alternating bands of dark (iron
oxide-rich) and light (silica-rich) layers.

Most studies agree that the origin of these rocks is related to the presence of acidic
solutions that altered the original host rock. The alternation of the bands, as well as the
other patterns observed in these rocks, have been described as Liesegang-like
patterning~\cite{loughnan_composition_1990,coward_how_2022,coward2023,liu_coefficients_2023}; however, the extent to which these bands conform to the spacing relations characteristic of classic Liesegang bands remains debated. The possibility that zebra rocks might be Turing patterns was
explored by Coward in his doctoral thesis~\cite{coward_how_2022}. However, in his conclusions
he indicates that he did not find Turing instabilities  in the reactions studied. The
mechanism thus remains an open question.

\subsubsection{Zebra textures}
 \label{sec:Ztextures}

\begin{figure}
\centering
\includegraphics{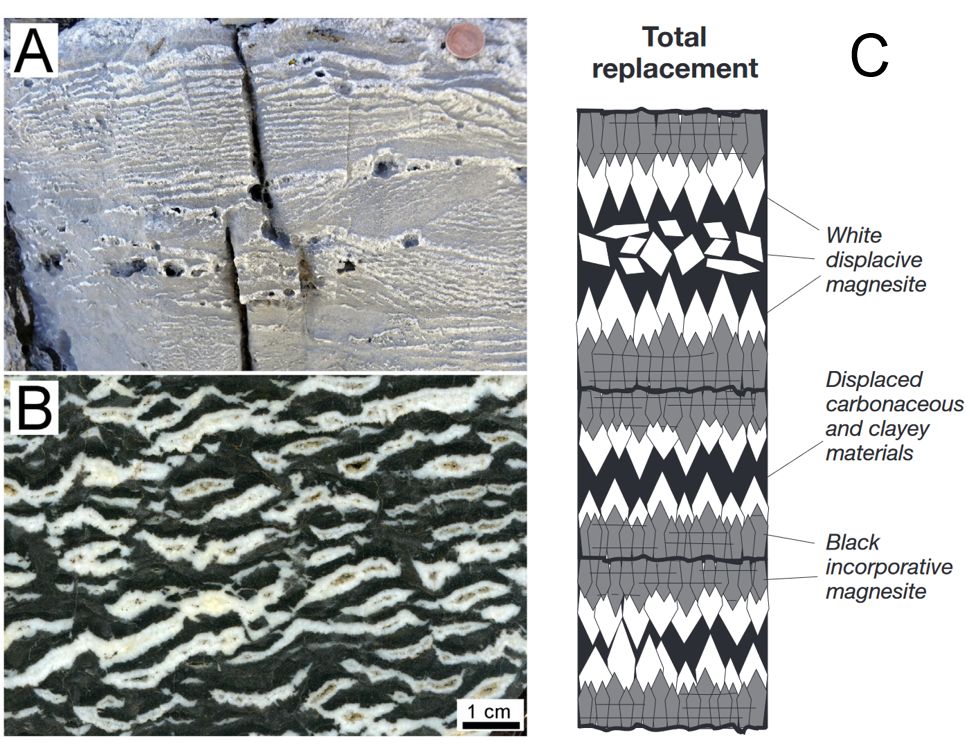}
\caption{\label{fig:zebratextures}Zebra textures.
A: Zebra dolomite outcrop and
B: zebra dolomite hand-polished  specimen. 
C: Schematic representation of the zebra texture in Eugui magnesites, Spain.}
\par\smallskip{\footnotesize\noindent\raggedright\emph{Sources:} A, B:~\citeA{wallace_zebra_2018}, C:~\citeA{lugli_petrography_2000}. \par}
\end{figure}

Zebra textures (Fig.~\ref{fig:zebratextures}) are very common in carbonate rocks occurring in a
variety of geological settings, such as Mississippi valley-type deposits and hydrothermal
dolomite deposits~\cite{fontbote_genesis_1990,morrow_zebra_2014,wallace_zebra_2018}. In all
these settings, the host rock is carbonate, while the mineralization producing the zebra
texture may consist of carbonates (e.g.,~dolomite and magnesite) or non-carbonate minerals
(e.g.,~sphalerite and fluorite). These textures are characterized by a rhythmic alternation
of light and dark bands of similar thickness. Generally, the bands are millimetric and,
although they can be parallel to stratification, they more commonly dip at a small
angle~\cite{wallace_zebra_2018}. These structures have been classified as examples of
self-organized patterns~\cite{fontbote1993self,merino_genesis_2006}.

The origin of these textures has historically been attributed to three main mechanisms:
dissolution~\cite{fontbote_genesis_1990,morrow_zebra_2014},
fracturing~\cite{park_dolomite_1938,swennen_genesis_2003}, and
metasomatism--replacement--recrystallization~\cite{lugli_petrography_2000,kelka_zebra_2017}.
Other mechanisms have also been proposed on an ad hoc
basis~\cite{krug_morphological_1996,arne_internal_1991,merino_genesis_2006}. Recently,
however, a unifying theory for their origin has been proposed that combines the three main
mechanisms. This theory, proposed by~\citeA{wallace_zebra_2018}, posits co-located processes:
mineral replacement, and dissolution/void generation. These two processes together imply that
in the hollows where new crystals are growing, crystallization can generate sufficient
pressure to fracture the host. These fractures may then be dissolved and filled by new
cement. The net result  is the formation of the zebra banding.

\subsection{Large and cold}

Near-surface processes with modest energy inputs nonetheless organize matter on centimetre to metre scales. Latent-heat release and fluid flow sculpt icicles and stalactites, salt-crust convection draws hexagons across dry lakes, and impact-like target patterns punctuate evaporitic flats. Shrinkage and weathering generate polygonal mud cracks, honeycomb tafoni, desert-rose efflorescences and vertical solution pipes or wormholes, all spaced tens of centimetres or more despite their low-temperature setting.

\subsubsection{Icicles, stalactites, and stalagmites}
\label{sec:icles}

\begin{figure}
\centering
\includegraphics{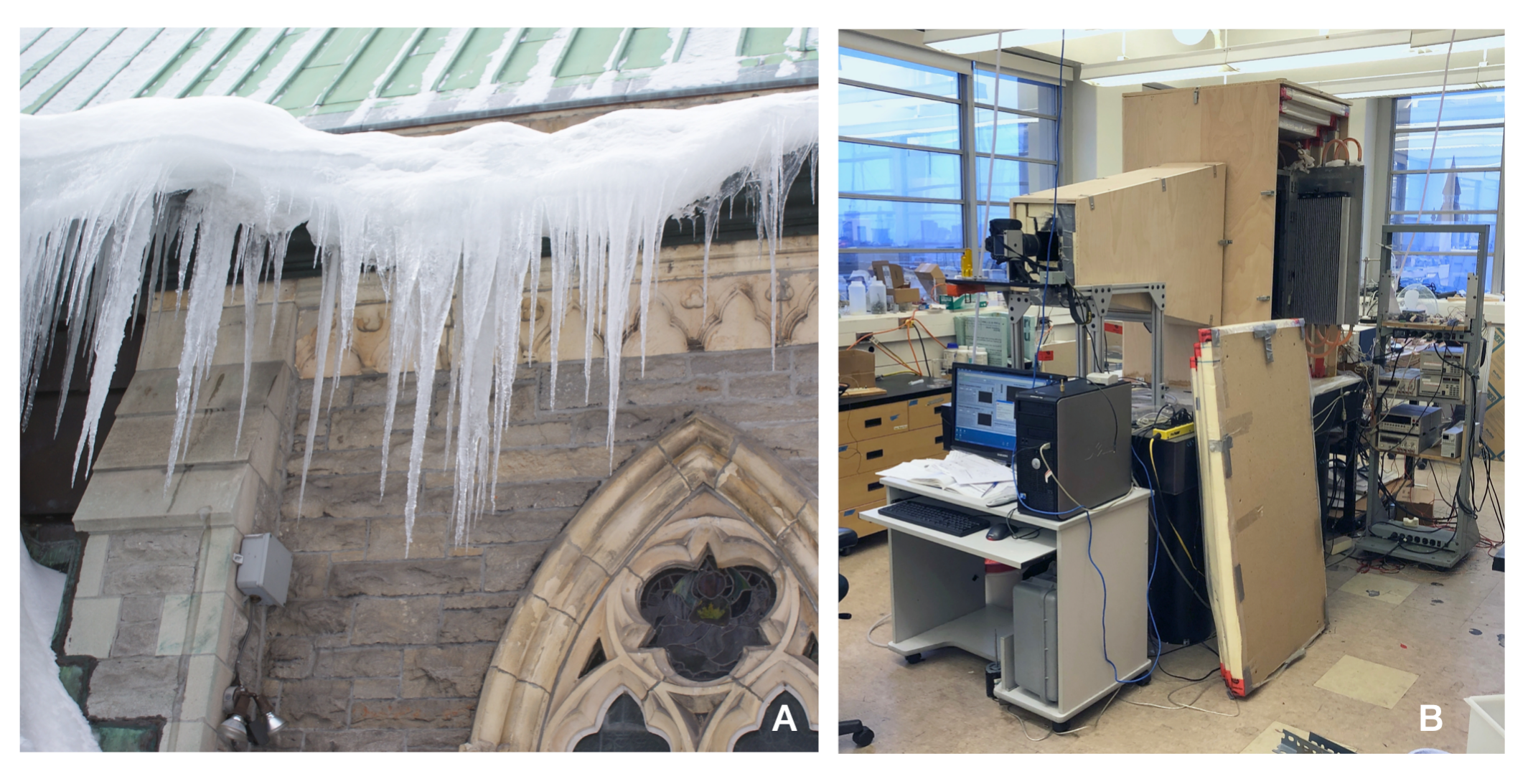}
\caption{\label{fig:icicle_motivational}Natural and laboratory icicles.
(A) Natural icicles hanging from a church roof in Montr\'eal. Both conical and drapery-like formations can be seen.  
(B) The laboratory icicle growing machine, showing the insulated, temperature controlled,
refrigerated box with one side opened, and its computer controlled digital camera and
associated electronics. Temperature controlled feed water of known composition is introduced
from the top of the box onto a slowly rotating support, while drip-off water is collected
below.  The humidity of the air inside the box is measured and may be turbulently stirred.
Using edge detection of images taken with slow rotation, the entire 3D shape of the icicle
can be measured as a function of growth
time~\cite{chen2011,chen2013,chen_thesis,icicle_inclusions,ladan2021wetting,ladan_thesis}.
Data from these experiments are published in the \textit{Icicle
Atlas}~\cite{icicle_atlas_home}. }
\par\smallskip{\footnotesize\noindent\raggedright\emph{Images:} Stephen Morris.\par}
\end{figure}

An \textit{icicle} is the familiar elongated, tapering structure of ordinary ice that forms when water
drips into freezing air from a point of support. Since ice formation requires the removal of
latent heat, we can infer that heat transfer is most efficient at the fast growing tip and
slower on the sides~\cite{makkonen1988}.  The shape forms a substrate for the subsequent flow
of water, some of which freezes while some drips off; the dripping water assists the heat
transfer, causing faster growth at the tip. Examples of natural icicles are shown in
Fig.~\ref{fig:icicle_motivational}A. Very similar fluid-mechanical transport is presumably
involved in the formation of \textit{stalactites}, which are calcium carbonate deposits that
hang from  roofs of caves. Here, the solid CaCO$_3$ is deposited when CO$_2$
comes out of solution~\cite{Short2005,shortphysfluids,camporeale}, but the water is not
consumed.  Both icicles and stalactites usually have a single dripping tip, but may sometimes
develop multiple branches~\cite{chen2011}. Other common morphologies are also observed, such
as sheet-like \emph{drapery} formations in which the water flows only along a leading edge,
or \emph{flow stone}, in which water flows over a broad surface~\cite{meakin2010}.

The suffix \emph{-icle} has also been applied to other elongated tapering structures, such as
\textit{snoticles} (also known as \textit{snottites}), \textit{brinicles}  and
\textit{rusticles}. Snoticles~\cite{snoticle} are drips of mucus-like microbial mats that
hang from cave roofs, resembling small stalactites. Brinicles~\cite{teston2024experimental}
are ice structures that form on the underside of sea ice when plumes of dense, supercooled
brine descend from channels in the ice, causing the surrounding sea water to freeze into an
elongated pipe. Rusticles~\cite{rusticle} are long, iron-oxide rich  structures that form on
shipwrecks. 
 In all cases, gravity- or buoyancy-driven fluid flows interact with the evolving shape,
producing long, downward-pointing structures. Submerged, elongated pipe-like structures are
also formed by chemical gardens (Sec.~\ref{sec:osmotic})~\cite{brinicle}; these are generally upward pointing, with
positive rather than negative buoyancy (Fig.~\ref{fig:osmosis}).

\begin{figure}
\centering
\includegraphics{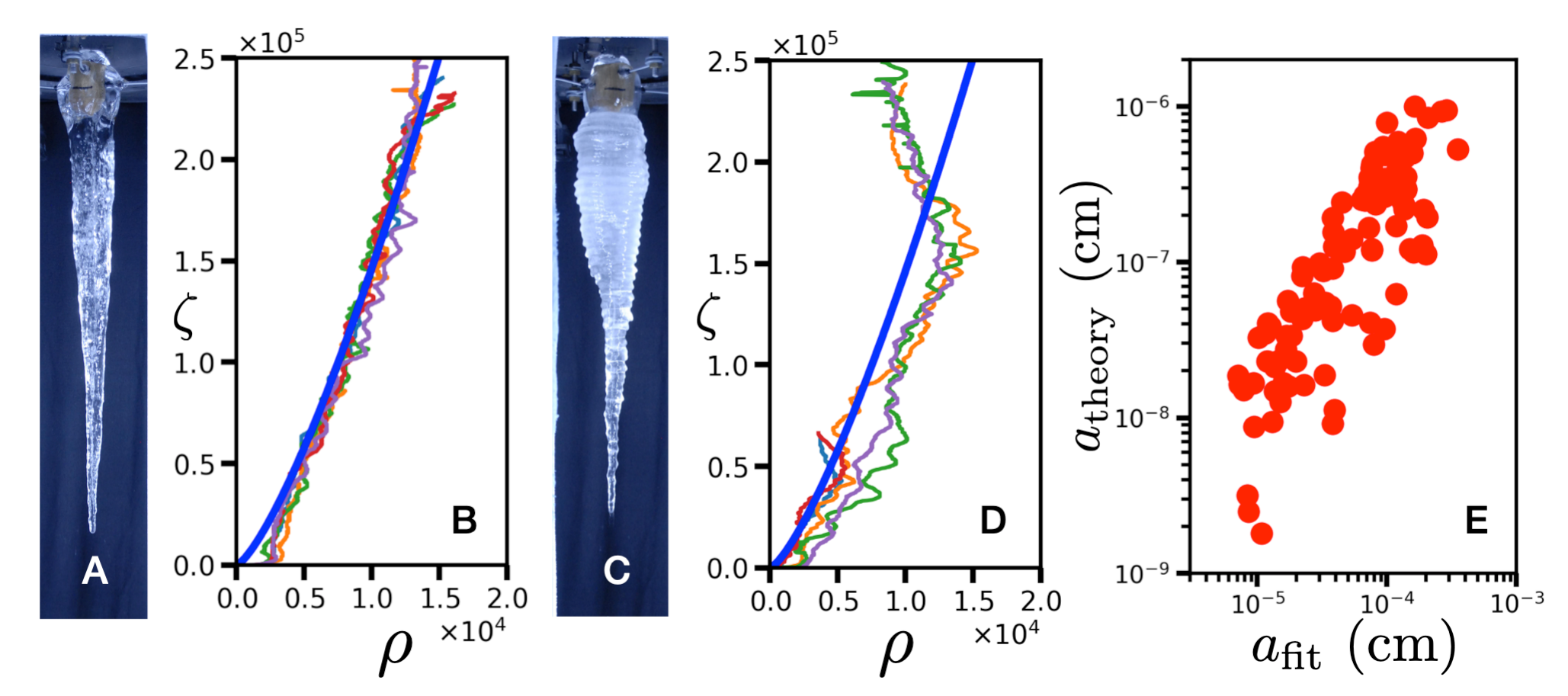}
\caption{\label{fig:icicle_shape}Universal icicle shape.
(A) and (C) show the final states of two icicles grown from different water sources under
otherwise identical conditions, using the apparatus shown in
Fig.~\ref{fig:icicle_motivational}(B): (A) was made with distilled
water~\cite{icicle_atlas_run_m7q10}; (C) was made with Toronto tap
water~\cite{icicle_atlas_run_m7q12}.  The temperature of the stirred air was
$-6.8${\degree}C, the input mass flux of water was 2.6~g/min, and the input water
temperature was 3.2{\degree}C. These results show clearly that even very small
concentrations of impurities have a profound effect on icicle shapes~\cite{chen2011}. Using
76 icicles grown under a variety of conditions and water sources, edge detected shapes were
fit to Eq.~\eqref{eq_shapesolution}, using the dimensionless variables given in
\eqref{dimvar}, with $a=a_{\rm fit}$ as a fitting parameter. (B) Shows the best five fits,
while (D) shows the worst five fits, as measured by reduced $\chi^2$. (E) shows a
comparison of $a_{\rm theory}$ to $a_{\rm fit}$ for a series of 106 icicles, grown using controlled
concentrations of NaCl as an impurity~\cite{chen_thesis}. }
\end{figure}

 Unlike stalactites and most other geological formations, icicles are unusually amenable to
laboratory growth studies.  Fig.~\ref{fig:icicle_motivational}B shows an apparatus for growing
icicles under controlled conditions in the laboratory.  Detailed experiments on
icicles~\cite{chen2011,chen2013,ladan2021wetting,icicle_inclusions,menno2023} have revealed a
wealth of subtle phenomena.  Analogous laboratory experiments on stalactites are somewhat
impractical owing to the longer time-scale.   
  
 Both icicles~\cite{short2006} and stalactites~\cite{Short2005,shortphysfluids} have been
theoretically predicted to have universal shapes, independent of growing conditions.
According to the ``platonic'' icicle theory, assuming axial symmetry, the radius $R$
and distance from the tip $z$ may be scaled by a common length $a$, so that
if the shape is expressed in dimensionless coordinates,
\begin{equation}
\rho = R/a~,~~~~\zeta = z/a,
\label{dimvar}
\end{equation}
then the predicted universal icicle shape is given by
\begin{equation}
\label{eq_shapesolution}
\rho(\zeta) = \frac{4}{3} \bigl(\zeta^{\frac{1}{2}}+2\bigr) \sqrt{\zeta^{\frac{1}{2}}-1}~~\sim \zeta^{3/4}~~{\rm for}~{\rm small}~a.
\end{equation}
The shape is universal because all the physical parameters are absorbed into $a$, which is given by
\begin{equation}
\label{eq_shapea}
a = a_{\rm theory} = \frac{g\beta_{A}(\Delta T_{A})^{5}}{\nu_{A}^{2}} \biggl( \frac{\Lambda_{A}}{L_{f}\rho_{I}Cv_{t}} \biggr)^{4},
\end{equation}
where $g$ is the acceleration due to gravity, $\beta_{A}$ is the volumetric thermal expansion coefficient of air, $\nu_{A}$ is the kinematic viscosity of air, $\Delta T_{A}$ is the difference between the surface temperature and the ambient temperature, $\Lambda_{A}$ is the thermal conductivity of air, $L_{f}$ is the latent heat of fusion of water and $\rho_{I}$ is the density of ice. $C$ is a dimensionless constant of order unity and $v_{t}$ is the speed of the growing tip. The theory relies on a number of simplifying assumptions, in addition to axial symmetry.  The growth rate of ice is assumed to be limited by the heat transfer across the rising thermal boundary layer that surrounds the icicle. The boundary layer is assumed to be laminar. The advection of heat by the film of flowing water on the sides of the icicle is neglected.  Pure water is assumed.  

The universal theory of icicle shapes has been thoroughly tested
experimentally~\cite{chen2011,chen_thesis,menno2023}. Results of one such
study~\cite{chen2011,chen_thesis} are summarized in  Fig.~\ref{fig:icicle_shape}. It is found
that the most important parameter affecting icicle morphology is the purity of the feed
water. Even the very low level of impurity in ordinary tap water is enough to produce
drastically non-universal shapes, as in  Fig.~\ref{fig:icicle_shape}C,D; on the other hand,
icicles grown from nominally distilled water can be close to the universal shape, as in
Fig.~\ref{fig:icicle_shape}A,B. The agreement with the universal shape seen in
Fig.~\ref{fig:icicle_shape}B is impressive considering that only one fit parameter, $a_{\rm fit}$
was used, and the slope of the fit line is independent of the parameter. However, icicles
grown from ultra-pure Milli-Q water~\cite{menno2023} have a ``dripping candle'' shape which
is poorly described by the universal shape. Thus, universal ``platonic'' icicles are only
observed for nearly pure, but not ultra-pure, water.  These very low levels of impurity are
ignored by the theory, but will of course be present in any natural icicle. Several other
assumptions of the universal shape theory are also violated; the air motion was turbulent,
rather than laminar, for example.  Furthermore, the physical interpretation of $a_{\rm theory}$ is
obscure, given its very small typical value of ${\sim} 10^{-7}~$cm, taking $C=1$ in
Eq.~\eqref{eq_shapea}.  But the values of the fit parameter $a_{\rm fit}$ are also very small,
and do appear to be positively correlated with the expected values of $a_{\rm theory}$, as shown
in  Fig.~\ref{fig:icicle_shape}E.  The magnitude discrepancy between $a_{\rm fit}$ and $a_{\rm theory}$
can be removed by taking $C = 0.2$  in Eq.~\eqref{eq_shapea}.  Thus, we may say that the
universal shape theory is, at best, a qualified success.

\begin{figure}
\centering
\includegraphics{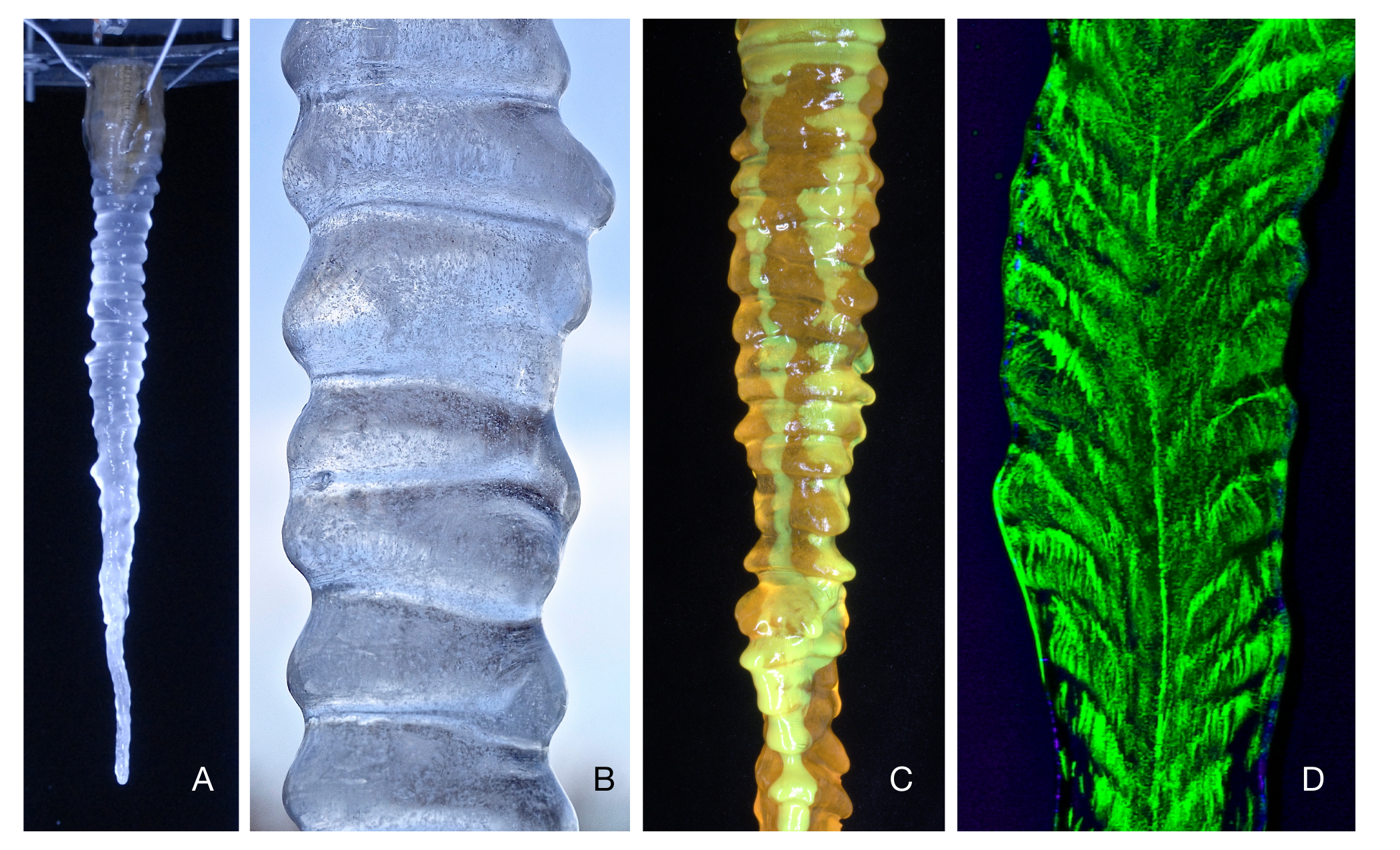}
\caption{\label{fig:icicle_ripples}Rippled icicles: 
(A) laboratory grown icicle~\cite{icicle_atlas_run_120906}, made from distilled water with
320~ppm NaCl as an impurity. 
(B) close up of the ripples on a natural icicle. In all cases, the ripple wave-length is close to 1~cm (image: S. W. Morris). 
(C)  a lab-grown icicle with 519~ppm sodium fluorescein dye as the impurity.  The dye glows
green in the liquid phase, but appears orange when trapped in inclusions in the ice. The
partial wetting of the surface is evident (Image: CC-BY Ladan and
Morris~\cite{ladan2021wetting}). 
(D)  a vertical section of a lab-grown icicle near 0~\textdegree C with 171~ppm sodium
fluorescein.  Here, the dye in the inclusions glows green. The chevron pattern of the
inclusions tracks the slow upward motion of the ripples during growth. The chevrons have a
substructure of crescents~\cite{icicle_inclusions}.}
\end{figure}

\begin{figure}
\centering
\includegraphics{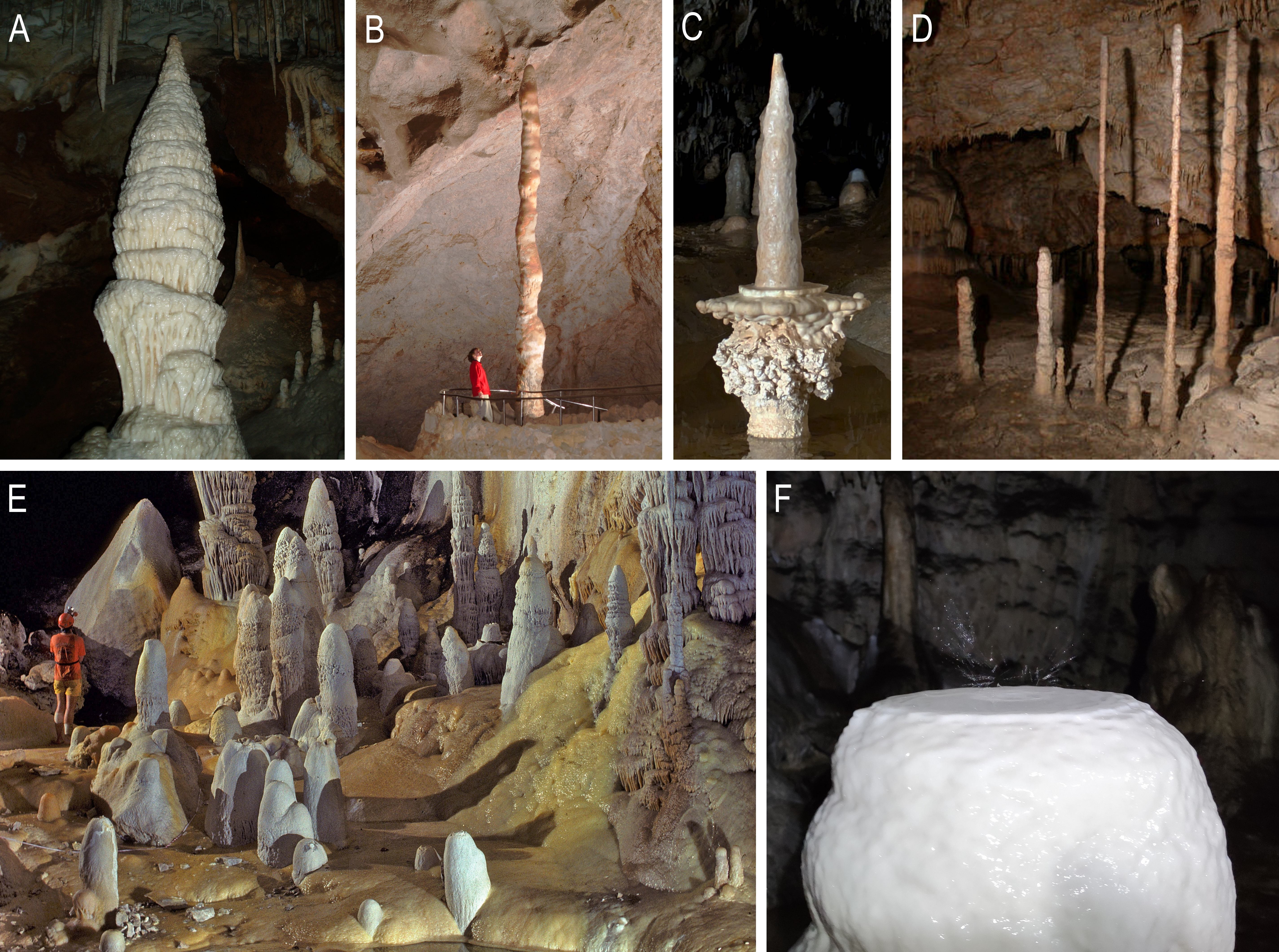}
\caption{\label{fig:stalagmites}(A) ``Minaret'' stalagmite in the Jenolan Caves, NSW, Australia; (B) ``Witch's Finger'' columnar stalagmite in Carlsbad Caverns, USA; (C) ``Candlestick'', conical stalagmite in Sloupsko-\v so\v s\r uvsk\'e caves, Czechia; (D) long columnar stalagmites in Katarynske caves, Czechia; (E) columnar stalagmites in Lechuguilla Cave, New Mexico, USA (F) flat-top stalagmites in Postojna Cave, Slovenia.}
\par\smallskip{\footnotesize\noindent\raggedright\emph{Images:} A: Bellman (public domain), B: Peter Jones, National Park Service, USA (public domain) C: Piotr Szymczak, D: Jochen Duckeck (public domain), E: Dave Bunnell/Under Earth Images (CC BY-SA 2.5), F: Courtesy of Matej Lipar.\par}
\end{figure}

\begin{figure}
\centering
\includegraphics[width=\linewidth,height=0.78\textheight,keepaspectratio]{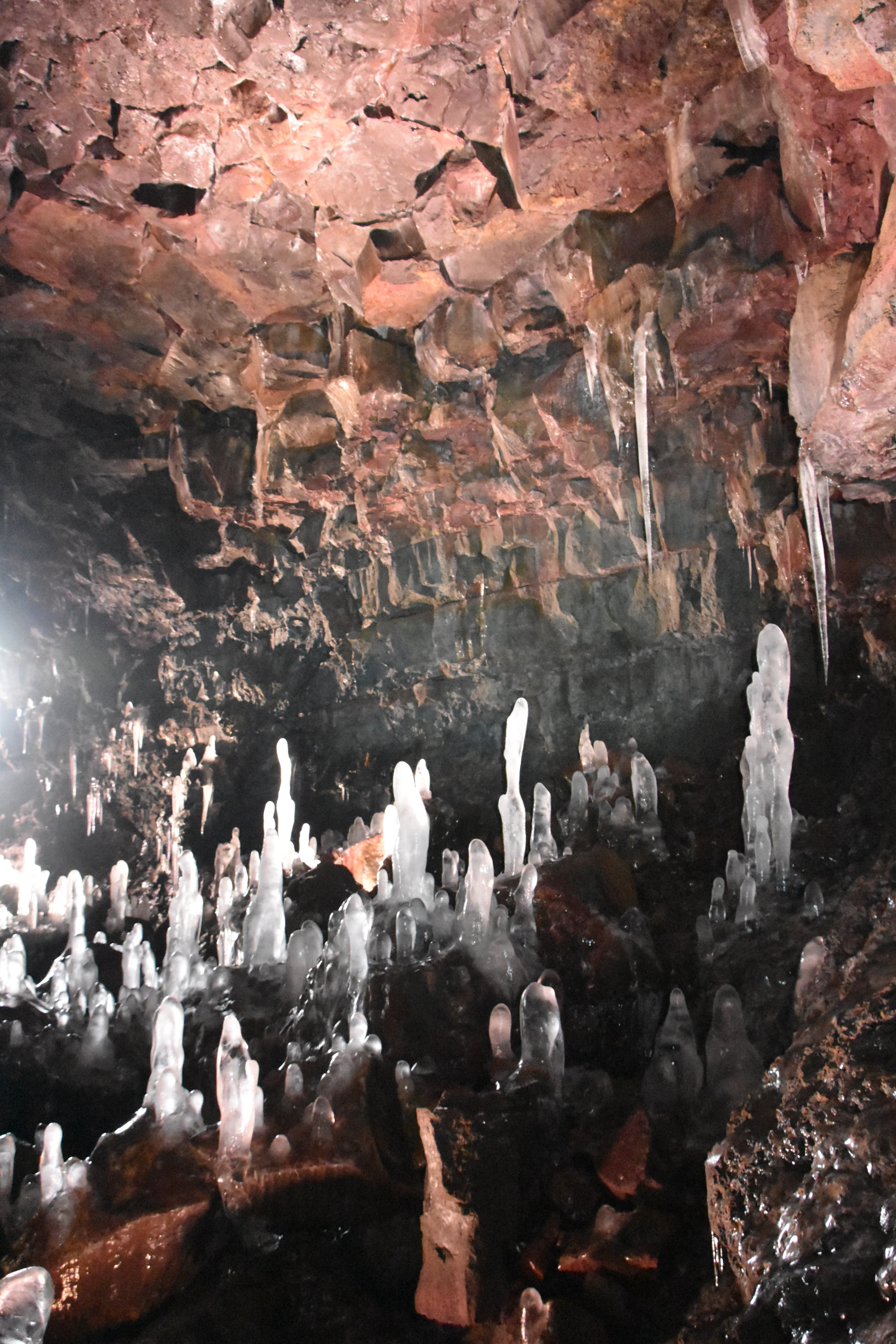}
\caption{\label{fig:cave}Icicles, hanging from the ceiling, and their rare cousins, the ice version of stalagmites, on the floor, in a lava tube cave in Iceland. The columnar jointing of the lava (Sec.~\ref{sec:basalt}) may also be appreciated.}
\par\smallskip{\footnotesize\noindent\raggedright\emph{Image:} Julyan Cartwright.\par}
\end{figure}

In addition to their overall pointy shape, icicles and stalactites may also exhibit ripple
patterns about their circumference.  These ripples have a near-universal wave-length of about
1~cm, independent of the growing conditions~\cite{ogawa2002,ueno2010a,camporeale,chen2013}.
Examples of rippled icicles are shown in  Fig.~\ref{fig:icicle_ripples}. Ripples are another
manifestation of impurities present in the feed water~\cite{chen2013,chen_thesis}.  The
effect of impurities on the wetting properties of the ice turns out to be
crucial~\cite{ladan2021wetting,ladan_thesis}. It is counterintuitive, but ultra-pure water
does not wet ultra-pure ice very well~\cite{demmenie2025}.  On an ultra-pure
icicle~\cite{menno2023}, water does not coat the whole icicle but rather descends in discrete
sliding drops that hardly wet the ice at all.  Even extremely low levels of impurity suffice
to increase the wetting of the ice surface and change the overall morphology of the resulting
icicle~\cite{chen2011,chen2013,ladan2021wetting}.  At moderate, but still very low, levels of
impurity, ripples emerge on the surface during growth~\cite{chen2013}, which are presumed to
be due to a morphological instability. Even natural icicles which form from atmospheric
precipitation are often impure enough to be in this regime, as in
Fig.~\ref{fig:icicle_ripples}B.  Ripples are observed to travel upward slowly during growth. As
the impurity concentration is increased, ripples grow faster and to larger amplitude and
eventually become disordered, but their wave-length always remains very close to 1~cm.  
 
 When a fluorescent dye is used as the impurity, it becomes possible to visualize the surface
flow over the icicle~\cite{ladan2021wetting} and the fate of impurities trapped in the
ice~\cite{icicle_inclusions}. The most well-formed ripples are found in an impurity
concentration regime in which the ice is only partially wetted by the liquid phase, which
descends in branching rivulets; see  Fig.~\ref{fig:icicle_ripples}C. At high
concentrations~\cite{ladan2021wetting,menno2023}, liquid coverage becomes complete and the
ripples are very disordered. A substantial fraction of the impurity concentration ends up
trapped in small liquid inclusions inside the ice; these inclusions are roughly spherical and
$100~\rmmu\mathrm{m}$ in size.  They form clusters that make a distinct pattern of chevrons following
the peaks of the ripples~\cite{icicle_inclusions}. Each chevron has a substructure of
crescents, which are discrete layers rich in inclusions.  These crescents are just visible in
Fig.~\ref{fig:icicle_ripples}D. At high concentrations, the ice becomes completely spongy with
liquid inclusions.
 
At present, none of these phenomena are quantitatively described by linear stability theories
for the onset of icicle ripples~\cite{ogawa2002,ueno2010a,worster2024}. These theories assume
complete wetting and do not treat the role of impurities at all.  When generalized to include
impurities~\cite{ladan2021wetting,ladan_thesis}, they fail to predict the wave-length of the
ripples. The rippling instability thus depends on the impurity dependence of the
non-equilibrium partial wetting of the ice, a process that is very poorly
understood~\cite{huerreARFM,demmenie2025}.    The process by which impurities are trapped to
become inclusions is also not understood. A linear stability theory of ripples on stalactites
has been proposed~\cite{camporeale}, which uses an analogous assumption of complete wetting,
but much less is known about stalactite growth dynamics. Sectioned stalactites do not exhibit
anything analogous to the pattern of inclusions in icicles. It would be interesting to
examine the path of water descending an actively growing stalactite using a dye tracer to
probe the role of wetting in that case.  While icicles and stalactites are superficially
similar, they differ in many important details.

A somewhat more complex but analogous argument for the universal shape of stalactites leads to the same universal scaling as for icicles $\rho(\zeta) \sim \zeta^{3/4}$ for $\zeta \gg a$.
For stalactites, however, this ideal outer shape follows from different physics~\cite{Short2005,shortphysfluids}. In icicles, growth is constrained by removal of
latent heat through an air-side thermal boundary layer; in stalactites, growth is controlled
within the gravity-driven water film, where the carbonate deposition rate is limited by the
net conversion of carbonic acid to aqueous $\text{CO}_2$. Consequently, the prefactor differs:
icicles depend on ventilation and air properties (via heat transfer), whereas stalactites
depend on drip rate, water $\text{pCO}_2$, temperature, and film rheology. In both cases the
invariant (shape-preserving) outer solution erases tip details, while the amplitude retains
sensitivity to the ambient environment.

Stalagmites, their floor-growing counterparts, do not mirror stalactites. Stalactites are
typically slender and elongated, whereas stalagmites are generally
bulkier (Fig.~\ref{fig:stalagmites}). This morphological asymmetry indicates different
controlling mechanisms. Stalagmite growth is well described by surface-reaction--limited
precipitation in a thin film that has largely equilibrated by the time droplets strike the
tip~\cite{Franke1965,Romanov2008,Szymczak2025}.
Under form-preserving upward translation, the analytic ``ideal stalagmite'' solutions form a one-parameter family indexed by the Damk\"ohler number $\text{Da}=k \pi R^2/Q$,
where $k$ is a precipitation rate constant, $R$ is the radius of the stalagmite, and $Q$ the volumetric drip flux. Three end-member morphologies arise and are observed in caves~\cite{Szymczak2025}: flat-top ($\text{Da}>1$), columnar ($\text{Da}=1$), and conical ($\text{Da}<1$). Selection reflects droplet delivery near the apex: when the fall height is large, impact and splash spread wetting over a finite radius $R_c$, producing a flat cap that joins smoothly to the outer profile (flat-tops). When the wetted radius is negligible ($R_c \ll R$), the columnar limit is selected. Conical stalagmites have pointed tips; their innermost region is well fit by an inverted-stalactite profile, but farther out the shape steepens as surface-reaction control dominates. Field photographs and CT sections show that these analytic profiles fit natural examples using radii alone as free parameters.

Ice stalagmites, Fig.~\ref{fig:cave} have recently been studied experimentally and theoretically in the context of growth by the solidification of impacting water droplets on a cold substrate~\cite{Papa2025}, with regime transitions controlled by substrate temperature, drop discharge, and a critical height above which unfrozen water accumulates at the tip and develops fingered, star-shaped forms.

\subsubsection{Solution pipes, wormholes, and replacement fingers}
\label{sec:pipes}

\begin{figure}
\centering
\includegraphics{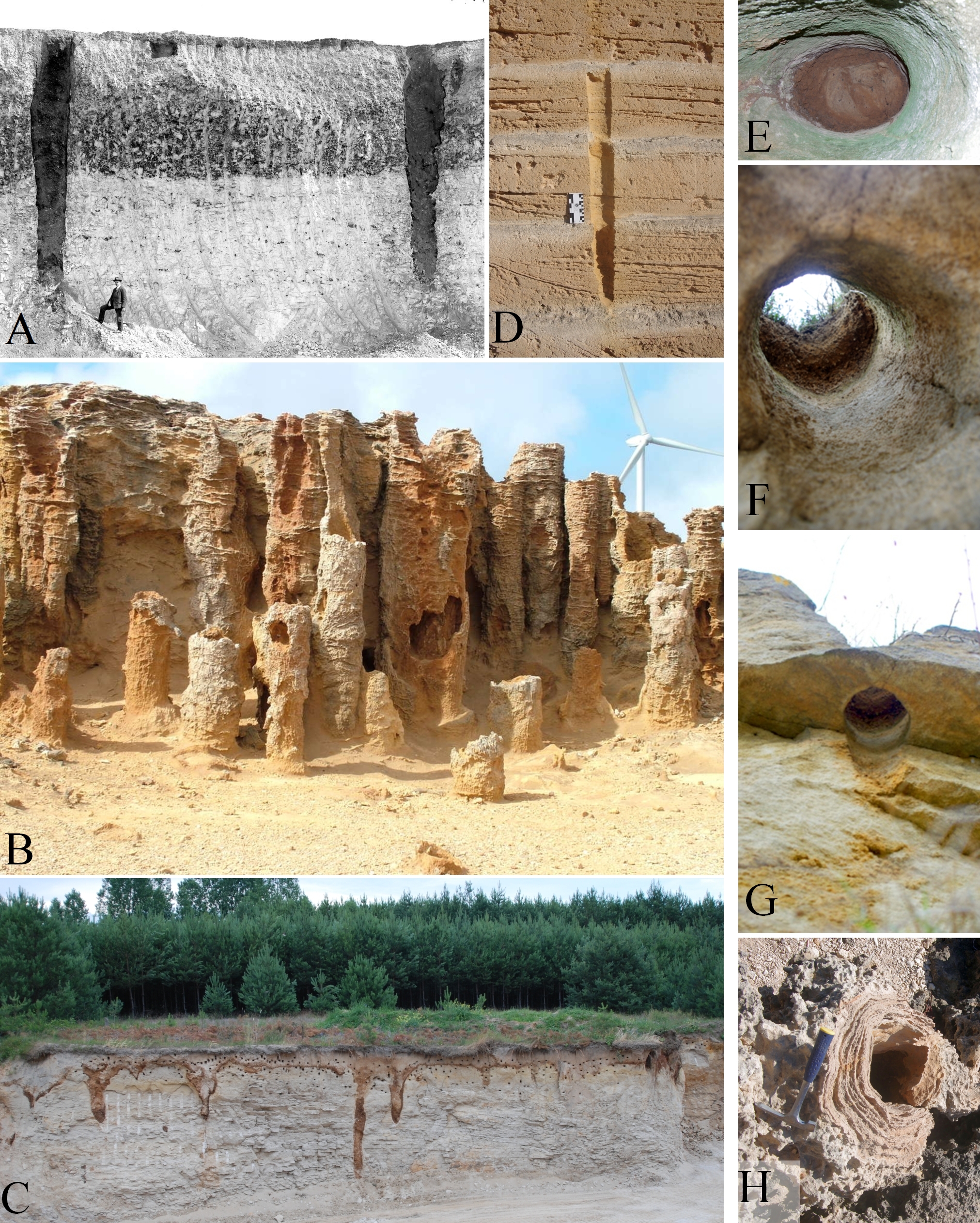}
\caption{\label{fig:Pipes}Solution pipes. 
A: pipes in Cretaceous chalk (Swanscombe, UK),    
B: pipes in Quaternary calcarenite (Cape Bridgewater, Australia), 
C,E--G: pipes in Miocene calcarenite, (Smerdyna quarry, Poland), 
D:  pipes in Neogene calcarenite (Guilderton quarry, Australia), 
H: pipe in Pleistocene calcarenite at
Cape Perron, Perth, Australia.
}
\par\smallskip{\footnotesize\noindent\raggedright\emph{Images:} 
A: J. Rhodes, courtesy of British Geological Survey; source: BGS GeoScenic (image P201858),
B and D--H, Piotr Szymczak,
C: courtesy of \L ukasz U\.zarowicz, Warsaw
University of Life Sciences, Poland.\par}
\end{figure}

Solution pipes,  Fig.~\ref{fig:Pipes}, are vertical, finger-like structures found in the epikarst zone of porous calcareous rocks. They vary in size from a few centimetres to a few metres in diameter, but are generally less than 1~m in diameter with variable lengths, the deepest reach 100~m. They also vary in shape;  some are more conical, tapering slowly towards the tip, whereas others have almost constant cross-section and a cigar-shaped termination. Their cross-sections are almost perfectly circular and very smooth (Fig.~\ref{fig:Pipes}E--H), such that one may have the impression that they were artificially drilled rather than formed by natural processes.

\begin{figure}
\centering
\includegraphics{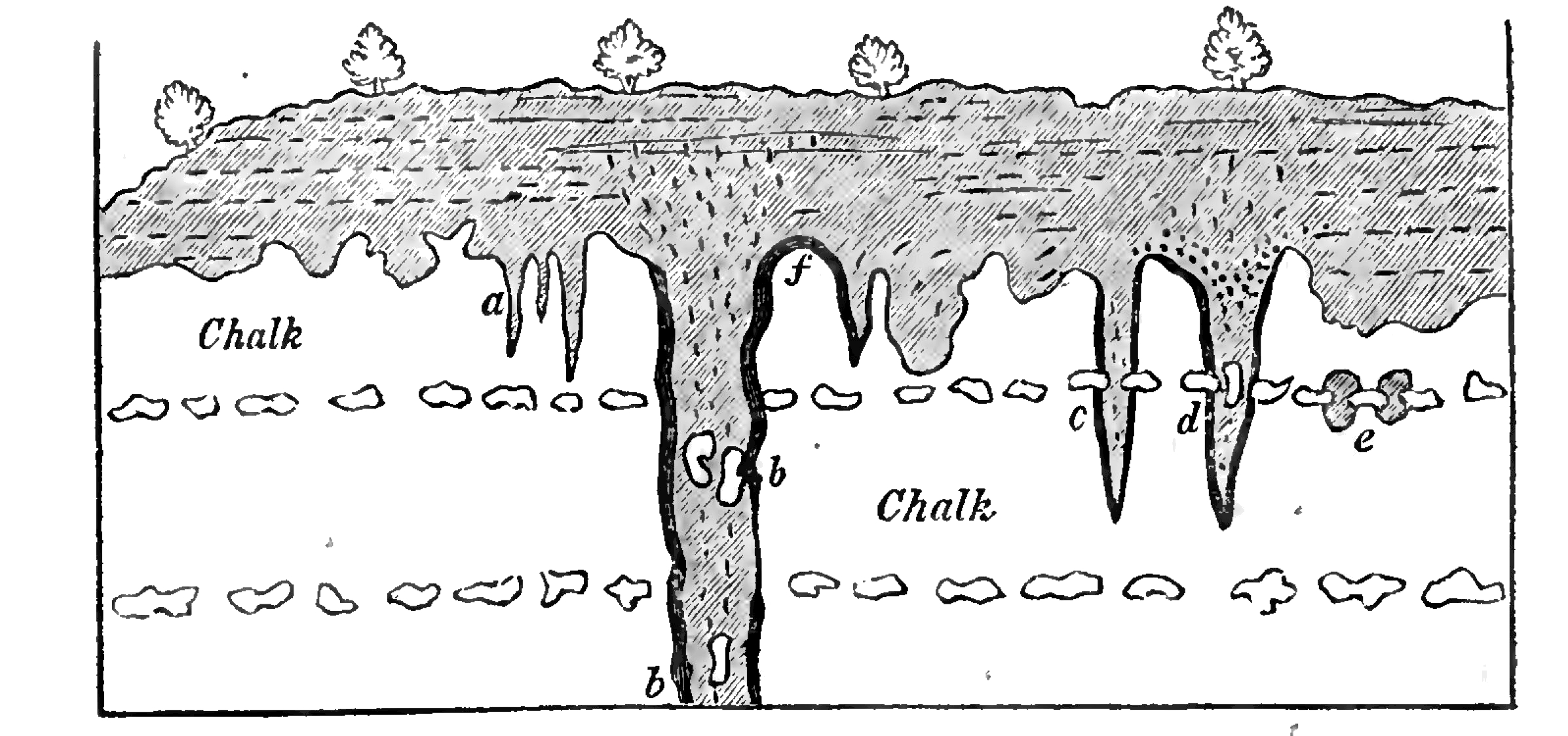}
\caption{\label{fig:Lyell}Solution pipes in  chalk at Eaton near Norwich, UK, as sketched
by~\citeA{Lyell1839}.}
\end{figure}

One of the first to describe in detail the morphology of solution pipes was Charles Lyell,
the father of modern geology. His 1839 paper \emph{On the tubular cavities filled with gravel
and sand called sand-pipes in the chalk near Norwich}~\cite{Lyell1839} remains to this day a
great example of deductive reasoning based on observations from nature. Lyell describes a
group of solution pipes in chalk strata (Fig.~\ref{fig:Lyell}), and observes that they are
elongated with lengths up to 20~m and round in cross-section. He also comments that they do
not merge, tend to stay separated from one another, and that they are not guided by the
fractures.  When the smaller pipes cross the horizontal layers of flint nodules (c \& d in
Fig.~\ref{fig:Lyell}), the nodules tend to stay in place, sticking out of the pipe, whereas in
larger pipes (b in  Fig.~\ref{fig:Lyell}) the nodules are found somewhat lower than their
original position. All of that leads him to the hypothesis that the pipes are solutional in
origin, that the chalk has been removed by the corroding action of water charged with acid,
in which the flint nodules, being insoluble, were left in situ in the smaller pipes after the
calcareous matrix had been dissolved. He also concludes from the manner in which the large
detached flints were dispersed through the contents of the widest pipes that the excavation
and filling of the pipe were gradual and contemporaneous processes. He further comments that
the fact that piping always proceeds under the cover of sand or gravel and never on the
exposed rock may be related to the fact that these upper layers provide space for the water
to focus.

Not all researchers agreed with Lyell. In particular,~\citeA{Trimmer1845} expressed the
opinion that pipes originate from the mechanical action of water, similar to potholes near
waterfalls. However, as argued by~\citeA{Prestwich1855}, this would imply that they were
empty during their formation, and thus all the insoluble material, like the flint nodules in
Fig.~\ref{fig:Lyell}, should be found at their base, which is not the case. Additionally, the
pipes do not seem to contain grinding materials required for pothole erosion (e.g.,~resistant
pebbles), Furthermore, some of the pipes have a huge aspect ratio (even 100:1), much larger
than any known potholes, making it hard to imagine the vertices of such an elongation
actively carving pipes.

The pipes in tropical and Mediterranean climates differ from those in temperate regions, as
they feature hard rims cemented with calcite (Fig.~\ref{fig:Pipes}B). In some areas, the eroded
matrix around the pipes reveals a reversed landscape, resembling a forest of vertical pipes
standing closely together. The resemblance of these pipes to tree trunks has historically led
to their misidentification as petrified trees~\cite{Bretz1960,Boutakoff1963}. This model
suggested that decaying tree trunks, rapidly buried by aeolian sand, could form conduits for
groundwater, leading to the precipitation of calcrete and forming a solid cast around the
original trunk.

However, this hypothesis was challenged by several researchers~\cite{Herwitz1993,Grimes2004},
who noted that the close spacing of the pipes, less than 0.5~m in some places, is too dense
for a forest, the bases of the pipes are rounded hemispheres without paleosol horizons or
branching root structures~\cite{Grimes2004}, and some pipes reach depths up to 20~m and are
unbranched vertical cylinders, which would imply rather peculiar, column-like trees. Despite
these objections, interpretive signs erected at Cape Bridgewater (Fig.~\ref{fig:Pipes}B) in the
early 2000s still described this as a petrified forest~\cite{Grimes2004}.

Nevertheless, it seems that the only hypothesis that has withstood scrutiny posits that the
rimmed pipes in tropical climates are solutional in
origin~\cite{Lundberg1995,DeWaele2011,Lipar2015,lipar2021}, i.e.,~their genesis is similar to
that postulated by Lyell for the chalk pipes in England, with the only difference being the
rim cementation triggered by high evaporation rates in tropical climates. Formation of the
pipes would then be linked with the reactive--infiltration instability (Sec.~\ref{VFRI}): small
inhomogeneities in the porous matrix 
tend to focus the flow, which is
followed by enhanced dissolution, which eventually transforms initial
inhomogeneities into a mature pipe.

A question remains regarding the sources of the large quantities of water that carved the
pipes. While most researchers suggest that it is CO$_2$-charged rainwater acting over
millennia, an alternative hypothesis exists for pipes formed in colder areas, which links
their formation with deglaciation.~\citeA{Morawiecka1997},~\citeA{Walsh2001}
and~\citeA{dobrowolski2015} studied pipes in Miocene calcarenite in Poland that could have
formed subglacially under the cover of a continental glacier till. Piping there is restricted
to areas beneath a till cover, and pipes did not develop where till is absent. This links the
formation of the pipes to the deglaciation at the end of the Elsterian period, when large
volumes of cold water were suddenly released into the fully saturated subglacial aquifer. In
such conditions karstification becomes very intense~\cite{lauritzen2013}. 
Related structures include
tubular concretions of authigenic carbonates that can preserve the subsurface plumbing network of  methane seep systems;  Sec.~\ref{sec:mudvolcanoes}.

An interesting issue related to the formation of solution pipes, and more broadly, to the
formation of self-organized structures in geological systems, is whether every structure
necessarily needs a precursor. A geologist would naturally lean towards an affirmative answer
to this question: since something is here now, there must have been a cause for it to form
here, rather than elsewhere. Physics, on the other hand, points to positive feedback
mechanisms underlying the onset of instabilities, which can produce ordered structures from
any small disturbance. In the case of solution pipes, this discussion has been ongoing for
several decades. Many studies suggested that pre-existing heterogeneities in rock or soil
have caused the flow to focus in a particular spot. The factors invoked included animal
burrows~\cite{Doerr2000,Devitt2002}, surface hollows, potholes or
cracks~\cite{Coetze1975,Lundberg1995,Grimes2004,Grimes2009,DeWaele2011,Lipar2015}, stemflow
and rootflow~\cite{Herwitz1993,Mitchell1995,Johnson2006,Lipar2015}. On the other hand,
numerical models~\cite{Upadhyay2015,Petrus2016,lipar2021} suggest that dissolution patterns
are largely insensitive to the initial conditions in the rock matrix. Inhomogeneities do
impact the time-scale and length-scale over which the channels are first
observed~\cite{Hanna1998,Cheung2002,Maheshwari2013,Kalia2009,Maheshwari2013a}, however in a
large enough system naturally growing instability eventually overwhelms any local
perturbation~\cite{Upadhyay2015}.

Recent experiments and theory show that solution fingers and pipes can evolve into
form-preserving bodies that simply translate without changing shape: after a brief transient
near the tip, the entire body profile becomes time-invariant and advances at nearly constant
speed~\cite{Zukowski2025}. In competitions, the winner maintains invariance while the loser
fattens and stalls, which helps to understand why a dominant pipe maintains a uniform
circular bore and why mergers are uncommon. Invariance thus acts as a shape attractor in
reactive infiltration: once the invariant regime is reached, geometry is fixed by a small set
of controls (flow and transport properties), and the tip merely translates. A practical
corollary is that the far-field side slope of the invariant body is set by those controls, so
measuring it, together with radius, provides a direct route to infer
discharge~\cite{Zukowski2025}. In this way, pipe outlines serve as paleo-flow gauges without
requiring system-specific chemistry.

\begin{figure}
\centering
\includegraphics{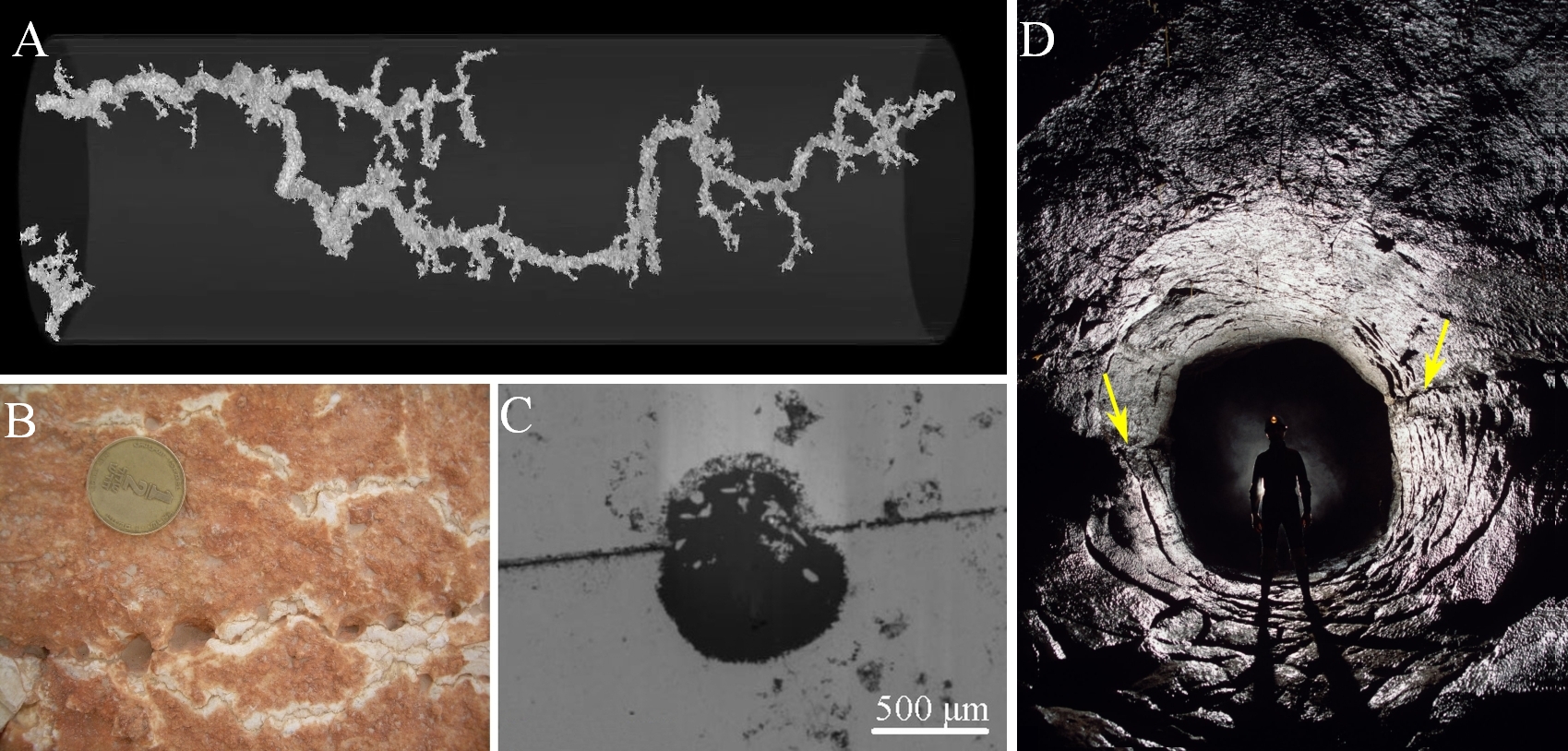}
\caption{\label{fig:Caves}Wormhole formation in porous rocks and fractures. 
A: tomographic image of a wormhole etched by carbonic acid in limestone
core. 
B: A dissolutionally
enlarged stylolite with several wormholes in a limestone wall of Mitzpe Ramon quarry, Israel.
C: Wormhole at the outlet of a fractured
limestone core. 
D: A conduit in a phreatic cave in Dan yr Ogof, Swansea Valley, South Wales. The yellow arrows mark the initial
fissure.}
\par\smallskip{\footnotesize\noindent\raggedright\emph{Sources and images:} A:~\citeA{Cooper2023}, B: courtesy of Pawel Kondratiuk, University of Warsaw,  C: courtesy of Linda Luquot,
University of Montpellier, D: courtesy of Brendan Marris, Dudley Caving Club.\par}
\end{figure}

Another manifestation of reactive-infiltration instability is wormholing, in which dissolution fingers form when a pressurized reactive solution is pushed through porous or fractured rock~\cite{Hoefner1988,Fredd1998,Cooper2023}. The distinction between solution pipes and wormholes is somewhat arbitrary, but in general the term solution pipes is used for gravity-driven dissolution fingers formed in the vadose zone near the surface and extending downwards whereas wormholes appear in fully saturated flow driven by external pressure gradients.

 Wormholes are relatively easy to obtain in experiments (Fig.~\ref{fig:Caves}A), and they have been used for at least 100~years by petroleum engineers to enhance oil and gas production by increasing the permeability of the rock. The shapes of the wormholes formed during acidization depend strongly on the flow rate, with more conical, smoother wormholes forming at lower flow rates, and highly ramified, tortuous wormholes appearing at high flow rates. Long, thin (dominant) wormholes, formed at intermediate flow rates, are the most effective for petroleum engineering, since they minimize the acid required for a given increase in permeability.
Estimates of wormhole growth rates and, in particular, the so-called breakthrough time, the moment when the longest wormhole reaches the outlet of the system, are crucial for a number of geotechnical problems. These include risk assessment of potential leakage of sequestered carbon dioxide, safety of dam sites in soluble rocks, risk of catastrophic ground subsidence due to solutional widening of fractures, and the danger of water seepage into toxic waste repositories.

Instabilities in fracture dissolution were first discovered by numerical
simulations~\cite{Hanna1998} and later confirmed by
experiments~\cite{Detwiler2003,Gouze2003}. Linear stability analysis shows that fractures are
even more unstable to dissolution than porous media~\cite{Szymczak2011}, where a minimum
wavelength is needed for unstable growth~\cite{Ortoleva1987b}. However, fracture dissolution
is almost always unstable~\cite{Szymczak2012,Starchenko2016a}, and intense wormholing is to
be expected (Sec.~\ref{sec:instabil}). The onset of a reactive-infiltration instability has been
observed in a microfluidic fracture system, with  good agreement with theoretical predictions
of the wavelength of the initial instability~\cite{Osselin2016}.

Wormholing is the driving force behind cave formation in karst
systems~\cite{Szymczak2011,Hanna1998,Dreybrodt1990}, and cave conduits are beautiful
manifestations of nonuniform dissolution. Caves are initiated along fractures and bedding
planes, which have a quasi two-dimensional structure, while the mature cave is almost always
a system of pipe-like conduits (Fig.~\ref{fig:Caves}D). 
Wormholes are formed across a range of length-scales. 
 Fig.~\ref{fig:Caves}B shows cm-size wormholes growing along a stylolite in a limestone block, while  Fig.~\ref{fig:Caves}C shows a mm-scale wormhole emerging in fracture dissolution experiments.

\begin{figure}
\centering
\includegraphics{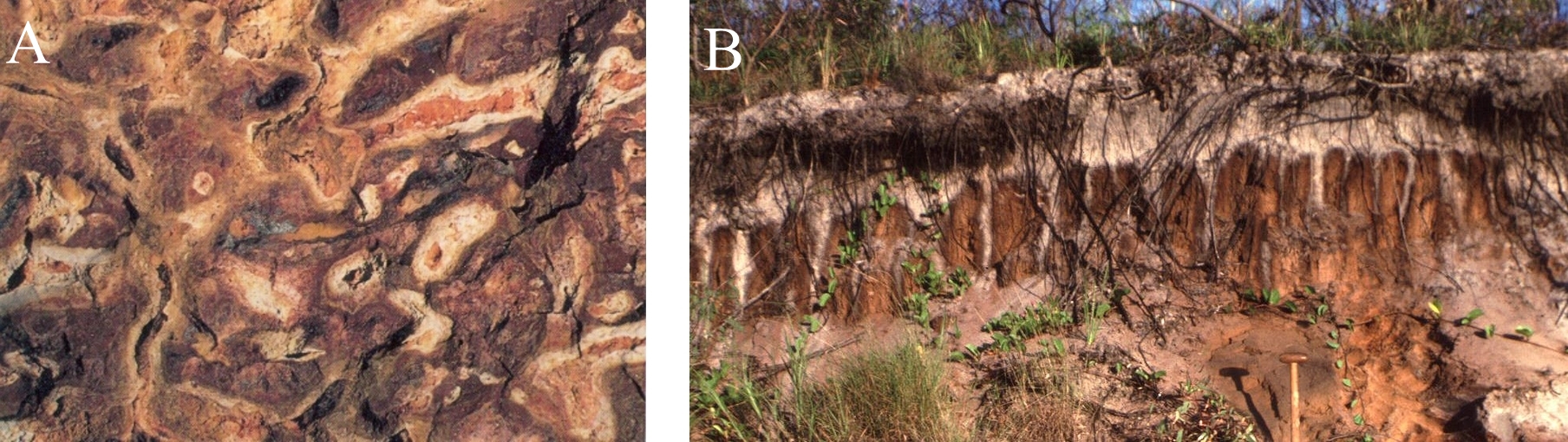}
\caption{\label{fig:laterites}A: close-up view, which spans about 1 dm vertically, of typical laterite showing tubular structures surrounded by various iron oxides~\cite{barge2015chemical}. B: Podzol tongues, K'gari Island, Australia.  
}
\par\smallskip{\footnotesize\noindent\raggedright\emph{Sources and images:} A: Werner
Schellmann,  ISRIC - World Soil Information~\cite{barge2015chemical}, B: courtesy of Ken G. Grimes.\par}
\end{figure}

Wormholes provide a way in which aggressive solutions can penetrate deeply into limestone
formations. A simple estimate of the penetration length of CO$_2$-charged water in uniform fractures gives
$l_p < 1$~m, meaning limestone caves should not exist at all~\cite{White1961}. One
possible resolution of this paradox is the sharp drop in the dissolution rate of ${\rm CaCO}_3$
near saturation~\cite{White1977}. However, wormhole formation offers a simpler
explanation~\cite{Szymczak2011}; flow in the spontaneously formed conduits (Fig.~\ref{fig:Caves})
is several orders of magnitude larger than the initially uniform flow across the fracture
width, and the penetration length increases accordingly. Numerical
simulations~\cite{Starchenko2016a} show that highly unsaturated solutions can penetrate
deeply into wormholes etched in fracture surfaces.

Dissolution processes acting at the surface also produce a variety of patterns. Karren, or lapies, are small- to medium-scale, intricate surface features in karst
landscapes, formed by the dissolution of soluble bedrock like limestone, dolomite, or gypsum
by acidic water. These miniature, sculpted landforms --- including grooves, ridges, pits, and
channels --- often appear on exposed rock surfaces forming landscapes known as karrenfields or
limestone pavements~\cite{gines2009karst}. Physical modelling has also addressed rillenkarren, parallel grooves typically formed on steep slopes~\cite{glew1980simulation,guerin2020streamwise}.

\begin{figure}
\centering
\includegraphics[width=\linewidth,height=0.78\textheight,keepaspectratio]{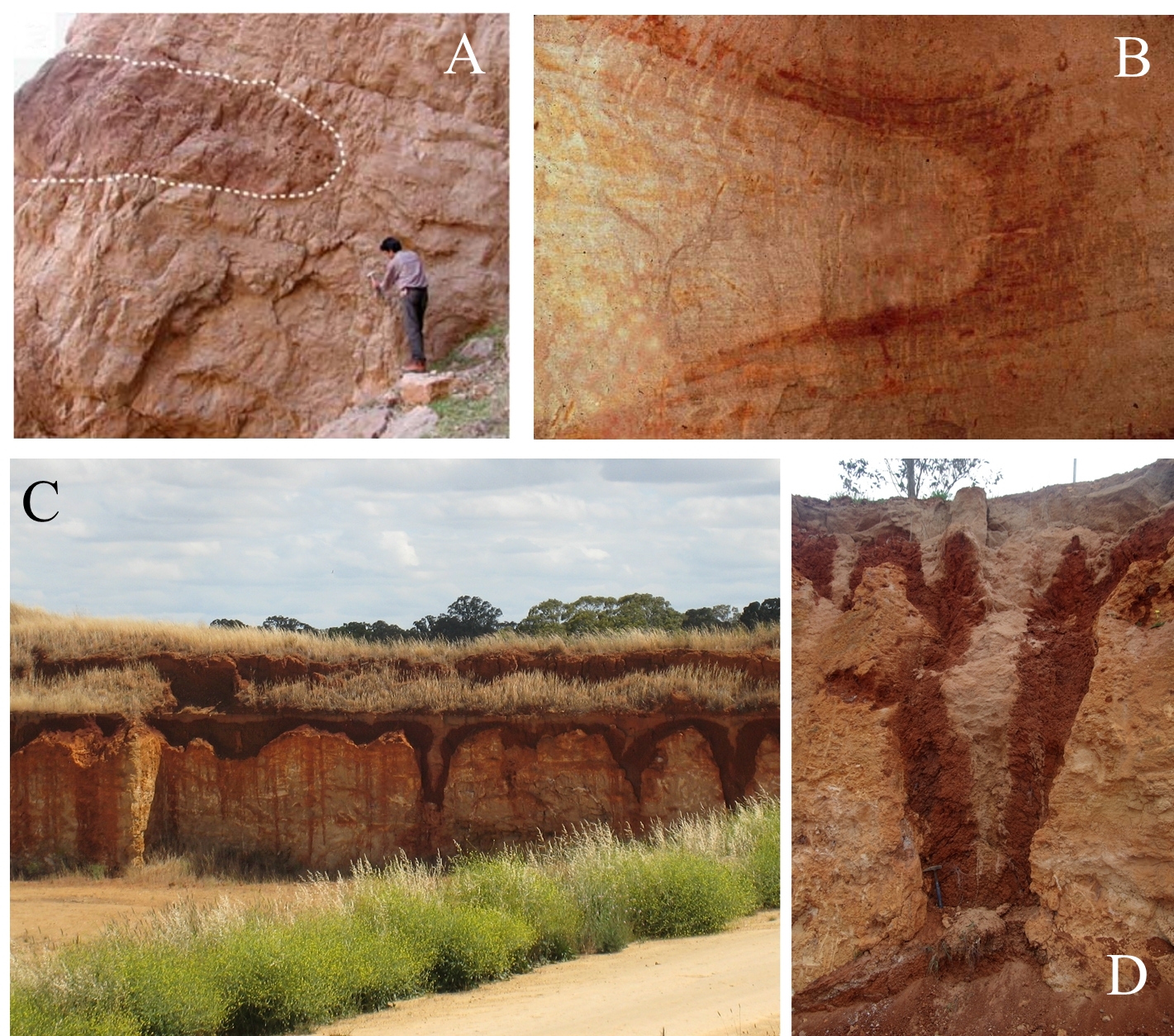}
\caption{\label{fig:Fingers}Replacement fingers. A: A dolomite finger (marked with white
dashed line) in limestone outcrop~\cite{Sharp2010}; B: Uranium roll with uraninite precipitation around its tip;
C: Terra Rossa fingers over limestone at Greatstone Winery near Coonawarra, South Australia; D: 
Closeup of one of the pipes, showing the terra rossa layer (dark red) and more recent siliceous sands overlying it.}
\par\smallskip{\footnotesize\noindent\raggedright\emph{Sources and images:} A: \citeA{Sharp2010}, 
 B: C. L. Van Alstine (U. S. Atomic Energy
Commission), courtesy of  Robert Gregory (Wyoming State Geological Survey),
C: courtesy of Les Sampson (Claremont Wines, South Australia), 
D: Piotr Szymczak.\par}
\end{figure}

Another instance of tubular void formation occurs in laterites~\cite{nahon1992ferruginous}.
Laterites are soils with the characteristics of geological soft matter, being soft yet cohesive like clay. They are often used as building material in the subtropical
regions where they are abundant. As  Fig.~\ref{fig:laterites}A shows, they commonly contain
tubular voids,  attributed to ``interstitial, chemically active pore liquids or gases
contained within the rock body, or introduced from external
sources''~\cite{aleva1994laterites}. The iron chemistry involved may be related to that of
laboratory chemical gardens and the physics to osmotic processes
(Sec.~\ref{sec:osmotic})~\cite{barge2015chemical}.

 While solution pipes and wormholes are associated with the dissolution of rock, replacement
fingers are connected with dissolution--precipitation processes, where the primary mineral is
replaced by a secondary one~\cite{Korzhinskii1968,Putnis2009,Kondratiuk2017}. Examples
include dolomitization (Fig.~\ref{fig:Fingers}A), which involves the replacement of limestone
(calcium carbonate) with dolomite (calcium magnesium carbonate). Since the molar volume of
dolomite is less than that of calcite, the primary mineral in limestone, dolomitization can
create additional pore space if the process is not accompanied by significant compaction or
cementation. The higher permeability of the dolomite then triggers instability, which can
lead to the formation of dolomite
fingers~\cite{Koeshidayatullah2020,Centrella2021,Merino2011}. Another example is the
formation of uranium rolls (Fig.~\ref{fig:Fingers}B), in which uraninite precipitates at the
redox front that separates oxidized rock from reduced rock. The oxidized rock is more porous
than the reduced one because the redox reactions produce sulphuric acid that dissolves
potassium feldspars. Increased porosity again results in instability leading to the formation
of characteristic fingers with a thin band~\cite{Dewynne1993}. Finally,
Figs.~\ref{fig:Fingers}C \& D show the terra-rossa fingers above limestone. An
intriguing hypothesis on their genesis was put forward by Merino and Banerjee~\cite{Merino2008}, who proposed that terra-rossa is formed by authigenic replacement of the underlying limestone at
a reaction front. The hydrogen ions produced in such a reaction dissolve the rock
matrix, triggering a reactive-infiltration instability which should result in intense
piping~\cite{Kondratiuk2017}.

Another manifestation of reactive-infiltration instability are   \emph{podzol tongues} in
sandy soils (Fig.~\ref{fig:laterites}B) -  downward-extending fingers of bleached eluvial horizon
material that penetrate into the underlying illuvial horizon, giving an irregular, wavy
boundary that may reach deep into the
subsoil~\cite{Lundstrom2000_PodzolReview,Schaetzl2020_Tongues}. Podzol tongues are a product of the podzolization process, which involves intense leaching of
upper horizons by organic acids, leading to translocation of dissolved organic matter, iron,
and aluminum downward in the soil~\cite{Lundstrom2000_PodzolReview}. In a typical podzol, the
surface organic layer and topsoil produce acidic compounds that percolate into the mineral
soil. These acids chelate Fe and Al from the sand and silt particles, removing pigments and
clay from the upper layer~\cite{Skjemstad1992_PodzolsII}. The result is a bleached, ash--grey
albic horizon from which metals and clays have been eluviated. As the leachate moves
downward, it encounters conditions (higher pH, reactive minerals, or saturation zones) where
the organic complexes break down and deposit their load. This forms a darker spodic horizon
enriched with illuvial humus, Fe, and Al, often appearing as a reddish-brown or black
layer~\cite{Lundstrom2000_PodzolReview}. 

Tongue formation occurs when this leaching-illuviation process is spatially non-uniform.
Instead of a level, horizontal E/B boundary, certain vertical zones experience more intensive
leaching, causing the eluvial horizon to protrude downward in finger- or tongue-shaped
extensions~\cite{Bourgault2015_HBEF}. Essentially, podzol tongues mark preferred flow
pathways where water and acids penetrate
deeper~\cite{Schaetzl2020_Tongues,Sommer2001_LateralPodzol_Sandstone}.
Chemically, the material within a tongue is highly leached: it is mostly quartz sand with
coatings removed. The edges and bottoms of a tongue tell the story of where mobilized solutes
redeposit: the centre is filled with bleached sand, the periphery is fringed by orange-brown
iron (hydr)oxide coatings, and the base of the tongue often accumulates iron-organic
compounds, forming a ferruginous or humus-rich halo at the tongue's
terminus~\cite{Schaetzl2020_Tongues}. This indicates that as acidic leachate travels down a
tongue, some iron diffuses laterally and precipitates along the margins, while further down
the remaining iron and organic matter precipitate at the tongue's end where flow diminishes
or soil chemistry changes. Thus, a podzol tongue can be viewed as a vertical conduit of
eluviation surrounded by a zone of illuviation (iron/humus staining) at its
margins~\cite{Thompson1992_PodzolsI,Bourgault2015_HBEF,Jankowski2014_LateralPodzol_Poland}.

In essence, a tongue is self-reinforcing: as eluviation strips coatings and fines, the column becomes more porous and permeable, capillary retention drops, and Darcy flow focuses into the same low-resistance path. The higher advective flux delivers more ligands and oxidants, accelerating Fe--Al stripping and deepening the bleached conduit---i.e.,~a classic reactive-infiltration instability driven by hydraulic focusing. At the same time, illuviation at the tongue rim lowers permeability in the surroundings, which further confines flow to the conduit. Structural features (roots, burrows) are not required to trigger the growth -  when present, they merely bias spacing and orientation. Growth saturates only when rim cementation chokes the path, reactants are depleted, or a textural/capillary break diverts flux laterally.

\subsubsection{Dry salt lakes and convection}
\label{sec:saltpan}

\begin{figure}
\centering
\includegraphics{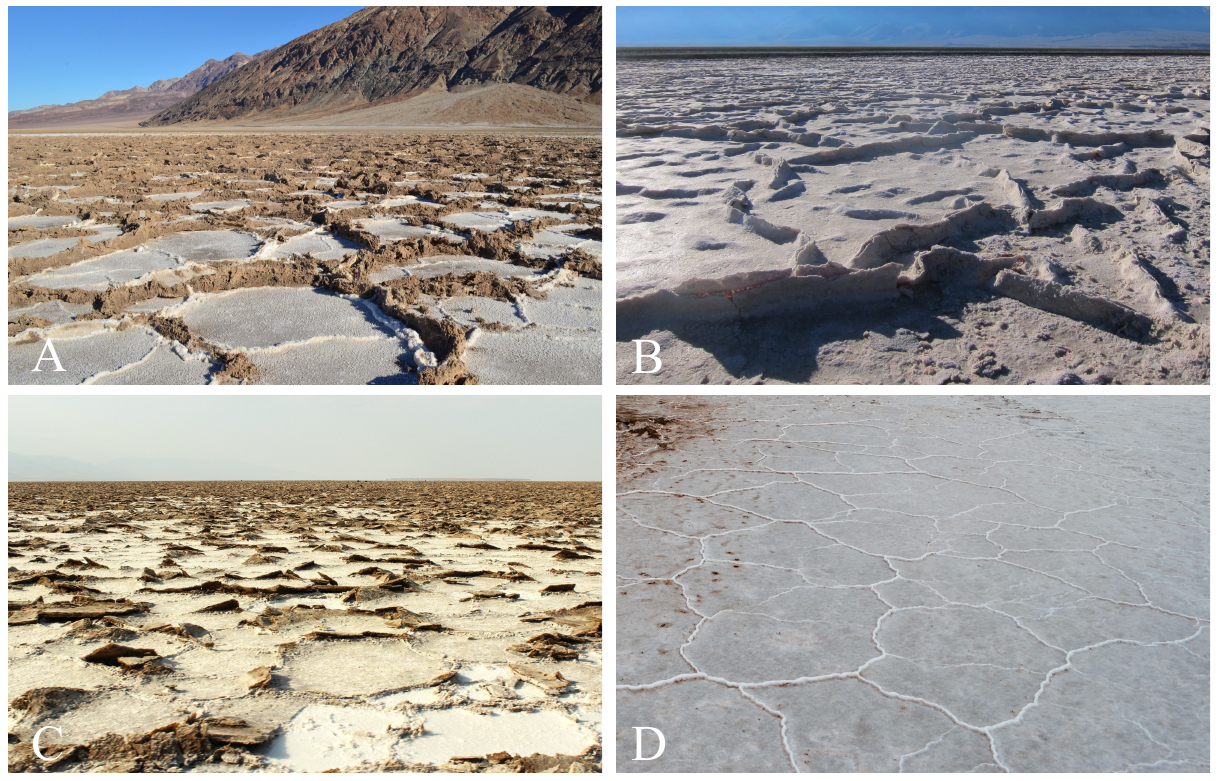}
\caption{\label{fig:saltpan}Polygonal patterns of salt frequently appear in dry salt lakes and evaporite pans.  Examples include at
A: Badwater Basin, Death Valley, California, 
B: Owens Lake, California, 
C: Danakil salt pan, Ethiopia  and 
D: the Skeleton Coast, Namibia.  In all cases, a salt crust develops on the surface of the pan, with narrow raised ridges defining the edges of the polygons.}
\par\smallskip{\footnotesize\noindent\raggedright\emph{Images:} A,B,D: Lucas Goehring; C: Electra Kotopoulou.\par}
\end{figure}

Dry salt lakes, also called evaporite pans, saline flats or playas, are found in arid
environments where evaporation of groundwater exceeds
precipitation~\cite{Lowenstein1985,Briere2000,Dixon2009}.  A common situation is a terminal
or endorheic valley with no outflow route for water, such as the case of Death Valley,
California, shown in  Fig.~\ref{fig:saltpan}A.  Evaporites can develop from the decline and
disappearance of a natural lake, like Owens Lake,  Fig.~\ref{fig:saltpan}B, which dried up after
the Owens River was diverted into aqueducts near Los
Angeles~\cite{Tyler1997,Groeneveld2010,Bertenthal2021}.  Large natural evaporite pans are
also found in arid basins including the Danakil salt plain, Ethiopia,  Fig.~\ref{fig:saltpan}C.
In coastal settings, similar features called sabkha can result from seawater evaporation, as
in  Fig.~\ref{fig:saltpan}D.  

Striking patterns of closed polygonal shapes, bounded by narrow ridges, often appear and
decorate the surface of dry lakebeds.  Here, as groundwater evaporates over time, it leaves
behind any dissolved salt that it had been carrying.  These salts accumulate at the surface
of the soil, forming into a solid salt crust, in which the patterns develop.   The polygons
typically have a diameter of a few (1--10) m, separated by 1--10~cm high ridges or
raised features~\cite{Nield2015,Lasser2020,Lasser2023}.   The surface patterns show dynamics
on time-scales of only a few months~\cite{Nield2015,Lasser2023}. 

Although beautiful, dry salt lakes produce mineral-rich dust through
erosion~\cite{Prospero2002,Klose2019}, which is detrimental to air quality and human health
and a major source of uncertainty in modelling climate sensitivity~\cite{Kok2023}. Changes in
water use policy and climate are exacerbating these problems worldwide, as saline lakes like
the Aral Sea, Dead Sea and Great Salt Lake are receding, leaving salt flats
behind~\cite{Wurtsbaugh2017}.  In this context, Owens Lake is seen as a case study for
developing remediation strategies and dust control at dry lakes, given its situation as a
man-made dry lake that was at one time the largest source of hazardous aerosols (PM-10) in
the United States~\cite{Bertenthal2021}.  

Early interpretations of the polygonal patterns in the salt crusts of dry lakes focused on a
range of mechanical instabilities of the surface crust itself, such as
fracture~\cite{Krinsley1970} or buckling~\cite{Christiansen1963}.  However, more recently an
appreciation of the hidden fluid dynamics of dry salt lakes has developed into an explanation
of the remarkably well-ordered surface shapes of this pattern~\cite{Lasser2021,Lasser2023}.
Evaporite pans are maintained by evaporation of groundwater, and the groundwater table at
active pans is usually very shallow, close enough to the surface to ensure good connection
through capillary action~\cite{Lowenstein1985,Tyler1997,Briere2000}.   Dissolved salt
concentrates in the groundwater near the surface, forming a boundary layer of salt-rich,
denser fluid.  Measurements of the density and salinity of groundwater have demonstrated this
layer, for example, in the upper ${\sim}$1~m of Owens Lake~\cite{Tyler1997,Lasser2023}.
This scenario of denser fluid resting atop a deep reservoir of less saline, and thus more
buoyant, groundwater, is inherently unstable.  

Since dry salt lakes occur in desert areas, thermal fluctuations might also be expected to
contribute to their dynamics.  In contrast to the more well-known double-diffusive fingering
structures (Sec.~\ref{sec:double-diffusive}) that produce stratification and `salt-fingers' in
the ocean~\cite{Huppert1981,Schmitt1995}, however, thermal buoyancy does not appear to be
relevant to dry lake convection: based on temperature logs and groundwater sampling at Owens
Lake, the density contrast due to salinity is estimated to be two orders of magnitude larger
than that caused by thermal expansion~\cite{Lasser2021}.   A significant contribution,
instead, might arise from the effects of temperature on the hydration states of some salts.
For example, sodium sulphate can swell in volume by up to 300\% when it changes from an
anhydrous mineral (thenardite) to a hydrated one (mirabilite)~\cite{Flatt2014}.  This phase
transition occurs at temperatures and relative humidities routinely experienced in dry lake
environments, but its effects on the salt crusts have not yet been explored.

\begin{figure}
\centering
\includegraphics{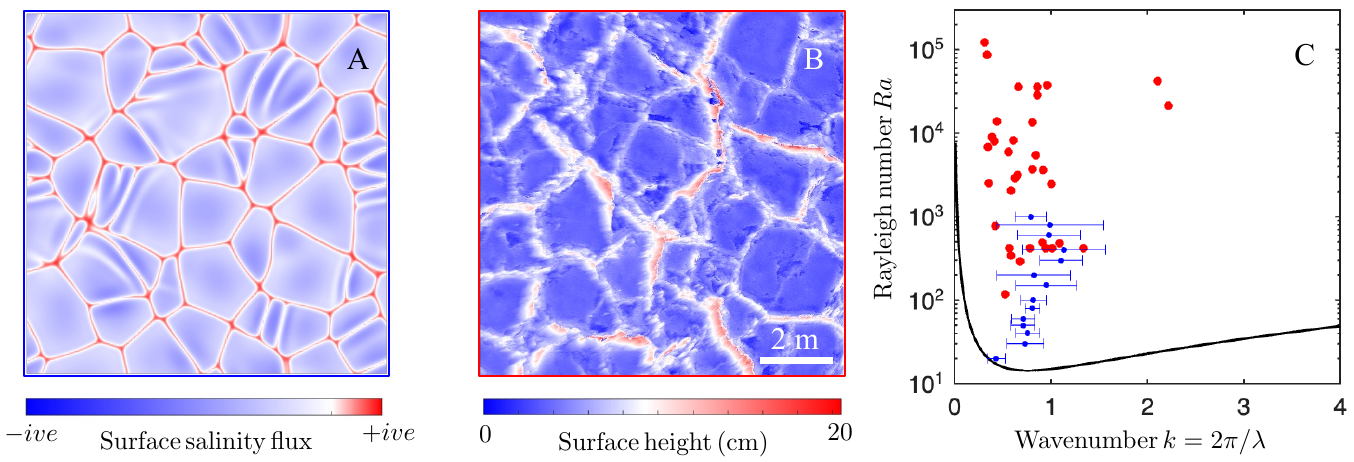}
\caption{\label{fig:saltpan2} Characteristic patterns and length-scale selection in dry salt lakes.  
A: Results of a simulation of the dry salt lake problem given by
Eqs.~\eqref{eq:mass}--\eqref{eq:darcy}, and a constant through-flow driven by surface
evaporation.  Colours give the pattern of surface salinity flux, with red indicating areas of
enhanced salt crust growth, and blue indicating where crust growth is inhibited.  The surface
shown has size 12$\pi$.  In dimensional terms, for a natural length of 0.15
m~\citep{Lasser2023}, this corresponds to a square of approximately 6~m in size.  (Image:
Matthew Threadgold and C\'edric Beaume.)
B: Surface height profile map of the natural salt crust at Badwater Basin~\citep{Lasser2021}. 
C: The dominant wave-number $k = 2\pi/\lambda$ of the patterns arising from dry salt lake models
(blue) and field data (red).  Linear stability analysis predicts the neutral stability curve
(solid black line), above which the model is unstable.  Data reproduced
from~\cite{Lasser2023}.}
\end{figure}

The soil of a dry salt lake can be considered as a porous medium, saturated in water.
Density-driven convection in porous media is a well-studied topic, and the subject of several
recent, detailed reviews~\cite{Hewitt2020,Nield2017}.  In the context of a dry lake, a
simplified model considers evaporation at a rate $E$ occurring at the upper surface
of the ground, while water is recharged from below, or from some distant reservoir, at a rate
just sufficient to balance the evaporative losses~\cite{Wooding1997a,Lasser2021}.  These
boundary conditions lead to the problem of one-sided convection with a through-flow of fluid.
Dissolved salt accompanies the flow, but is left behind by evaporation, accumulating above
the soil surface in a salt crust, and near the surface in salt-rich fluid.   In dimensionless
terms, the system of equations taken to model this system is
\begin{eqnarray}
\nf&\nabla\cdot\mathbf{u} = 0,
    \label{eq:mass}\\
\nf& \displaystyle\frac{\partial S}{\partial t}+\mathbf{u}\cdot\nabla S=\nabla^2 S,
    \label{eq:salinity}\\
\nf& \mathbf{u}=-\nabla p+Ra\,S \hat{z},
    \label{eq:darcy}
\end{eqnarray}
which describes the flow of an incompressible fluid in a porous medium, with some velocity $\mathbf{u}$, accompanied by the transport of dissolved salt that moves along with the fluid flow, and by diffusion.  The salt concentration of the groundwater is represented by the salinity $S$. This is bounded by the background salt concentration present in the reservoir source ($S=0$), and the saturation limit ($S=1$) at which dissolved salt will precipitate out as a solid.  Salinity increases the density of the water, and these buoyancy forces are represented on the right hand side of Eq.~\eqref{eq:darcy} in the Boussinesq approximation, along with any other contributions to the fluid pressure $p$.

The system of equations given above was first studied for the analogous but thermally-driven
case of a geyser~\cite{Wooding1960}, and later adapted to the setting of a dry salt
lake~\cite{Wooding1997a,Wooding1997b}.  In this model, the stability of the salt-rich,
near-surface boundary layer is controlled by a dimensionless Rayleigh number, $Ra=\kappa\Delta\rho g/\mu E$.
Here, $\kappa$ is the permeability of the soil, $\Delta\rho$ is the density contrast
between reservoir fluid and salt-saturated water, $g$ is the magnitude of
acceleration due to gravity,  $\mu$ is the viscosity of water, and $E$ is the
evaporation rate in terms of a volume flux of water.  The Rayleigh number describes the
balance between diffusive and advective effects.  It is also the ratio of the speed at which
a large denser plume of fluid will fall under its own weight, to the speed of the flow moving
upward through the soil to replace water lost to evaporation.  The steady-state boundary
layer of the dry salt lake model is unstable to convection when $Ra$ exceeds about 7 when the
upper surface is modelled with a uniform pressure boundary condition~\cite{Wooding1960}, or
14.3 for the case of a uniform vertical flow~\cite{Homsy1976}. Analogue laboratory
experiments in Hele-Shaw cells have confirmed the critical value for
convection~\cite{Wooding1997a}, and explored the longer-term convective
dynamics~\cite{Lasser2023}.  Similarly, linear stability analyses have been made for an
initially uniform lake, to investigate the time needed to build up a sufficiently thick
boundary layer that will be unstable to
convection~\cite{Lasser2021,Bringedal2022,Bringedal2025}. At several dry lakes and sabkhas
around the world, Rayleigh numbers have been estimated to be between about
$10^2$--$10^5$~\cite{vanDam2009,Lasser2023}, showing that groundwater should be
actively convecting at these sites. Indeed, large, denser plumes of salt-rich water have been
directly observed near the surface of sabkhas using electrical conductivity
measurements~\cite{vanDam2009,Stevens2009}.  

Simulations of the groundwater flows in dry salt lakes have shown that the pattern of the
convective cells that develops near the surface of a modelled lakebed strongly resembles the
patterns seen in natural salt crust~\cite{Lasser2021,Lasser2023}, as demonstrated in
Fig.~\ref{fig:saltpan2}A,B. These models also predict that the typical size of convective
features is governed by a characteristic scale given by the ratio of the diffusion constant
of salts in the groundwater, to the volumetric evaporation rate.  These predictions, along
with data extracted from surface profilometry of dry salt lakes, are shown in
Fig.~\ref{fig:saltpan2}C.  Since the typical diffusivity of salts in water is around
10$^{-9}~m/s^2$, while the groundwater evaporation rates measured in active dry salt lakes are
of order 0.1~mm/day or 10$^{-9}~{\rm m/s}$, the characteristic scale that emerges in the model
system is of order 1~m in most cases, close to what is observed at places like Badwater
Basin, Owens Lake, and Sua Pan~\cite{Lasser2023}.   Furthermore, the models can predict rates
of salt crust growth by estimating the salt flux into the surface from a balance between
advection and diffusion.  In this context, the polygonal salt ridges should appear over
downwelling sheets of salt-rich water, as has been confirmed by mapping out the groundwater
salinity through direct sampling~\cite{Lasser2023}.  Under the field conditions typical of
Owens Lake, convective transport of salt has been estimated to contribute to growth rates
about 1~mm/month, which is comparable to the growth rates observed at similar dry lakes in
Botswana~\cite{Nield2015}.  Since more salt should enter the crust above the salt-rich
downwelling flows, these models also predict that salt ridges should grow about twice as fast
as the rest of the crust~\cite{Lasser2023}.  

As explored further in  Sec.~\ref{sec:roses}, the growth and internal structure of salt crusts
can be quite complex.  A better understanding of how water and salt move through the crust,
and of the feedback between crust growth and local evaporation, are likely to be the next
steps in further developing predictive models of dry salt lakes.  The effects of an
unsaturated capillary fringe have also recently begun to be
investigated~\cite{Liu2023,Bringedal2025}, as have the effects of more efficient convective
flows within any cracks in the ground, as might be caused for example by desiccation or a
cycling of water availability~\cite{Haque2024}.

\subsubsection{Desert roses and other efflorescence structures}
\label{sec:roses}

\begin{figure}
\centering
\includegraphics{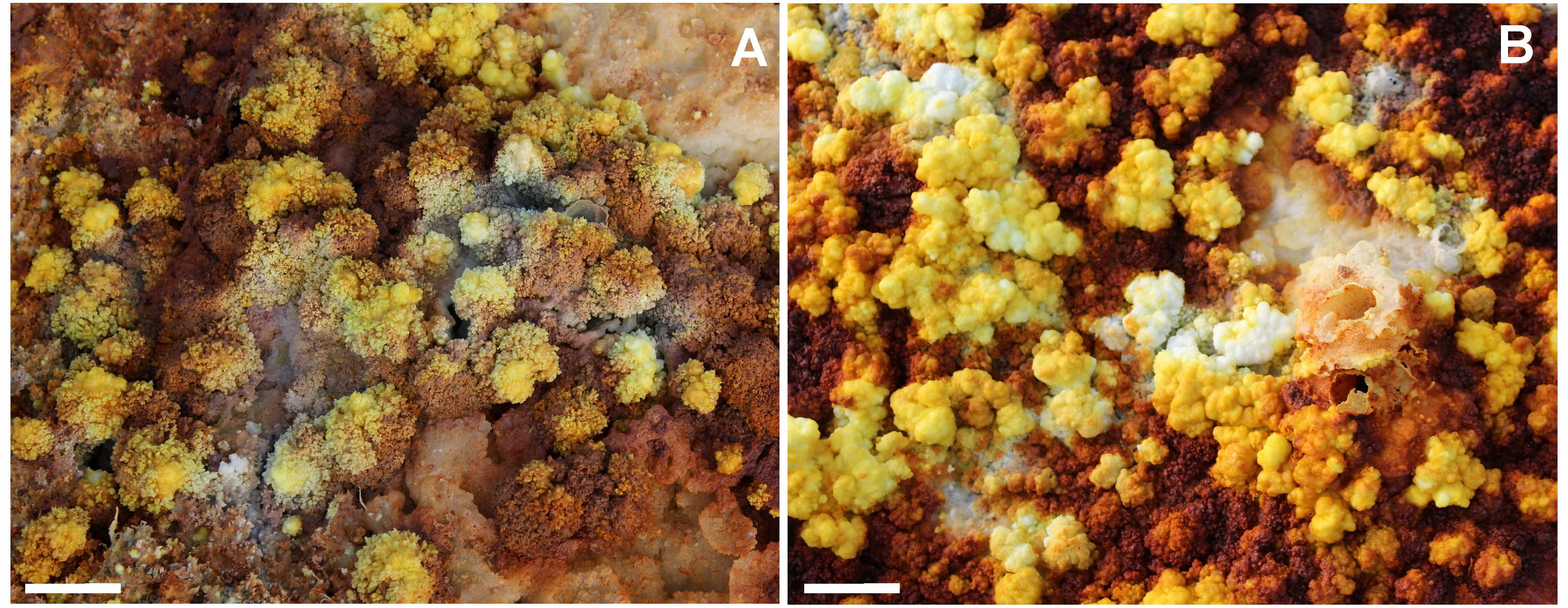}
\caption{\label{fig:Dallol_efflorescence_1}Salt efflorescence at the acidic hydrothermal
field of Dallol, Ethiopia~\cite{Kotopoulouetal.2019}. Dendritic (A) and cauliflower (B)
efflorescence of gypsum 
(CaSO$_4$$\cdot$ 2H$_2$O), sylvite (KCl), carnallite (KCl$\cdot$MgCl$_2$$\cdot$
6H$_2$O), jarosite (KFe$_3$(SO$_4$)$_2$(OH)$_6$) and
iron-oxide minerals. Scale bars are 3~cm.}
\par\smallskip{\footnotesize\noindent\raggedright\emph{Source:} \citeA{Kotopoulouetal.2019}.\par}
\end{figure}

\begin{figure}
\centering
\includegraphics[width=\linewidth,height=0.78\textheight,keepaspectratio]{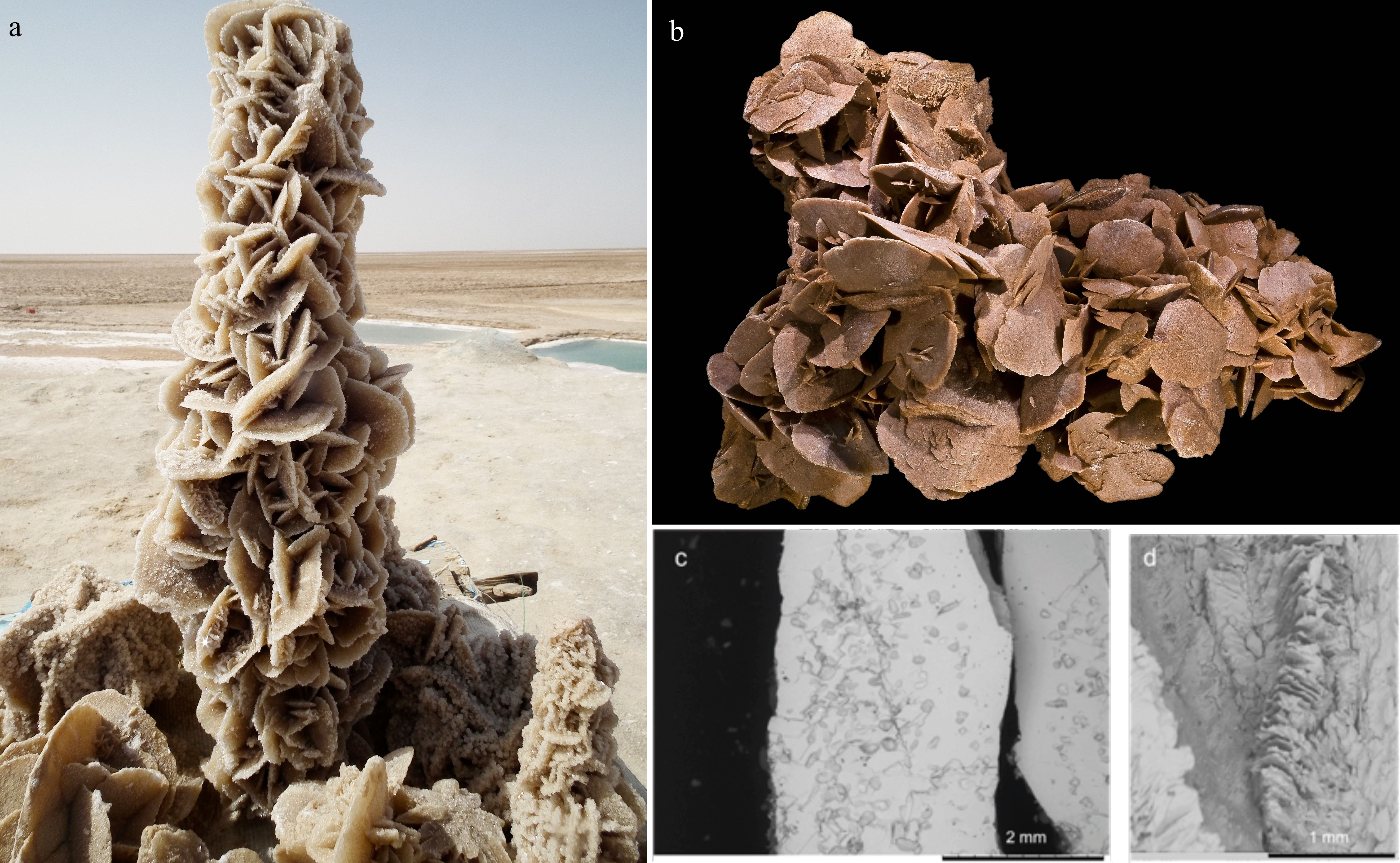}
\caption{\label{fig:creeping2}
Desert roses formed by crystallization viewed at different scales. 
A: Large desert rose formation in the Tunisian desert. 
B: Saharan gypsum desert rose from Tunisia (47x33 cm). 
C: SEM image of sand grains entrapped between two broken crystalline petals. 
D: SEM image of the internal structure of one petal showing a self-organized crystallization at an even smaller scale. }
\par\smallskip{\footnotesize\noindent\raggedright\emph{Images:} A: Laura Pe\~na; CC BY-SA 3.0, B: Didier Descouens; CC BY-SA 4.0; 
C--D: Noushine Shahidzadeh.\par}
\end{figure}

The presence of salts in natural environments can create spectacular landscapes but can also pose significant threats to ecosystems and historical
artefacts~\cite{pitman2002global,cooke1968salt,shahidzadeh2010damage}. With environmental
fluctuations such as rainfall or temperature and relative humidity changes, salt present in
granular materials --- consolidated or unconsolidated --- such as stones or soil can dissolve and
form salt solutions that can flow. Consequently, the salts can be transported and precipitate
when the water  evaporates at the surface~\cite{desarnaud2015drying}. This is the origin of
salt creeping, a phenomenon in which salt crystals continue to precipitate far from an
evaporating salt solution
boundary~\cite{washburn2002creeping,huang1976creeping,hird2016migration,qazi2019salt}. Salt
creeping can initiate and grow on flat surfaces but also on top of porous materials such as
soil, sand and stones; this is called efflorescence;
Fig.~\ref{fig:Dallol_efflorescence_1}~\cite{desarnaud2015drying,qazi2019drying,veran2012discrete}.
The latter can, for example, occur with ground water in soils or at salt lakes. Salt creeping
can  result in spectacular structures in natural environments, such as the formation of
desert roses in arid regions,  Fig.~\ref{fig:creeping2}A,B, and salt crystallization motifs near
the Black and Dead Sea coasts. In  addition, salt creep can lead to soil salinization,
vegetation decline, and water quality issues, impacting biodiversity~\cite{pitman2002global}.

Salt creeping is a general phenomenon that happens for different salts under conditions of
low relative humidity and is affected by the presence of impurities and other
microcrystallites that provide secondary nucleation sites near the evaporation
front~\cite{qazi2019salt}.  As salt crystals are wettable by their own salt solutions, such
multiple nucleation sites provide a platform for a film of salt solution to spread even
further over the newly formed microcrystallites, well beyond the initial evaporation
front~\cite{hird2016migration,qazi2019salt}. This in turn enlarges the evaporative area,
which leads to a faster precipitation, causing an exponential increase in the precipitating
crystal mass in time. This self-amplifying process results in spectacular three-dimensional
crystal networks at macroscopic distances from the salt solution source.  Microscopic
analysis of the 3D network of such efflorescence reveals an internal porous structure of the
crystalline salt network with a fractal structure characterized by a self-organized,
self-similar arrangement of salt crystals, as we discuss in  Sec.~\ref{sec:capillary}. 

Laboratory and field observations on different salts suggest that macroscopic patterning varies between salts because different minerals have different crystal structures, reflecting the seven crystal systems. Desert roses typically consist of
calcium sulphate (gypsum) or barium sulphate (barite) with sand grains (SiO$_2$)
interspersed between the crystalline structures. When investigated in detail, desert roses
reveal a spectacular self-organized structure: large petals, smaller petals in between these
and even smaller ones between those, in a fractal organization. A close-up of the structure
by SEM imaging,  Fig.~\ref{fig:creeping2}C,D, reveals that each petal itself is composed of
parallel smaller leaflet crystals, emphasizing the universality of this pattern and the
internal porous structure of the 3D salt network~\cite{Wijnhorst2023}. These  structures can grow to 1 to 3~m in height. 

Salt creeping is therefore both universal and self-amplifying, typically initiated by the presence of salt crystals or impurities. It hinges on the formation of numerous crystallites under low humidity conditions, with impurities often serving as nucleation centres or salt mixtures catalysing this spectacular self-organized process.
 Addressing salt creep is vital for preserving our geo- and cultural
heritage~\cite{brocx2010geoheritage}, calling for further research and interdisciplinary
collaboration to develop sustainable solutions for their conservation.
 On the other hand,  when salt crystallization occurs at the pore scale within the porous
materials it can induce weathering and damage outdoor historic and archaeological sites such
as Petra in Jordan~\cite{heinrichs2008diagnosis}, or Pharaonic sandstone monuments in Luxor,
Egypt~\cite{fitzner2003weathering},  and more generically our outdoor  built cultural
heritage.

\subsubsection{Salt weathering, honeycomb weathering, and tafoni}
\label{sec:honeycomb}

\begin{figure}
\centering
\includegraphics{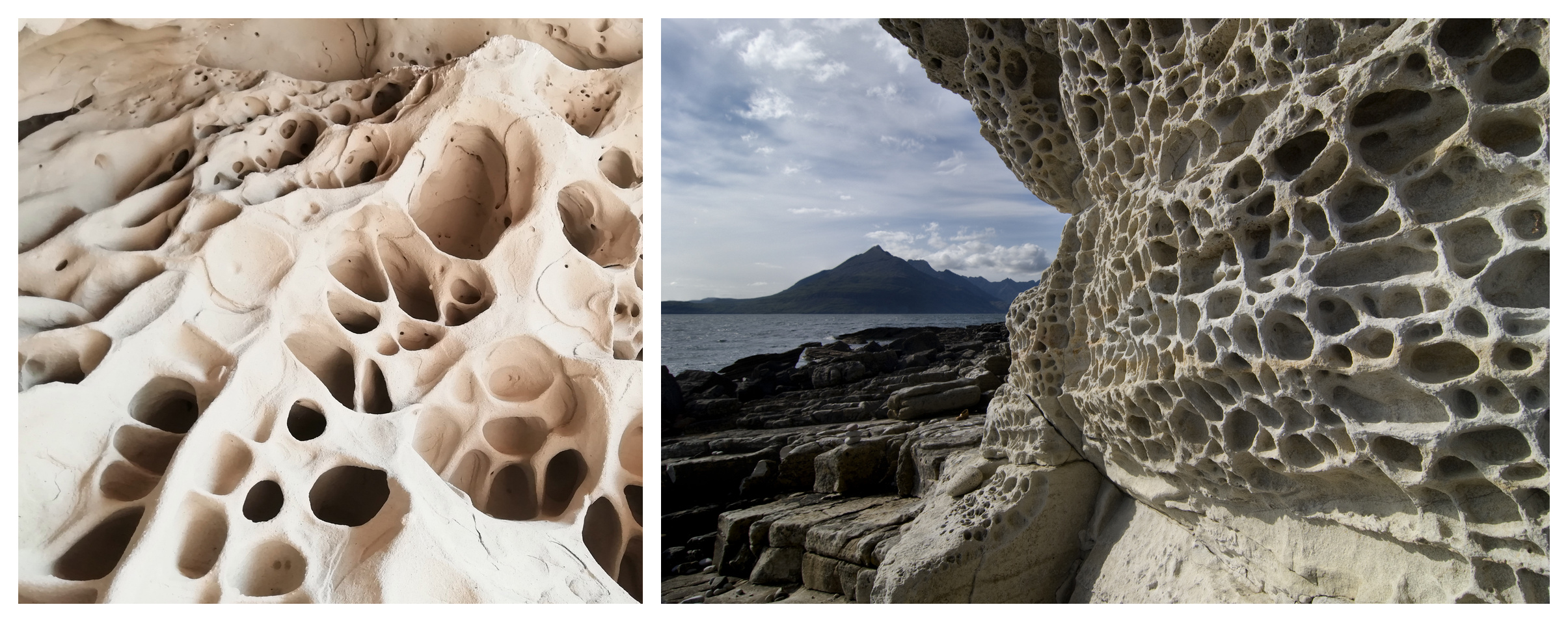}
\caption{\label{fig:tafoni}Salt weathering: 
honeycomb/tafoni patterns.
Left:  Le Grotte di Soprasasso, Bolognese Apennines, Italy,
Right:  Elgol, Isle of Skye, UK.}
\par\smallskip{\footnotesize\noindent\raggedright\emph{Images:} Left:   Elena Tartaglione; CC-BY-SA-4.0;
Right:   Kalense;
CC-BY-SA-3.0.\par}
\end{figure}

Salt weathering, haloclasty, honeycomb weathering, and tafoni,  Fig.~\ref{fig:tafoni}, are terms that describe a set of related geological processes and features.
Salt weathering~\cite{wellman1965salt,doehne2002salt}, also known as haloclasty, is a suite
of physical weathering processes in which dissolved salts infiltrate pore and fracture
networks and then crystallize and/or hydrate, generating stresses within rock pores or
fractures. This phenomenon is particularly common in arid and semi-arid regions where
evaporation rates are high, and salt solutions can become concentrated, as well as in coastal
environments where salt spray is present. In both settings, saline solutions enter the rock;
concentration by drying or cooling leads to nucleation and growth of salt crystals, and
crystal growth or hydration cycles exert crystallization pressures on pore walls. Over time,
this stress field causes the rock to break apart, leading to granular disintegration,
flaking, and the emergence of characteristic surface relief.

Honeycomb (alveolar, cavernous) weathering is characterized by small, closely spaced cavities
resembling a honeycomb pattern on rock surfaces~\cite{mustoe1982origin,rodriguez1999origins}.
It is often found in coastal and desert environments where salt spray or other saline
moisture sources are present and is typically associated with sedimentary rocks, such as
sandstones, which have high porosity. Salt weathering is widely regarded as a dominant driver
of honeycomb formation, acting together with microclimatic gradients and case-hardening of
surface crusts, though additional processes may contribute. 

Tafoni are rounded, cave-like features found on the surfaces of granular rocks such as
sandstone, granite, and basalt. Tafoni structures can range from small pits to large
hollowed-out spaces. There is no clear consensus on the threshold between honeycomb
weathering and tafoni~\cite{groom2015defining}. Usage in the literature is primarily
morphological:  \emph{honeycomb} is typically used for centimetre-scale cellular pits and
\emph{tafoni} for decimetre- to metre-scale alcoves, with overlap between categories.  There
are a number of modelling studies of honeycombs and tafoni~\cite{huinink2004simulating}. In
addition to salt-weathering feedbacks,
a Turing-type mechanism has been proposed for the formation of these honeycomb/tafoni features
\cite{mcbride2004origin}, although its applicability likely depends on lithology, transport
regimes, and environmental forcing.

\subsubsection{Sedimentary crack patterns}
\label{sec:crackpatterns}

\begin{figure}
\centering
\includegraphics{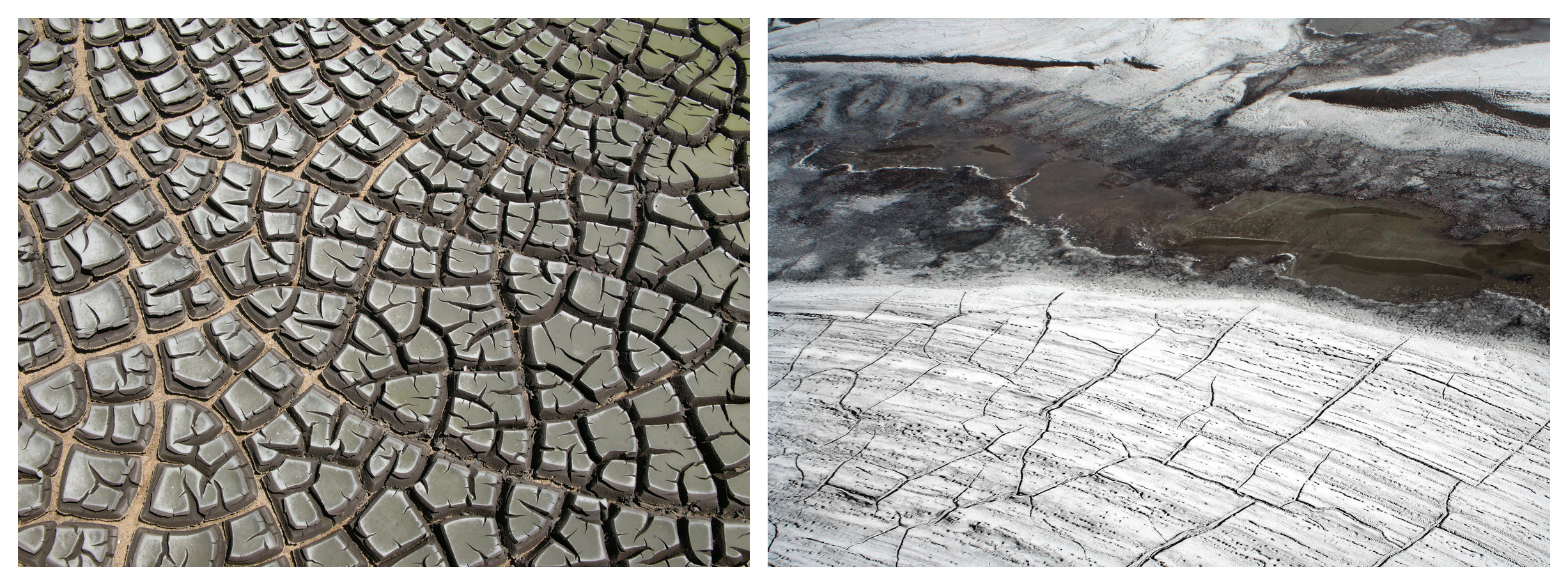}
\caption{\label{fig:sedimentary_cracking}
Sedimentary cracking patterns:
Left: cm-scale desiccation cracks in dried sludge,
Right: hm-scale ice-wedge polygons in permafrost.
}
\par\smallskip{\footnotesize\noindent\raggedright\emph{Images:} Left: 
 Hannes Grobe; CC-BY-SA-2.5;
Right:  
 Brocken Inaglory; CC-BY-SA-3.0.\par}
\end{figure}

Fractures in soft sedimentary layers arise when the material accumulates tensile strain from
drying, cooling, or differential shear. Once nucleated, cracks propagate and link to relieve
the stress, assembling plan-view networks;
Fig.~\ref{fig:sedimentary_cracking}~\cite{weinberger1999initiation}. The resulting morphologies
range from polygonal meshes --- often near-hexagonal or orthogonal --- to branched, hierarchical
patterns. Outcomes depend on composition, moisture content, layer thickness, substrate
adhesion, and the drying/cooling history~\cite{plummer1981shrinkage}. Their geometry is
amenable to scaling descriptions: stress build-up and release set characteristic spacings and
size distributions, and many fracture systems display scale-invariant
statistics~\cite{bonnet2001scaling}.
Two canonical examples illustrate the range. In desiccating mud
(Fig.~\ref{fig:sedimentary_cracking}, left), evaporation drives surface shrinkage; tensile
stresses concentrate near the surface, cracks nucleate, and the network expands as drying
proceeds~\cite{zhao2014growth}. In permafrost (Fig.~\ref{fig:sedimentary_cracking}, right),
repeated winter cooling produces thermal-contraction cracking that, over years to millennia,
organizes into ice-wedge polygons~\cite{lachenbruch1962mechanics}.

Beyond these, several sedimentary settings develop crack networks through internally driven
volumetric strain under burial, dewatering, or early cementation. Polygonal fault systems in
fine-grained marine tiers form during early burial and dewatering, producing layer-bound,
near-planar meshes that are largely independent of far-field tectonic stress. Spacing scales
with tier thickness and mechanical layering; modelling and seismic mapping document a robust
wave-length---thickness relation and arrest at mechanical
boundaries~\cite{Cartwright2003,Hansen2004}. 
Subaqueous, syneresis cracks record shrinkage and dewatering in cohesive muds below water.
Unlike classic mudcracks, they may be discontinuous, cut lamination at high angles, and show
meniscate voids; proposed drivers include salinity shocks (deflocculation), slow compaction,
and earthquake-induced dewatering with sand/silt injection that later fills the
fissures~\cite{Pratt1998,Mcmahon2017}.
In peritidal carbonates, early cementation and displacive crystal growth generate in-plane
expansion of thin crusts; the result is buckling, ridge-and-polygon or \emph{teepee}
topography, and associated breccias. The dominant wave-length reflects crust thickness,
stiffness, and substrate coupling, and in some successions rapid, earthquake-triggered
failure has been inferred~\cite{Kendall1987,Pratt2002}.
Lastly, organic-rich shales develop bedding-parallel and cross-cutting microfracture networks
during thermal maturation. Imaging and experiments show cracks nucleating near kerogen
domains as internal pressure rises during conversion and then coalescing into percolating
pathways; spacing and connectivity reflect kerogen distribution, mechanical heterogeneity,
and maturation kinetics~\cite{Han2022,Kobchenko2011}. 

These examples fit naturally into a scaling viewpoint: in burial-compaction tiers and cemented crusts the dominant wave-length scales with an active-layer thickness $H$ (tier or crust thickness), whereas in maturation-driven shales it additionally reflects kerogen patch size and pressure-generation rates; in all cases, boundary conditions and mechanical layering set arrest lines and pattern coarsening pathways.

\subsubsection{Geodes}
\label{sec:geodes}

\begin{figure}
\centering
\includegraphics{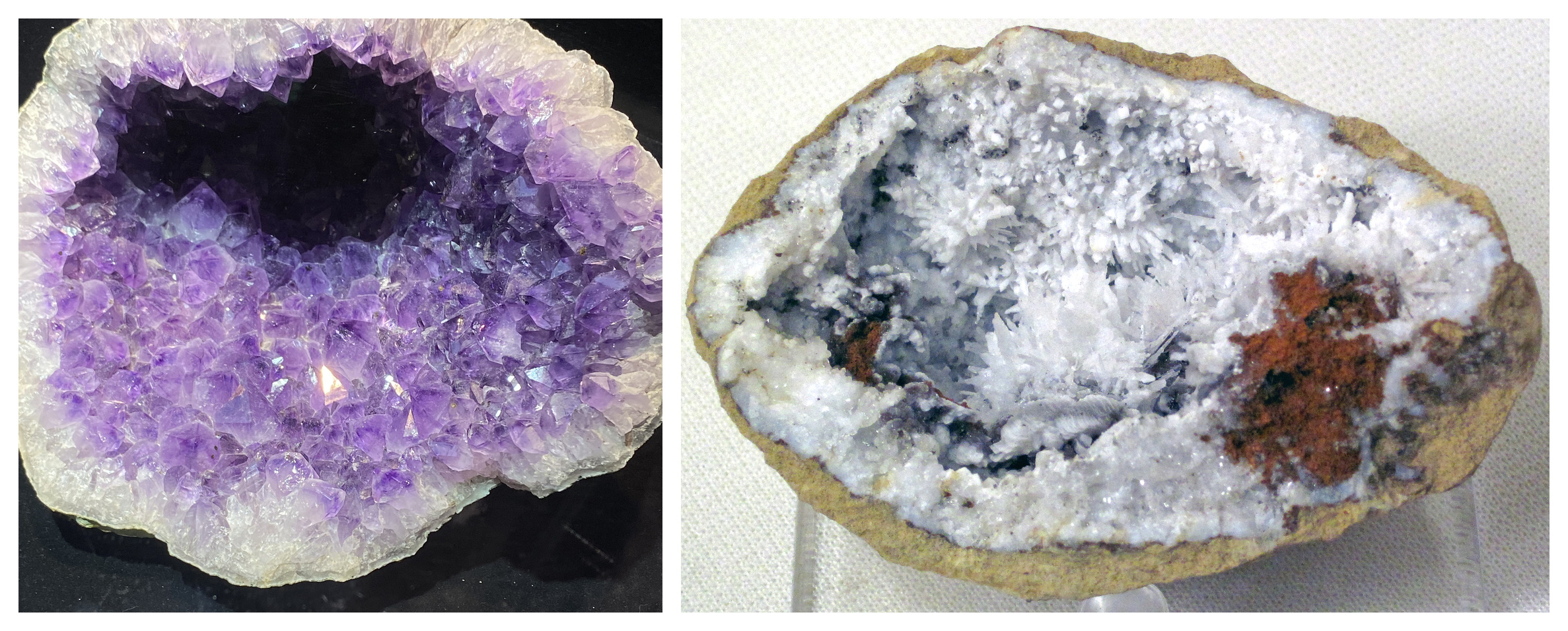}
\caption{\label{fig:geodes}Geodes.
Left: 
 Amethyst geode,
Right: 
 Geode with aragonite, Bloomington, Indiana, USA.
}
\par\smallskip{\footnotesize\noindent\raggedright\emph{Images:} Left: Madeleinemcc; CC-BY-SA-3.0;
Right:  James St. John; CC-BY-2.0.\par}
\end{figure}

Geodes,  Fig.~\ref{fig:geodes}, are hollow, spherical or oval rock cavities lined with
inward-facing crystals or mineral deposits~\cite{bassler1908formation,rakovan2017word}. They
can range in size from a few centimetres to over a metre in diameter. The largest known
geode, located in Pulp\'\i, Almer\'\i a, Spain,  measures approximately 8~m long, 2~m wide,
and 2~m high~\cite{Fernandez2006Pulpi}.
Geodes occur in a variety of geological environments; they
are commonly found in areas with volcanic activity and in sedimentary rock regions where water has caused dissolution of rocks, such as limestone and shale. Their hollow interiors are often lined with minerals such as quartz, amethyst, calcite, agate, and other silicate or carbonate minerals.

The formation of geodes occurs in two major stages: cavity creation and mineral deposition.
Geodes begin as hollow cavities within rocks, which can form through several processes:
In volcanic environments, molten lava can trap gas bubbles as it cools. These bubbles, or vesicles, form empty spaces within the solidifying rock. Over time, these cavities serve as the initial space for mineral deposition.
In sedimentary environments, cavities can form due to the dissolution of minerals (e.g.,~limestone or gypsum) by groundwater. Acidic groundwater can dissolve portions of the rock, leaving behind voids that become the basis for geode formation.

Once a cavity exists, the second stage involves mineral deposition on the inner walls. 
Groundwater or hydrothermal fluids, often rich in dissolved ions, infiltrate the cavity. Over time, the minerals precipitate and crystallize along the inner walls of the cavity.
As mineral-laden water continues to flow through the cavity, crystals slowly grow inward. Depending on the local geochemistry, different minerals precipitate, creating colourful and varied geodes. Quartz is the most common mineral, but other varieties, such as amethyst (a purple form of quartz), calcite, and agate, are also frequent.

Geodes often display banding of layers of different minerals, as changes in water chemistry or temperature over time cause sequential deposition of different minerals. This can result in intricate patterns of colours and textures.
Quartz geodes are the most common and feature clear or white quartz crystals lining the cavity. Some may also contain amethyst (purple quartz) or smoky quartz (grey-brown quartz).
Calcite is another common mineral found in geodes, often forming larger, blockier crystals compared to quartz.
Some geodes have a lining of banded agate, which forms concentric layers of microcrystalline quartz around the inner walls (Sec.~\ref{sec:agates}).

\begin{figure}
\centering
\includegraphics[width=\linewidth,height=0.78\textheight,keepaspectratio]{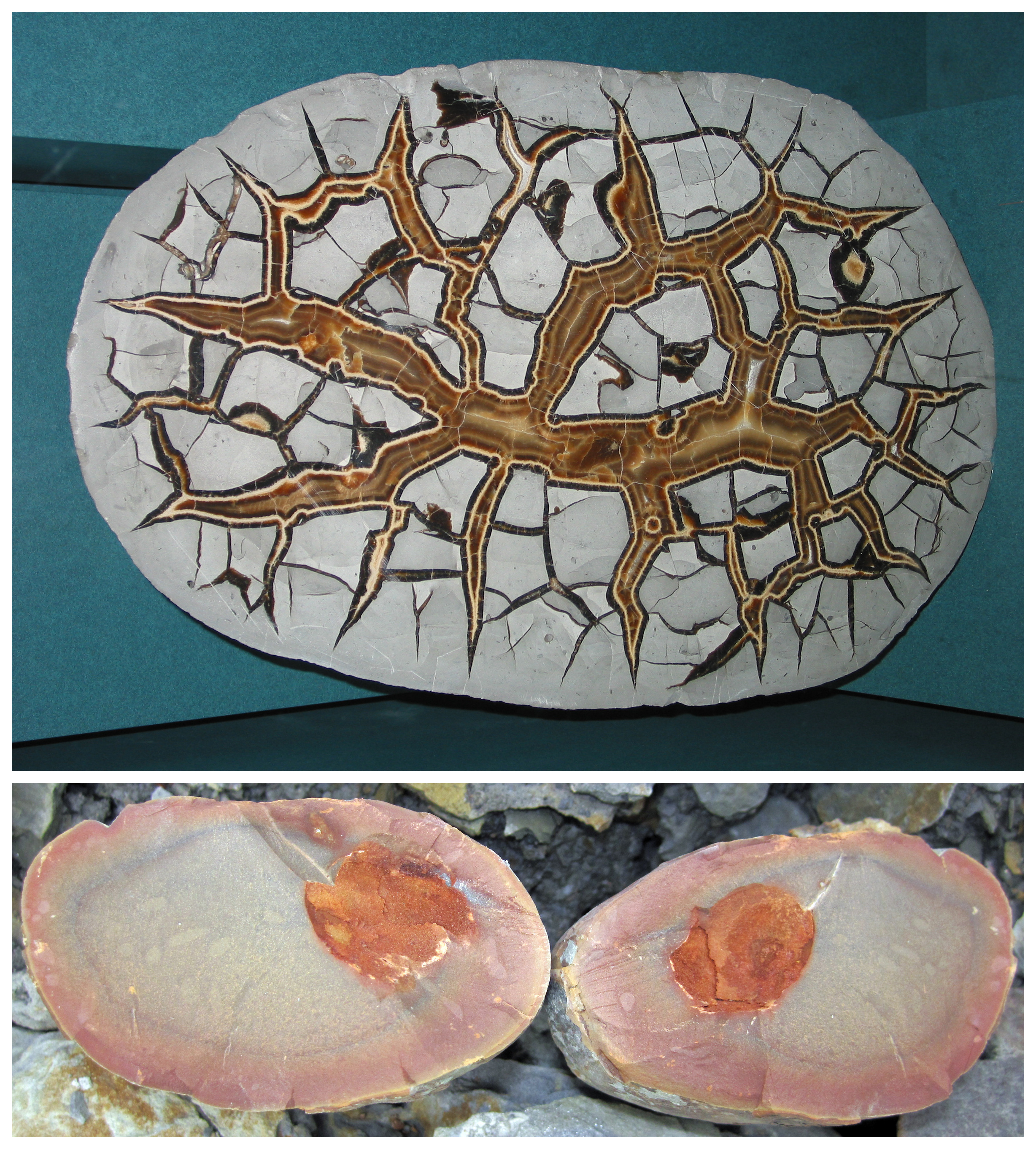}
\caption{\label{fig:concretion_nodules}Concretions and nodules.
Above: Septarian concretion,
below: Siderite (ironstone) nodule, Ohio, USA.
}
\par\smallskip{\footnotesize\noindent\raggedright\emph{Images:} Above: Keith Pomakis; CC-BY-SA-2.5,
Below:  James St. John, CC-BY-2.0.\par}
\end{figure}

\subsubsection{Concretions and nodules}
\label{sec:concretions}

Concretions and nodules (Fig.~\ref{fig:concretion_nodules}) are secondary mineral features found in sedimentary rocks, but they differ in origin and internal structure. Terminology varies in the literature; in this review, \emph{concretion} refers to early-diagenetic mineral precipitation around a nucleus --- often with radial or concentric fabrics --- whereas \emph{nodule} denotes a diagenetic replacement body lacking concentric growth. Both are rounded or irregularly shaped mineral masses, typically harder than the surrounding rock, and occur in a range of geological settings.

Concretions~\cite{raiswell2000mudrock,seilacher2001concretion,marshall2013carbonate} are
hard, compact masses of mineral material that form by the precipitation of minerals around a
nucleus or core, typically in sedimentary rocks like sandstone, shale, or limestone.
Concretions form when minerals precipitate from groundwater that percolates through sediment. These minerals accrete around a nucleus, which can be an organic object, like a fossil, shell, or plant fragment, or an inorganic particle, such as a grain of sand.
As minerals continue to precipitate, the concretion grows outward, often developing concentric layering. The growth occurs syndepositionally (during deposition) or early diagenetically (soon after burial), before the surrounding sediment fully lithifies.
Common minerals that precipitate to form concretions include calcite, silica (quartz or chalcedony), iron oxides (such as hematite), and siderite (iron carbonate).
Concretions are often spherical or elliptical in shape, but they can also be irregular. They vary in size from a few centimetres to several metres across.
Concretions often exhibit a radial  internal structure, with mineral layers emanating out from the nucleus. This distinguishes them from nodules, which lack this  concentric growth pattern.
Concretions are typically harder than the surrounding sedimentary rock, so they tend to be  preferentially exposed as the host rock weathers.

Septarian concretions couple early carbonate growth with later shrinkage fracturing and
crack-seal infill. A millimetre-scale precipitation front builds a stiffening carbonate body
around a nucleus while the host mud remains weak and
permeable~\cite{marshall2013carbonate,RaiswellFisher2000_MudrockConcretionsReview,Seilacher2001_Morphodynamics}.
As the concretion densifies, internal volumetric strain and pore-fluid evolution trigger
tensile failure of the still-weak interior. Proposed drivers include dewatering/shrinkage of
clay-rich gels (syneresis-style strain), localized pore-pressure transients, differential
compaction, and even shaking by earthquakes, all of which naturally produce polygonal or
radial crack
meshes~\cite{Astin1986_SeptarianCrackFormation,Hounslow1997_LocalPorePressureSeptaria,Pratt2001_SeptarianEarthquakes,HendryEtAl2006_BacterialPhysicalChemical}.
The resulting septaria act as short-lived high-permeability conduits; later fluids
precipitate sparry calcite (often ferroan), ankerite or barite in crack-seal fashion,
commonly in multiple generations that post-date the micritic
matrix~\cite{Scotchman1991_KimmeridgeSeptariaGeochem,AstinScotchman1988_KimmeridgeSeptariaDiagenesis,HudsonEtAl2001_OxfordClaySeptariaFluids,DeCraenSwennenKeppens1998_BoomClaySeptaria,GreenPirrie2001_OxfordianSeptaria,MolinaReolid2024_BariteSeptaria}.
Stable-isotope, cathodoluminescence and clumped-isotope studies document distinct fluid
histories for matrix versus septarian cements and repeated opening--sealing
cycles~\cite{HudsonEtAl2001_OxfordClaySeptariaFluids,PaxtonEtAl2021_ClumpedIsotopesSeptaria}.

Nodules~\cite{boggs2009petrology} are  mineral masses that form by post-depositional
(diagenetic) replacement of the surrounding sediment or rock, rather than by mineral
precipitation around a core. Nodules form through diagenetic replacement, where the original sedimentary material is replaced by minerals as groundwater circulates through the rock. This process often occurs 
after lithification.
Common minerals found in nodules include flint (a variety of microcrystalline quartz), chert, pyrite, phosphate, and gypsum. The replacement process can result in nodules that are chemically different from the surrounding rock.
Nodules tend to be irregular in shape and are often found scattered within the host rock. They can vary significantly in size, similar to concretions, from small pebbles to large boulders.
Nodules typically lack the concentric layering that concretions possess. Their internal structure is more uniform, often reflecting the mineral that replaced the original material.
Nodules are often found in sedimentary rocks like limestone, chalk, and shale. For example, flint nodules are commonly found in chalk formations.

Quantitative descriptions of concretion growth treat the body as a moving precipitation front
(concretions) or a coupled dissolution--precipitation front (replacement nodules), embedded
in a porous medium governed by transport and reaction. Classical field syntheses constrain
the geometry and timing
windows~\citep{raiswell2000mudrock,seilacher2001concretion,marshall2013carbonate}, while
forward models for iron--oxide systems couple transport with pH/redox buffering to reproduce
rinds and banding~\citep{Chan2007} and link to observed rhythmic
features~\citep{Potter2011,Katsuta2024}.

\begin{figure}
\centering
\includegraphics{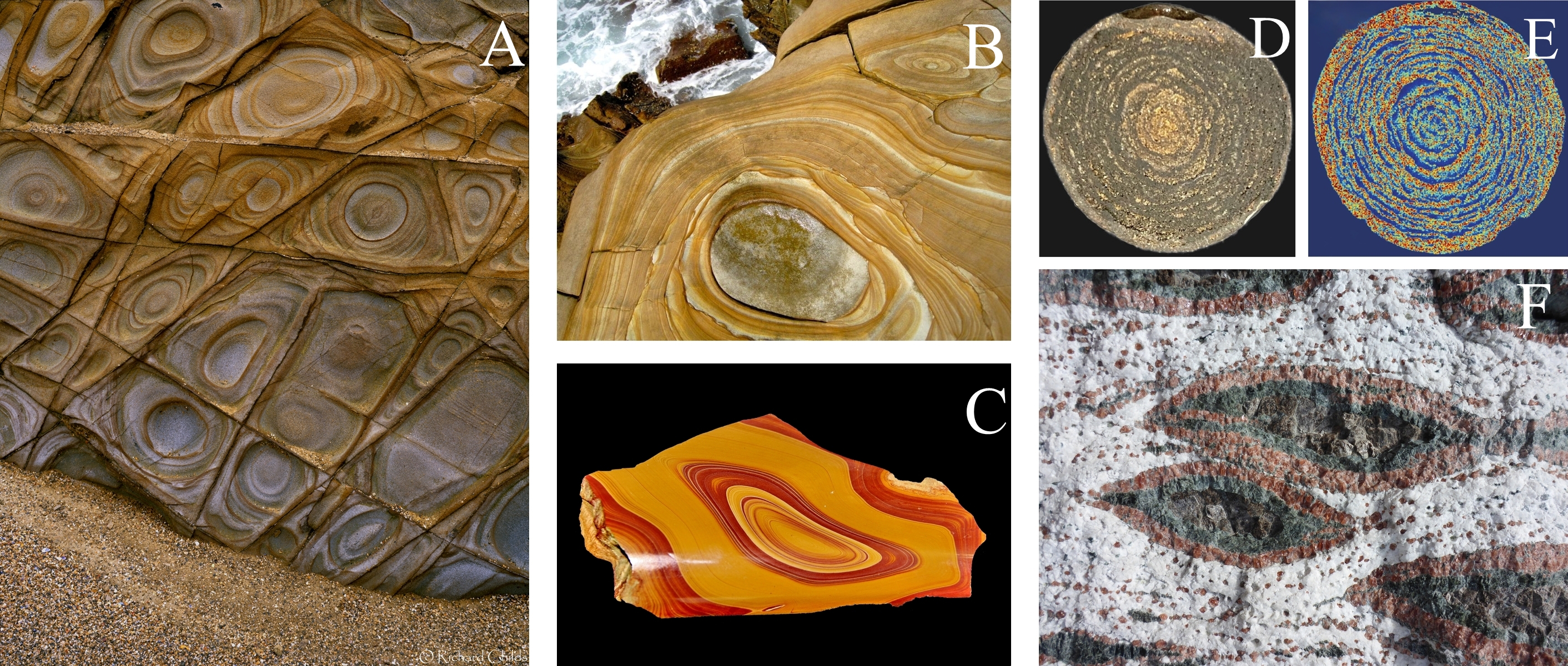}
\caption{\label{fig:target}Geological target patterns. A: in sandstones at Widemouth Bay, Bude, England. B: at Bouddi National Park, New South Wales, Australia. C: in volcano-sedimentary rock, Gobi desert. 
D: Layered moqui marble, an iron oxide concretion from the Navajo Sandstone in the
southwestern United States, with a radius of approximately 1.2~cm~\cite{Katsuta2024}. E: Iron
distribution within the concretion shown in panel D~\cite{Katsuta2024}. F: Reaction corona in
metamorphosed anorthosite: the original igneous rock was dominated by plagioclase with
orthopyroxene; the green clinopyroxene and garnet grew during subsequent metamorphism.
}
\par\smallskip{\footnotesize\noindent\raggedright\emph{Images:} A: Courtesy of   Richard Childs, 
B: Courtesy of    Evelyn Ward, 
C:  Dmitry Demezhko, distributed via imaggeo.egu.eu on the Creative Commons license, 
D-E: \citeA{Katsuta2024},
 F: Courtesy of Chris Clark.\par}
\end{figure}

\subsubsection{Target patterns}
\label{sec:targetpatterns}

Geological target patterns --- nested, approximately concentric bands around a nucleus --- occur in many rock types (Fig.~\ref{fig:target}) and are conspicuous in agates (Sec.~\ref{sec:agates}). In much of the geological literature these rings are casually labelled \emph{Liesegang} by analogy with laboratory precipitation experiments (Sec.~\ref{sec:liesegang}). We reserve Liesegang rings for the reaction--diffusion mechanism \emph{sensu stricto}: a diffusing reactant drives repeated exceedance of a precipitation threshold, producing quasi-periodic bands with characteristic spacing laws and front-controlled chemistry
\cite{Henisch1991,Nabika2020}. Where such diagnostics are absent, \emph{target} or
\emph{concentric banding} are  neutral morphological terms. Recent reviews emphasize this
mechanistic definition and its tests (spacing progression, front kinematics, band
chemistry)~\cite{nabika2019pattern}.

 Geological examples of target morphology include rhythmically banded iron-oxide concretions
(Sec.~\ref{sec:concretions}), e.g.,~the layered end-members in the Navajo Sandstone \emph{moqui
marbles} (Fig.~\ref{fig:target}D),  which show millimetre-scale, concentric Fe-oxide bands
throughout the body~\cite{Potter2011,Katsuta2024}. In metamorphic settings, reaction coronas
can likewise present repeated, compositionally alternating shells around relict grains
(Fig.~\ref{fig:target}E); when this rhythmic alternation is present, the resulting concentric
banding constitutes a target-like analogue, with layer successions and growth captured by
diffusion--thermodynamic rim-growth models~\cite{Joesten1977,Ashworth1990}. A broader family
of iron-oxide banding on outcrops --- ranging from irregular stains to well-ordered concentric
banding --- highlights how transport--reaction feedbacks in similar materials can yield both
highly ordered targets and less ordered banding, depending on boundary conditions and
supply~\cite{Chan2007}.

\subsection{Small and hot}

High heat or high stress acts on micro-scales, forging patterns measured in millimetres to a few centimetres. Banded sphalerite layers, the agate-lined cavities of thunder eggs, and orbicular granitoid shells record episodic crystallization in magmatic settings, while stylolite teeth and spiral ``snowball'' garnets capture differential stress during metamorphism. Menilite nodules, clathrite cages and other earthquake-fluidized or hydrothermal growths complete this hot-yet-small collection.

\subsubsection{Stylolites}
\label{sec:stylolites}

\begin{figure}
\centering
\includegraphics{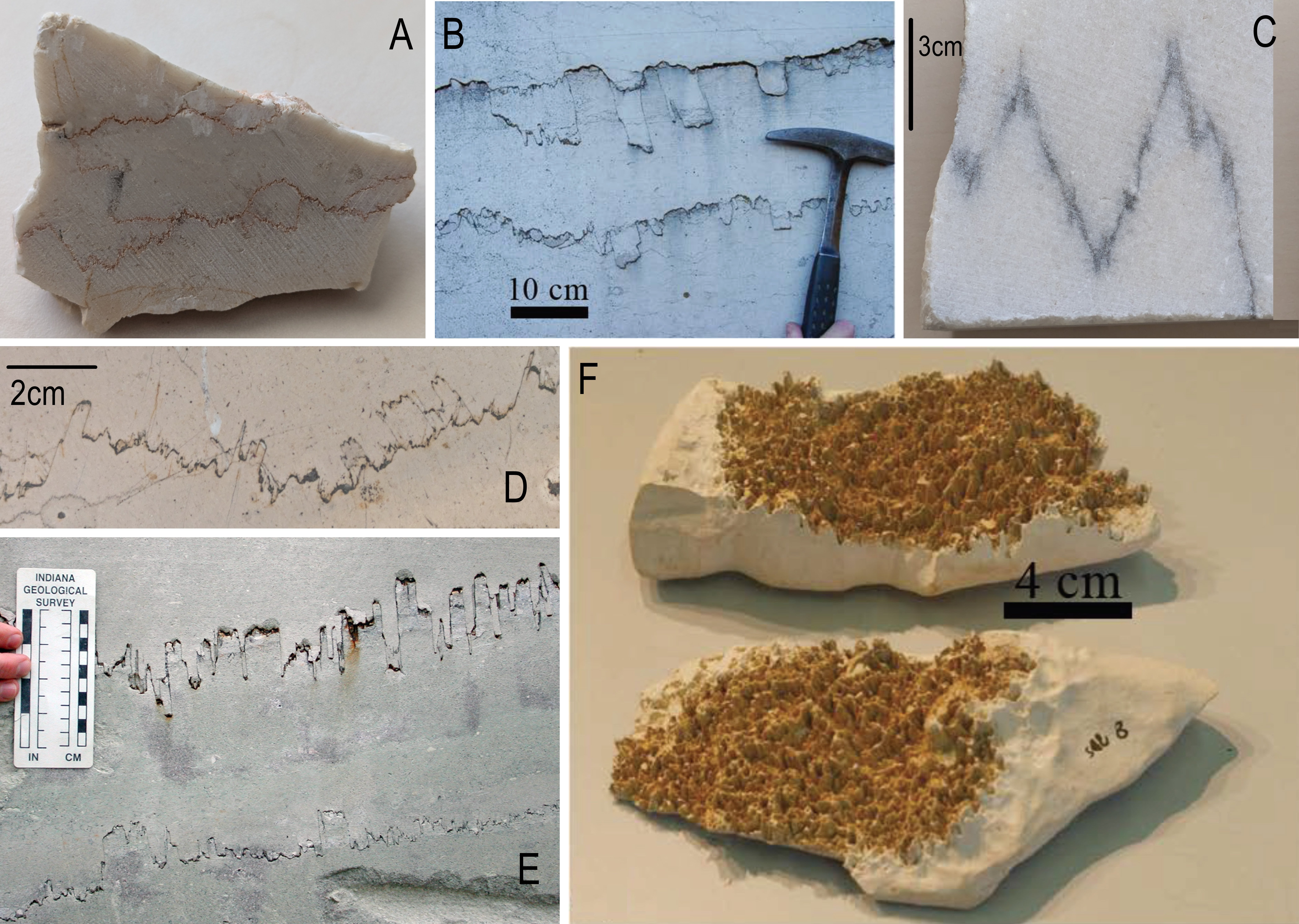}
\caption{\label{fig:stylolites}Stylolites.
A: Stylolite in limestone from Mitzpe Ramon, Israel; the red colour comes from clay residue. 
B: Stylolites in  limestone from Burgundy, France. 
C: Stylolites in marble from Dolni Morava quarry, Czechia.  
D: Stylolites in limestone from Morawica, Poland.
E: Stylolites in the Salem Limestone from Bloomington, Indiana, USA. 
F:  Surface of a stylolite in  limestone from Vercors, France.}
\par\smallskip{\footnotesize\noindent\raggedright\emph{Sources and images:} A, C, D: Piotr Szymczak, 
B, F:~\citeA{Toussaint2018},
E: Michael C. Rygel; CC-BY-SA-3.0.\par}
\end{figure}

 Stylolites~\cite{park1968stylolites},  Fig.~\ref{fig:stylolites}, are serrated, interlocking
dissolution surfaces commonly found in sedimentary rocks, especially carbonates and
sandstones~\cite{Stockdale1922,Shaub1939}. They form under compressive stress in the Earth's
upper crust by a process of pressure solution: minerals dissolve at stressed grain contacts
and re-precipitate in pore spaces, causing a net volume loss~\cite{Toussaint2018}.  The
resulting structure is a rough, tooth-like interface enriched in insoluble residues (clays,
organics) that were left behind as the host mineral dissolved~\cite{Stockdale1922}.
Stylolites play a significant role in rock deformation and diagenesis: they accommodate
strain, influence fluid flow, and record stress history. Their characteristic ``tooth''
geometry (pointing in the direction of maximum compressive stress) reflects the underlying
physics of their growth.  Stylolites can act as conduits for fluid migration, influencing
fluid flow and replacement reactions~\cite{koehn2016new}, helping transport fluids and
precipitating minerals in the fractures~\cite{heap2014stylolites,bruna2019stylolites}.

At the heart of stylolite formation is a thermodynamic driving force for dissolution under stress. A mineral under a normal stress $\sigma_n$ at a contact has a higher chemical potential or solubility than the same mineral under less stress. To first order, the change in chemical potential due to stress is
\begin{equation}
\delta \mu = \Omega \sigma_n,
\end{equation}
where $\Omega$ is the molar volume of the solid. This means that material at highly
stressed contact points can dissolve into the pore fluid, increasing the local ion
concentration, and then diffuse away, precipitating in regions of lower stress; e.g.,~open
pore space. In essence, the system tends towards lower free energy by removing material from
stressed locations. This thermodynamic principle --- sometimes called the pressure solution or
stress-induced dissolution mechanism --- was recognized early on as the origin of
stylolites~\cite{Weyl1959,Heald1959}. 
Indeed, a flat dissolution surface cannot persist under non-uniform stress: even minute
undulations raise the normal stress on the peaks and lower it in the valleys, so the peaks
dissolve faster while the valleys are protected. This positive feedback, an
Asaro--Tiller--Grinfeld-type instability~\cite{Asaro1972,Grinfeld1986,Kassner2001},
continually amplifies the roughness until the strain--energy gain from removing stressed
material is balanced by the surface-energy cost of creating new interface.

 A stylolite can be viewed mechanically as a sort of anticrack; the opposite of a tensile
crack. Instead of opening and creating void space, material is removed by dissolution and the
opposing faces move together.~\citet{Fletcher1981} formalized this idea, noting that
stylolite teeth focus compressive stress at their tips, much as crack tips focus tensile
stress. In an elastic host, a corrugated interface under compression bears higher normal
stress on protrusions; if those protrusions dissolve, the interface advances locally and bulk
shortening accrues as compaction strain. In some models the stylolite seam is treated as a
thin, soft (residuum-filled) layer embedded in an elastic medium; the associated stress
redistribution can even produce small tensile zones just beyond tips, occasionally nucleating
microcracks or veins.

The anticrack picture is a useful qualitative guide, but by itself does not guarantee
indefinite propagation under compression. In simulations by~\citet{Aharonov2009}, purely
stress-driven pressure-solution defects do not self-propagate laterally: calculated stresses
are low along the flanks, suppressing dissolution there, and continued tip dissolution
progressively smears out the tip stress concentration, which limits further advance.
Sustained lateral growth therefore requires an additional mechanism that maintains high tip
stresses and/or enhances flank dissolution; e.g.,~clay-assisted pressure solution together
with material heterogeneity that pins parts of the interface and promotes tooth formation. 
 Clays can enhance pressure solution by maintaining a thin water film and providing ionic pathways. Spring-network and related heterogeneous models show that a homogeneous medium tends to dissolve uniformly, whereas weak zones (``clay patches'') trigger localization into stylolite planes. 
In nature, one often sees stylolites initiating where a shale parting or a marl layer is present within limestone; exactly what the clay-enhanced dissolution theory would suggest.

The development of a stylolite's characteristic jagged morphology can be viewed through the
lens of interface dynamics and roughening. Initially, a roughly planar dissolution seam
localizes, perhaps along a bedding plane or a thin clay layer. As pressure solution proceeds,
any slight geometric irregularity on that interface will tend to amplify because of the
stress-driven instability described earlier.  The corresponding evolution equation for the
stylolite height on a 2D interface $(x,y)$
reads~\cite{schmittbuhl2004,bonnetier2009,rolland2012modeling}
\begin{equation}
\partial_t h(x,t) = v_0 + \nu\,\mathrm{PV}\!\int_{\mathbb R} \frac{h(x,t)-h(y,t)}{(x-y)^2}\,\mathrm dy  -  \kappa\,\partial_x^2 h(x,t) + \eta(x,h),
\label{eq:SRGT}
\end{equation}
where the Cauchy principal-value integral $\mathrm{PV}\!\int$ is the nonlocal elastic term, $\kappa>0$ captures stabilizing surface energy, $v_0$ sets the mean advance from kinetics,  $\nu$ collects together a combination of physical constants, and $\eta$ is quenched heterogeneity, arising from the spatial variability of the material properties composing the
grain~\cite{rolland2012modeling}. In Fourier space,
\begin{equation}
\partial_t \hat h(k,t) = \bigl[\nu\,\svert k\svert  - \kappa k^2\bigr]\,\hat h(k,t) + \hat\eta(k),
\end{equation}
so long wave-lengths are elastically unstable ($\propto \svert k\svert $) while short wave-lengths are stabilized by capillarity ($\propto k^2$), implying a fastest-growing wave-number $k_{\max}=\nu/(2\kappa)$ and the corresponding length-scale $L^{\star} = 4 \pi \kappa/\nu$.

Stylolite surfaces exhibit fractal characteristics, indicating scale-invariant roughness over
certain ranges. Geometrical analyses of natural stylolites in limestone have found that the
height profiles are self-affine with different Hurst exponents at different
scales~\cite{rolland2012modeling,schmittbuhl2004}. Specifically, a three-regime scaling is
often observed in sedimentary stylolites:
At small scales of tens to hundreds of microns, comparable to grain size, the interface is relatively smooth with a high Hurst exponent $H \approx 1.1 \pm 0.1$. This implies strong correlation and gentle variations at the grain-scale; the stylolite does not look very jagged under the microscope, partly because grain heterogeneity pins the interface.
At intermediate scales of 0.1--10~mm up to centimetres, a roughening regime appears with
$H \approx 0.5$--0.6. This regime is interpreted as the result of the stress-driven instability:
the interface wanders in a statistically rough manner governed by elastic interactions and
dissolution kinetics, largely independent of individual grain structure. At large scales of
several cm and above, sedimentary stylolites often show a saturation of roughness  with Hurst
exponent effectively 0, and  a flat power spectrum~\cite{BenItzhak2012}. Thus there is an
upper cut-off to the self-affine behaviour; beyond a certain length, the amplitude of
stylolite teeth no longer grows with scale.  Tectonic stylolites formed by tectonic stress,
not just burial, can show more anisotropic scaling: the roughness exponent differs in the
direction parallel to bedding versus perpendicular, reflecting stress anisotropy in the plane
of the stylolite~\cite{Ebner2010}. For example, a stylolite in folded limestone might have
longer, more spine-like teeth in the vertical direction, due to vertical stress, but more
irregular smaller serrations along the horizontal direction, due to lateral confining stress
being lower. 

The existence of scaling laws for stylolites has practical implications. One major result
from recent studies is that the crossover length between the small-scale and
intermediate-scale regimes encodes the stress that was present during stylolite
formation~\cite{Koehn2012,rolland2012modeling,Toussaint2018}. This crossover length is
essentially the critical wave-length of the instability, $L^{\star}$,  determined by the
balance of surface energy and elastic strain energy. 
In effect, stylolites can serve as a gauge of past stress. This has been demonstrated by comparing stylolite-derived stress values with independent estimates; e.g.,~calcite twinning paleostress. Additionally, the height of the largest teeth, related with the plateau of the roughness spectrum, correlates with the total amount of dissolution/compaction that occurred. These insights show how a deep understanding of interface roughening dynamics and scaling laws not only explains stylolite shapes but also unlocks quantitative geological information from them. 

Because stylolites typically form over millions of years under geological conditions,
reproducing them in the laboratory is challenging. Nonetheless, experimental efforts have
been made to simulate stylolite formation in accelerated time, usually by using more reactive
minerals or higher temperatures.~\citet{Gratier2005} conducted a landmark experiment in which
they compacted a layer of quartz sand in a pressure vessel at 350~{\degree}C and
${\sim}$50 MPa differential stress. Over the course of a few weeks to months, they
successfully grew miniature stylolites at the contacts between the quartz grains.
High-resolution imaging revealed tiny tooth-like dissolution features on the order of
$\rmmu$m in height that qualitatively resemble natural stylolite shapes. This experiment
confirmed that stylolites can form via pressure solution under controlled conditions, and
provided direct observation of the process: small dissolution pits initiated at random defect
points and gradually coalesced into a rough dissolution seam, with grains becoming truncated
and material re-precipitated in pore space. The geometry of the experimental stylolites
showed an interplay between heterogeneity and mechanical constraints, consistent with theory.
For instance, peaks tended to form where there were harder grains or impurities that
dissolved slower, and the overall amplitude of the teeth was limited by the need to
remain within the thin sample layer. Despite these successes, lab-grown stylolites are
inherently limited in scale. The experimental stylolite teeth were at most on the scale of
one or two grain diameters in height --- tens of microns --- whereas natural stylolites can have
teeth from mm up to cm high penetrating many grain diameters. The time scale is the limiting
factor: to grow large stylolites requires dissolving a significant volume, which is extremely
slow at laboratory time-scales unless one uses unfeasibly high temperatures or very soluble
minerals. 

 Ongoing work continues to refine both the experiments and the models. Stylolites, once merely a curious serration in rocks, are now appreciated as natural laboratories of far-from-equilibrium thermodynamics and fractal growth, where the interplay of stress, chemistry, and time produces complexity from simplicity.

\begin{figure}
\centering
\includegraphics[width=\linewidth,height=0.78\textheight,keepaspectratio]{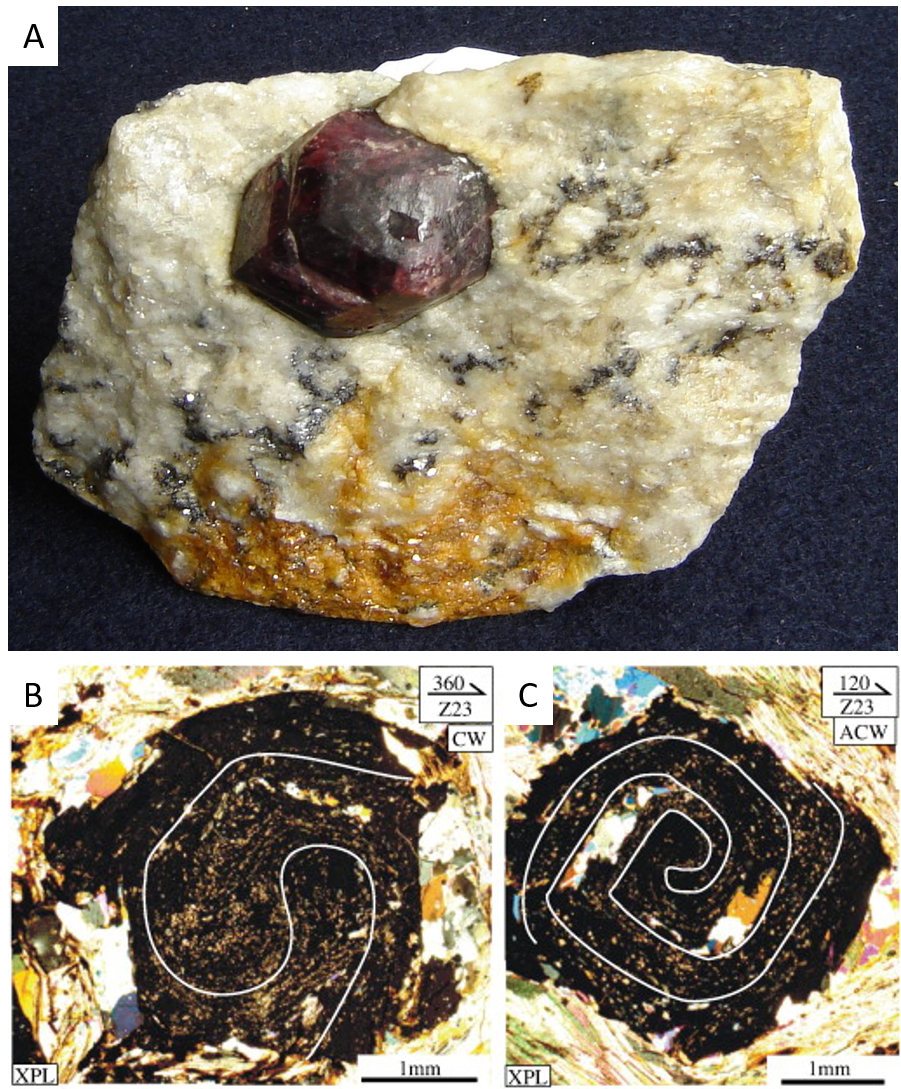}
\caption{\label{fig:porphyroblasts}
Porphyroblasts.
A: Garnet porphyroblast (3~cm width) in a gneiss matrix. 
B,C: Crossed polars optical microscope images of spiral garnets showing (B) clockwise  and
(C) anticlockwise asymmetry.
}
\par\smallskip{\footnotesize\noindent\raggedright\emph{Images and sources:} A: Eurico Zimbres (FGEL/UERJ) and Tom Epaminondas (mineral collector), CC-BY-SA, B, C: 
\citeA{Shah2011Foliation}\par}
\end{figure}

\subsubsection{Porphyroblasts and spiral garnets}
\label{sec:porphyroblasts}

Porphyroblasts (Fig.~\ref{fig:porphyroblasts}) are large mineral crystals that grow within a finer-grained matrix in metamorphic rocks. These crystals form during metamorphism, the process whereby rocks recrystallize under elevated temperature and pressure. Common minerals that grow as porphyroblasts include garnet, staurolite, and andalusite.
Porphyroblasts begin to form when specific minerals nucleate during metamorphism. As metamorphic conditions (temperature and pressure) increase, these minerals grow larger than the surrounding matrix owing to differences in mineral stability and growth kinetics.
Porphyroblasts commonly develop in reaction zones, where minerals become unstable and new, more stable minerals grow. The growth of porphyroblasts reflects the rock's chemical and physical evolution during metamorphism.

As they grow, porphyroblasts can trap small fragments of the surrounding rock, known as inclusions, preserving a record of the rock's history prior to their formation. 
During metamorphism, rocks are often subjected to deformation, such as folding or shearing. This can cause porphyroblasts to rotate within the matrix.
Under simple shear, relatively rigid porphyroblasts may undergo rotation within a flowing matrix. 
Alternatively, under some conditions porphyroblasts may grow essentially without rotation. Some garnet porphyroblasts exhibit spiral-shaped inclusion trails, known as spiral garnets or snowball garnets, which were once thought to indicate rotation. However, this spiral pattern can also arise from progressive overgrowth while the external foliation reorients, so that successive inclusions are entrained at slightly different azimuths.

There has been a long-standing debate in geology over porphyroblast
rotation~\cite{Fay2008,Bons2009}.
The question of whether porphyroblasts rotate or not is clearly one that falls within the domain of physics, in particular fluid dynamics. A classical result due to Einstein states that, in a Newtonian fluid undergoing simple shear at low Reynolds number, a suspended sphere rotates at half the shear rate,  $\dot{\gamma}/2$. This has been used to argue that porphyroblasts must rotate as the rocks  shear about them. However, that result applies only to Newtonian fluids, whereas rocks deform as complex non-Newtonian media. 
For non-Newtonian fluids,~\citet{brunn} and~\citet{gauthier} showed theoretically and
experimentally that the rotation rate is the same as in a Newtonian fluid in the case of an
isolated sphere in a second-order fluid (i.e.,~at low shear rates). On the other hand,
simulations of~\citet{DAvino2008}  and experiments of~\citet{Snijkers2009} found reduced
rotation at higher shear rates relative to the Newtonian case.
\citet{michele} showed that for sufficiently high rates of shear, spherical particles of size
60--$70~\rmmu$m suspended within a non-Newtonian fluid self-organize into string-like
structures oriented in the direction of flow. They observed, moreover, that the particle
rotation rate was strongly reduced relative to the Newtonian case;  it decreased further as
interparticle spacing diminished and practically vanished when the spheres lined up, as if
the string of particles were a long rigid thread.
\citet{Won2004} and~\citet{Scirocco} showed that the formation of these chains increases with
the degree of shear thinning of the fluid and with the shear rate.~\citet{Hwang} performed
numerical simulations with several different non-Newtonian fluids and showed a transition in
particle structures in a sequence: random, clustering, clustered string, and string
formation, as the solvent viscosity decreases and the Weissenberg number increases.
Thus the physics of non-Newtonian fluids predicts a rotation of a single particle much lower
than in the Newtonian case, and the formation of strings of particles that behave as
quasi-rigid structures. Note the similarity of these string-like structures to the millipede
structures of porphyroblasts~\cite{bell2007progressive}.

\subsubsection{Banded sphalerite}
\label{sec:sphalerite}

\begin{figure}
\centering
\includegraphics{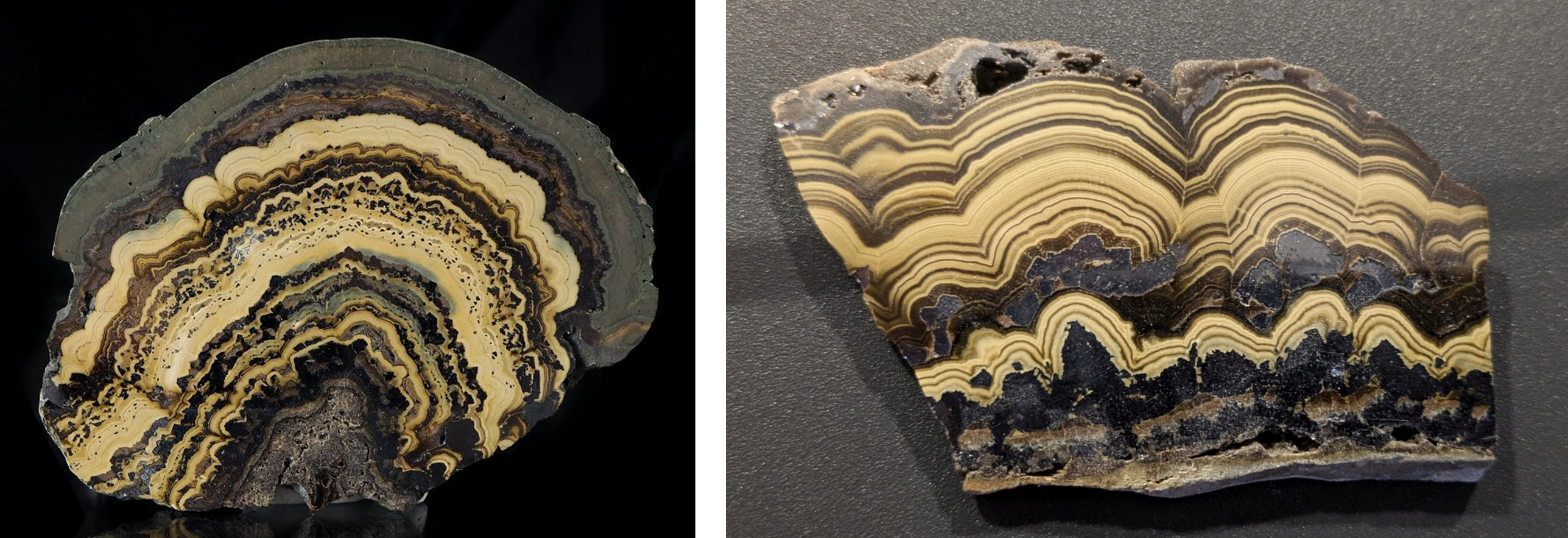}
\caption{\label{fig:Blende}Banded sphalerite. Banded precipitation in hydrothermal ore deposition, Schalenblende and Zincblende Pomorzany, Poland.}
\par\smallskip{\footnotesize\noindent\raggedright\emph{Images:} Left: Courtesy of Przemys\l aw Budzy\'nski; Right: Piotr Szymczak.\par}
\end{figure}

An example of a layered structure with an elusive origin is banded sphalerite, a prominent zinc ore,  Fig.~\ref{fig:Blende},
which often has light coloured ZnS rich cores surrounded by sub-mm nested, sequentially
coloured, bands of ZnS crystals that are arranged in cm-sized clusters~\cite{Katsev2001}.

In a similar fashion to the agate problem (Sec.~\ref{sec:agates}), both intrinsic and extrinsic
mechanisms have been proposed to explain the formation of these patterns. In particular, the
origin of banding has been sought in pressure--temperature fluctuations, where shifts in
these conditions during sphalerite formation alter the solubility of various elements. For
instance,~\citeA{Zhu2018} demonstrate that sphalerite deposits show variations in zinc and
sulphur isotopes, which correlate with changes in fluid temperatures over time. This results
in the rhythmic deposition of sphalerite layers with slightly different compositions,
contributing to band formation.
Other models~\cite{Cathles1983,Yardley1991} suggest episodic fluid flow during hydrothermal
mineralization, with long hiatuses separating periods of rapid crystallization. 
Variations in the concentration of chemical elements in successive hydrothermal infiltrations
can lead to alternating layers of different compositions, creating banding. A common
criticism of these models is that attributing each of the tens or hundreds of thousands of
layers to an episodic event, such as a crack-seal cycle, seems to violate the principle of
Ockham's razor~\cite{shore_oscillatory_1996}. 

Intrinsic models, on the other hand, suggest that the origin of the sphalerite banding is a process of self-organization and
that it was formed in a single event of oscillatory
zoning~\cite{Oen1980,LHeureux2000,Katsev2001,Katsev2001b} (Sec.~\ref{sec:zoning}). A pioneering
model of this kind~\cite{Oen1980} attributes oscillatory zoning in sphalerite to cyclic
supersaturation and diffusion-controlled concentration gradients during  crystallization.
There is a feedback loop here occurring during crystallization when two or more different
cations compete for incorporation into a mineral structure. Initially, one cation is
preferentially incorporated into the growing crystal while the other accumulates in the
surrounding fluid. Once the concentration of the excluded cation reaches a threshold, it
begins to be incorporated into the crystal, which results in a banding
pattern~\cite{Benedetto2005}.

A somewhat more complex mechanism has been proposed by L'Heureux, Katsev, and Fowler~\citeA{LHeureux2000,Katsev2001,Katsev2001b}. Their models combines the effects of
geochemical reactions, crystal growth, dissolution, and ripening under far-from-equilibrium
conditions. An important role in the final formation of the banded structure is played by a
self-propagating sequence of growth and dissolution events, known as coarsening
waves~\cite{Feeney1983,Boudreau1995}. The mechanism here is a form of Ostwald ripening
(Sec.~\ref{sec:ostwald}): as larger particles grow, they deplete the solute in their vicinity,
causing smaller particles to dissolve and the solute to diffuse towards the larger particles.
As shown by~\citeA{Feeney1983} and~\citeA{Boudreau1995}, this feedback mechanism leads to the
formation of periodic structures or bands, which propagate as waves through the medium. These
waves represent regions of alternating high and low particle concentration, coarsening over
time as the system evolves towards a more stable state. In the case of sphalerite patterns,
this process can occur over geologically short time-scales: although the formation of the
entire cm-scale sphalerite cluster may take thousands of years, the band pattern itself is
predicted to form on a time scale of months~\cite{Katsev2001}. This opens the possibility
that intrinsic and extrinsic models may work in conjunction, with intrinsic mechanisms
responsible for sub-mm scale features and external conditions controlling cm-scale
variability.

Lastly, it is worth noting that sphalerite banding has also been linked to microbial
activity~\cite{Kucha2010}, with microglobular sphalerite textures interpreted as fossil
microbial mats of sulphate-reducing bacteria, as evidenced by sulphur isotope data.

\begin{figure}
\centering
\includegraphics{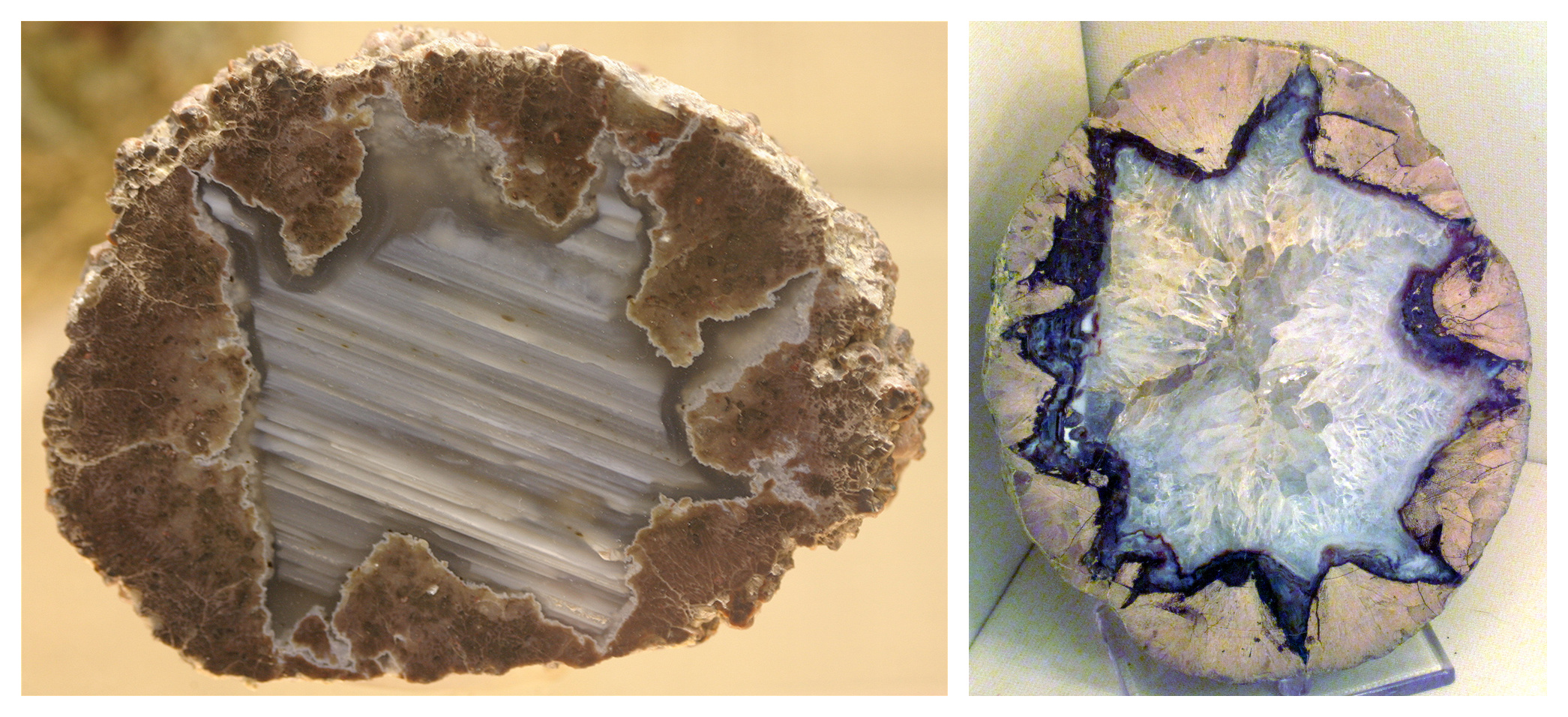}
\caption{\label{fig:thunder_egg}
Thunder eggs.
Left: Agate thunder egg, 
Right: Quartz thunder egg, Oregon, USA.
}
\par\smallskip{\footnotesize\noindent\raggedright\emph{Images:} Left: Cacophony; CC-BY-SA-3.0,2.5,2.0,1.0,
     Right: James St. John, CC-BY-2.0.
\par}
\end{figure}

\subsubsection{Thunder eggs}
\label{sec:thundereggs}

Thunder eggs or lithophysae~\cite{breitkreuz2013spherulites}  are spherical to ovoid cavities
(Fig.~\ref{fig:thunder_egg}) typically found within rhyolitic and other silicic volcanic
rocks~\cite{roots1952thunder}. They may look externally similar to geodes
(Sec.~\ref{sec:geodes}), but thunder eggs have a distinct origin and internal structures. They
are often filled with minerals such as agate~\cite{kile2002occurrence}, quartz, chalcedony,
or opal~\cite{bryan1963later}.
Thunder eggs form in silicic volcanic rocks, particularly rhyolites and tuffs, during or
shortly after emplacement~\cite{zarembo1991rare}. Their formation is commonly described as
two main stages: cavity formation and mineral infill. Initial cavities within solidified
volcanic rock generally form from trapped gas bubbles in the viscous magma or through
shrinkage cracks as the lava cools and contracts. An alternative or complementary mechanism
involves spherulitic growth, in which silicate minerals like quartz or feldspar crystallize
radially from a central nucleus, creating a ball-like structure. These structures may then
develop voids or fractures that later become filled with minerals.
After the cavities form, mineral-rich solutions, often silica-laden, permeate the volcanic rock. Over time, chalcedony, agate, quartz, or opal precipitate from these solutions and fill the hollow centres of the thunder eggs.

\subsubsection{Orbicules}
\label{sec:orbicule}

\begin{figure}
\centering
\includegraphics{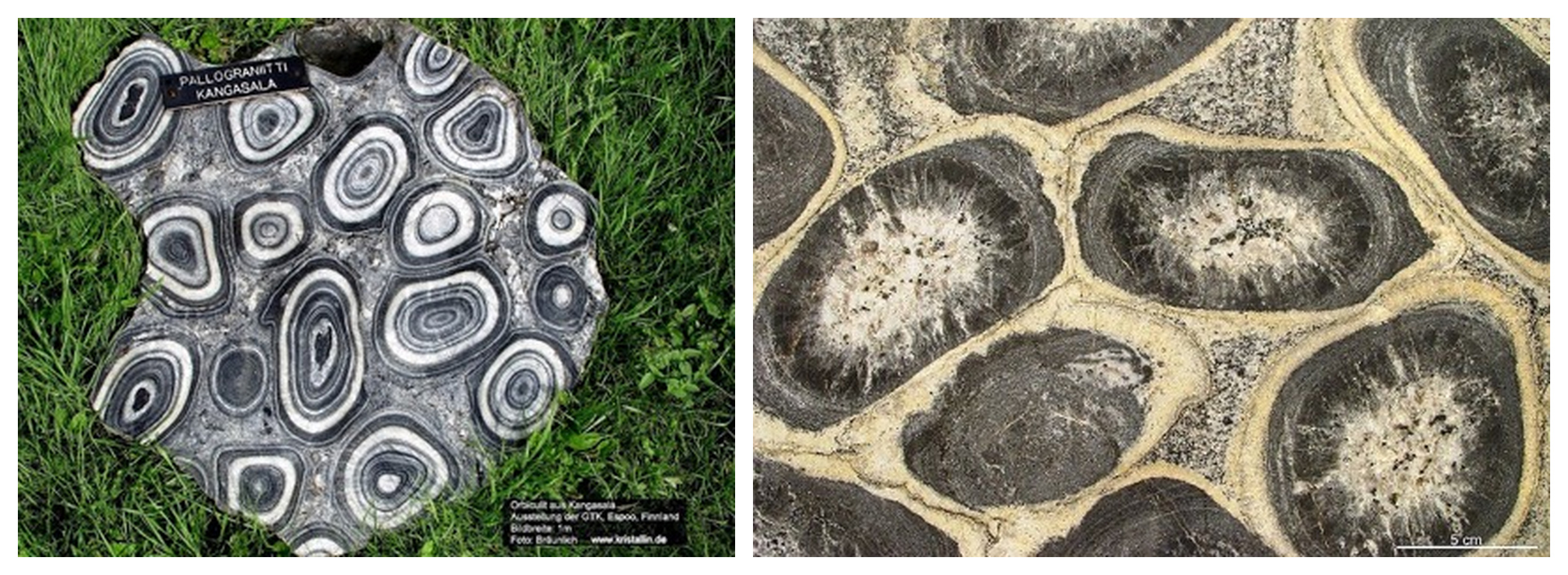}
\caption{\label{fig:orbicular}Orbicules in  granitoid rocks.
}
\par\smallskip{\footnotesize\noindent\raggedright\emph{Images:} kristallin.de; CC-BY-SA 3.0.\par}
\end{figure}

Orbicular granitoids  Fig.~\ref{fig:orbicular}, are uncommon rocks of localized occurrence that have attracted the attention of geologists for at least a
century~\cite{sederholm1928orbicular}. Their defining feature is the presence of orbicules, spheroidal
bodies with concentric shells, comprising layers of mafic minerals, primarily phyllosilicate,
and felsic minerals, predominantly quartz and feldspar~\cite{Ortoleva1994}. These layers
comprise elongated crystals growing radial or tangential to the
shells~\cite{leveson1966orbicular,vernon1985possible}. The orbicules seem to form around a
grain in a cooling magma chamber, although alternative theories about a metasomatic origin
have been proposed in the past; see references in~\citeA{vernon1985possible}. The diameter of
these orbicular structures can range from less than 5~cm (ultrabasic rocks), 5--10~cm (basic
rocks) and up to 40~cm for granites~\cite{lahti_orbicular_2005}. Finland has the highest number of reported occurrences, although rocks of this kind are also found elsewhere. 

\begin{figure}
\centering
\includegraphics{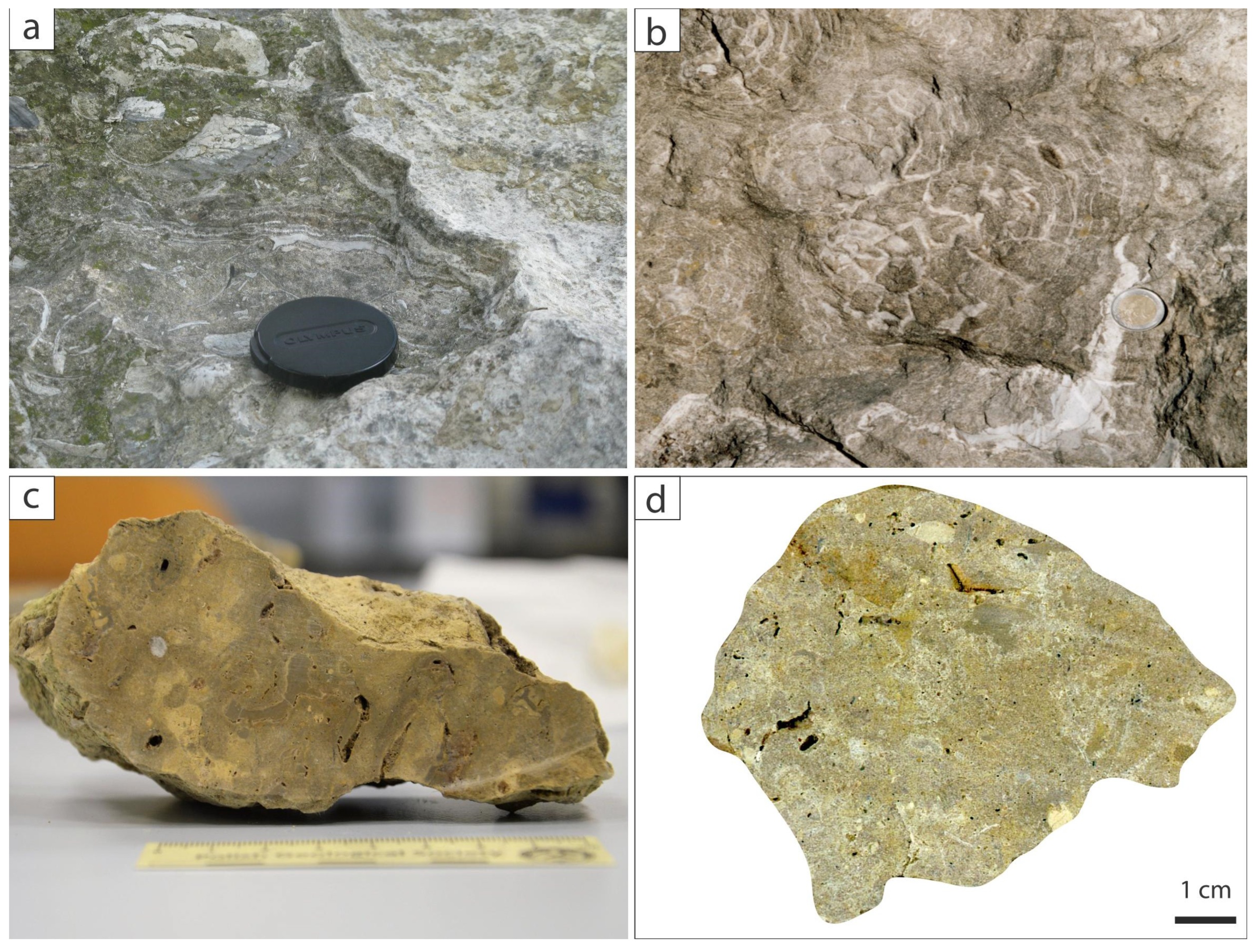}
\caption{\label{fig:clathrites}
Clathrite structures. 
A: Thin layered structures surrounding a carbonate breccia with shell fragments. 
B: Irregular network (mainly radial to concentric) of carbonate-filled veins. C,D: Vacuolar,
spongy and vuggy-like fabrics: cavities have various shapes, empty or filled with carbonate
cements and/or coarser sediments and coquina debris.}
\par\smallskip{\footnotesize\noindent\raggedright\emph{Source:} \citeA{argentino2019evidences}.\par}
\end{figure}

\subsubsection{Clathrites}
\label{sec:clathrites}

Clathrites,  Fig.~\ref{fig:clathrites}, are methane-derived precipitates formed by the
transformation of clathrates into carbonates. Clathrates, clathrate hydrates, or gas hydrates
are partially solid structures composed of water molecules arranged in crystalline cages.
Within these structures, small molecules, usually non-polar gases, are confined within the
ice-like polyhedral water cages. Clathrates can evolve into clathrites over time. This
replacement is directly linked to the transformation of clathrates into carbonates, while
partially preserving and mirroring their patterns and mesoscale
structures.~\citeA{kennett2000relationship} coined the term clathrate in 2000 to describe
mesoscale structures showing  soft-sediment deformation associated with gas hydrate formation
or dissociation. These structures are found within massive gas-hydrate layers and sometimes
exhibit surface morphologies created by a direct mirroring of clathrates bubble structures.
Examples of sites where clathrites have been observed include the Gulf of
Mexico~\cite{smith2014methane,suess2014marine}, north Italian
Apennines~\cite{argentino2019evidences}, Mediterranean Sea mud
volcanoes~\cite{aloisi2000methane}, and Cascadia in northwestern North  America~\cite{greinert2001gas}.

\subsubsection{Menilites}
\label{sec:menilites}

\begin{figure}
\centering
\includegraphics{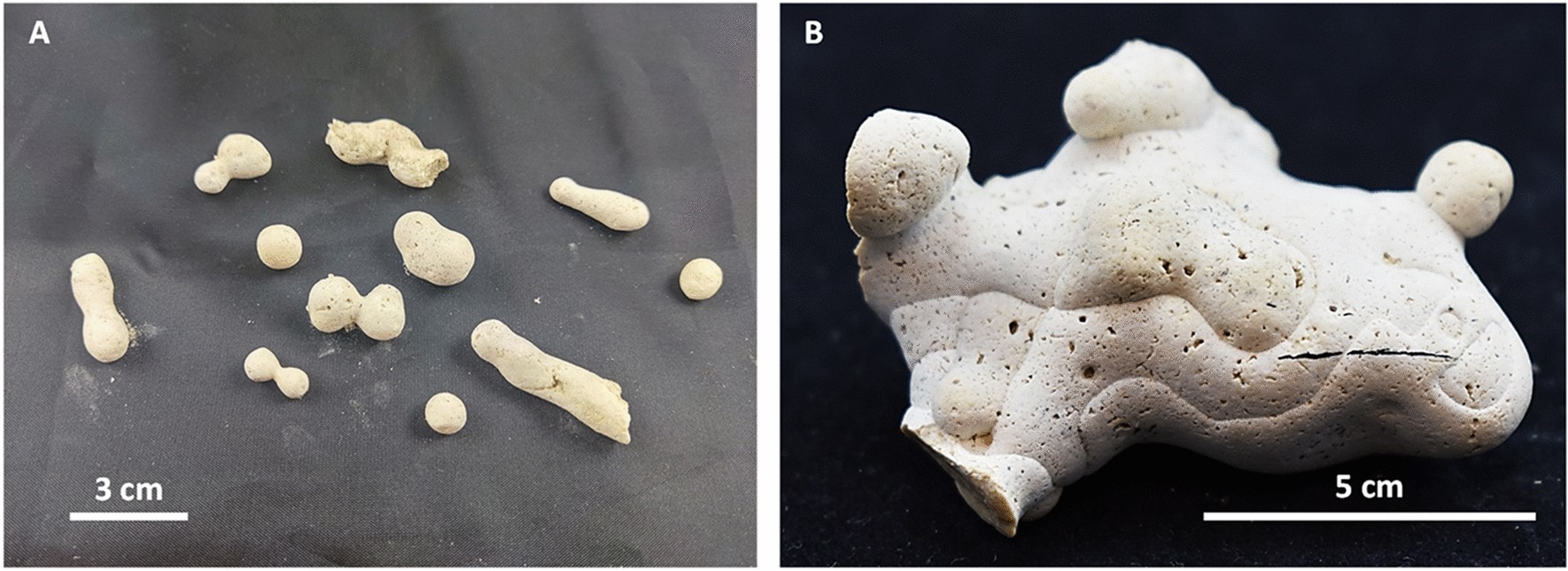}
\caption{\label{fig:menilite}Menilites with simple (A) and complex (B) shapes.}
\par\smallskip{\footnotesize\noindent\raggedright\emph{Source:} \citeA{pimentel_mineralogical_2024}.\par}
\end{figure}

Menilites are rocks of botryoidal appearance composed mainly of opal (Sec.~\ref{sec:opals}) and
dolomite~\cite{molina_grande_idolos_1980,pimentel_mineralogical_2024}. They occur mainly in
simple centimetric forms such as spheres, dumbbells or rods (Fig.~\ref{fig:menilite}A). However,
the most interesting are the larger specimens ($>$~5~cm) with more complex shapes
(Fig.~\ref{fig:menilite}B). Recently, it has been proposed that the formation of these rocks may
be related to  paleoseismic events~\cite{pimentel_mineralogical_2024}. A fluid phase is
involved in their formation, developing within porous strata above the water table and
becoming fluidized by earthquake shaking that drives upward water flow. The density contrast
between these nodules (opaline, less dense) and the bedrock (marls and limestones, denser)
implies that, under wet conditions, an osmotic-pressure difference (Sec.~\ref{sec:osmotic}) can
develop if the opal-carbonate interface is considered a membrane (i.e.,~an elastic
interface). If inflowing water ruptures this membrane, the pressurized internal fluid is
extruded, forming the characteristic shapes of these rocks.

\subsection{Large and hot}

Here we meet metre-scale structures powered by intense thermal or mechanical budgets. Cooling magmas fracture into columnar joints, lightning fuses sand into hollow fulgurites, buoyant hydrothermal fluids build towering chimneys, and deep-crystal stress forges gneissic foliation. Mud volcanoes, pockmarks, seeps and earthquake-driven seismites round out the set, all tapping high heat flow or transient tectonic energy to pattern rock and sediment over decimetres to metres.

\subsubsection{Basalt columns}
\label{sec:basalt}

Appearing as a striking pattern of roughly hexagonal pillars of rock, columnar joints are
famous from landscapes like those of the Giant's Causeway in Northern Ireland, and the
Devil's Postpile in California.  At these places, basalt has broken into a colonnade
consisting of regular, prismatic columns of rock, each a few tens of centimetres across.  The
cracks that delimit the edges of these columns were formed as the basalt cooled from molten
lava, millions of years ago.  Subsequently exposed by erosion, the evocative scenery of
columnar jointing, as shown in  Fig.~\ref{fig:basalt_columns}, has attracted scientific attention
for centuries~\cite{Foley1694}. The origins of columnar joints in the thermal contraction of
a cooling lava flow have been recognized since at least the 19th century~\cite{Mallet1875},
and recent research has focused on explaining the origins of the regular pattern of cracks
that form the joints, as well as the physics that determines the size of the columns. 

\begin{figure}
\centering
\includegraphics{gr73}
\caption{\label{fig:basalt_columns}
Columnar jointing in lava and starch.  
A: At Staffa, Scotland, a well-formed colonnade sits atop an ash layer, and is surmounted by an entablature of thinner, more disordered columns. The colonnade cooled from the ash-contact upwards, while the entablature cooled from above. 
B: A cross-section of columns at the Giant's Causeway, N. Ireland, highlights their polygonal shape. 
C:  Striae, appearing as short bands on the face of any column, record the stick--slip motion of the column's formation. Plumose (left/right-leaning feathery lines, marked by arrows) also indicate the direction in which each crack advance occurred. 
D: When a slurry of corn starch is dried, it also cracks into columnar joints. The sample here is shown inverted, with the base of the colonnade exposed, to better display the hexagonal pattern that developed.}
\par\smallskip{\footnotesize\noindent\raggedright\emph{Images:} Lucas Goehring.\par}
\end{figure}

Although best known in the form of columnar
basalt~\cite{Ryan1978,Long1986,Aydin1988,Goehring2008,Phillips2013}, columnar joints occur in
a wide variety of materials. They are common in other lava types, including rhyolite and
andesite~\cite{Spry1962,Degraff1987}, and have been reported in igneous rocks on
Mars~\cite{Milazzo2009,Milazzo2012} and the moon~\cite{Xiao2015,BasilevskyY2015}. Columns can
also be found in sandstone~\cite{Summer1995} and chalk~\cite{Weinberger2019} where they are
thought to result from cooling and contraction after some intense thermal treatment, for
example by the invasion of a nearby dyke or sill; for similar reasons, columns are sometimes
found in smelter slag, coal, and rapidly quenched glass~\cite{Degraff1987,French1925}.  

Columnar joints in lava typically form when a lava flow cools slowly over time.  The lava
could be emplaced as a massive flood basalt, covering thousands of square
kilometres~\cite{Long1986,Degraff1987,Goehring2008} or as a pond or stream that formed after
a smaller volcanic eruption, as is common in Hawaii~\cite{Peck1968,Ryan1978}.  As the
initially fluid lava cools, it solidifies.  The details of this process will depend on the
specific chemistry of a lava, with a series of distinct minerals crystallizing from the melt
over a range of temperatures. However, to a good approximation, there will be a change from a
fluid-like to a solid-like rheology, corresponding to a glass transition or liquidus--solidus
transition around 800--{900~{\degree}C}~\cite{Peck1968,Lore2001,Gottsmann2004,Lamur2018}.
Below this temperature the lava will behave as an elastic solid, and accumulate tensile
stress due to thermal contraction.  This situation sets up a stress gradient in the lava,
mirroring the thermal gradient, which drives fracture.

Since a molten lava flow will cool inwards from its boundaries, the initial cracks will
appear at the surfaces of the flow.  This process has been observed, for example, in the
cooling of fresh lava lakes from the Kilauea volcano~\cite{Peck1968}.  As time and cooling
proceed, the cracks from the surface will grow deeper.  They will develop into a network of
crack tips following just behind the solidification front, where the cooling lava supports an
elastic stress.  This crack network effectively carves out the columnar features as it
advances~\cite{Mallet1875,Spry1962}.  The motions of the crack tips are not smooth, however,
but rather the cracks that will form the column faces advance intermittently.  When any crack
grows, its active tip will move into warmer conditions.  As a result, the tensile forces
driving the fracture become weaker, and it will soon slow down and halt.  After sufficient
thermal stress has then had time to build up again, the crack can be reactivated, grow
forward, and halt once more.  As shown in  Fig.~\ref{fig:basalt_columns}C, this stick--slip
motion is recorded by a roughly periodic banding along the faces of the columns; features
called striae that look similar to chisel marks~\cite{Ryan1978}.  Within each band, the crack
surface is smoothest under the more brittle conditions of the crack initiation, and becomes
rougher as the crack advances into warmer rock and slows down~\cite{Ryan1978,Degraff1987}.
Combined with slight mismatches in the orientation of subsequent striae, and with plumose
markings --- faint feathery lines that develop perpendicular to the direction of crack
propagation --- these periodic variations in texture can be used to determine the direction in
which a colonnade developed~\cite{Ryan1978,Degraff1987,Budkewitsch1994,Goehring2008}. For
further details, see~\citeA{Pollard1988} who review the geology and petrology of columnar
joints, and explain how to interpret the surface features (striae, plumose, hackle) on column
faces. 

The directional growth of columnar joints also provides the route for columns to take on
well-ordered polygonal shapes.  Near the flow margin, the fracture network that defines the
columns looks similar to the contraction cracks that can be seen in dried mud or clay
(Sec.~\ref{sec:crackpatterns})~\cite{Peck1968,Aydin1988,Budkewitsch1994}.  This pattern is
dominated by cracks intersecting at T-shaped junctions, which is the result of the sequential
fragmentation of the surface: later cracks will curve to intersect earlier ones at right
angles, to maximize their energy release rate~\cite{Bohn2005}.    As the columns grow, the
incremental advances that leave the striae features allow the crack network to change: each
crack advance can be slightly misaligned with the previous step.  Over time, the crack
intersections that form the corners of the columns will then shift towards Y-shaped
junctions.  In lava flows, this can be seen within the first 1--2~m of a flow
margin~\cite{Aydin1988}.  The Y-junctions are energetically favoured in situations where the
advancing cracks are guided or triggered by the locations of previous features, but where the
sequence in which the cracks form is randomized in each cycle of fracture
growth~\cite{Goehring2013}.~\citeA{Goehring2014} review in more detail this process of the
ordering of hexagonal fracture patterns.  

\begin{figure}
\centering
\includegraphics{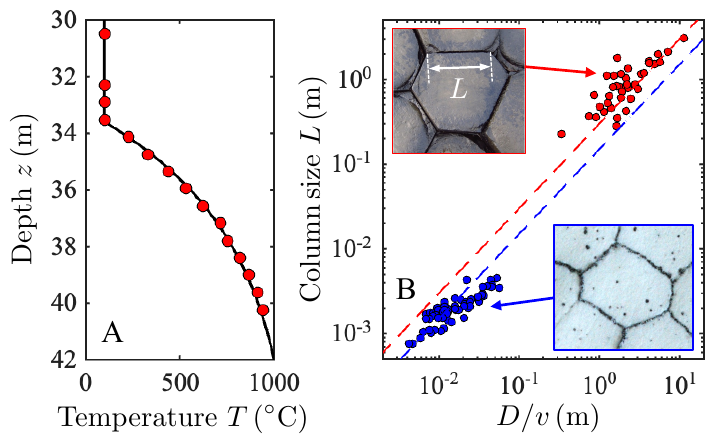}
\caption{\label{fig:scaling_columns}
Scaling of columnar joints.  
A: Temperature measurements (red points) from boreholes at Kilauea Iki, Hawaii, are well-fit
by stationary solutions to the advection--diffusion problem (black line); data reproduced
from~\citeA{Hardee1980}, \citet{Helz2020}, borehole KI75-1. 
B: Testing of the thermoelastic/poroelastic model of columnar joint scaling.  The size of the
columnar features, $L$, can be taken as the distance between adjacent vertices of a
column.  The corresponding advection--diffusion length, 
$D/v$, can be estimated for the
cases of cooling lava columns (red points) and drying starch columns (blue points).  Both
cases show a simple scaling of $Pe = vL/D \sim 1$. Data reproduced from~\citeA{Goehring2009}.
}
\end{figure}

The manner in which a colonnade grows means that it records how a lava flow cools.  In
particular, the columns extend along the direction of cooling, and perpendicular to isotherms
at the cracking front, near to the solidus or glass-transition
temperature~\cite{Mallet1875,Spry1962,Budkewitsch1994,Kattenhorn2008}.  Some common
structures formed in this way include fans of columns, rosettes and chevrons, in addition to
the more typical horizontal colonnade~\cite{Spry1962}.   A fan can form as columns radiate
away from a strong cooling site, for example. The fracture network of a developing colonnade
is highly permeable, allowing water and gas to circulate within the cracks, enhancing
cooling~\cite{Hardee1980,Long1986,Budkewitsch1994,Goehring2008,Lamur2018}.  Clear evidence
for this comes from borehole measurements in the cooling Kilauea Iki lava
lake~\cite{Hardee1980}, which for many years maintained a near-constant {100~{\degree}C}
temperature from the surface to within about ten metres of the still-molten regions of the
lake, see  Fig.~\ref{fig:scaling_columns}.  It is thought that water percolating through the
cracks boils near the molten lava, then flows away as steam to condense near the flow
margins~\cite{Hardee1980,Ryan1981,Budkewitsch1994,Goehring2008}.  The efficiency of this
two-phase convective mechanism means that a relatively constant cooling rate can be
maintained over the  periods of time --- years or decades --- that a regular colonnade takes to
develop.   

A large lava flow will tend to spread out into a thin but wide layer, which cools
simultaneously from above and below.  This leads to the common arrangement of an upper
colonnade, which cooled from the top down, and a lower colonnade, which cooled from the
bottom up.   In many landscapes multiple flow units, each from a different volcanic event,
can be layered on top of each other; there are at least four such layers visible in the
cliffs above the Giant's Causeway, for example~\cite{Tomkeieff1940}.  More complex,
multi-tiered structures have also been reported within a single flow unit, such as at the
Columbia Plateau~\cite{Long1986,Degraff1987}.  Finally, the upper colonnade is frequently
replaced by a layer of narrow and much more randomly oriented columns, a structure called
entablature, as in the top layer of  Fig.~\ref{fig:basalt_columns}A.  It is still not clear why
or how entablature forms, although it has been implicated with enhanced and localized
cooling~\cite{Long1986,Forbes2014}.~\citeA{Hamada2020} give a recent, detailed discussion of
this curious open problem.

The rate at which the lava cools also affects the size of the columns, with faster cooling
generally leading to smaller
columns~\cite{Ryan1978,Grossenbacher1995,Budkewitsch1994,Goehring2008,Goehring2009}.  The
striae provide the best evidence of this connection, which is also supported by analysis of
the crystal texture~\cite{Long1986}.   In particular, the thickness or height of the striae
provide insight into the cooling rate of the lava, and there is a strong correlation between
the size of the striae, and the size of the columns on which they
appear~\cite{Grossenbacher1995,Degraff1993,Goehring2008,Lamur2018}. This would be the case if
the column size is inversely proportional to the cooling rate of the lava when the columns
formed.

A simple physical model of the elastic stresses generated as a lava lake cools can summarize
much of our current understanding of the formation of columnar joints~\cite{Goehring2009}.
Above the boiling point of water, heat will only be transported by diffusion, with some
effective thermal diffusivity $D$.  Attention can thus focus on the temperature
$T$ of a slab or layer of rock, lying between this isotherm, and that of the
glass-transition temperature.  This region represents a cooling front that is advancing at
some speed $v$ into the lava lake.  By transforming to a coordinate system that
moves with the cooling front, heat transport is modelled by a one-dimensional
advection--diffusion equation,
\begin{equation}
\partial_t T = v \partial_z T+ D \partial_{zz} T,
\label{eq:lava_diff}
\end{equation}
for depth $z$ and time $t$.  Away from the surface, the temperature profiles across the cooling front will evolve towards a steady-state solution in the co-moving reference frame.  This problem can then be made dimensionless by scaling the system by a characteristic length-scale $L$, such that
\begin{equation}
Pe \partial_z T+ \partial_{zz} T = 0,
 \label{eq:lava_pe}
\end{equation}
where $Pe = vL/D$ is the P\'eclet number, which represents the relative strengths of
advection and diffusion.   In support of this model, borehole measurements of the cooling of
the Kilauea Iki lava lake in Hawaii show good agreement with the solutions of
Eq.~\eqref{eq:lava_pe}, as first reported by~\cite{Hardee1980} and shown in
Fig.~\ref{fig:scaling_columns}A.  

As lava cools, thermal contraction leads to the development of tensile stress, which is
relieved by fracture, as has  been described  above.   The key point here is that the crack
tips, and their driving force for fracture, are confined to a thin layer near the
solidification front.  This elastic layer is the direct result of thermal contraction and its
thickness will be inherited from the thermal profile of the lava, obeying the same scaling as
Eq.~\eqref{eq:lava_pe}: faster cooling, or weaker diffusion, will result in a thinner, more
compact front.   When a thin, brittle layer cracks, each crack will release the stress in its
immediate vicinity and the crack spacing will tend to saturate when the average distance
between cracks is a small multiple of the layer thickness~\cite{Bai2000,Yin2010}.  This is
also the case when a layer of mud, paint, or slurry dries: the final crack spacing is a few
times the layer thickness~\cite{Groisman1994,Hull1999,Shorlin2000}.  By identifying the
column size as the representative length-scale, $L$, of the thermoelastic problem,
this model of columnar joint formation predicts that all colonnades should grow under
conditions of some fixed $Pe$, of order one.  In other words, that the column size
$L$ is proportional to, and of the same order as, the advection--diffusion length
$D/v$.   This prediction is shown in  Fig.~\ref{fig:scaling_columns}B, along with field
data from~\citeA{Goehring2009} that is best fit by $Pe = 0.3$.  The observed scaling also
explains the earlier finding that a column's size, on average, is inversely proportional to
the cooling rate, and hence $v$, of a colonnade.

The development and validation of this model of columnar joints has been greatly aided by
analogue laboratory experiments involving colonnades produced in drying starch; see
Fig.~\ref{fig:basalt_columns}D.  The interpretation of such studies relies on the  mathematical
analogy between the theories of poroelasticity and thermoelasticity, i.e.,~the equivalence
between the elastic stresses generated by drying and by cooling,
respectively~\cite{Goehring2009}.  This correspondence is demonstrated in
Fig.~\ref{fig:scaling_columns}B, where data from dried starch colonnades are plotted against a
best-fit line of $Pe=0.15$.  The first detailed experiments of~\citeA{Muller1998}
highlighted the value of drying starch as an analogue system, although there are occasional
mentions of it in earlier literature~\cite{Huxley1881,French1925}.  M\"uller's work was
followed by a series of quantitative studies that showed how to control column size by
adjusting the drying rate of the starch~\cite{Toramaru2004}; tracked how the pattern ordered
with depth~\cite{Goehring2005}; linked column growth to the passage of a sharp drying
front~\cite{Mizuguchi2005}; and provided numerical simulations of crack arrays advancing
along with the drying front~\cite{Nishimoto2007}.  Plumose structures on the larger joints of
dried starch~\cite{Muller2000} and other pastes~\cite{Sakaguchi2022} confirmed the
interpretation of their appearance in lava joints. Three-dimensional images of dried starch
colonnades, obtained by X-ray
tomography~\cite{Muller1998,Mizuguchi2005,Goehring2005,Goehring2006,Crostack2012,Hamada2020},
also confirmed how columns grow normal to the direction of drying, analogous to cooling, and
led to the current appreciation of how the column size depends inversely on the drying or
cooling rate.  These analogue experiments, and the mathematical analogy between cooling and
drying that enables a strong interpretation of their results, are reviewed in detail in
chapter 7 of~\citeA{Bacchin2018}, and chapter 9 of~\citeA{GoehringBook}.   

More recently, research on columnar jointing has progressed into a number of promising
additional areas.  Detailed mapping of the topology and geometry of columnar joints has led
to the development of simple, low-dimensional metrics that can distinguish between a wide
variety of crack patterns with distinct origins~\cite{Domokos2020,Roy2021,Domokos2024}.  The
origins of entablature remain enigmatic, but innovative lines of experiment~\cite{Hamada2020}
have begun to unravel their nature.  The related concept of \emph{master cracks}, joints that
enable faster local cooling, is also being
explored~\cite{Forbes2014,Moore2019,Hamada2020,Akiba2021}.  With more practical aims,
research has been made into how the strongly anisotropic texture of columnar joints affects
their hydrological and mechanical properties, such as their relative permeability to fluid
flow along different directions~\cite{Vasseur2019,Chao2020,Niu2023}, and their strength and
resistance to deformation~\cite{Que2023}.  These applications address questions such how best to
 utilize basalt formations for carbon geosequestration~\cite{Jayne2019,Wu2021}.

\subsubsection{Hydrothermal vents}
\label{sec:hydrothermalvents}

\begin{figure}
\centering
\includegraphics{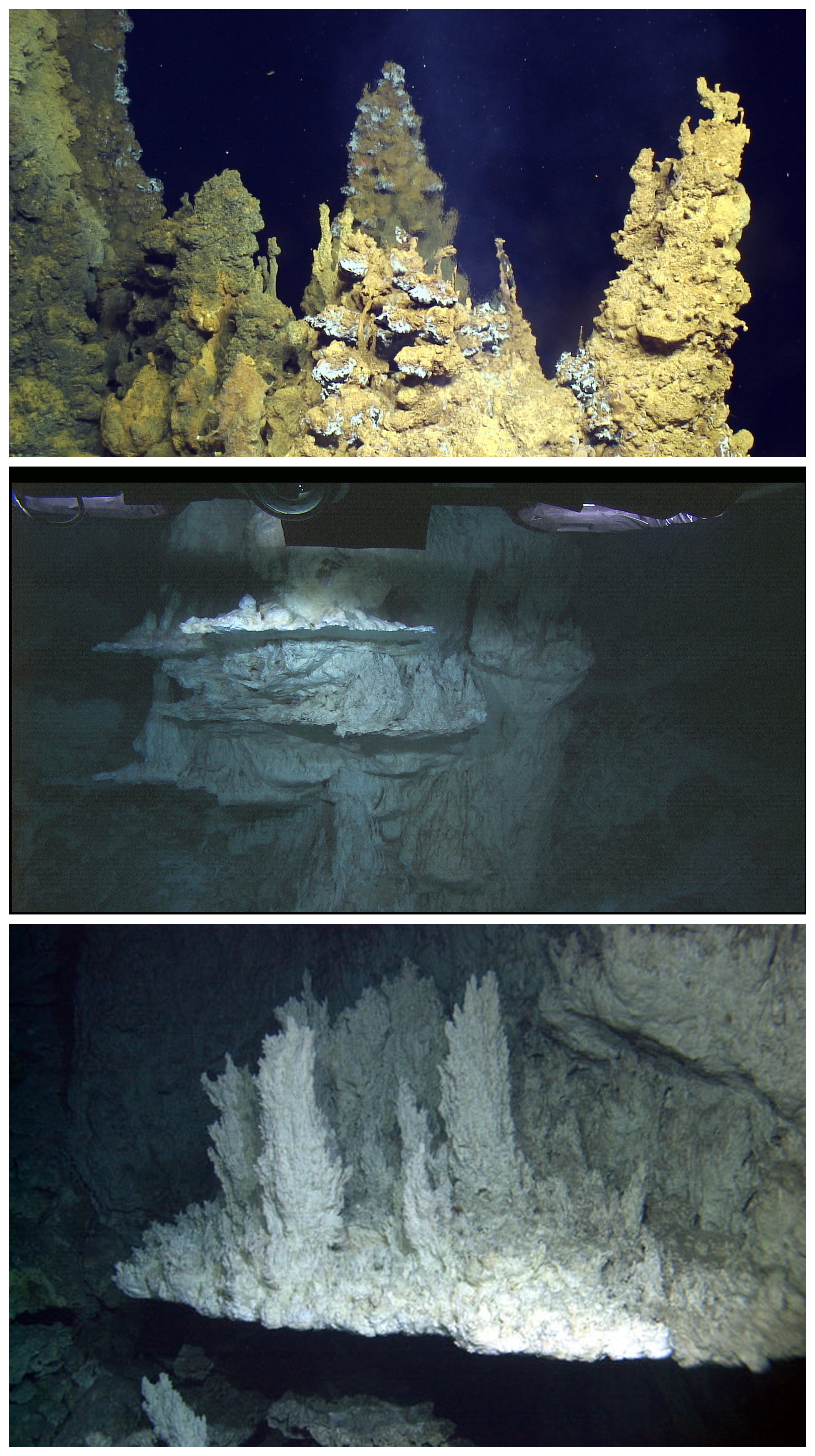}
\caption{\label{fig:hydrothermal_vents}Hydrothermal vent structures: chimneys and flanges. \\
Images: Submarine Ring of Fire 2014 - Ironman, NSF/NOAA, CC-BY-SA-2.0; NOAA, public domain; NSF, public domain.}
\end{figure}

Hydrothermal vents are fissures of the oceanic or continental crust that emit hot water (up to 400{\degree}C), typically rich in metals and metalloids. The majority of the known hydrothermal vents are found along mid-ocean ridges (65\%), in back-arc basins (22\%), on volcanic arcs (12\%), and on intraplate volcanoes (1\%) \cite{baker2004global,hannington2002global}. Submarine hydrothermal vents form as seawater percolates down through cracks in the oceanic crust, gets heated by magmatic sources, reacts with the surrounding rocks/sediments leaching metals, and rises back to the seafloor rich in magmatic gases. When the hydrothermal water, meets the cold, oxygenated, slightly-basic pH seawater, minerals precipitate out forming the characteristic chimneys or smokers (Fig.~\ref{fig:hydrothermal_vents}). Black smokers tend to form atop oceanic spreading ridges when hot (up to > 400{\degree}C), sulfide-rich, acidic plumes vent out, primarily composed of iron sulphide minerals, responsible for the characteristic dark color. While, cooler (250--300{\degree}C) vents typically issuing from more distal, off-axis centers, that emit lighter-coloured plumes, rich in barium, calcium, and silica, are called white smokers \cite{von1990seafloor}.  In 2000 a new category of off-axis hydrothermal vents was discovered, known as the Lost City \cite{kelley2001off}, discharging low temperature ($< 91${\degree}C), alkaline fluids (pH 9--11), rich in methane, hydrogen and hydrocarbons, arising from the hydrothermal alteration of the host, ultramafic rocks (a process known as serpentinization).

\begin{figure}
\centering
\includegraphics{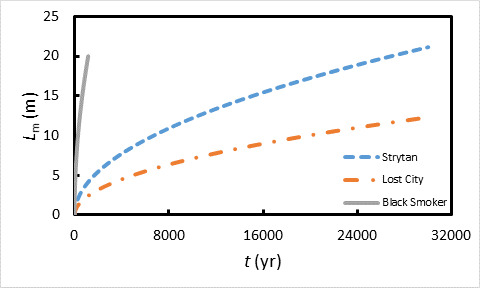}
\caption{\label{fig:hydrothermal_vents2}Fluid-mechanical model prediction of the growth of a hydrothermal vent
over time $t$ in terms of wall thickness $L_m$ compares black
smokers (grey) Strytan (blue) and Lost City Hydrothermal Field (orange)
vents.
}
\par\smallskip{\footnotesize\noindent\raggedright\emph{Source:} \citeA{gutierrez2024magnesium}.\par}
\end{figure}

The chemical and thermal differences between the vent types are reflected in differing physical growth mechanisms \cite{cardoso2017differing}. A black smoker attains a thickness of 7~cm in 5 days~\cite{black}, while a white smoker needs 2 years to achieve a thickness of 10 cm~\cite{white}. Thus transport across the wall for hot black-smoker vents should be slower than for cold white-smoker vents, all other conditions being equal, and so hot vents are more isolated from the surroundings. Radial tube growth can be driven by (1) thermal diffusion or (2) chemical diffusion; the former is faster and produces a solid wall more rapidly than the latter. Consequently, hot vents experience less contact between vent fluids and seawater (the chimney becomes solid quickly), whereas cold vents permit longer exchanges across the wall. These differences are seen in Fig.~\ref{fig:hydrothermal_vents2}, which shows the growth rates of different vent types~\cite{gutierrez2024magnesium}. In fact, low-temperature hydrothermal vents can be considered as natural analogues of far-from-equilibrium, self-organized, tubular precipitates best known as chemical gardens~\cite{barge2015chemical}. In a chemical-garden experiment, a soluble metal salt introduced into a solution containing a counter-anion, often silicate or carbonate, precipitates an insoluble salt~\cite{cartwright2011chemical}. The precipitate forms a cohesive membrane that envelopes the jet while still permitting ion transport and continued reaction~\cite{cartwright2002formation}. 

Aside from submarine hydrothermal systems, terrestrial hydrothermal vents also occur, with well-known examples in Yellowstone, USA \cite{hurwitz2014dynamics} and Dallol, Ethiopia \cite{Kotopoulouetal.2019,kotopoulou2025natural}. In this case they form as meteoric and/or groundwater seeps underground through cracks, is heated by shallow magmatic or geothermal sources, reacts with the surrounding rocks/sediments and rises up discharging hot water, rich in magmatic gases in the form of fumaroles, geysers and chimneys. 

Hydrothermal vents display certain characteristic patterns; chimneys and flanges are typical of submarine systems (Fig.~\ref{fig:hydrothermal_vents}), while fumaroles, geysers, chimneys, terraces and associated pools are common in terrestrial ones. The primary structures, chimneys, are tall, tubular structures formed by mineral precipitation that builds a sheath around the fluid plume as it emerges from below, and are primary structures at venting sites. The grow over time, as long as the hydrothermal spring remains active, thus can reach several meters in height. Their internal structure may consist of a single conduit through which the mineral-laden water flows, or it may be much more complex, comprising a multitude of micro- to nano-scale channels permitting slower, more diffuse flow through a porous matrix  (Fig.~\ref{fig:membrane}). Such flows are likely to display chaotic advection (Sec.~\ref{sec:chaotic}). Flanges are horizontal structures that project from the sides of chimneys, creating ledge-like features from which secondary chimneys may also grow upward. Flanges form by mineral deposition as the hydrothermal fluids cool and precipitate minerals. They often trap pockets of superheated water beneath them~\cite{turner1995laboratory}. Meanwhile, terracing is a classic example of self-organizing pattern formation, found commonly, but not exclusively, in hydrothermal systems. Terrace and associated nested pool formation is a complex process, controlled by a series of parameters such as hydrothermal flow dynamics, brine chemistry, geomorphology, mineral nucleation rates, presence of microorganisms and fault activity, among others \cite{meakin2010}. Interruptions of vent activity, shifts of the vent location, and/or of the flow directions and changes in vent chemistry, that commonly occur in hydrothermal systems, further shape the growth and morphology of the terraces and pools. We discuss terracing mechanisms, both biological and otherwise, in a separate review  \cite{geo_bio_review}.

Ever since their discovery \cite{corliss1979submarine}, hydrothermal vents have attracted scientific interest for various reasons. Firstly, they are one of the oldest and most important ore-forming processes on Earth that have produced some of the largest and most valuable ore deposits mined to date~\cite{tivey2007generation}. Secondly, hydrothermal vents support unique chemosynthetic ecosystems, including extremophilic bacteria and archaea, the discovery of which has pushed further the limits of life on Earth and beyond \cite{van2000ecology}. Thirdly, Lost City type vents that are associated with serpentinization~\cite{kelley2005serpentinite,sainz2018growth} and analogous shallow-sea systems at Strytan, Iceland~\cite{stanulla2017,gutierrez2024magnesium} and at Prony Bay, New Caledonia~\cite{monnin2014}, have been suggested as candidate sites for the origin of life on Earth~\cite{russell1997,Martin2007,Cartwright2019}. 
The hypothesis is that micro-compartments
in submarine alkaline hydrothermal vents separated by mineral membranes may have been where
the first proto-bio-chemistry emerged~\cite{Sander2011,Sojo2016}.

\subsubsection{Mud volcanoes, pockmarks and seeps}
\label{sec:mudvolcanoes}

Submarine seeps and mud volcanoes  are structures that form  part of the Earth's fluid
venting systems. These structures expel  fluids, including liquids, principally water but
also hydrocarbons, and gases (e.g.,~carbon dioxide and methane), from the subsurface to the
seafloor~\cite{judd2009seabed} or land surface~\cite{kopf2002significance}. 
Cold seeps are associated with the slow seepage of hydrocarbons, methane, and other fluids at or near ambient seafloor temperatures.
Hot seeps involve the expulsion of warmer fluids, often associated with hydrothermal systems.
Pockmarks are shallow depressions formed by the escape of fluids and gases from the seabed, resulting in crater-like morphologies.
Submarine mud volcanoes occur on the seafloor and are often linked to deep-seated overpressure zones, faulting, or hydrocarbon reservoirs.
Terrestrial mud volcanoes are found on land and are often associated with tectonic plate boundaries or sedimentary basins with high fluid pressure.

\begin{figure}
\centering
\includegraphics{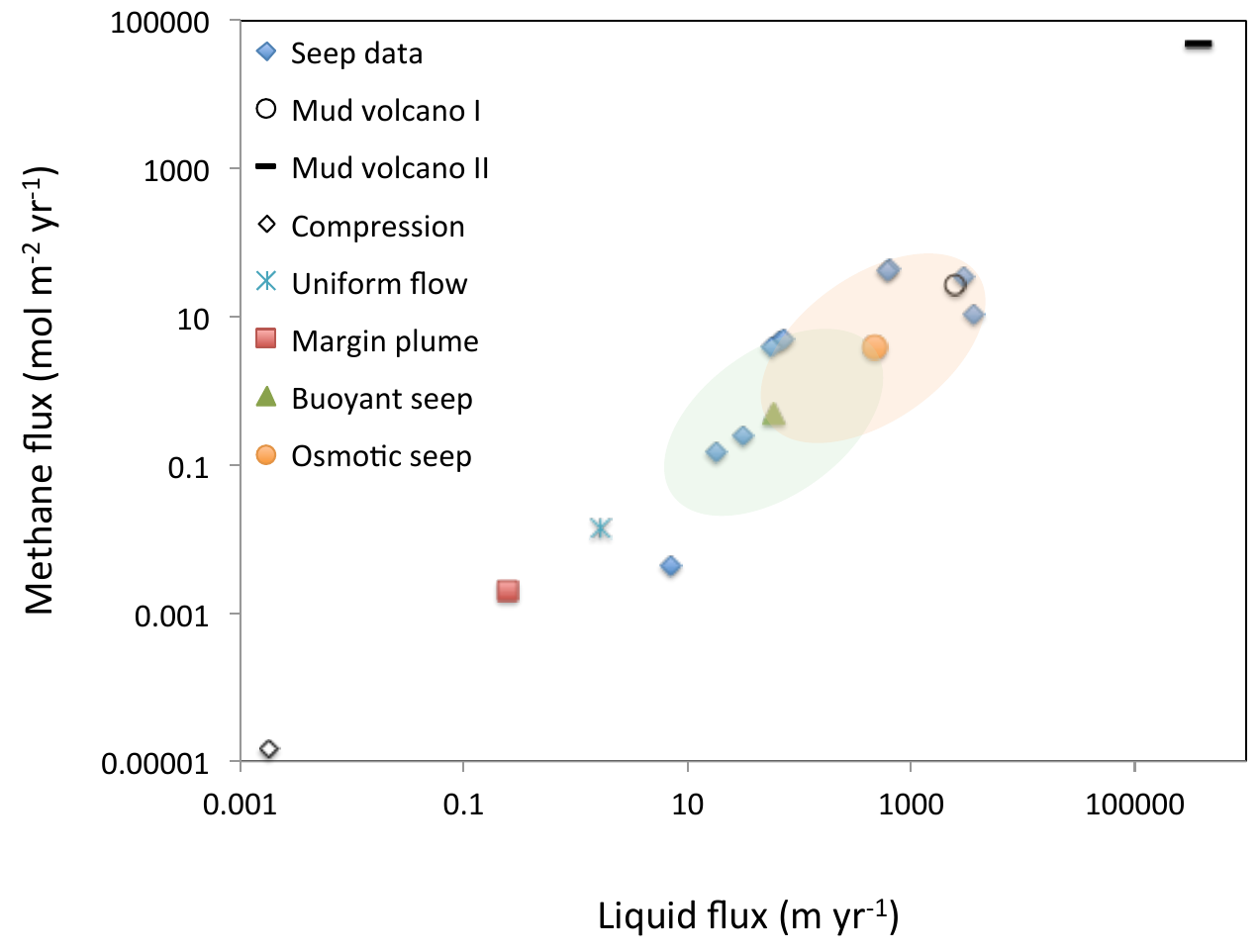}
\caption{\label{fig:mud_volcanoes2}
Dissolved methane flux plotted against liquid flux; comparison of field measurements at seeps and mud
volcanoes with  theoretical predictions for a
uniform source of solute and a margin heat plume, and a buoyant or
osmotic plume in a developed seep. The predictions are for a sediment
permeability of 10\tsup{\tmin12}~m$^2$ and an exit methane concentration of 8~mM; the
green and orange shaded ellipses represent the range of permeabilities
10\tsup{\tmin13}--10\tsup{\tmin11}~m$^2$ (along the major axis) and methane concentrations
0.6--126~mM (along the minor axis) for a buoyant and osmotic seep flow,
respectively. An estimate of efflux from sediment compression alone  is shown
as a baseline.}
\par\smallskip{\footnotesize\noindent\raggedright\emph{Source:} \citeA{cardoso2016increased}.\par}
\end{figure}

\begin{figure}
\centering
\includegraphics{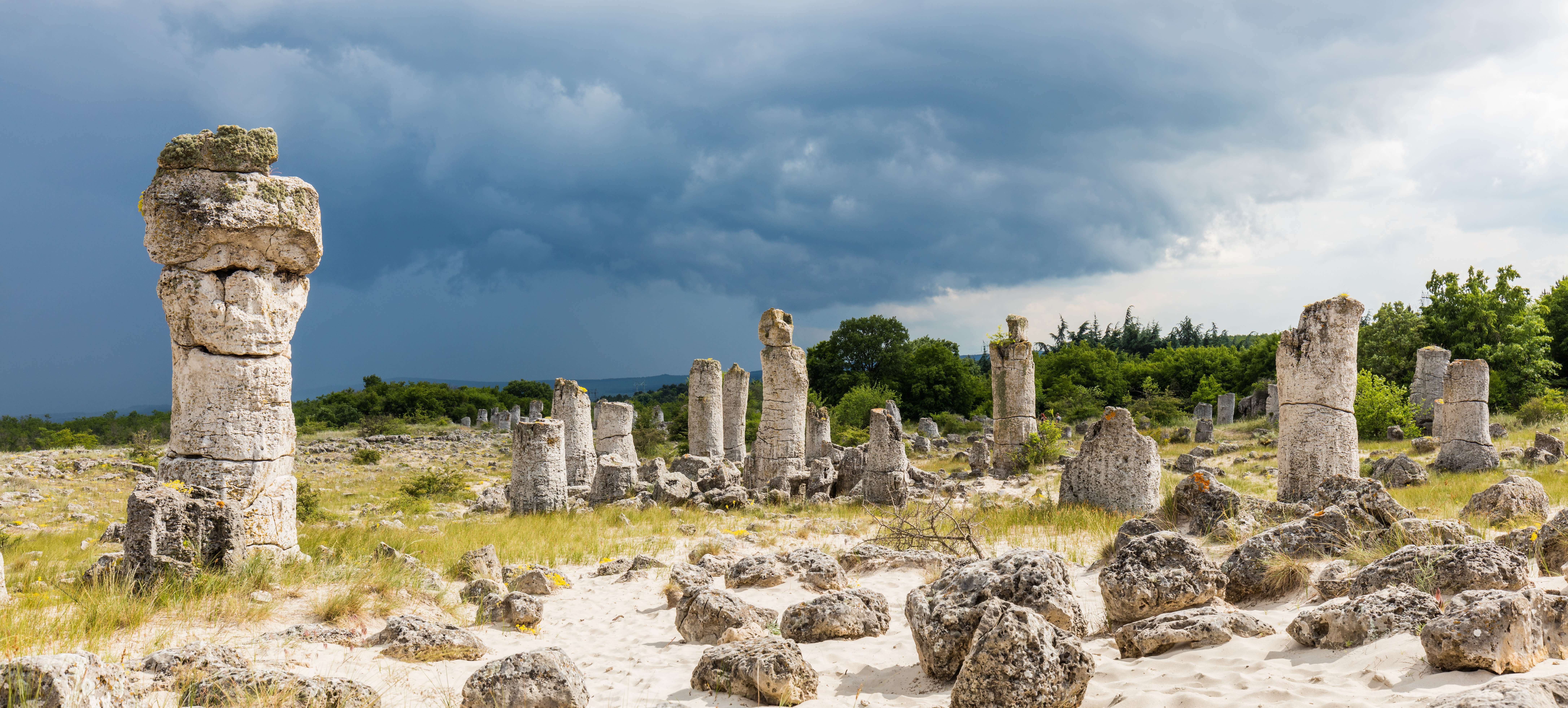}
\caption{\label{fig:mud_volcanoes}
Tubular concretions of authigenic carbonates, Pobiti Kamani, Varna, Bulgaria, preserve the subsurface plumbing network of an Early Eocene methane seep system in the Balkanides foreland 
\cite{de2009formation}.
}
\par\smallskip{\footnotesize\noindent\raggedright\emph{Image:} Diego Delso; CC-BY-SA-4.0.\par}
\end{figure}

Submarine seeps typically involve the upward migration of fluids through vertical conduits, such as fractures, faults, or permeable sedimentary layers. 
Mud volcanoes, both submarine and terrestrial, are characterized by the upward movement of fluidized sediments, mud, and gases through a central vertical conduit. 
These conduits allow fluids to move from deep subsurface reservoirs to the seafloor or land surface. 
The vertical conduits of mud volcanoes are typically associated with high-pressure environments, where fluid overpressure exceeds the strength of overlying sediments, leading to the formation of fractures and the eruption of materials.
The process is driven by buoyancy forces, where lighter fluids like hydrocarbons and gases
rise through denser surrounding sediments, and by osmotic
pressure~\cite{cardoso2016increased} (Fig.~\ref{fig:mud_volcanoes2}).
In some cases, fluids may migrate laterally along permeable layers before reaching the
seafloor~\cite{rocha2021formation}.
Horizontal migration is influenced by geological structures like unconformities or bedding
planes, as well as by self-sealing/self-capping processes~\cite{hovland2002self}.
 
Submarine seeps and mud volcanoes often lead to the precipitation of carbonate minerals, forming distinctive authigenic carbonate structures, such as mounds, tubes, or chimneys (Fig.~\ref{fig:mud_volcanoes}). 
As in many geological examples in this review, the formation mechanisms have not been confirmed experimentally. At least three aspects of their formation can, however, be identified. First, dissolution may play a role in some cases; as in the case of solution pipes (Sec.~\ref{sec:pipes}), its importance depends on the rock type and the pH of the fluid. Second, probably more common than dissolution is the washout of unconsolidated sediment by fluid flow in a self-concentrating process. Third, precipitation commonly occurs along the flow walls as CO$_2$ is converted into carbonate rock. This is frequently bacterially mediated.
Expelled materials from mud volcanoes, such as mud and brecciated rock fragments, can lithify over time, forming mudstone or breccia deposits. These rocks serve as evidence of past mud volcanic activity.
Mud volcanoes may leave behind diatreme-like structures or intrusive bodies of mudstone that
cut through surrounding sediments, providing clues to the ancient venting
processes~\cite{brown1990nature}.
In some cases, large carbonate structures known as chemoherms can form around seeps due to
the activity of chemosynthetic organisms~\cite{teichert2005chemoherms}. 
These features are preserved in the rock record and indicate past fluid seepage
events~\cite{conti1999miocene}.

\subsubsection{Fulgurites}
\label{sec:Fulgurites}

\begin{figure}
\centering
\includegraphics{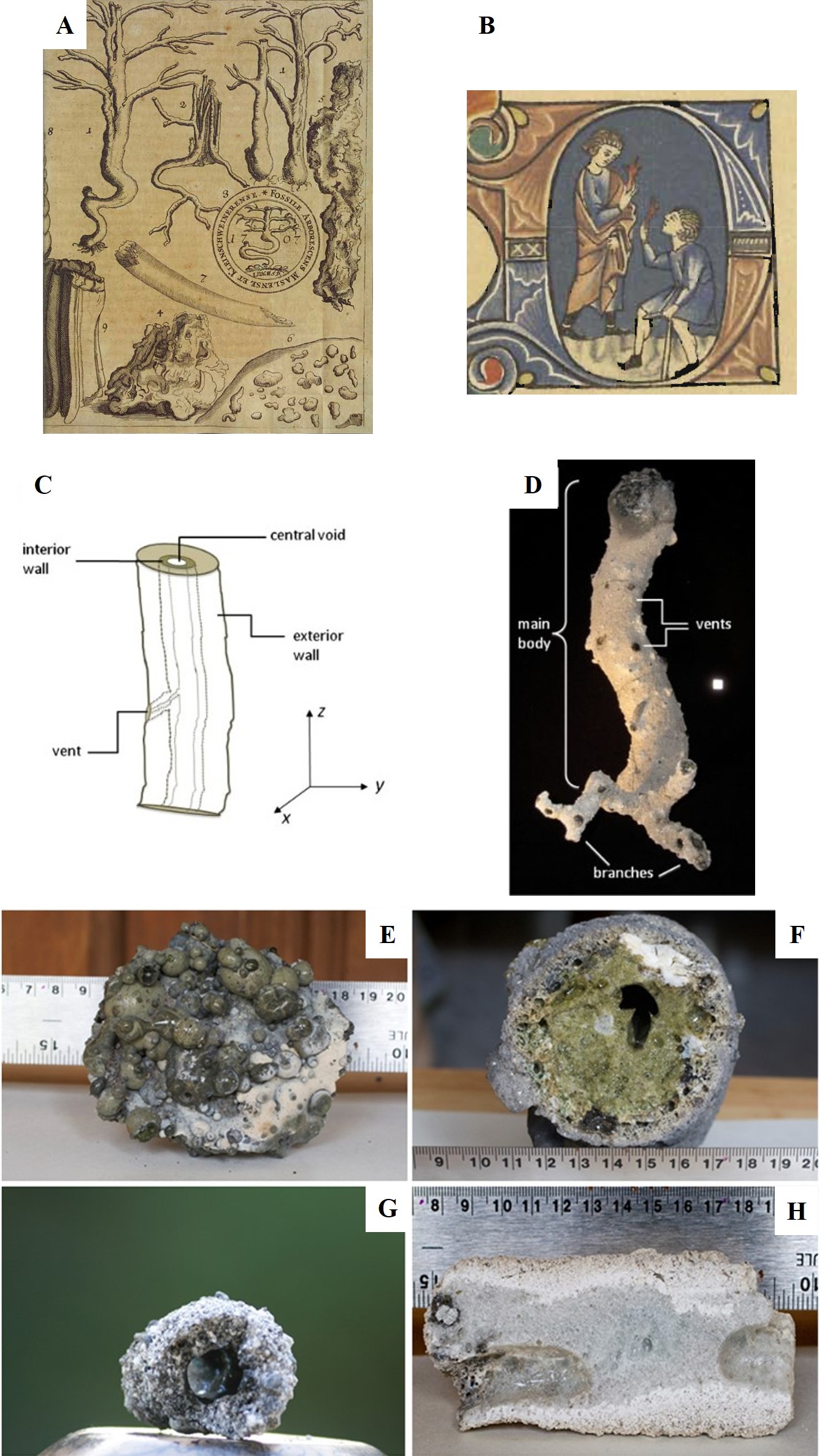}
\caption{\label{fulgurite}A: Fulgurite illustrations from the book written
by~\citeA{hermann_maslographia_1711}. B: Illustration of the lapidary of Alfonso X El
Sabio~\cite{alfonso_x_lapidario_1881}, showing rocks with an arborescent
aspect that resemble fulgurites.
C: Scheme of fulgurite structure~\cite{pasek_fulgurite_2012}. D: Lateral view of a natural
fulgurite~\cite{pasek_fulgurite_2012}. 
E--H: Fulgurites of ambiguous (natural or artificial/anthropogenic)
origin~\cite{pasek_forensics_2018}. An exogenic fulgurite can be seen in E, while in F--H
fulgurites with a hollow tube in the inner part are shown. }
\par\smallskip{\footnotesize\noindent\raggedright\emph{Sources:} A:~\citeA{hermann_maslographia_1711},
B:~\citeA{alfonso_x_lapidario_1881},
    C,D:~\citeA{pasek_fulgurite_2012},
E--H:~\citeA{pasek_forensics_2018}.\par}
\label{naturalfulgurite}\label{fulguriteOLD}
\end{figure}

Fulgurites are hollow glass tubes formed in sand, soil or rock by high-intensity electrical
discharges, with lengths ranging from centimetres to
metres~\cite{pasek_forensics_2018,grapes_pyrometamorphism_2010}. The first formal description
of fulgurites may have been by~\citeA{hermann_maslographia_1711} in the early 18th century
(Fig.~\ref{fulguriteOLD}A)~\cite{petty_origin_1936,gailliot_petrified_1980,carter_raman_2010},
but in  the 13th
century~\cite{garciaguinea_quartzofeldspathic_2009,martin-ramos_characterization_2019}, in
the lapidary of Alfonso X \emph{El Sabio}, King of Spain,  they may appear under the name
mazintarincan (Fig.~\ref{fulguriteOLD}B)~\cite{alfonso_x_lapidario_1881}.
They form when lightning rapidly heats the substrate via the high-current discharge,
producing glass. The substrate may be sand, clay soil, caliche, rock, or anthropogenic
material~\cite{pasek_forensics_2018,pasek_fulgurite_2012}. For a comprehensive survey of
artificial fulgurites, see~\citeA{pasek_forensics_2018}.

When lightning strikes, rapid heating generates voids and vesicles along the discharge path,
allowing volatile escape. These processes produce a  glassy silica wall of cylindrical or
elongated conical, usually hollow, morphology. This vitreous inner wall is surrounded by a
rough outer rind containing partially melted and unmelted grains
(Fig.~\ref{naturalfulgurite})~\cite{pasek_fulgurite_2012}. The origin of the central cavity has
been attributed classically to gas expansion that presses and fuses the sand as the lightning
travels through the ground~\cite{petty_origin_1936}, and, more recently, to sand vaporization
along the discharge path~\cite{pasek_forensics_2018}. The central hole, which may be single
or composed of multiple closely spaced voids, is aligned with the lightning propagation
direction ($z$-direction in  Fig.~\ref{naturalfulgurite}C). Secondary branches commonly
emanate from the primary tube. The primary branch is typically the widest, though seldom more
than a few centimetres in diameter~\cite{pasek_fulgurite_2012}.

Fulgurites have been divided into four categories, based on the ground
composition~\cite{pasek_fulgurite_2012}. Type I, sand fulgurites, are mainly composed of
silica glass, and develop a thin inner wall (interior wall in  Fig.~\ref{naturalfulgurite}C). If
present, the outer wall is also composed mainly of silica glass with traces of Al and Fe
oxides. Type II, clay fulgurites, form in soils composed of clays, sand and/or small rock
fragments. Both inner and outer walls are clearly visible in these fulgurites. This type is
wider and has larger glassy regions than type I. Type III, caliche fulgurites, form in desert
hardpan or caliche. Silica glass constitutes less than 10\% of the material and is
embedded within calcite. Type IV, rock fulgurites, form in bedrock; the glass walls are
surrounded by unmelted host rock. They often occur as narrow tubes on the rock surface.
In addition, a subcategory has been proposed: exogenic/droplet fulgurites (type
V)~\cite{pasek_fulgurite_2012}. This type forms when, after a very powerful lightning strike,
($>100$~GW), melted substrate is ejected into the atmosphere, solidifying in the air.
Rapid air cooling yields amorphous, droplet- or bubble-shaped morphologies that are generally
greenish in colour~\cite{pasek_fulgurite_2012,martin-ramos_characterization_2019}. These
fulgurites are mainly composed of glass and are often associated with Types II and IV
fulgurites~\cite{pasek_fulgurite_2012}.
It has been proposed that the phosphide mineral schreibersite, (Fe,Ni)$_3$P, which
can  form as spherulites in fulgurites, may have provided an important source of prebiotic
phosphorus on the early Earth, potentially facilitating the synthesis of  phosphorus-bearing
organic compounds~\cite{hess_lightning_2021}.

Similar discharge morphologies are also observed in insulating materials, providing a useful
counterpart to fulgurites. In this context, one speaks of Lichtenberg figures: intricate,
branching patterns formed by dielectric breakdown in insulators~\cite{Niemeyer1984}. They
develop on the surface or within materials such as wood, acrylic, or glass when subjected to
high voltage, and resemble fern-like or lightning-bolt shapes.  These patterns trace the path
of least resistance that the electrical current takes through the insulating material,
leaving permanent marks or indentations. Breakdown can occur either on the surface or within
the bulk, depending on material properties and field strength. Both Lichtenberg figures and
fulgurites preserve the geometry of electrical discharge paths, but in different media:
Lichtenberg figures record breakdown channels within solid dielectrics, while fulgurites
capture the same process in geological materials, vitrified by extreme heating.

\subsubsection{Foliation in gneiss}
\label{sec:gneiss}

\begin{figure}
\centering
\includegraphics{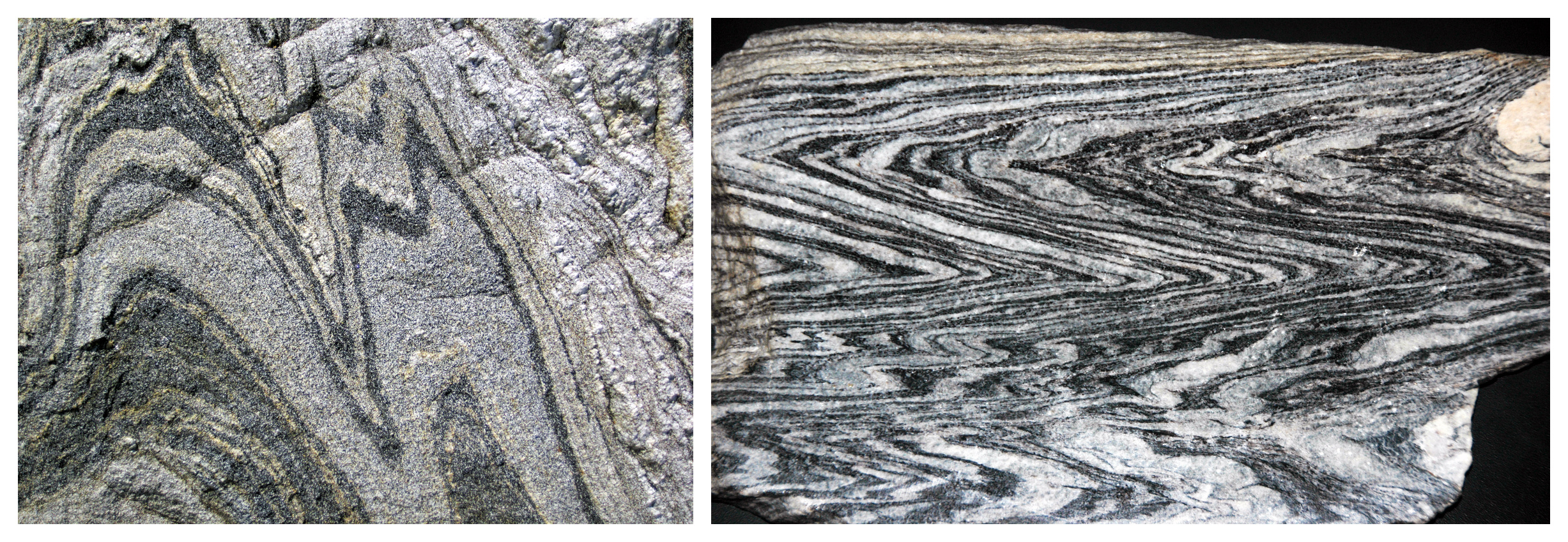}
\caption{\label{fig:gneiss}Foliation in gneiss.
Left: Gneiss foliation, Archean; Ennis Lake North roadcut, Madison County, Montana, USA;
Right: Chevron folds in gneiss, Precambrian; Medicine Bow Mountains, Wyoming, USA.
}
\par\smallskip{\footnotesize\noindent\raggedright\emph{Images:} James St. John; CC-BY-2.0.\par}
\end{figure}

 Foliation or banding in gneiss refers to the distinctive layering  observed in this
metamorphic rock,  Fig.~\ref{fig:gneiss}~\cite{passchier2012field}. Gneiss is formed through the
process of metamorphism, which involves the alteration of pre-existing rocks under high
pressure and temperature conditions. During this process, the minerals within the rock
recrystallize and reorganize, leading to the development of a foliation.
The foliation in gneiss is a result of the alignment of minerals or mineral layers parallel to the direction of pressure during metamorphism. This alignment gives gneiss its characteristic layered appearance, with alternating bands of different mineral compositions and colours. The banding is typically visible to the naked eye and can be straight or wavy.
Foliation in gneiss can be classified into two main types: compositional and structural. Compositional foliation refers to the variation in mineral composition across the rock, creating bands of different minerals such as quartz, feldspar, and mica. Structural foliation, on the other hand, is related to the deformation of the rock during metamorphism, resulting in the alignment of mineral crystals or the development of preferred orientations.
The presence of foliation in gneiss provides important information about its geological history and the conditions under which it formed. It indicates the intense pressure and temperature that the rock experienced during metamorphism. Additionally, the distinct layering in gneiss can be used to determine the direction and intensity of tectonic forces that acted on the rock.

\subsubsection{Seismites}
\label{sec:seismites}

The term \emph{seismite} was coined in 1969 by Seilacher to describe structures in the
Miocene Monterey Formation at Santa Barbara, California,
USA~\cite{seilacher_fault-graded_1969}. He described a regular sequence comprising, from top
to bottom, a liquefied zone, a debris zone, and a step zone, with gradational internal
contacts and a sharp upper boundary, and interpreted this succession as earthquake
generated~\cite{seilacher_fault-graded_1969}. Since then, numerous authors have used the term
\emph{seismite} to denote an earthquake-induced sedimentary
structure~\cite{alfaro_significance_2010,alfaro_soft-sediment_1997,montenat_seismites_2007,moretti_liquefaction_2011}.
However, the term is debated, because many purported seismites reflect sediment
liquefaction~\cite{shanmugam_seismite_2016}, and liquefaction can be triggered by multiple
mechanisms; up to 21 have been described, including earthquakes, sediment loading, salt
tectonics, and volcanic activity. Accordingly, some authors advocate the broader term
soft-sediment deformation structures (SSDS)~\cite{shanmugam_seismite_2016}. Settling this
terminological debate is beyond the scope of this review (see,
e.g.,~\citeA{shanmugam_seismite_2016}, which lists 27 SSDS types). Here we describe
structures that have been reported as earthquake-related and we  refer to them to as
seismites. 

\begin{figure}
\centering
\includegraphics{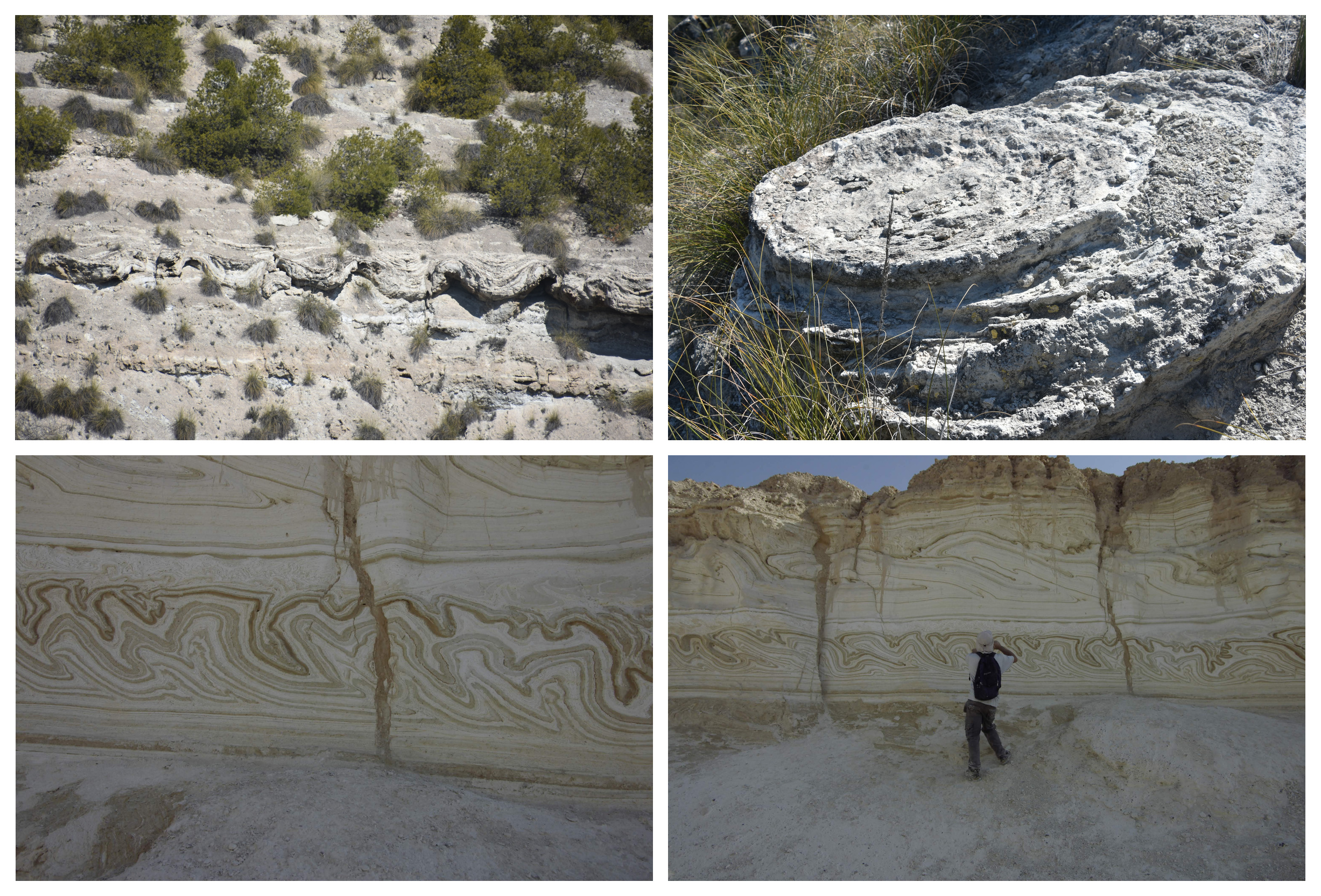}
\caption{\label{seismite}Seismites. Above: giant seismites of Galera, Granada, Spain; on the left an overview of the pillow-like structures, on the right a detail of one of the pillows, about two metres across. 
Below: seismites in Lake Lisan (Pleistocene Dead Sea) sediments, Israel.
}
\par\smallskip{\footnotesize\noindent\raggedright\emph{Images:} Above: Julyan Cartwright, below: Piotr Szymczak.\par}
\end{figure}

Seismites (Fig.~\ref{seismite})  occur all over the world, e.g.,~in the
USA~\cite{seilacher_fault-graded_1969,Jewell2004ancient},
Germany~\cite{Bachmann2005seismite}, and 
Israel~\cite{Kagan2018Integrated},    to name just a few. 
In the work of~\citet{rodriguez-pascua_soft-sediment_2000}, focused on the Miocene basins of
the Prebetic Zone,  Spain, the different SSDS found in the region are described and related
to earthquake magnitude. According to~\citet{rodriguez-pascua_soft-sediment_2000}, seismites
in this region fall into two groups: features formed without liquefaction --- typically linked
to earthquakes of $M < 5$ --- and those formed after sediment liquefaction, generally
requiring larger magnitudes. Non-liquefaction structures include loop bedding, disturbed
varved lamination, and mixed layers without fluidization. Liquefaction-related structures
include mushroom-like silts protruding into laminites, mixed layers with fluidization,
pseudonodules, sand dykes, pillow structures, and intruded and fractured gravels; metre-scale
features observed in shallow-lake, detrital settings.
Among these soft-sediment deformation structures, the most striking are pillow structures, of
which exceptional examples are found in the Baza Basin, Spain (Fig.~\ref{seismite}). These pillow
structures occur in deformed beds with thickness varying from a few cm to 2.5~m, the best
examples being 1.5--2~m thick and 2--4~m in width~\cite{alfaro_significance_2010}. These
structures were generated when a lower-density unit (clays and silt) was overlain by a higher
density unit (fine- and coarse-grained sands). When an earthquake occurred, the liquefaction
of the clays caused them to rise upwards, breaking the overlying layer, generating escape
structures and imparting the sand unit with a pillow-like
morphology~\cite{alfaro_significance_2010} in a geological manifestation of the
Rayleigh--Taylor instability (Sec.~\ref{sec:RT}). In some cases, lateral shear between mobilized
and overlying layers may also produce secondary Kelvin--Helmholtz billows (Sec.~\ref{sec:KH}),
adding a wavy or rolled appearance to the interface.

\section{Summary and outlook}

The physical mechanisms outlined in  Sec.~\ref{sec:physical} generate the mesoscale geological patterns compiled in  Sec.~\ref{sec:patterns}. They act on the material classes of  Sec.~\ref{sec:types}, which in many respects  align with soft-matter systems (Sec.~\ref{sec:soft}). In this sense, mesoscale geological pattern formation is mediated by soft-matter physics, with mesoscopic building blocks (grains, pores, films) assembling into structures that remain mechanically compliant and reconfigurable as they form. Understanding their formation mechanisms is essential for explaining geological self-organization and constraining the physicochemical and mechanical conditions in the environments where these patterns originate. This, in turn, is critical for reconstructing Earth's past environmental conditions and geodynamic processes, and for distinguishing complex abiotic patterns from potential biosignatures---especially when evidence is sought for ancient life on Earth and, potentially, on other planets.

One key aspect is \emph{self}-organization, i.e., the emergence of structure through internal feedbacks within a minimally defined system. Because ``the system'' can always be expanded  so that any pattern appears self-organized, we adopt the smallest system that closes the feedback loop and treat Earth's boundary conditions and large-scale drivers as external.

Although Earth scientists have long studied paradigmatic geological pattern-forming systems,
they have not always labelled them \emph{self-organized}. An oscillatory mesoscale mechanism
was first proposed by~\citeA{harloff1927},~\citeA{phemister1934}, and~\citeA{hills1936} to
explain plagioclase zoning. In the 1970s--80s, studies
by~\citeA{fisher1973},~\citeA{fraser1977},~\citeA{ortoleva1984} and~\citeA{ortoleva1987} explicitly
attributed some rock and mineral patterns to geochemical self-organization. Since then, ideas from nonlinear dynamics have spread from crystals and rocks to a wide range of geological systems. Interest in this area has recently grown for two reasons: (i) mounting evidence that geochemical
self-organization may have catalysed key steps in the transition from geochemistry to
biology~\cite{russell1994,hanczyc2007,mcglynn2012,cardoso2020,kotopoulou2020,garcia2020,kotopoulou2021};
and (ii) the realization that abiotic self-organized structures such as silicate--carbonate
biomorphs, though of limited geological relevance, can mimic biological morphologies and may therefore be misinterpreted ~\cite{garcia2003,geo_bio_review}.

\begin{figure}
\centering
\includegraphics{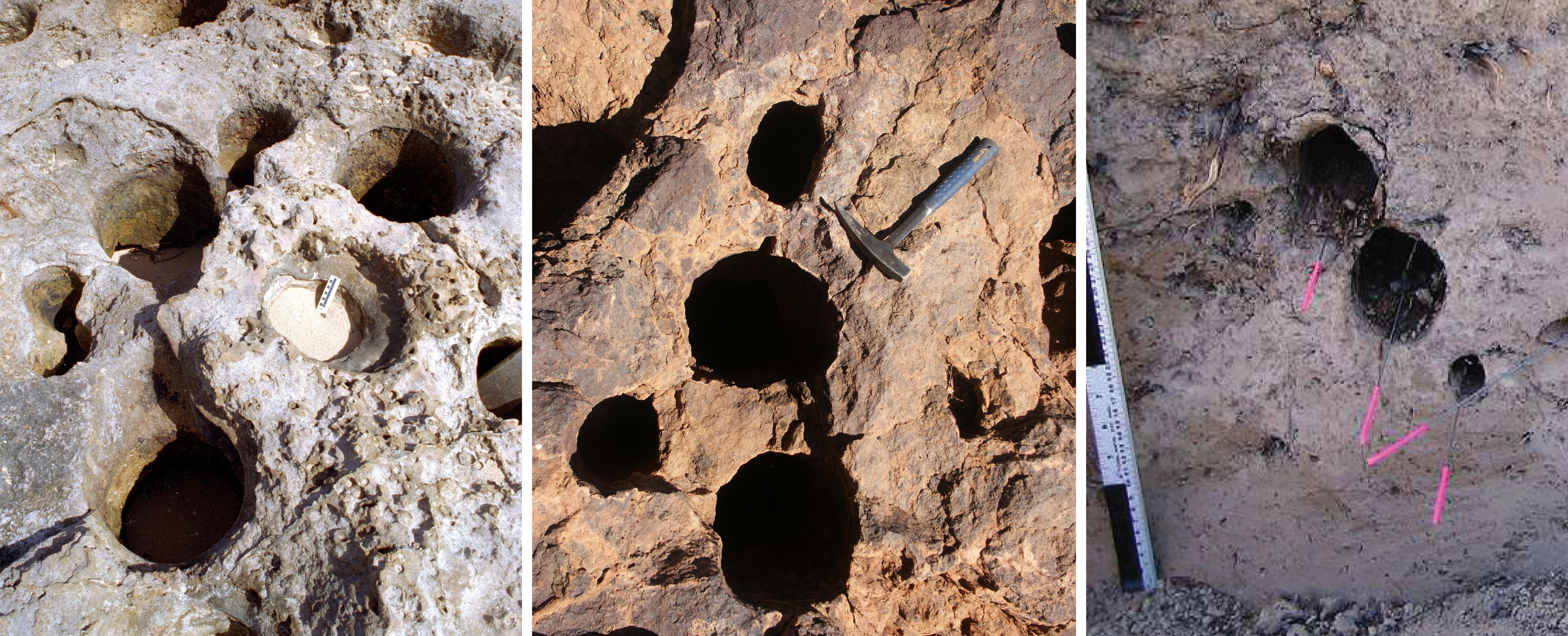}
\caption{\label{fig:soilpipes}
Equifinality of pipe-like forms produced by different processes. (Left) Karst solution pipes
in dune limestone, Canunda National Park, South Australia. (Centre) Pipe-like conduits in a
lateritized duricrust (lateritic pseudokarst), southern Queensland, Australia. (Right) Soil
pipes formed along decaying tree roots, near Troy, Idaho, USA~\cite{Leslie2013}. Similar
morphologies arise from distinct mechanisms: dissolution in carbonate rock (left), mechanical
erosion in soil (right) and a mixed chemical--mechanical regime in laterite (centre), which
is harder to classify~\cite{grimes2008laterite}.
}
\par\smallskip{\footnotesize\noindent\raggedright\emph{Images:} Left and centre: courtesy of Ken G. Grimes.\par}
\end{figure}

Many geological patterns are \emph{equifinal}: distinct mechanisms and environments can converge to nearly indistinguishable morphologies. Shape alone is therefore an ambiguous indicator of origin. Solution pipes in karst, for example, can resemble soil pipes in regolith even though one arises from dissolution along focused flow paths while the other reflects seepage-induced piping and mechanical removal (Fig.~\ref{fig:soilpipes}). Polygonal networks in columnar joints share pitch, node geometry, and arrested fronts with permafrost polygons, yet the active layer and energy source differ. Dendritic motifs in Mn-oxide films echo the branching statistics of drainage basins or electrochemical deposition, but the governing transport laws and kinetic thresholds are not the same. Banded textures --- Liesegang rhythms, oscillatory zoning in crystals, varve-like laminations, or travertine and silica-sinter layering --- likewise repeat at different scales under different couplings between transport, nucleation, and surface renewal. Even spherical or concentric bodies such as cannonball concretions, ooids, and accretionary lapilli share radius-selection logics while nucleation environments and growth regimes diverge; honeycomb (tafoni) textures produced by salt cycling or freeze--thaw can be indistinguishable at a glance; ring-like planforms tie together impact craters, volcanic ring dikes, and collapse dolines.

Terminology also drifts across communities and can obscure connections or suggest false contrasts. The same morphology may be labelled differently (for example, solution pipe versus soil pipe; chimney versus chemical garden), while a single term may span mechanisms (mudcrack versus polygonal terrain). To reduce ambiguity, we pair each morphology label with an explicit mechanism hypothesis, state the controlling parameter set using shared symbols and non-dimensional groups, and give the most common synonyms at first use. This keeps datasets comparable across geology, soft matter, and fluid physics while acknowledging that convergent forms need not imply identical drivers.

Overall, viewing these patterns through the lens of nonlinear dynamical systems yields constraints on the environmental and geodynamic conditions present during their formation. Such constraints aid both the reconstruction of Earth's geological history and the discrimination of abiotic from biologically influenced mesoscale pattern formation. More broadly, progress in physics has made it possible to move beyond describing geological patterns in terms of quantities such as fractal dimension, spacing laws, or banding statistics, and toward inferring the transport, kinetic, and mechanical processes responsible for their formation. We hope this review helps advance that aim.

 \section*{Acknowledgements}
 
This review paper originated at the workshop ``Geochemobrionics: Self-Organization in Geological Systems'' organized by Sean McMahon at the University of Edinburgh in 2022.  The workshop was funded by the \grantsponsor{European COST} (Cooperation in Science and Technology) programme, action number CA17120, Chemobrionics.
JHEC and CISD acknowledge support from the \grantsponsor{Spanish Ministerio de Ciencia, Innovaci\'on y Universidades} through grant \grantnumber{2}{PID2024-160443NB-I00}.
SFJ has received funding from the European Research Council (ERC) under the European Union's Horizon Europe research and innovation programme (grant agreement No 1101114969). SFJ also acknowledges support from Science Foundation Ireland (SFI Pathway award 22/PATH-S/10692) and from La Caixa Foundation (ID 100010434) and the European Union's Horizon 2020 research and innovation programme under the Marie Sk\l{}odowska-Curie grant agreement No 847648 (fellowship code LCF/BQ/PI21/11830015).
EK acknowledges funding from the \grantsponsor{European Union's Horizon 2020 research and innovation programme under the Marie Sk\l{}odowska-Curie} grant agreement No \grantnumber{3}{101031812} `NanoBioS'. CCL acknowledges funding by the Leverhulme Trust grant ECF-2023-202. PS acknowledges support by the National Science Centre (Poland) under CEUS-UNISONO grant 2020/02/Y/ST3/00121.

\bibliographystyle{elsarticle-num-names}
\bibliography{final.bib}

@book{Ortoleva1994,
title={Geochemical {Self-Organization}},
SerialNo={1},
author={P. J. Ortoleva},
 year={1994},
booktitle={Geochemical {Self-Organization}},
 orgname={Oxford University Press},
 city = {New York},
}

@book{Jamtveit1999,
title={Growth, Dissolution and Pattern Formation in Geosystems},
SerialNo={2},
author={Bj{\o}rn Jamtveit and Paul Meakin},
 year={1999},
booktitle={Growth, Dissolution and Pattern Formation in Geosystems},
 orgname={Springer},
 city = {Dordrecht},
 isbn={9780412832406, 9789401591799},
}

@article{Jamtveit2012,
SerialNo={3},
author={Jamtveit, B. and Hammer, {\O}.},
 year={2012},
title={Sculpting of Rocks by Reactive Fluids},
journal={{Geochem.} {Perspectiv}},
 volume={1},
 number={3},
  pages={341--481},
doi={10.7185/geochempersp.1.3},
}

@article{voigtlander2024,
SerialNo={4},
author={Voigtl{\"a}nder, Anne and Houssais, Morgane and Bacik, Karol A. and
  Bourg, Ian C. and Burton, Justin C. and Daniels, Karen E. and Datta, Sujit S. and
  {Del Gado}, Emanuela and Deshpande, Nakul S. and Devauchelle, Olivier and
  others},
 year={2024},
title={Soft matter physics of the ground beneath our feet},
journal={Soft Matter},
 volume={20},
 number={30},
  pages={5859--5888},
doi={10.1039/D4SM00391H},
}

@misc{geo_bio_review,
mypattern={[authors,atitle,midc][booktag,bktitle,yr,endbooktag,bibpages,bibdoi][post,myehost]},
SerialNo={5},
author={Julyan H. E. Cartwright and Charles S. Cockell and Julie G. Cosmidis
  and Silvia Holler and F. Javier Huertas and Sean F. Jordan and Pamela Knoll
  and Electra Kotopoulou and Corentin C. Loron and Sean McMahon and Anna
  Neubeck and Carlos Pimentel and C. Ignacio Sainz-D{\'\i}az and Piotr Szymczak},
 year={2025},
title={Self-assembled versus biological pattern formation in geology},
post={Preprint},
doi={10.48550/arXiv.2601.00323},
}

@article{de1992soft,
SerialNo={6},
author={{De Gennes}, Pierre-Gilles},
 year={1992},
title={Soft matter},
journal={Science},
 volume={256},
 number={5056},
  pages={495--497},
 orgname={American Association for the Advancement of Science},
doi={10.1126/science.256.5056.495},
}

@article{jerolmack2019,
SerialNo={7},
author={Douglas J. Jerolmack and Karen E. Daniels},
 year={2019},
title={Viewing {Earth}'s surface as a soft-matter landscape},
journal={Nature {Rev.} Phys.},
doi={10.1038/s42254-019-0111-x},
}

@book{nash2007geochemical,
SerialNo={8},
author={Nash, David J. and McLaren, Sue J.},
 year={2007},
title={Geochemical Sediments and Landscapes},
 series={RGS-IBG Book Series},
 orgname={Blackwell Publishing Ltd},
 city = {Oxford, UK},
doi={10.1002/9780470712917},
 isbn={9781405125192},
}

@book{budd2016introduction,
title={Autogenic dynamics and self-organization in sedimentary systems},
SerialNo={9},
 year={2016},
booktitle={Autogenic dynamics and self-organization in sedimentary systems},
 orgname={SEPM Society for Sedimentary Geology},
 editor={Budd, David A. and Hajek, Elizabeth A. and Purkis, Sam J.},
doi={10.2110/sepmsp.106},
}

@article{evans2019,
SerialNo={10},
author={Evans, Robert and Frenkel, Daan and Dijkstra, Marjolein},
 year={2019},
title={From simple liquids to colloids and soft matter},
journal={Phys. Today},
 volume={72},
 number={2},
  pages={38--39},
 orgname={AIP Publishing},
doi={10.1063/PT.3.4135},
}

@article{oehler1976,
SerialNo={11},
author={Oehler, John H.},
 year={1976},
title={Hydrothermal crystallization of silica gel},
journal={{Geol.} {Soc.} {Amer.} {Bull.}},
 volume={87},
 number={8},
  pages={1143--1152},
 orgname={Geological Society of America},
doi={10.1130/0016-7606(1976)87\lt 1143:HCOSG\gt 2.0.CO;2},
}

@article{howard2018,
SerialNo={12},
author={Howard, Charles Brian and Rabinovitch, Avinoam},
 year={2018},
title={A new model of agate geode formation based on a combination of
  morphological features and silica sol--gel experiments},
journal={Eur. J. Mineral.},
 volume={30},
 number={1},
  pages={97--106},
 orgname={E. Schweizerbart'sche Verlagsbuchhandlung Science Publishers},
doi={10.1127/ejm/2017/0029-2685},
}

@article{langer1989,
SerialNo={13},
author={Langer, J.S.},
 year={1989},
title={Dendrites, viscous fingers, and the theory of pattern formation},
journal={Science},
 volume={243},
 number={4895},
  pages={1150--1156},
 orgname={American Association for the Advancement of Science},
doi={10.1126/science.243.4895.1150},
}

@article{Mallet1875,
SerialNo={14},
author={Robert Mallet},
 year={1875},
title={On the origin and mechanism of production of the prismatic (or
  columnar) structure of basalt},
journal={Phil. Mag.},
 volume={50},
  pages={122--135 and 201--226},
doi={10.1080/14786447508641268},
}

@article{Spry1962,
SerialNo={15},
author={Alan Spry},
 year={1962},
title={The origin of columnar jointing, particularly in basalt flows},
journal={{J.} {Geol.} {Soc.} {Austral.}},
 volume={8},
 number={2},
  pages={191--216},
doi={10.1080/14400956208527873},
}

@article{Ryan1981,
SerialNo={16},
author={Michael P. Ryan and Charles G. Sammis},
 year={1981},
title={The glass transition in basalt},
journal={J. Geophys. Res.},
 volume={86},
  pages={9519--9535},
doi={10.1029/JB086iB10p09519},
}

@article{Goehring2008,
SerialNo={17},
author={L. Goehring and S.W. Morris},
 year={2008},
title={Scaling of columnar joints in basalt},
journal={J. Geophys. Res.},
 volume={113},
  pages={B10203},
doi={10.1029/2007JB005018},
}

@article{Monecke2023,
SerialNo={18},
author={Monecke, Thomas and Reynolds, T. James and Taksavasu, Tadsuda and
  Tharalson, Erik R. and Zeeck, Lauren R. and Guzman, Mario and Gissler, Garrett
  and Sherlock, Ross},
 year={2023},
title={Natural growth of gold dendrites within silica gels},
journal={Geology},
 volume={51},
  pages={189----192},
doi={10.1130/G48927.1},
}

@article{kirkpatrick2013,
SerialNo={19},
author={Kirkpatrick, J.D. and Rowe, C.D. and White, J.C. and Brodsky, E.E.},
 year={2013},
title={Silica gel formation during fault slip: \?{E}vidence from the rock record},
journal={Geology},
 volume={41},
 number={9},
  pages={1015--1018},
 orgname={Geological Society of America},
doi={10.1130/G34483.1},
}

@article{borhara2020,
SerialNo={20},
author={Borhara, Krishna and Onasch, Charles M.},
 year={2020},
title={Evidence for silica gel and its role in faulting in the {Tuscarora}
  \?{Sandstone}},
journal={J. Struct. Geol.},
 volume={139},
  artnum={104140},
 orgname={Elsevier},
doi={10.1016/j.jsg.2020.104140},
}

@article{davidovits1994,
SerialNo={21},
author={Davidovits, Joseph},
 year={1994},
title={Geopolymers: man-made rock geosynthesis and the resulting development
  of very early high strength cement},
journal={{J.} {Mater.} {Educ.}},
 volume={16},
  post={91--91},
 orgname={MATERIALS EDUCATION COUNCIL},
}

@article{kim2006,
SerialNo={22},
author={Kim, Daeik and Lai, Hsuan-Ting and Chilingar, George V. and Yen, Teh
  Fu},
 year={2006},
title={Geopolymer formation and its unique properties},
journal={{Environ.} {Geol.}},
 volume={51},
  pages={103--111},
 orgname={Springer},
doi={10.1007/s00254-006-0308-z},
}

@article{jaupart1989,
SerialNo={23},
author={Jaupart, Claude and Vergniolle, Sylvie},
 year={1989},
title={The generation and collapse of a foam layer at the roof of a basaltic
  magma chamber},
journal={J. Fluid Mech.},
 volume={203},
  pages={347--380},
 orgname={Cambridge University Press},
doi={10.1017/S0022112089001497},
}

@article{pal2003,
SerialNo={24},
author={Pal, Rajinder},
 year={2003},
title={Rheological behavior of bubble-bearing magmas},
journal={Earth Planet Sci. Lett.},
 volume={207},
 number={1--4},
  pages={165--179},
 orgname={Elsevier},
doi={10.1016/S0012-821X(02)01104-4},
}

@article{vasseur2020,
SerialNo={25},
author={Vasseur, J{\'e}r{\'e}mie and Wadsworth, Fabian B. and Dingwell, Donald
  B.},
 year={2020},
title={Permeability of polydisperse magma foam},
journal={Geology},
 volume={48},
 number={6},
  pages={536--540},
 orgname={Geological Society of America},
doi={10.1130/G47094.1},
}

@article{morris1930amygdules,
SerialNo={26},
author={Morris, F. K.},
 year={1930},
title={Amygdules and pseudo-amygdules},
journal={{Bull.} {Geol.} {Soc.} {Amer.}},
 volume={41},
 number={3},
  pages={383--404},
 orgname={Geological Society of America},
doi={10.1130/GSAB-41-383},
}

@book{yariv1979,
title={Geochemistry of Colloid Systems: \?{F}or Earth Scientists},
SerialNo={27},
author={Yariv, Shmuel and Cross, Harold},
 year={1979},
booktitle={Geochemistry of Colloid Systems: \?{F}or Earth Scientists},
 orgname={Springer Science \& Business Media},
}

@article{neelamma2022bentonite,
SerialNo={28},
author={Neelamma, M.K. and Holla, Sowmya R. and Selvakumar, M. and Chandran, P.
  Akhil and De, Shounak},
 year={2022},
title={Bentonite clay liquid crystals for high-performance supercapacitors},
journal={J. Electron. Mater.},
 volume={51},
 number={5},
  pages={2192--2202},
 orgname={Springer},
doi={10.1007/s11664-022-09469-y},
}

@article{jones1964,
SerialNo={29},
author={J. B. Jones and Sanders, J. V. and Segnit, E. Ralph},
 year={1964},
title={Structure of opal},
journal={Nature},
 volume={204},
 number={4962},
  pages={990--991},
 orgname={Nature Publishing Group UK London},
doi={10.1038/204990a0},
}

@article{Saunders2022,
SerialNo={30},
author={Saunders, James A.},
 year={2022},
title={Colloids and Nanoparticles: \?{I}mplications for Hydrothermal Precious
  Metal Ore Formation},
journal={SEG {Discover.}},
 number={130},
  pages={15--21},
 orgname={GeoScienceWorld},
doi={10.5382/SEGnews.2022-130.fea-01},
}

@article{zocher1925,
SerialNo={31},
author={Z{\"o}cher, H.},
 year={1925},
title={\"Uber freiwillige Strukturbildung in Solen},
journal={Z. {Anorgan.} Chem.},
 volume={147},
  pages={91--110},
}

@article{davidson2005,
SerialNo={32},
author={Davidson, Patrick and Gabriel, Jean-Christophe P.},
 year={2005},
title={Mineral liquid crystals},
journal={{Curr.} {Opin.} Colloid Interface {Sci.}},
 volume={9},
 number={6},
  pages={377--383},
 orgname={Elsevier},
doi={10.1016/j.cocis.2004.12.001},
}

@article{gabriel1996,
SerialNo={33},
author={Gabriel, Jean-Christophe P. and Sanchez, Cl{\'e}ment and Davidson,
  Patrick},
 year={1996},
title={Observation of nematic liquid-crystal textures in aqueous gels of
  smectite clays},
journal={J. Phys. Chem.},
 volume={100},
 number={26},
  pages={11139--11143},
 orgname={ACS Publications},
doi={10.1021/jp961088z},
}

@article{michot2006,
SerialNo={34},
author={Michot, Laurent J. and Bihannic, Isabelle and Maddi, Solange and
  Funari, S{\'e}rgio S. and Baravian, Christophe and Levitz, Pierre and
  Davidson, Patrick},
 year={2006},
title={Liquid-crystalline aqueous clay suspensions},
journal={{Proc.} {Natl} {Acad.} {Sci.}},
 volume={103},
 number={44},
  pages={16101--16104},
 orgname={National Acad Sciences},
doi={10.1073/pnas.0605201103},
}

@misc{mattievich2003,
SerialNo={35},
author={Mattievich, Enrico and Chadwick, J. and Cashion, John D. and Boas, John
  Frank and Clark, M. J. and Mackie, R. D.},
 year={2003},
title={Macroscopic ferronematic liquid crystals determine the structure of
  {Kimberley Zebra Rock}},
booktitle={27th Ann. Cond. Matt. Phys. Meet. Conf. Handbook},
}

@article{kelley2005serpentinite,
SerialNo={36},
author={Kelley, Deborah S. and Karson, Jeffrey A. and Fruh-Green, Gretchen L. and
  Yoerger, Dana R. and Shank, Timothy M. and Butterfield, David A. and Hayes, John
  M. and Schrenk, Matthew O. and Olson, Eric J. and Proskurowski, Giora and Mike
  Jakuba and Al Bradley and Ben Larson and Kristin Ludwig and Deborah Glickson
  and Kate Buckman and Alexander S. Bradley and William J. Brazelton and Kevin
  Roe, Mitch J. Elend and Ad{\'e}lie Delacour and Stefano M. Bernasconi and
  Marvin D. Lilley and John A. Baross and Roger E. Summons and Sean P. Sylva},
 year={2005},
title={A serpentinite-hosted ecosystem: the {Lost City} hydrothermal field},
journal={Science},
 volume={307},
 number={5714},
  pages={1428--1434},
 orgname={American Association for the Advancement of Science},
doi={10.1126/science.1102556},
}

@article{kharaka1973,
SerialNo={37},
author={Kharaka, Yousif K. and Berry, Frederick A. P.},
 year={1973},
title={Simultaneous flow of water and solutes through geological
  membranes---\?{I}. \?{Experimental} investigation},
journal={Geochim. Cosmochim. Acta},
 volume={37},
 number={12},
  pages={2577--2603},
 orgname={Elsevier},
doi={10.1016/0016-7037(73)90267-6},
}

@article{fritz1986,
SerialNo={38},
author={Fritz, Steven J.},
 year={1986},
title={Ideality of clay membranes in osmotic processes: a review},
journal={Clays Clay {Miner.}},
 volume={34},
  pages={214--223},
 orgname={Springer},
doi={10.1346/CCMN.1986.0340212},
}

@article{neuzil2000osmotic,
SerialNo={39},
author={Neuzil, C. E.},
 year={2000},
title={Osmotic generation of `anomalous' fluid pressures in geological
  environments},
journal={Nature},
 volume={403},
 number={6766},
  pages={182--184},
doi={10.1038/35003174},
}

@article{russell1994,
SerialNo={40},
author={Russell, Michael J. and Daniel, Roy M. and Hall, Allan J. and
  Sherringham, John A.},
 year={1994},
title={A hydrothermally precipitated catalytic iron sulphide membrane as a
  first step toward life},
journal={J. Mol. Evol.},
 volume={39},
  pages={231--243},
 orgname={Springer},
doi={10.1007/BF00160147},
}

@article{barge2015chemical,
SerialNo={41},
author={Barge, Laura M. and Cardoso, Silvana S. S. and Cartwright, Julyan H. E. and
  Cooper, Geoffrey J. T. and Cronin, Leroy and {De Wit}, Anne and Doloboff, Ivria J.
  and Escribano, Bruno and Goldstein, Raymond E. and Haudin, Florence and David
  E. H. Jones and Alan L. Mackay and Jerzy Maselko and Jason J. Pagano and J.
  Pantaleone and Michael J. Russell and C. Ignacio Sainz-D{\'\i}az and Oliver
  Steinbock and David A. Stone and Yoshifumi Tanimoto and Noreen L. Thomas},
 year={2015},
title={From chemical gardens to chemobrionics},
journal={Chem. Rev.},
 volume={115},
 number={16},
  pages={8652--8703},
 orgname={ACS Publications},
doi={10.1021/acs.chemrev.5b00014},
}

@article{zieg2005,
SerialNo={42},
author={Zieg, Michael J. and Marsh, Bruce D.},
 year={2005},
title={The {Sudbury Igneous Complex}: \?{V}iscous emulsion differentiation of a
  superheated impact melt sheet},
journal={{Geol.} {Soc.} {Amer.} {Bull.}},
 volume={117},
 number={11--12},
  pages={1427--1450},
 orgname={Geological Society of America},
doi={10.1130/B25579.1},
}

@article{gogoi2019,
SerialNo={43},
author={Gogoi, Bibhuti and Saikia, Ashima},
 year={2019},
title={The genesis of emulsion texture owing to magma mixing in the {Ghansura
  Felsic Dome} of the {Chotanagpur Granite Gneiss Complex} of eastern \?{India}},
journal={{Can.} {Miner.}},
 volume={57},
 number={3},
  pages={311--338},
 orgname={GeoScienceWorld},
doi={10.3749/canmin.1800064},
}

@article{cunningham1916iv,
SerialNo={44},
author={Cunningham-Craig, E. H.},
 year={1916},
title={The Origin of Oil-Shale},
journal={{Proc.} {Roy.} {Soc.} {Edin.}},
 volume={36},
 number={1--2},
  pages={44--86},
 orgname={Royal Society of Edinburgh Scotland Foundation},
doi={10.1017/S0370164600018125},
}

@article{cicconi2019,
SerialNo={45},
author={Cicconi, Maria Rita and Neuville, Daniel R.},
 year={2019},
title={Natural glasses},
journal={Springer {Handbook} {Glas.}},
  pages={771--812},
 orgname={Springer},
doi={10.1007/978-3-319-93728-1\_22},
}

@article{heide2011,
SerialNo={46},
author={Heide, Klaus and Heide, Gerhard},
 year={2011},
title={Vitreous state in nature---\?{O}rigin and properties},
journal={Geochemistry},
 volume={71},
 number={4},
  pages={305--335},
 orgname={Elsevier},
doi={10.1016/j.chemer.2011.10.001},
}

@article{ma2001micro,
SerialNo={47},
author={Ma, Chi and Gresh, Jennifer and Rossman, George R. and Ulmer, Gene C.
  and Vicenzi, Edward P.},
 year={2001},
title={Micro-analytical study of the optical properties of rainbow and sheen
  obsidians},
journal={{Can.} {Miner.}},
 volume={39},
 number={1},
  pages={57--71},
 orgname={Mineralogical Association of Canada},
doi={10.2113/gscanmin.39.1.57},
}

@article{pasek_fulgurite_2012,
SerialNo={48},
author={Pasek, M. A. and Block, Kristin and Pasek, Virginia},
 year={2012},
title={Fulgurite morphology: a classification scheme and clues to formation},
journal={{Contrib.} {Miner.} {Petrol.}},
 volume={164},
 number={3},
  pages={477--492},
doi={10.1007/s00410-012-0753-5},
 issn={1432-0967},
}

@article{boswell1951,
SerialNo={49},
author={Boswell, P. G. H.},
 year={1951},
title={The Trend of Research on the Rheotropy of Geological Materials},
journal={{Sci.} {Prog.} (1933-)},
 volume={39},
 number={156},
  pages={608--622},
 orgname={JSTOR},
}

@article{biot1961,
SerialNo={50},
author={Biot, Maurice Anthony},
 year={1961},
title={Theory of folding of stratified viscoelastic media and its implications
  in tectonics and orogenesis},
journal={{Geol.} {Soc.} {Amer.} {Bull.}},
 volume={72},
 number={11},
  pages={1595--1620},
 orgname={Geological Society of America},
doi={10.1130/0016-7606(1961)72[1595:TOFOSV]2.0.CO;2},
}

@article{matthes1953quicksand,
SerialNo={51},
author={Matthes, Gerard H.},
 year={1953},
title={Quicksand},
journal={Sci. Am.},
 volume={188},
 number={6},
  pages={97--104},
 orgname={JSTOR},
}

@article{khaldoun2005liquefaction,
SerialNo={52},
author={Khaldoun, A. and Eiser, E. and Wegdam, G. H. and Bonn, Daniel},
 year={2005},
title={Liquefaction of quicksand under stress},
journal={Nature},
 volume={437},
 number={7059},
  post={635--635},
 orgname={Nature Publishing Group UK London},
doi={10.1038/437635a},
}

@article{kadau2009living,
SerialNo={53},
author={Kadau, Dirk and Herrmann, Hans J. and Andrade, Jos{\'e} S. and
  Ara{\'u}jo, Asc{\^a}nio D. and Bezerra, Luiz J. C. and Maia, Luis P.},
 year={2009},
title={Living quicksand},
journal={{Granular} Matter},
 volume={11},
  pages={67--71},
 orgname={Springer},
doi={10.1007/s10035-008-0117-z},
}

@article{griffiths2000dynamics,
SerialNo={54},
author={Griffiths, Ross W.},
 year={2000},
title={The dynamics of lava flows},
journal={Annu. Rev. Fluid Mech.},
 volume={32},
 number={1},
  pages={477--518},
 orgname={Annual Reviews 4139 El Camino Way, PO Box 10139, Palo Alto, CA
  94303-0139, USA},
doi={10.1146/annurev.fluid.32.1.477},
}

@article{shalev2012,
SerialNo={55},
author={Shalev, Eyal and Lyakhovsky, Vladimir},
 year={2012},
title={Viscoelastic damage modeling of sinkhole formation},
journal={J. Struct. Geol.},
 volume={42},
  pages={163--170},
 orgname={Elsevier},
doi={10.1016/j.jsg.2012.05.010},
}

@article{rubin1993,
SerialNo={56},
author={Rubin, Allan M.},
 year={1993},
title={Dikes vs.\ diapirs in viscoelastic rock},
journal={Earth Planet Sci. Lett.},
 volume={117},
 number={3--4},
  pages={653--670},
 orgname={Elsevier},
doi={10.1016/0012-821X(93)90109-M},
}

@article{veveakis2021,
SerialNo={57},
author={Veveakis, Manolis and Poulet, Thomas},
 year={2021},
title={A note on the instability and pattern formation of shrinkage cracks in
  viscoplastic soils},
journal={{Geomechan.} Energy Environ.},
 volume={25},
  artnum={100198},
 orgname={Elsevier},
doi={10.1016/j.gete.2020.100198},
}

@article{burnley2013,
SerialNo={58},
author={Burnley, P.C.},
 year={2013},
title={The importance of stress percolation patterns in rocks and other
  polycrystalline materials},
journal={{Nature} {Commun.}},
 volume={4},
 number={1},
  pages={2117},
 orgname={Nature Publishing Group UK London},
doi={10.1038/ncomms3117},
}

@book{andreotti2013granular,
title={Granular Media: Between Fluid and Solid},
SerialNo={59},
author={Andreotti, Bruno and Forterre, Yo{\"e}l and Pouliquen, Olivier},
 year={2013},
booktitle={Granular Media: Between Fluid and Solid},
 orgname={Cambridge University Press},
}

@inbook{gravish2016entangled,
SerialNo={60},
author={Gravish, Nick and I. Goldman, Daniel},
 year={2016},
title={Entangled Granular Media},
booktitle={Fluids, Colloids and Soft Materials},
  pages={341--354},
 orgname={John Wiley \& Sons, Ltd},
doi={10.1002/9781119220510.ch17},
 isbn={9781119220510},
}

@article{zheng2021mechanics,
SerialNo={61},
author={Zheng, Hu and Niu, Wenqing and Mao, Wuwei and Li, Lihui and Wang, Fawu
  and Huang, Yu},
 year={2021},
title={Mechanics of granular material and the application in engineering
  geology},
journal={{J.} {Engin.} {Geol.}},
 volume={29},
 number={1},
  pages={12--24},
doi={10.13544/j.cnki.jeg.2021-0017},
}

@article{turing1952chemical,
SerialNo={62},
author={Turing, Alan Mathison},
 year={1952},
title={The chemical basis of morphogenesis},
journal={Phil. {Trans.} R. {Soc.} {London} B},
 volume={237},
  pages={37--72},
doi={10.1098/rstb.1952.0012},
}

@book{Murray2002,
title={Mathematical Biology: \?{I}. \?{An} Introduction},
SerialNo={63},
author={Murray, James D.},
 year={1989},
booktitle={Mathematical Biology: \?{I}. \?{An} Introduction},
 orgname={Springer},
}

@article{ball2015forging,
SerialNo={64},
author={Ball, Philip},
 year={2015},
title={Forging patterns and making waves from biology to geology: a commentary
  on {Turing} (1952)`The chemical basis of morphogenesis'},
journal={Phil. {Trans.} R. {Soc.} {London} B},
 volume={370},
 number={1666},
  artnum={20140218},
 orgname={The Royal Society},
doi={10.1098/rstb.2014.0218},
}

@article{mcbride2004origin,
SerialNo={65},
author={McBride, Earle F. and Picard, M. Dane},
 year={2004},
title={Origin of honeycombs and related weathering forms in {Oligocene Macigno
  Sandstone}, {Tuscan} coast near {Livorno, \?{Italy}}},
journal={Earth Surf. Process. Landf.},
 volume={29},
 number={6},
  pages={713--735},
 orgname={Wiley Online Library},
doi={10.1002/esp.1065},
}

@article{hammer2008calcite,
SerialNo={66},
author={Hammer, {\O} and Dysthe, D.K. and Lelu, B. and Lund, H. and Meakin, P. and
  Jamtveit, B.},
 year={2008},
title={Calcite precipitation instability under laminar, open-channel flow},
journal={Geochim. Cosmochim. Acta},
 volume={72},
 number={20},
  pages={5009--5021},
 orgname={Elsevier},
doi={10.1016/j.gca.2008.07.028},
}

@article{meron1992pattern,
SerialNo={67},
author={Meron, Ehud},
 year={1992},
title={Pattern formation in excitable media},
journal={Phys. Rep.},
 volume={218},
 number={1},
  pages={1--66},
 orgname={Elsevier},
doi={10.1016/0370-1573(92)90098-K},
}

@article{zhabotinsky1991history,
SerialNo={68},
author={Zhabotinsky, Anatol M.},
 year={1991},
title={A history of chemical oscillations and waves},
journal={Chaos: {Interdisc.} {J.} Nonlinear {Sci.}},
 volume={1},
 number={4},
  pages={379--386},
 orgname={American Institute of Physics},
doi={10.1063/1.165848},
}

@article{cartwright2012crystal,
SerialNo={69},
author={Cartwright, Julyan H. E. and Checa, Antonio G. and Escribano, Bruno and
  Ignacio Sainz-Diaz, C.},
 year={2012},
title={Crystal growth as an excitable medium},
journal={{Phil.} {Trans.} {Roy.} {Soc.} A},
 volume={370},
 number={1969},
  pages={2866--2876},
 orgname={The Royal Society Publishing},
doi={10.1098/rsta.2011.0600},
}

@article{burridge1967model,
SerialNo={70},
author={Burridge, Robert and Knopoff, Leon},
 year={1967},
title={Model and theoretical seismicity},
journal={Bull. Seismol. Soc. Am.},
 volume={57},
 number={3},
  pages={341--371},
 orgname={The Seismological Society of America},
doi={10.1785/BSSA0570030341},
}

@article{cartwright1997burridge,
SerialNo={71},
author={Cartwright, Julyan H. E. and {Hern{\'a}ndez-Garc{\'\i}a}, Emilio and Piro,
  Oreste},
 year={1997},
title={Burridge-{Knopoff} models as elastic excitable media},
journal={Phys. Rev. Lett.},
 volume={79},
 number={3},
  pages={527},
 orgname={APS},
doi={10.1103/PhysRevLett.79.527},
}

@article{rothman2019characteristic,
SerialNo={72},
author={Rothman, Daniel H.},
 year={2019},
title={Characteristic disruptions of an excitable carbon cycle},
journal={{Proc.} {Natl} {Acad.} {Sci.}},
 volume={116},
 number={30},
  pages={14813--14822},
 orgname={National Acad Sciences},
doi={10.1073/pnas.1905164116},
}

@article{pelletier2004spiral,
SerialNo={73},
author={Pelletier, Jon D.},
 year={2004},
title={How do spiral troughs form on {Mars}?},
chsep={\chsep[atitle]{~}},
journal={Geology},
 volume={32},
 number={4},
  pages={365--367},
 orgname={Geological Society of America},
doi={10.1130/G20228.2},
}

@article{Jablczynski1926,
SerialNo={74},
author={Jab{\l}czy{\'n}ski, K.},
 year={1926},
title={{\"Uber \?{Lie}segang-Ringe}},
journal={Kolloid-Zeitschrift},
 volume={40},
 number={1},
  pages={22--28},
}

@article{matalon1955liesegang,
SerialNo={75},
author={Matalon, R. and Packter, A.},
 year={1955},
title={The {\?{Lie}segang} phenomenon. \?{I}. \?{S}ol protection and diffusion},
journal={{J.} Colloid {Sci.}},
 volume={10},
 number={1},
  pages={46--62},
 orgname={Elsevier},
doi={10.1016/0095-8522(55)90076-3},
}

@book{Runge1855,
title={{Der Bildungstrieb der Stoffe : veranschaulicht in selbstst\"andig
  gewachsenen Bildern (Fortsetzung der Musterbilder)}},
SerialNo={76},
author={Friedlieb Ferdinand Runge},
 year={1855},
booktitle={{Der Bildungstrieb der Stoffe : veranschaulicht in selbstst\"andig
  gewachsenen Bildern (Fortsetzung der Musterbilder)}},
 orgname={Selbstverlag},
 city = {Oranienburg},
}

@book{Ord1879,
title={On the Influence of Colloids upon Crystalline Form and Cohesion},
SerialNo={77},
author={Ord, W. M.},
 year={1879},
booktitle={On the Influence of Colloids upon Crystalline Form and Cohesion},
 orgname={E. Stanford},
 city = {London},
}

@article{liesegang1896uber,
SerialNo={78},
author={Liesegang, Raphael E.},
 year={1896},
title={Uber einige eigenschaften von gallerten},
journal={Naturwissensch Wochenschr},
 volume={11},
  pages={353--362},
}

@book{henisch2014periodic,
title={Periodic Precipitation},
SerialNo={79},
author={Henisch, Heinz K.},
 year={2014},
booktitle={Periodic Precipitation},
 orgname={Elsevier},
}

@inbook{LHeureux1999,
SerialNo={80},
author={{L'Heureux}, Ivan and Fowler, Anthony D.},
 year={1999},
title={Branching and oscillatory patterns in plagioclase and
  {Mississippi}-valley type sphalerite deposits},
booktitle={Growth, Dissolution and Pattern Formation in Geosystems},
  pages={85--108},
 orgname={Springer},
doi={10.1007/978-94-015-9179-9\_4},
}

@article{nabika2019pattern,
SerialNo={81},
author={Nabika, Hideki and Itatani, Masaki and Lagzi, Istv{\'a}n},
 year={2019},
title={Pattern formation in precipitation reactions: \?{T}he {\?{Lie}segang}
  phenomenon},
journal={Langmuir},
 volume={36},
 number={2},
  pages={481--497},
 orgname={ACS Publications},
doi={10.1021/acs.langmuir.9b03018},
}

@article{Ostwald1897,
SerialNo={82},
author={Ostwald, Wilhelm},
 year={1897},
title={{Besprechung der Arbeit von \?{Lie}segangs ``A-Linien''}},
journal={Z. Phys. Chem.},
 volume={22},
  pages={289--330},
}

@article{Prager1956,
SerialNo={83},
author={Prager, S.},
 year={1956},
title={Periodic precipitation},
journal={J. Chem. Phys.},
 volume={25},
 number={2},
  pages={279--283},
doi={10.1063/1.1742871},
}

@article{Lifshitz1961,
SerialNo={84},
author={Lifshitz, Ilya M. and Slyozov, Vitaly V.},
 year={1961},
title={The kinetics of precipitation from supersaturated solid solutions},
journal={{J.} {Phys.} {Chem.} Solids},
 volume={19},
 number={1--2},
  pages={35--50},
 orgname={Elsevier},
doi={10.1016/0022-3697(61)90054-3},
}

@article{Boudreau1995,
SerialNo={85},
author={Boudreau, A. E.},
 year={1995},
title={Crystal aging and the formation of fine-scale igneous layering},
journal={{Miner.} {Petrol.}},
 volume={54},
 number={1},
  pages={55--69},
 orgname={Springer Wien},
doi={10.1007/BF01162758},
}

@article{krug_morphological_1999,
SerialNo={86},
author={Krug, Hans-J{\"u}rgen and Brandtst{\"a}dter, Hermann},
 year={1999},
title={Morphological Characteristics of {\?{Lie}segang} Rings and Their
  Simulations},
journal={{J.} {Phys.} {Chem.} A},
 volume={103},
 number={39},
  pages={7811--7820},
doi={10.1021/jp991092l},
 issn={1089-5639},
}

@article{dayeh_transition_2014,
SerialNo={87},
author={Dayeh, Malak and Ammar, Manal and Al-Ghoul, Mazen},
 year={2014},
title={Transition from rings to spots in a precipitation reaction--diffusion
  system},
journal={RSC {Adv.}},
 volume={4},
 number={104},
  pages={60034--60038},
doi={10.1039/C4RA11223G},
 issn={2046-2069},
}

@article{papp_fine_2020,
SerialNo={88},
author={Papp, P. and Bohner, B. and T{\'o}th, {\'A}gota and Horv{\'a}th, D.},
 year={2020},
title={Fine tuning of pattern selection in the cadmium-hydroxide-system},
journal={J. Chem. Phys.},
 volume={152},
 number={9},
  artnum={094906},
doi={10.1063/1.5144292},
 issn={1089-7690},
}

@article{cartwright1999pattern,
SerialNo={89},
author={Cartwright, Julyan H. E. and Garc{\'i}a-Ruiz, Juan Manuel and
  Villacampa, Ana I},
 year={1999},
title={Pattern formation in crystal growth: {\?{Lie}segang} rings},
journal={Comput. Phys. Comm.},
 volume={121},
  pages={411--413},
 orgname={Elsevier},
doi={10.1016/S0010-4655(99)00370-7},
}

@article{reeder_oscillatory_1990,
SerialNo={90},
author={Reeder, Richard J. and Fagioli, Richard O. and Meyers, William J.},
 year={1990},
title={Oscillatory zoning of {Mn} in solution-grown calcite crystals},
journal={Earth Sci. Rev.},
 volume={29},
 number={1},
  pages={39--46},
doi={10.1016/0012-8252(0)90026-R},
 issn={0012-8252},
}

@article{prieto_nucleation_1997,
SerialNo={91},
author={Prieto, M. and Fern{\'a}ndez-Gonz{\'a}lez, A. and Putnis, A. and
  Fern{\'a}ndez-D{\'i}az, L.},
 year={1997},
title={Nucleation, growth, and zoning phenomena in crystallizing
  \?{({Ba,Sr}){\?{CO}}3, {Ba}({\?{SO}}4,{CrO}4), ({Ba,Sr}){\?{SO}}4, and ({Cd,\?{Ca}}){\?{CO}}3} solid
  solutions from aqueous solutions},
journal={Geochim. Cosmochim. Acta},
 volume={61},
 number={16},
  pages={3383--3397},
doi={10.1016/S0016-7037(97)00160-9},
 issn={0016-7037},
}

@article{ling_nanospectroscopy_2018,
SerialNo={92},
author={Ling, Florence T. and Hunter, Heather A. and Fitts, Jeffrey P. and
  Peters, Catherine A. and Acerbo, Alvin S. and Huang, Xiaojing and Yan, Hanfei
  and Nazaretski, Evgeny and Chu, Yong S.},
 year={2018},
title={Nanospectroscopy Captures Nanoscale Compositional Zonation in Barite
  Solid Solutions},
journal={{Sci.} {Rep.}},
 volume={8},
 number={1},
  pages={13041},
doi={10.1038/s41598-018-31335-3},
}

@article{ginibre_crystal_2007,
SerialNo={93},
author={Ginibre, Catherine and W{\"o}rner, Gerhard and Kronz, Andreas},
 year={2007},
title={Crystal Zoning as an Archive for Magma Evolution},
journal={Elements},
 volume={3},
 number={4},
  pages={261--266},
doi={10.2113/gselements.3.4.261},
}

@article{shore_oscillatory_1996,
SerialNo={94},
author={Shore, Mark and Fowler, Anthony D.},
 year={1996},
title={Oscillatory zoning in minerals; a common phenomenon},
journal={{Can.} {Miner.}},
 volume={34},
 number={6},
  pages={1111--1126},
}

@misc{Matsson2008,
  author    = {Matsson, J.},
  title     = {A Student Project On \?{Rayleigh} \?{Benard} Convection},
  booktitle = {2008 Annual Conference \& Exposition Proceedings},
  address   = {Pittsburgh, Pennsylvania},
  month     = jun,
  year      = {2008},
  doi       = {10.18260/1-2--3591}
}

@article{wollkind1982kelvin,
SerialNo={95},
author={Wollkind, David J. and Alexander, J. Iwan D.},
 year={1982},
title={Kelvin--\?{Helmholtz} instability in a layered \?{Newton}ian fluid model of the
  geological phenomenon of rock folding},
journal={SIAM J. Appl. Math.},
 volume={42},
 number={6},
  pages={1276--1295},
 orgname={SIAM},
doi={10.1137/0142089},
}

@article{fletcher1977folding,
SerialNo={96},
author={Fletcher, Raymond C.},
 year={1977},
title={Folding of a single viscous layer: exact infinitesimal-amplitude
  solution},
journal={Tectonophysics},
 volume={39},
 number={4},
  pages={593--606},
 orgname={Elsevier},
doi={10.1016/0040-1951(77)90155-X},
}

@book{getling1998rayleigh,
title={{Rayleigh}-\?{B\'enard} Convection: Structures and Dynamics},
SerialNo={97},
author={Getling, Alexander V.},
 year={1998},
series={{Rayleigh}-\?{B\'enard} Convection: Structures and Dynamics},
 volume={vol. 11},
 orgname={World Scientific},
}

@article{bodenschatz2000recent,
SerialNo={98},
author={Bodenschatz, Eberhard and Pesch, Werner and Ahlers, Guenter},
 year={2000},
title={Recent developments in \?{Rayleigh}-\?{B\'enard} convection},
journal={Annu. Rev. Fluid Mech.},
 volume={32},
 number={1},
  pages={709--778},
 orgname={Annual Reviews 4139 El Camino Way, PO Box 10139, Palo Alto, CA
  94303-0139, USA},
doi={10.1146/annurev.fluid.32.1.709},
}

@inbook{busse1989fundamentals,
SerialNo={99},
author={Busse, Friedrich H.},
 year={1989},
title={Fundamentals of Thermal Convection},
booktitle={Mantle Convection: Plate Tectonics and Global Dynamics},
 series={Fluid Mechanics of Earth and Planets},
 volume={vol. 4},
  pages={23--95},
 orgname={Gordon and Breach Science Publishers},
 city = {New York},
 isbn={978-2-88124-691-3},
}

@article{anderson2002plate,
SerialNo={100},
author={Anderson, Don L.},
 year={2002},
title={Plate tectonics as a far-from-equilibrium self-organized system},
journal={Plate {Boundary} Zones},
 volume={30},
  pages={411--425},
 orgname={American Geophysical Union Washington, DC},
doi={10.1029/GD030p0411},
}

@book{Wager1968,
title={Layered Igneous Rocks},
SerialNo={101},
author={Wager, L. R. and Brown, G. M.},
 year={1968},
booktitle={Layered Igneous Rocks},
 orgname={Oliver and Boyd},
 city = {Edinburgh},
}

@article{Cawthorn1994,
SerialNo={102},
author={Cawthorn, R. G.},
 year={1994},
title={Layered igneous rocks: 25 years after {Wager and Brown}},
journal={South {Afr.} {J.} {Geol.}},
 volume={97},
 number={3},
  pages={313--331},
doi={10.10520/AJA10120750\_509},
}

@article{Thy2023,
SerialNo={103},
author={Thy, P. and Tegner, C. and Lesher, C. E.},
 year={2023},
title={{Petrology of the Skaergaard Layered Series}},
journal={GEUS {Bull.}},
 volume={51},
  artnum={e8327},
doi={10.34194/geusb.v51.8327},
}

@article{Kessler2003,
SerialNo={104},
author={Kessler, M. A. and Werner, B. T.},
 year={2003},
title={Self-Organization of Sorted Patterned Ground},
journal={Science},
 volume={299},
 number={5605},
  pages={380--383},
doi={10.1126/science.1077300},
}

@article{Hallet2013,
SerialNo={105},
author={Hallet, Bernard},
 year={2013},
title={Stone circles: form and soil kinematics},
journal={{Phil.} {Trans.} {Roy.} {Soc.} A},
 volume={371},
 number={2004},
  artnum={20120357},
doi={10.1098/rsta.2012.0357},
}

@article{gibbons2023demonstrating,
SerialNo={106},
author={Gibbons, Melissa M. and Muldoon, Dillon and Khalil, Imane},
 year={2023},
title={Demonstrating the \?{Kelvin}-\?{Helmholtz} Instability Using a Low-Cost
  Experimental Apparatus and Computational Fluid Dynamics Simulations},
journal={Fluids},
 volume={8},
 number={12},
  pages={318},
 orgname={MDPI},
doi={10.3390/fluids8120318},
}

@article{kull1991theory,
SerialNo={107},
author={Kull, Hans-J{\"o}rg},
 year={1991},
title={Theory of the {\?{Rayleigh}-Taylor} instability},
journal={Phys. Rep.},
 volume={206},
 number={5},
  pages={197--325},
 orgname={Elsevier},
doi={10.1016/0370-1573(91)90153-D},
}

@article{whitehead1975dynamics,
SerialNo={108},
author={{Whitehead, Jr.}, John A. and Luther, Douglas S.},
 year={1975},
title={Dynamics of laboratory diapir and plume models},
journal={J. Geophys. Res.},
 volume={80},
 number={5},
  pages={705--717},
 orgname={Wiley Online Library},
doi={10.1029/JB080i005p00705},
}

@article{trusheim1957halokinese,
SerialNo={109},
author={Trusheim, Ferdinand},
 year={1957},
title={\"Uber Halokinese und ihre Bedeutung f\"ur die strukturelle
  Entwicklung Norddeutschlands},
journal={{Z.} {Deutsche} {Geol.} {Gesellschaft}},
  pages={111--151},
 orgname={Schweizerbart'sche Verlagsbuchhandlung},
}

@article{Hudec2007,
SerialNo={110},
author={Hudec, Martin R. and Jackson, Michael P. A.},
 year={2007},
title={Terra infirma: \?{U}nderstanding salt tectonics},
journal={Earth Sci. Rev.},
 volume={82},
 number={1--2},
  pages={1--28},
doi={10.1016/j.earscirev.2007.01.001},
}

@article{Davison2000,
SerialNo={111},
author={Davison, I. and Alsop, G. I. and Evans, N. G. and Safaricz, M.},
 year={2000},
title={Overburden deformation patterns and mechanisms of salt diapir
  penetration in the {Central Graben, North Sea}},
journal={{Mar.} {Petrol.} {Geol.}},
 volume={17},
 number={5},
  pages={601--618},
doi={10.1016/S0264-8172(00)00011-8},
}

@article{de1996evolution,
SerialNo={112},
author={{De Silva}, I. P. D. and Fernando, H. J. S. and Eaton, F. and Hebert, D.},
 year={1996},
title={Evolution of {\?{Kelvin}-\?{Helmholtz}} billows in nature and laboratory},
journal={Earth Planet Sci. Lett.},
 volume={143},
 number={1--4},
  pages={217--231},
 orgname={Elsevier},
doi={10.1016/0012-821X(96)00129-X},
}

@article{heifetz2005soft,
SerialNo={113},
author={Heifetz, Eyal and Agnon, Amotz and Marco, Shmuel},
 year={2005},
title={Soft sediment deformation by {Kelvin \?{Helmholtz} Instability}: \?{A} case
  from {Dead Sea} earthquakes},
journal={Earth Planet Sci. Lett.},
 volume={236},
 number={1--2},
  pages={497--504},
 orgname={Elsevier},
doi={10.1016/j.epsl.2005.04.019},
}

@article{ryan2012coils,
SerialNo={114},
author={Ryan, A. J. and Christensen, Philip R.},
 year={2012},
title={Coils and polygonal crust in the {Athabasca Valles} region, {Mars}, as
  evidence for a volcanic history},
journal={Science},
 volume={336},
 number={6080},
  pages={449--452},
 orgname={American Association for the Advancement of Science},
doi={10.1126/science.1219437},
}

@article{thomas2013formation,
SerialNo={115},
author={Thomas, Katherine and Herminghaus, Stephan and Porada, Hubertus and
  Goehring, Lucas},
 year={2013},
title={Formation of {Kinneyia} via shear-induced instabilities in microbial
  mats},
journal={{Phil.} {Trans.} {Roy.} {Soc.} A},
 volume={371},
 number={2004},
  artnum={20120362},
 orgname={The Royal Society Publishing},
doi={10.1098/rsta.2012.0362},
}

@article{herminghaus2016kinneyia,
SerialNo={116},
author={Herminghaus, Stephan and Thomas, Katherine Ruth and Aliaskarisohi,
  Saeedeh and Porada, Hubertus and Goehring, Lucas},
 year={2016},
title={Kinneyia: a flow-induced anisotropic fossil pattern from ancient
  microbial mats},
journal={{Frontier} {Mater.}},
 volume={3},
  pages={30},
 orgname={Frontiers Media SA},
doi={10.3389/fmats.2016.00030},
}

@article{fabbri2017fluid,
SerialNo={117},
author={Fabbri, Stefania and Li, Jian and Howlin, Robert P. and Rmaile, Amir
  and Gottenbos, Bart and {De Jager}, Marko and Starke, E. Michelle and Aspiras,
  Marcelo and Ward, Marilyn T. and Cogan, Nicholas G. and Paul Stoodley},
 year={2017},
title={Fluid-driven interfacial instabilities and turbulence in bacterial
  biofilms},
journal={{Environ.} {Microbiol.}},
 volume={19},
 number={11},
  pages={4417--4431},
 orgname={Wiley Online Library},
doi={10.1111/1462-2920.13883},
}

@article{schatz2001,
SerialNo={118},
author={Schatz, Michael F. and Neitzel, G. Paul},
 year={2001},
title={Experiments on thermocapillary instabilities},
journal={Annu. Rev. Fluid Mech.},
 volume={33},
  pages={93--127},
 orgname={Annual Reviews},
doi={10.1146/annurev.fluid.33.1.93},
}

@article{craster2009dynamics,
SerialNo={119},
author={Craster, Richard V. and Matar, Omar K.},
 year={2009},
title={Dynamics and stability of thin liquid films},
journal={Rev. Mod. Phys.},
 volume={81},
 number={3},
  pages={1131--1198},
 orgname={APS},
doi={10.1103/RevModPhys.81.1131},
}

@article{baron2022melting,
SerialNo={120},
author={Baron, Marzena A. and Fiquet, Guillaume and Morard, Guillaume and
  Miozzi, Francesca and Esteve, Im{\`e}ne and Doisneau, B{\'e}atrice and
  Pakhomova, Anna S. and Ricard, Yanick and Guyot, Fran\c{c}ois},
 year={2022},
title={Melting of basaltic lithologies in the \?{Earth's} lower mantle},
journal={Phys. Earth Planet. Inter.},
 volume={333},
  artnum={106938},
 orgname={Elsevier},
doi={10.1016/j.pepi.2022.106938},
}

@article{rouwet2017sedimentation,
SerialNo={121},
author={Rouwet, Dmitri and Iorio, Marta},
 year={2017},
title={The sedimentation of {Suminagashi}-like floating clay patterns at {El}
  {Chich\'on} crater lake ({Chiapas, \?{Mexico}})},
journal={{Geol.} {Soc.} {London} {Special} {Pub.}},
 volume={437},
 number={1},
  pages={153--161},
 orgname={The Geological Society of London},
doi={10.1144/SP437.9},
}

@article{rehman2018role,
SerialNo={122},
author={Rehman, F. and Singh, O.P.},
 year={2018},
title={Role of \?{Rayleigh} numbers on characteristics of double diffusive salt
  fingers},
journal={Heat Mass Transf.},
 volume={54},
 number={11},
  pages={3483--3492},
 orgname={Springer},
doi={10.1007/s00231-018-2385-4},
}

@article{toramaru2012numerical,
SerialNo={123},
author={Toramaru, Atsushi and Matsumoto, Mitsuo},
 year={2012},
title={Numerical experiment of cyclic layering in a solidified binary eutectic
  melt},
journal={{J.} {Geophys.} {Res.}: Solid Earth},
 volume={117},
 number={B2},
 orgname={Wiley Online Library},
doi={10.1029/2011JB008204},
}

@article{Huppert1981,
SerialNo={124},
author={Huppert, Herbert E. and Turner, J. Stewart},
 year={1981},
title={Double-diffusive convection},
journal={J. Fluid Mech.},
 volume={106},
  pages={299--329},
doi={10.1017/S0022112081001614},
}

@article{Schmitt1995,
SerialNo={125},
author={Raymond W. Schmitt},
 year={1995},
title={The Ocean's Salt Fingers},
journal={Sci. Am.},
 volume={272},
 number={5},
  pages={70--75},
}

@book{radko2013double,
title={Double-Diffusive Convection},
SerialNo={126},
author={Radko, Timour},
 year={2013},
booktitle={Double-Diffusive Convection},
 orgname={Cambridge University Press},
}

@article{huppert1984double,
SerialNo={127},

author={Huppert, Herbert E. and Sparks, R. Stephen J.},
 year={1984},
title={Double-diffusive convection due to crystallization in magmas},
journal={{Annu.} {Rev.} Earth {Planet.} {Sci.}},
 volume={12},
  pages={11},
doi={10.1146/annurev.ea.12.050184.000303},
}

@article{turner1986komatiites,
SerialNo={128},
author={Turner, J. Stewart and Huppert, Herbert E. and Sparks, R. Stephen J.},
 year={1986},
title={Komatiites \?{II}: \?{E}xperimental and theoretical investigations of
  post-emplacement cooling and crystallization},
journal={J. Petrol.},
 volume={27},
 number={2},
  pages={397--437},
 orgname={Oxford University Press},
doi={10.1093/petrology/27.2.397},
}

@article{hansen1990nonlinear,
SerialNo={129},
author={Hansen, Ulrich and Yuen, David A.},
 year={1990},
title={Nonlinear physics of double-diffusive convection in geological
  systems},
journal={Earth Sci. Rev.},
 volume={29},
 number={1--4},
  pages={385--399},
 orgname={Elsevier},
doi={10.1016/0012-8252(90)90050-6},
}

@article{bischoff1989salinity,
SerialNo={130},
author={Bischoff, James L. and Rosenbauer, Robert J.},
 year={1989},
title={Salinity variations in submarine hydrothermal systems by layered
  double-diffusive convection},
journal={{J.} {Geol.}},
 volume={97},
 number={5},
  pages={613--623},
 orgname={University of Chicago Press},
doi={10.1086/629338},
}

@book{charlier2015layered,
title={Layered Intrusions},
SerialNo={131},
 year={2015},
booktitle={Layered Intrusions},
 orgname={Springer},
 editor={Charlier, Bernard and Namur, Olivier and Latypov, Rais and Tegner,
  Christian},
}

@article{kerr1982layered,
SerialNo={132},
author={Kerr, R. C. and Turner, J. S.},
 year={1982},
title={Layered convection and crystal layers in multicomponent systems},
journal={Nature},
 volume={298},
 number={5876},
  pages={731--733},
 orgname={Nature Publishing Group UK London},
doi={10.1038/298731a0},
}

@article{naslund1996mechanisms,
SerialNo={133},
author={Naslund, H. R. and McBirney, A. R.},
 year={1996},
title={Mechanisms of formation of igneous layering},
journal={{Dev.} {Petrol.}},
 volume={15},
  pages={1--43},
 orgname={Elsevier},
doi={10.1016/S0167-2894(96)80003-0},
}

@article{rogerson2000patterns,
SerialNo={134},
author={Rogerson, M. A. and Cardoso, S. S. S.},
 year={2000},
title={Patterns of bubble desorption during the solidification of a
  multicomponent melt},
journal={J. Fluid Mech.},
 volume={419},
  pages={263--282},
 orgname={Cambridge University Press},
doi={10.1017/S0022112000001208},
}

@book{nield2017convection,
title={Convection in Porous Media},
SerialNo={135},
author={Nield, Donald A. and Bejan, Adrian},
 year={2017},
booktitle={Convection in Porous Media},
 orgname={Springer},
}

@article{imhoff1988experimental,
SerialNo={136},
author={Imhoff, Paul T. and Green, Theodore},
 year={1988},
title={Experimental investigation of double-diffusive groundwater fingers},
journal={J. Fluid Mech.},
 volume={188},
  pages={363--382},
 orgname={Cambridge University Press},
doi={10.1017/S002211208800076X},
}

@article{anderson2020convective,
SerialNo={137},
author={Anderson, Daniel M. and Guba, Peter},
 year={2020},
title={Convective phenomena in mushy layers},
journal={Annu. Rev. Fluid Mech.},
 volume={52},
 number={1},
  pages={93--119},
 orgname={Annual Reviews},
doi={10.1146/annurev-fluid-010719-060332},
}

@article{Shahidzadeh2015Salt,
SerialNo={138},
author={Shahidzadeh, Noushine and Schut, Marthe F. L. and Desarnaud, Julie and
  Prat, Marc and Bonn, Daniel},
 year={2015},
title={Salt stains from evaporating droplets},
journal={{Sci.} {Rep.}},
 volume={5},
 number={1},
doi={10.1038/srep10335},
}

@article{qazi2019salt,
SerialNo={139},
author={Qazi, M. J. and Salim, H. and Doorman, C. A. W. and Jambon-Puillet, E.
  and Shahidzadeh, N.},
 year={2019},
title={Salt creeping as a self-amplifying crystallization process},
journal={{Sci.} {Adv.}},
 volume={5},
 number={12},
  pages={eaax1853},
 orgname={American Association for the Advancement of Science},
doi={10.1126/sciadv.aax1853},
}

@article{Wijnhorst2024,
SerialNo={140},
author={Wijnhorst, R. and {Van der Sloot}, F. and Pel, L. and Shahidzadeh, N.},
 year={2024},
title={Effect of evaporative surface area on salt efflorescence and
  subflorescence formation in a given porous material},
journal={{Phys.} {Rev.} {Appl.}},
 volume={21},
 number={6},
  artnum={064055},
doi={10.1103/PhysRevApplied.21.064055},
}

@article{desarnaud2015drying,
SerialNo={141},
author={Desarnaud, Julie and Derluyn, Hannelore and Molari, Luisa and {de
  Miranda}, Stefano and Cnudde, Veerle and Shahidzadeh, Noushine},
 year={2015},
title={Drying of salt contaminated porous media: \?{E}ffect of primary and
  secondary nucleation},
journal={J. Appl. Phys.},
 volume={118},
 number={11},
 orgname={AIP Publishing},
doi={10.1063/1.4930292},
}

@article{Saffman1958,
SerialNo={142},
author={Saffman, P. G. and Taylor, Geoffrey},
 year={1958},
title={The Penetration of a Fluid into a Porous Medium or Hele-Shaw Cell
  Containing a More Viscous Liquid},
journal={{Proc.} R. {Soc.} {London} A.},
 volume={245},
 number={1242},
  pages={312--329},
doi={10.1098/rspa.1958.0085},
}

@article{Lovoll2004,
SerialNo={143},
author={L{\o}voll, Grunde and M{\'e}heust, Yves and Toussaint, Renaud and
  Schmittbuhl, Jean and M{\aa}l{\o}y, Knut J{\o}rgen},
 year={2004},
title={Growth activity during fingering in a porous {Hele-Shaw} cell},
journal={Phys. Rev. E},
 volume={70},
  artnum={026301},
 orgname={APS},
doi={10.1103/PhysRevE.70.026301},
}

@article{Zukowski2025,
SerialNo={144},
author={{\.Z}ukowski, Stanis{\l}aw and Magni, Silvana and Osselin, Florian and
  Dutka, Filip and Cooper, Max and Ladd, Anthony J.C. and Szymczak, Piotr},
 year={2025},
title={Invariant Forms of Dissolution Fingers},
journal={Phys. Rev. Lett.},
 volume={134},
 number={9},
  artnum={094101},
 orgname={APS},
doi={10.1103/PhysRevLett.134.094101},
}

@article{mcduff2010,
SerialNo={145},
author={D. R. McDuff and C. E. Shuchart and S. K. Jackson and D. Postl and J.
  S. Brown},
 year={2010},
title={Understanding Wormholes in Carbonates: \?{U}nprecedented Experimental
  Scale and \?{{\?{3}}-D} Visualization},
journal={{J.} {Petrol.} {Technol.}},
 volume={62},
  pages={78--81},
doi={10.2118/134379-MS},
}

@article{Hill1952,
SerialNo={146},
author={Hill, S.},
 year={1952},
title={Channeling in packed columns},
journal={Chem. Eng. Sci.},
 volume={1},
 number={6},
  pages={247--253},
 orgname={Elsevier},
doi={10.1016/0009-2509(52)87017-4},
}

@article{Chuoke1959,
SerialNo={147},
author={Chuoke, R. L. and {Van Meurs}, P. and {van der Poel}, C.},
 year={1959},
title={The instability of slow, immiscible, viscous liquid-liquid
  displacements in permeable media},
journal={{Trans.} AIME},
 volume={216},
 number={01},
  pages={188--194},
 orgname={SPE},
doi={10.2118/1141-G},
}

@article{Ortoleva1987b,
SerialNo={148},
author={P. Ortoleva and J. Chadam and E. Merino and A. Sen},
 year={1987},
title={Geochemical Self-Organization {\?{II}}: \?{T}he Reactive-Infiltration
  Instability},
journal={Am. J. Sci.},
 volume={287},
  pages={1008--1040},
doi={10.2475/ajs.287.10.1008},
}

@article{Hinch1990,
SerialNo={149},
author={E. J. Hinch and B. S. Bhatt},
 year={1990},
title={Stability of an acid front moving through porous rock},
journal={J. Fluid Mech.},
 volume={212},
  pages={279--288},
doi={10.1017/S0022112090001963},
}

@article{Szymczak2012,
SerialNo={150},
author={P. Szymczak and A. J. C. Ladd},
 year={2012},
title={Reactive infiltration instabilities in rocks. \?{Fracture} dissolution},
journal={J. Fluid Mech.},
 volume={702},
  pages={239--264},
doi={10.1017/jfm.2012.174},
}

@article{Szymczak2014,
SerialNo={151},
author={P. Szymczak and A. J. C. Ladd},
 year={2014},
title={Reactive infiltration instabilities in rocks. \?{Part {\?{2}}}. \?{Dissolution}
  of a porous matrix},
journal={J. Fluid Mech.},
 volume={738},
  pages={591--630},
doi={10.1017/jfm.2013.586},
}

@article{Groves1994a,
SerialNo={152},
author={C. G. Groves and A. D. Howard},
 year={1994},
title={Early Development of Karst Systems. {\?{I}. \?{P}referential} Flow Path
  Enlargement Under Laminar Flow},
journal={Water Resour. Res.},
 volume={30},
  pages={2837--2846},
doi={10.1029/94WR01303},
}

@article{Hanna1998,
SerialNo={153},
author={R. B. Hanna and H. Rajaram},
 year={1998},
title={Influence of Aperture Variability on Dissolutional Growth of Fissures
  in Karst Formations},
journal={Water Resour. Res.},
 volume={34},
  pages={2843--2853},
doi={10.1029/98WR01528},
}

@article{Szymczak2011,
SerialNo={154},
author={P. Szymczak and A. J. C. Ladd},
 year={2011},
title={The initial stages of cave formation: \?{B}eyond the one-dimensional
  paradigm},
journal={Earth Planet Sci. Lett.},
 volume={301},
  pages={424--432},
doi={10.1016/j.epsl.2010.10.026},
 issn={0009-2509},
}

@article{lipar2021,
SerialNo={155},
author={Lipar, Matej and Szymczak, Piotr and White, Susan Q. and Webb, John A.},
 year={2021},
title={Solution pipes and focused vertical water flow: \?{G}eomorphology and
  modelling},
journal={Earth Sci. Rev.},
 volume={218},
  artnum={103635},
 orgname={Elsevier},
doi={10.1016/j.earscirev.2021.103635},
}

@article{Hoefner1988,
SerialNo={156},
author={M. L. Hoefner and H. S. Fogler},
 year={1988},
title={Pore Evolution and Channel Formation During Flow and Reaction in Porous
  Media},
journal={AIChE J.},
 volume={34},
  pages={45--54},
doi={10.1002/aic.690340107},
}

@article{Fredd1998,
SerialNo={157},
author={C. N. Fredd and H. S. Fogler},
 year={1998},
title={Influence of Transport and Reaction on Wormhole Formation in Porous
  Media},
journal={AIChE J.},
 volume={44},
  pages={1933--1949},
doi={10.1002/aic.690440902},
}

@book{Economides2000,
SerialNo={158},
author={M. J. Economides and K. G. Nolte},
 year={2000},
 title={Reservoir Stimulation},
 publisher={Wiley},
 city = {Chichester, UK},
}

@article{Charalambous2021,
SerialNo={159},
author={Charalambous, Andreas N.},
 year={2021},
title={Water well acidization revisited: includes oil and geothermal well
  perspectives},
journal={{Quart.} {J.} {Engin.} {Geol.} {Hydrogeol.}},
 volume={54},
 number={3},
  pages={qjegh2020--071},
 orgname={The Geological Society of London},
doi={10.1144/qjegh2020-071},
}

@article{Sutra2017,
SerialNo={160},
author={Sutra, Emilie and Spada, Matteo and Burgherr, Peter},
 year={2017},
title={Chemicals usage in stimulation processes for shale gas and deep
  geothermal systems: a comprehensive review and comparison},
journal={{Renew.} {Sustain.} Energy {Rev.}},
 volume={77},
  pages={1--11},
 orgname={Elsevier},
doi={10.1016/j.rser.2017.03.108},
}

@article{Toussaint2018,
SerialNo={161},
author={Toussaint, Renaud and Aharonov, Einat and Koehn, Daniel and Gratier,
  J-P and Ebner, Martin and Baud, Patrick and Rolland, Alexandra and Renard,
  Francois},
 year={2018},
title={Stylolites: \?{A} review},
journal={J. Struct. Geol.},
 volume={114},
  pages={163--195},
doi={10.1016/j.jsg.2018.05.003},
}

@article{Witten1981,
SerialNo={162},
author={Witten, T.A. and Sander, Leonard M.},
 year={1981},
title={Diffusion-limited aggregation, a kinetic critical phenomenon},
journal={Phys. Rev. Lett.},
 volume={47},
 number={19},
  pages={1400},
doi={10.1103/PhysRevLett.47.1400},
}

@article{cardoso2014geochemistry,
SerialNo={163},
author={Cardoso, Silvana S. S. and Andres, Jeanne T. H.},
 year={2014},
title={Geochemistry of silicate-rich rocks can curtail spreading of carbon
  dioxide in subsurface aquifers},
journal={{Nature} {Commun.}},
 volume={5},
 number={1},
  pages={5743},
 orgname={Nature Publishing Group UK London},
doi={10.1038/ncomms6743},
}

@article{kang2014pore,
SerialNo={164},
author={Kang, Peter K and {De Anna}, Pietro and Nunes, Joao P and Bijeljic,
  Branko and Blunt, Martin J and Juanes, Ruben},
 year={2014},
title={Pore-scale intermittent velocity structure underpinning anomalous
  transport through \?{{\?{3}}-D} porous media},
journal={Geophys. Res. Lett.},
 volume={41},
 number={17},
  pages={6184--6190},
 orgname={Wiley Online Library},
doi={10.1002/2014GL061475},
}

@article{perugini2003chaotic,
SerialNo={165},
author={Perugini, Diego and Poli, Giampiero and Mazzuoli, Roberto},
 year={2003},
title={Chaotic advection, fractals and diffusion during mixing of magmas:
  evidence from lava flows},
journal={J. Volcanol. Geotherm. Res.},
 volume={124},
 number={3--4},
  pages={255--279},
 orgname={Elsevier},
doi={10.1016/S0377-0273(03)00098-2},
}

@article{aref1984stirring,
SerialNo={166},
author={Aref, Hassan},
 year={1984},
title={Stirring by chaotic advection},
journal={J. Fluid Mech.},
 volume={143},
  pages={1--21},
 orgname={Cambridge University Press},
doi={10.1017/S0022112084001233},
}

@article{cartwright1999introduction,
SerialNo={167},
author={Cartwright, Julyan H. E. and Feingold, Mario and Piro, Oreste},
 year={1999},
title={An introduction to chaotic advection},
journal={{Mix.}: Chaos {Turbulen.}},
  pages={307--342},
 orgname={Springer},
doi={10.1007/978-1-4615-4697-9\_13},
}

@article{aref2017frontiers,
SerialNo={168},
author={Aref, Hassan and Blake, John R. and Budi{\v s}i{\'c}, Marko and
  Cardoso, Silvana S S and Cartwright, Julyan H. E. and Clercx, Herman J H and {El
  Omari}, Kamal and Feudel, Ulrike and Golestanian, Ramin and Gouillart,
  Emmanuelle and GertJan F. {van Heijst} and Tatyana S. Krasnopolskaya and Yves
  Le Guer and Robert S. MacKay and Vyacheslav V. Meleshko and Guy Metcalfe and
  Igor Mezi{\'c} and Alessandro P. S. {de Moura} and Oreste Piro and Michel F. M.
  Speetjens and Rob Sturman and Jean-Luc Thiffeault and Idan Tuval},
 year={2017},
title={Frontiers of chaotic advection},
journal={Rev. Mod. Phys.},
 volume={89},
 number={2},
  artnum={025007},
 orgname={APS},
doi={10.1103/RevModPhys.89.025007},
}

@article{metcalfe2010partially,
SerialNo={169},
author={Metcalfe, Guy and Lester, Daniel and Ord, Alison and Kulkarni,
  Pandurang and Trefry, Mike and Hobbs, Bruce E and Regenauer-Lieb, Klaus and
  Morris, Jeffery},
 year={2010},
title={A partially open porous media flow with chaotic advection: towards a
  model of coupled fields},
journal={{Phil.} {Trans.} {Roy.} {Soc.} A},
 volume={368},
 number={1910},
  pages={217--230},
 orgname={The Royal Society Publishing},
doi={10.1098/rsta.2009.0198},
}

@article{lester2016chaotic,
SerialNo={170},
author={Lester, Daniel R. and Dentz, Marco and {Le Borgne}, Tanguy},
 year={2016},
title={Chaotic mixing in three-dimensional porous media},
journal={J. Fluid Mech.},
 volume={803},
  pages={144--174},
 orgname={Cambridge University Press},
doi={10.1017/jfm.2016.486},
}

@article{lester2016chaotic2,
SerialNo={171},
author={Lester, D.R. and Trefry, M.G. and Metcalfe, Guy},
 year={2016},
title={Chaotic advection at the pore scale: \?{M}echanisms, upscaling and
  implications for macroscopic transport},
journal={Adv. Water Resour.},
 volume={97},
  pages={175--192},
 orgname={Elsevier},
doi={10.1016/j.advwatres.2016.09.007},
}

@article{lester2012mechanics,
SerialNo={172},
author={Lester, Daniel R. and Ord, Alison and Hobbs, Bruce E.},
 year={2012},
title={The mechanics of hydrothermal systems: {\?{II}}. \?{Fluid} mixing and chemical
  reactions},
journal={Ore Geol. Rev.},
 volume={49},
  pages={45--71},
 orgname={Elsevier},
doi={10.1016/j.oregeorev.2012.08.002},
}

@article{trefry2019temporal,
SerialNo={173},
author={Trefry, Michael G. and Lester, Daniel R. and Metcalfe, Guy and Wu,
  Junhong},
 year={2019},
title={Temporal fluctuations and poroelasticity can generate chaotic advection
  in natural groundwater systems},
journal={Water Resour. Res.},
 volume={55},
 number={4},
  pages={3347--3374},
 orgname={Wiley Online Library},
doi={10.1029/2018WR023864},
}

@article{schoofs1999chaotic,
SerialNo={174},
author={Schoofs, Stan and Spera, Frank J. and Hansen, Ulrich},
 year={1999},
title={Chaotic thermohaline convection in low-porosity hydrothermal systems},
journal={Earth Planet Sci. Lett.},
 volume={174},
 number={1--2},
  pages={213--229},
 orgname={Elsevier},
doi={10.1016/S0012-821X(99)00264-2},
}

@article{oberst2018detection,
SerialNo={175},
author={Oberst, Sebastian and Niven, Robert K. and Lester, Daniel R. and Ord,
  Alison and Hobbs, Bruce and Hoffmann, Norbert},
 year={2018},
title={Detection of unstable periodic orbits in mineralising geological
  systems},
journal={Chaos: {Interdisc.} {J.} Nonlinear {Sci.}},
 volume={28},
 number={8},
 orgname={AIP Publishing},
doi={10.1063/1.5024134},
}

@inbook{yuen1992strongly,
SerialNo={176},
author={Yuen, David A. and Malevsky, Andrei V.},
 year={1992},
title={Strongly chaotic {\?{Newton}ian} and non-{\?{Newton}ian} mantle convection},
booktitle={Chaotic Processes in the Geological Sciences},
  pages={71--88},
 orgname={Springer},
doi={10.1007/978-1-4684-0643-6\_4},
}

@article{perugini2008virtual,
SerialNo={177},
author={Perugini, Diego and Petrelli, Maurizio and Poli, Giampiero},
 year={2008},
title={A virtual voyage through {{{\?{3D}}}} structures generated by chaotic mixing of
  magmas and numerical simulations: a new approach for understanding spatial
  and temporal complexity of magma dynamics},
journal={{Vis.} {Geosci.}},
 volume={13},
 number={1},
  pages={1--24},
 orgname={Springer},
doi={10.1007/s10069-006-0004-x},
}

@article{de2015chaotic,
SerialNo={178},
author={{De Campos}, Cristina P.},
 year={2015},
title={Chaotic flow patterns from a deep plutonic environment: a case study on
  natural magma mixing},
journal={Pure Appl. Geophys.},
 volume={172},
  pages={1815--1833},
 orgname={Springer},
doi={10.1007/s00024-014-0940-6},
}

@article{petrelli2016effects,
SerialNo={179},
author={Petrelli, Maurizio and {El Omari}, Kamal and {Le Guer}, Yves and Perugini,
  Diego},
 year={2016},
title={Effects of chaotic advection on the timescales of cooling and
  crystallization of magma bodies at mid crustal levels},
journal={{Geochem.} {Geophys.} Geosystems},
 volume={17},
 number={2},
  pages={425--441},
 orgname={Wiley Online Library},
doi={10.1002/2015GC006109},
}

@article{cardoso2014dynamics,
SerialNo={180},
author={Cardoso, Silvana S. S. and Cartwright, Julyan H. E.},
 year={2014},
title={Dynamics of osmosis in a porous medium},
journal={{Roy.} {Soc.} Open {Sci.}},
 volume={1},
 number={3},
  artnum={140352},
doi={10.1098/rsos.140352},
}

@article{cardoso2016increased,
SerialNo={181},
author={Cardoso, Silvana S. S. and Cartwright, Julyan H. E.},
 year={2016},
title={Increased methane emissions from deep osmotic and buoyant convection
  beneath submarine seeps as climate warms},
journal={{Nature} {Commun.}},
 volume={7},
 number={1},
  pages={13266},
 orgname={Nature Publishing Group UK London},
doi={10.1038/ncomms13266},
}

@article{rocha2021formation,
SerialNo={182},
author={Rocha, Luis A M and Guti{\'e}rrez-Ariza, Carlos and Pimentel, Carlos
  and S{\'a}nchez-Almazo, Isabel and Sainz-D{\'\i}az, C. Ignacio and Cardoso,
  Silvana S S and Cartwright, Julyan H E},
 year={2021},
title={Formation and structures of horizontal submarine fluid conduit and
  venting systems associated with marine seeps},
journal={{Geochem.} {Geophys.} Geosystems},
 volume={22},
 number={11},
  post={e2021GC009724},
 orgname={Wiley Online Library},
doi={10.1029/2021GC009724},
}

@article{morgado2024osmosis,
SerialNo={183},
author={Morgado, Ana M O and Rocha, Luis A M and Cartwright, Julyan H. E. and
  Cardoso, Silvana S S},
 year={2024},
title={Osmosis drives explosions and methane release in {Siberian}
  permafrost},
journal={Geophys. Res. Lett.},
 volume={51},
 number={18},
  post={e2024GL108987},
doi={10.1029/2024GL108987},
}

@book{glauber_furni_1646,
title={Furni novi philosophici, sive {Descriptio} artis distillatori{\ae} nov{\ae};
  nec non spirituum, olcorum, florum, aliorumque medicamentorum illus
  beneficio, facilim\^a qu\^adam \& perculiari vi\'a \`e vegetabilibus, animalibus
  \& mineralibus conficiendorum \& quidem magno cum lucro; agens quoque de
  illorum usu t\`am chymico qu\`am medico, edita \& publicata in gratiam
  veritatis studiosorum},
SerialNo={184},
author={Glauber, Johann Rudolf},
 year={1646},
booktitle={Furni novi philosophici, sive {Descriptio} artis distillatori{\ae} nov{\ae};
  nec non spirituum, olcorum, florum, aliorumque medicamentorum illus
  beneficio, facilim\^a qu\^adam \& perculiari vi\'a \`e vegetabilibus, animalibus
  \& mineralibus conficiendorum \& quidem magno cum lucro; agens quoque de
  illorum usu t\`am chymico qu\`am medico, edita \& publicata in gratiam
  veritatis studiosorum},
 orgname={prostant apud J. Janssonium},
 city = {Amsterodami},
}

@article{Pimentel2023Chemobrionics,
SerialNo={185},
author={Pimentel, Carlos and Zheng, Mingchuan and Cartwright, Julyan H. E. and
  SainzD{\'i}az, C. Ignacio},
 year={2023},
title={Chemobrionics Database: \?{C}ategorisation of Chemical Gardens According to
  the Nature of the Anion, Cation and Experimental Procedure},
journal={ChemSystemsChem},
 volume={5},
 number={4},
  artnum={e202300002},
 orgname={Wiley},
doi={10.1002/syst.202300002},
}

@article{cardoso2017differing,
SerialNo={186},
author={Cardoso, Silvana S. S. and Cartwright, Julyan H. E.},
 year={2017},
title={On the differing growth mechanisms of black-smoker and {Lost City}-type
  hydrothermal vents},
journal={{Proc.} {Roy.} {Soc.} A: {Math.} {Phys.} {Engin.} {Sci.}},
 volume={473},
 number={2205},
  artnum={20170387},
 orgname={The Royal Society Publishing},
doi={10.1098/rspa.2017.0387},
}

@article{sainz2018growth,
SerialNo={187},
author={Sainz-D{\'\i}az, C. Ignacio and Escamilla-Roa, Elizabeth and
  Cartwright, Julyan H E},
 year={2018},
title={Growth of Self-Assembling Tubular Structures of Magnesium Oxy/Hydroxide
  and Silicate Related With Seafloor Hydrothermal Systems Driven by
  Serpentinization},
journal={{Geochem.} {Geophys.} Geosystems},
 volume={19},
 number={8},
  pages={2813--2822},
 orgname={Wiley Online Library},
doi={10.1029/2018GC007594},
}

@article{Cartwright2019,
SerialNo={188},
author={Julyan H. E. Cartwright and Michael J. Russell},
 year={2019},
title={The origin of life: the submarine alkaline vent theory at 30},
journal={Interface {Focus}},
 volume={9},
doi={10.1098/rsfs.2019.0104},
 issn={20428901},
}

@article{brinicle,
SerialNo={189},
author={Cartwright, Julyan H. E. and Escribano, Bruno and Gonzalez, Diego L.
  and Sainz-Diaz, C. Ignacio and Tuval, Idan},
 year={2013},
title={Brinicles as a Case of Inverse Chemical Gardens},
journal={Langmuir},
 volume={29},
 number={25},
  pages={7655--7660},
doi={10.1021/la4009703},
}

@article{teston2024experimental,
SerialNo={190},
author={Test{\'o}n-Mart{\'\i}nez, Sergio and Barge, Laura M and Eichler, Jan
  and Sainz-D{\'\i}az, C. Ignacio and Cartwright, Julyan H E},
 year={2024},
title={Experimental modelling of the growth of tubular ice brinicles from
  brine flows under sea ice},
journal={Cryosphere},
 volume={18},
 number={5},
  pages={2195--2205},
 orgname={Copernicus Publications G\"ottingen, Germany},
doi={10.5194/tc-18-2195-2024},
}

@article{vance2019self,
SerialNo={191},
author={Vance, Steven D and Barge, Laura M and Cardoso, Silvana S S and
  Cartwright, Julyan H E},
 year={2019},
title={Self-assembling ice membranes on \?{Europa}: brinicle properties, field
  examples, and possible energetic systems in icy ocean worlds},
journal={Astrobiology},
 volume={19},
 number={5},
  pages={685--695},
doi={10.1089/ast.2018.1826},
}

@article{russell1997,
SerialNo={192},
author={M. J. Russell and A. J. Hall},
 year={1997},
title={The emergence of life from iron monosulphide bubbles at a submarine
  hydrothermal redox and \?{pH} front},
journal={{J.} {Geol.} {Soc.}},
 volume={154},
  pages={377--402},
 orgname={J Geol Soc London},
doi={10.1144/gsjgs.154.3.0377},
 issn={0016-7649},
}

@article{cardoso2020,
SerialNo={193},
author={Cardoso, Silvana S. S. and Cartwright, Julyan H. E. and {\v C}ejkov{\'a},
  Jitka and Cronin, Leroy and {De Wit}, Anne and Giannerini, Simone and
  Horv{\'a}th, Dezs{\H{o}} and Rodrigues, Al{\'\i}rio and Russell, Michael J
  and Sainz-D{\'\i}az, C. Ignacio and T{\'o}th, {\'A}.},
 year={2020},
title={Chemobrionics: \?{F}rom self-assembled material architectures to the origin
  of life},
journal={{Artific.} Life},
 volume={26},
 number={3},
  pages={315--326},
doi={10.1162/artl\_a\_00323},
}

@article{ding2019intrinsic,
SerialNo={194},
author={Ding, Yang and Cartwright, Julyan H. E. and Cardoso, Silvana S. S.},
 year={2019},
title={Intrinsic concentration cycles and high ion fluxes in self-assembled
  precipitate membranes},
journal={Interface {Focus}},
 volume={9},
 number={6},
  artnum={20190064},
doi={10.1098/rsfs.2019.0064},
}

@article{ding2024dynamics,
SerialNo={195},
author={Ding, Yang and Cardoso, Silvana S. S. and Cartwright, Julyan H. E.},
 year={2024},
title={Dynamics of the osmotic lysis of mineral protocells and its avoidance
  at the origins of life},
journal={Geobiology},
 volume={22},
 number={4},
  artnum={e12611},
doi={10.1111/gbi.12611},
}

@article{cartwright2012beyond,
SerialNo={196},
author={Cartwright, Julyan H. E. and Mackay, Alan L.},
 year={2012},
title={Beyond crystals: the dialectic of materials and information},
journal={{Phil.} {Trans.} {Roy.} {Soc.} A: {Math.} {Phys.} {Engin.} {Sci.}},
 volume={370},
 number={1969},
  pages={2807--2822},
doi={10.1098/rsta.2012.0106},
}

@article{cartwright2016dna,
SerialNo={197},
author={Cartwright, Julyan H. E. and Giannerini, Simone and Gonz{\'a}lez, Diego
  L},
 year={2016},
title={{\?{DNA}} as information: at the crossroads between biology, mathematics,
  physics and chemistry},
journal={{Phil.} {Trans.} {Roy.} {Soc.} A},
 volume={374},
 number={2063},
  artnum={20150071},
doi={10.1098/rsta.2015.0071},
}

@article{cartwright2024information,
SerialNo={198},
author={Cartwright, Julyan H. E. and {\v C}ejkov{\'a}, Jitka and Fimmel, Elena
  and Giannerini, Simone and Gonzalez, Diego Luis and Goracci, Greta and
  Gr{\'a}cio, Clara and Houwing-Duistermaat, Jeanine and Mati{\'c}, Dragan and
  Mi{\v s}i{\'c}, Nata{\v s}a and others},
 year={2024},
title={Information, Coding, and Biological Function: \?{T}he Dynamics of Life},
journal={{Artific.} Life},
 volume={30},
 number={1},
  pages={16--27},
doi={10.1162/artl\_a\_00432},
}

@article{witten1983diffusion,
SerialNo={199},
author={Witten, Thomas A. and Sander, Leonard M.},
 year={1983},
title={Diffusion-limited aggregation},
journal={Phys. Rev. B},
 volume={27},
 number={9},
  pages={5686},
 orgname={APS},
doi={10.1103/PhysRevB.27.5686},
}

@article{Meakin1983,
SerialNo={200},
author={Paul Meakin},
 year={1983},
title={Diffusion-controlled flocculation: \?{T}he effects of attractive and
  repulsive interactions},
journal={J. Chem. Phys.},
 volume={79},
  pages={2426--2429},
doi={10.1063/1.446051},
}

@article{Garik1985,
SerialNo={201},
author={Peter Garik},
 year={1985},
title={Anisotropic growth of diffusion-limited aggregates},
journal={Phys. Rev. A},
 volume={32},
  pages={1275--1278},
doi={10.1103/PhysRevA.32.1275},
}

@article{Erlebacher1993,
SerialNo={202},
author={J. Erlebacher and P. C. Searson and K. Sieradzki},
 year={1993},
title={Computer simulation of dense-branching patterns},
journal={Phys. Rev. Lett.},
 volume={71},
  pages={3311--3314},
doi={10.1103/PhysRevLett.71.3311},
}

@book{Meakin1998,
title={Fractals, Scaling and Growth Far from Equilibrium},
SerialNo={203},
author={Paul Meakin},
 year={1998},
booktitle={Fractals, Scaling and Growth Far from Equilibrium},
 orgname={Cambridge University Press},
 city = {Cambridge, UK},
}

@article{Halsey2000,
SerialNo={204},
author={Halsey, Thomas C.},
 year={2000},
title={Diffusion-Limited Aggregation: \?{A} Model for Pattern Formation},
journal={Phys. Today},
 volume={53},
 number={11},
  pages={36--41},
doi={10.1063/1.1333284},
}

@article{Mineev-Weinstein2000,
SerialNo={205},
author={M. Mineev-Weinstein and P. B. Wiegmann and A. Zabrodin},
 year={2000},
title={Integrable Structure of Interface Dynamics},
journal={Phys. Rev. Lett.},
 volume={84},
 number={24},
  pages={5106--5109},
doi={10.1103/PhysRevLett.84.5106},
}

@article{Mineev-Weinstein2008,
SerialNo={206},
author={Mark Mineev-Weinstein and Mihai Putinar and Razvan Teodorescu},
 year={2008},
title={Random Matrices in {{\?{2D}}}, \?{Laplacian} Growth and Operator Theory},
journal={J. Phys. A},
 volume={41},
 number={26},
  artnum={263001},
doi={10.1088/1751-8113/41/26/263001},
}

@article{khan2016piezoelectric,
SerialNo={207},
author={Khan, Asif and Abas, Zafar and Kim, Heung Soo and Oh, Il-Kwon},
 year={2016},
title={Piezoelectric thin films: an integrated review of transducers and
  energy harvesting},
journal={Smart Mater. Struct.},
 volume={25},
 number={5},
  artnum={053002},
 orgname={IOP Publishing},
doi={10.1088/0964-1726/25/5/053002},
}

@article{pishvar2020foundations,
SerialNo={208},
author={Pishvar, Maya and Harne, Ryan L.},
 year={2020},
title={Foundations for soft, smart matter by active mechanical metamaterials},
journal={{Adv.} {Sci.}},
 volume={7},
 number={18},
  artnum={2001384},
 orgname={Wiley Online Library},
doi={10.1002/advs.202001384},
}

@article{voisey2024gold,
SerialNo={209},
author={Voisey, Christopher R. and Hunter, Nicholas J.R. and Tomkins, Andrew G.
  and Brugger, Jo{\"e}l and Liu, Weihua and Liu, Yang and Luzin, Vladimir},
 year={2024},
title={Gold nugget formation from earthquake-induced piezoelectricity in
  quartz},
journal={{Nature} {Geosci.}},
  pages={1--6},
 orgname={Nature Publishing Group UK London},
doi={10.1038/s41561-024-01514-1},
}

@article{zhang_redox_2015,
SerialNo={210},
author={Zhang, Zhaorui and Wang, Zhenni and He, Shengnan and Wang, Chaoqi and
  Jin, Mingshang and Yin, Yadong},
 year={2015},
title={Redox reaction induced {Ostwald} ripening for size- and shape-focusing
  of palladium nanocrystals},
journal={{Chem.} {Sci.}},
 volume={6},
 number={9},
  pages={5197--5203},
doi={10.1039/C5SC01787D},
 issn={2041-6539},
}

@article{wagner_theorie_1961,
SerialNo={211},
author={Wagner, Carl},
 year={1961},
title={Theorie der {Alterung} von {Niederschl\"agen} durch {Uml\"osen}
  ({Ostwald}-{Reifung})},
journal={{Z.} {Elektrochem.} Berichte {Bunsengesellschaft} {Physik} {Chemie}},
 volume={65},
 number={7--8},
  pages={581--591},
doi={10.1002/bbpc.19610650704},
}

@article{lifshitz_kinetics_1961,
SerialNo={212},
author={Lifshitz, I. M. and Slyozov, V. V.},
 year={1961},
title={The kinetics of precipitation from supersaturated solid solutions},
journal={{J.} {Phys.} {Chem.} Solids},
 volume={19},
 number={1},
  pages={35--50},
doi={10.1016/0022-3697(61)90054-3},
}

@article{van_westen_effect_2018,
SerialNo={213},
author={van Westen, Thijs and Groot, Robert D.},
 year={2018},
title={Effect of Temperature Cycling on {Ostwald} Ripening},
journal={{Cryst.} Growth {Des.}},
 volume={18},
 number={9},
  pages={4952--4962},
doi={10.1021/acs.cgd.8b00267},
}

@article{caccin2015framework,
SerialNo={214},
author={Caccin, Marco and Li, Zhenwei and Kermode, James R and De Vita,
  Alessandro},
 year={2015},
title={A framework for machine-learning-augmented multiscale atomistic
  simulations on parallel supercomputers},
journal={Int. J. Quantum Chem.},
 volume={115},
 number={16},
  pages={1129--1139},
 orgname={Wiley Online Library},
doi={10.1002/qua.24952},
}

@article{eberl1990ostwald,
SerialNo={215},
author={Eberl, Dennis D. and {\'S}rodo{\'n}, Jan and Kralik, Martin and Taylor,
  Bruce E. and Peterman, Zell E.},
 year={1990},
title={Ostwald ripening of clays and metamorphic minerals},
journal={Science},
 volume={248},
 number={4954},
  pages={474--477},
doi={10.1126/science.248.4954.474},
}

@article{morse1988ostwald,
SerialNo={216},
author={Morse, John W. and Casey, William H.},
 year={1988},
title={Ostwald processes and mineral paragenesis in sediments},
journal={Am. J. Sci.},
 volume={288},
 number={6},
  pages={537--560},
doi={10.2475/ajs.288.6.537},
}

@article{cabane2005experimental,
SerialNo={217},
author={Cabane, Hugues and Laporte, Didier and Provost, Ariel},
 year={2005},
title={An experimental study of {Ostwald} ripening of olivine and plagioclase
  in silicate melts: implications for the growth and size of crystals in
  magmas},
journal={{Contrib.} {Miner.} {Petrol.}},
 volume={150},
  pages={37--53},
doi={10.1007/s00410-005-0002-2},
}

@article{hastie2021transport,
SerialNo={218},
author={Hastie, E.C.G. and Schindler, M. and Kontak, D.J. and Lafrance, B.},
 year={2021},
title={Transport and coarsening of gold nanoparticles in an orogenic deposit
  by dissolution--reprecipitation and {Ostwald} ripening},
journal={{Commun.} Earth {Environ.}},
 volume={2},
 number={1},
  pages={57},
doi={10.1038/s43247-021-00126-6},
}

@article{cartwright2007ostwald,
SerialNo={219},
author={Cartwright, Julyan H. E. and Piro, Oreste and Tuval, Idan},
 year={2007},
title={Ostwald ripening, chiral crystallization, and the common-ancestor
  effect},
journal={Phys. Rev. Lett.},
 volume={98},
 number={16},
  artnum={165501},
doi={10.1103/PhysRevLett.98.165501},
}

@book{anderson2017fracture,
title={Fracture Mechanics: Fundamentals and Applications},
SerialNo={220},
author={Anderson, Ted L.},
 year={2017},
booktitle={Fracture Mechanics: Fundamentals and Applications},
 edition={fourth ed.},
 orgname={CRC Press},
}

@book{broberg1999cracks,
title={Cracks and Fracture},
SerialNo={221},
author={Broberg, K. Bertram},
 year={1999},
booktitle={Cracks and Fracture},
 orgname={Elsevier},
}

@article{meyer2000crack,
SerialNo={222},
author={Meyer, D. and Br{\"u}ckner-Foit, M.},
 year={2000},
title={Crack interaction modelling},
journal={Fatigue Fract. Eng. Mater. Struct.},
 volume={23},
 number={4},
  pages={315--323},
 orgname={Wiley Online Library},
doi={10.1046/j.1460-2695.2000.00283.x},
}

@article{sun2021state,
SerialNo={223},
author={Sun, Yanan and Edwards, Michael G. and Chen, Bin and Li, Chenfeng},
 year={2021},
title={A state-of-the-art review of crack branching},
journal={Eng. Fract. Mech.},
 volume={257},
  artnum={108036},
 orgname={Elsevier},
doi={10.1016/j.engfracmech.2021.108036},
}

@article{hakim2005crack,
SerialNo={224},
author={Hakim, Vincent and Karma, Alain},
 year={2005},
title={Crack path prediction in anisotropic brittle materials},
journal={Phys. Rev. Lett.},
 volume={95},
 number={23},
  artnum={235501},
 orgname={APS},
doi={10.1103/PhysRevLett.95.235501},
}

@article{shcherbakov2003damage,
SerialNo={225},
author={Shcherbakov, R. and Turcotte, D. L.},
 year={2003},
title={Damage and self-similarity in fracture},
journal={Theor. Appl. Fract. Mech.},
 volume={39},
 number={3},
  pages={245--258},
 orgname={Elsevier},
doi={10.1016/S0167-8442(03)00005-3},
}

@article{tarasovs2014self,
SerialNo={226},
author={Tarasovs, Sergejs and Ghassemi, Ahmad},
 year={2014},
title={Self-similarity and scaling of thermal shock fractures},
journal={Phys. Rev. E},
 volume={90},
 number={1},
  artnum={012403},
 orgname={APS},
doi={10.1103/PhysRevE.90.012403},
}

@article{bavzant1993scaling,
SerialNo={227},
author={Ba{\v z}ant, Zden\v{e}k P.},
 year={1993},
title={Scaling laws in mechanics of failure},
journal={{J.} {Engin.} {Mech.}},
 volume={119},
 number={9},
  pages={1828--1844},
 orgname={American Society of Civil Engineers},
doi={10.1061/(ASCE)0733-9399(1993)119:9(1828)},
}

@article{laubach2019role,
SerialNo={228},
author={Laubach, Stephen E. and Lander, R.H. and Criscenti, Louise J. and Anovitz,
  Lawrence M. and Urai, J.L. and Pollyea, Ryan M. and Hooker, John N. and Narr,
  Wayne and Evans, Mark A. and Kerisit, Sebastien N. and J. E. Olson and T.
  Dewers and D. Fisher and R. Bodnar and B. Evans and P. Dove and L. M. Bonnell
  and M. P. Marder and L. Pyrak-Nolte},
 year={2019},
title={The role of chemistry in fracture pattern development and opportunities
  to advance interpretations of geological materials},
journal={Rev. Geophys.},
 volume={57},
 number={3},
  pages={1065--1111},
 orgname={Wiley Online Library},
doi={10.1029/2019RG000671},
}

@article{jamtveit2009reaction,
SerialNo={229},
author={Jamtveit, Bj{\o}rn and Putnis, Christine V and Malthe-S{\o}renssen,
  Anders},
 year={2009},
title={Reaction induced fracturing during replacement processes},
journal={{Contrib.} {Miner.} {Petrol.}},
 volume={157},
  pages={127--133},
 orgname={Springer},
doi={10.1007/s00410-008-0324-y},
}

@article{Walger2009,
SerialNo={230},
author={Walger, Eckart and Matthe{\ss}, Georgvon and Seckendorff, Volker and
  Liebau, Friedrich},
 year={2009},
title={The formation of agate structures: models for silica transport, agate
  layer accretion, and for flow patterns and flow regimes in infiltration
  channels},
journal={Neues {Jahrbuch} Mineralogie-Abhandlungen},
  pages={113--152},
 orgname={Schweizerbart'sche Verlagsbuchhandlung},
doi={10.1127/0077-7757/2009/0141},
}

@INPROCEEDINGS{Jackson2005,
SerialNo={231},
author={Jackson, Brian},
 year={2005},
title={Structures and micro-structures in {Scottish} agates},
booktitle={Symposium on Agate and Cryptocrystalline Quartz, Golden, Colorado},
  pages={89--94},
}

@article{Moxon2020,
SerialNo={232},
author={Moxon, T. and Palyanova, G.},
 year={2020},
title={Agate genesis: \?{A} continuing enigma},
journal={Minerals},
 volume={10},
 number={11},
  pages={953},
 orgname={MDPI},
doi={10.3390/min10110953},
}

@article{Goetze2020,
SerialNo={233},
author={G{\"o}tze, J. and M{\"o}ckel, R. and Pan, Y.},
 year={2020},
title={Mineralogy, geochemistry and genesis of agate---\?{A} review},
journal={Minerals},
 volume={10},
 number={11},
  pages={1037},
 orgname={MDPI},
doi={10.3390/min10111037},
}

@article{Heaney1995,
SerialNo={234},
author={Heaney, P.J. and Davis, A.M.},
 year={1995},
title={Observation and origin of self-organized textures in agates},
journal={Science},
 volume={269},
 number={5230},
  pages={1562--1565},
 orgname={American Association for the Advancement of Science},
doi={10.1126/science.269.5230.1562},
}

@article{Cady1998,
SerialNo={235},
author={Cady, S.L. and Wenk, H.-R. and Sintubin, M.},
 year={1998},
title={Microfibrous quartz varieties: characterization by quantitative \?{X-ray}
  texture analysis and transmission electron microscopy},
journal={{Contrib.} {Miner.} {Petrol.}},
 volume={130},
  pages={320--335},
 orgname={Springer},
doi={10.1007/s004100050368},
}

@article{Frondel1985,
SerialNo={236},
author={Frondel, C.},
 year={1985},
title={Systematic compositional zoning in the quartz fibers of agates},
journal={Am. Miner.},
 volume={70},
 number={9--10},
  pages={975--979},
 orgname={Mineralogical Society of America},
}

@article{French2013,
SerialNo={237},
author={French, M. W. and Worden, R. H. and Lee, D. R.},
 year={2013},
title={Electron backscatter diffraction investigation of length-fast
  chalcedony in agate: implications for agate genesis and growth mechanisms},
journal={Geofluids},
 volume={13},
 number={1},
  pages={32--44},
 orgname={Wiley Online Library},
doi={10.1111/gfl.12006},
}

@article{Jones1952,
SerialNo={238},
author={Jones, Francis T.},
 year={1952},
title={Iris agate},
journal={{Amer.} {Miner.}: {J.} Earth {Planet.} {Mater.}},
 volume={37},
 number={7--8},
  pages={578--587},
 orgname={Mineralogical Society of America},
}

@article{Frondel1978,
SerialNo={239},
author={Frondel, Clifford},
 year={1978},
title={Characters of quartz fibers},
journal={Am. Miner.},
 volume={63},
 number={1--2},
  pages={17--27},
 orgname={Mineralogical Society of America},
}

@article{Landmesser1984,
SerialNo={240},
author={Landmesser, Michael},
 year={1984},
title={Das problem der Achatgenese},
journal={Mitt. Pollichia},
 volume={72},
  pages={5--137},
}

@article{Liesegang1910,
SerialNo={241},
author={Liesegang, R.E.},
 year={1910},
title={Die Entstehung der Achate},
journal={Zentralblatt {Miner.}},
 volume={11},
  pages={593--597},
}

@article{hsu2016fordite,
SerialNo={242},
author={Hsu, Tao and Lucas, Andrew},
 year={2016},
title={Fordite from the \?{Corvette} assembly plant},
journal={Gems Gemology},
 volume={52},
 number={1},
  pages={87--88},
 orgname={GEMOLOGICAL INST AMER 5345 ARMADA DR, CARLSBAD, CA 92008 USA},
}

@book{nova2021bestiary,
title={The Bestiary of the Anthropocene},
SerialNo={243},
author={Nova, Nicolas and Maigret, Nicolas and Roszkowska, Maria},
 year={2021},
booktitle={The Bestiary of the Anthropocene},
 orgname={Onomatop\'ee},
}

@book{Farrington1927,
title={Agate: Physical Properties and Origin, Archaeology and Folklore},
SerialNo={244},
author={Farrington, Oliver Cummings and Laufer, Berthold},
 year={1927},
booktitle={Agate: Physical Properties and Origin, Archaeology and Folklore},
 orgname={Field Museum of Natural History},
}

@article{Schalm2011,
SerialNo={245},
author={Schalm, O. and Proost, K. and {De Vis}, K. and Cagno, S. and Janssens, K. and
  Mees, F. and Jacobs, P. and Caen, J.},
 year={2011},
title={Manganese staining of archaeological glass: the characterization of
  {Mn}-rich inclusions in leached layers and a hypothesis of its formation},
journal={Archaeometry},
 volume={53},
 number={1},
  pages={103--122},
 orgname={Wiley Online Library},
doi={10.1111/j.1475-4754.2010.00534.x},
}

@article{Haidinger1848,
SerialNo={246},
author={Haidinger, W.},
 year={1848},
title={\"Uber die Metamorphose der Gebirgsarten},
journal={Ber. \"Uber Die {Mittheilungen} Von Freunden Naturwissenschaften Wien},
 volume={4},
  pages={103--134},
}

@article{Noeggerath1849,
SerialNo={247},
author={Noeggerath, J.},
 year={1849},
title={Sendschreiben an den k.k. wirklichen {Bergrath und Prof.}, {Herrn} {W.
  Haidinger in Wien}, \"uber die {Achat Mandeln in den Melaphyren}},
journal={Verh. Naturhist. Vereins {Preuss} {Rheinland} U. Westphalens},
  pages={243--260},
}

@article{Kenngott1851,
SerialNo={248},
author={Kenngott, J.G.A.},
 year={1851},
title={{\"Uber die Achatmandeln in den Melaphyren, namentlich \"uber die
  von Theiss in Tirol}},
journal={Naturwiss. Abh. Gesammelt Durch {Subscription} Herausgeg. W. Haidinger},
 volume={4},
  pages={71--104},
}

@article{Reusch1864,
SerialNo={249},
author={Reusch, E.},
 year={1864},
title={{\"Uber den Agat}},
journal={{Annal} {Phys.} {Chem.}},
 volume={123},
  pages={94--114},
}

@article{Harris1989,
SerialNo={250},
author={Harris, Chris},
 year={1989},
title={Oxygen-isotope zonation of agates from {Karoo} volcanics of the
  {\?{Skeleton Coast}, Namibia}},
journal={Am. Miner.},
 volume={74},
 number={3--4},
  pages={476--481},
 orgname={Mineralogical Society of America},
}

@book{Liesegang1914,
title={{Die Achate}},
SerialNo={251},
author={Liesegang, R. E.},
 year={1914},
booktitle={{Die Achate}},
 orgname={Springer},
}

@book{Pabian1994,
title={{Banded Agates, Origins and Inclusions}},
SerialNo={252},
author={Pabian, Roger K. and Zarins, Andrejs},
 year={1994},
booktitle={{Banded Agates, Origins and Inclusions}},
 orgname={University of Nebraska-Lincoln},
}

@article{Wang1990,
SerialNo={253},
author={Wang, Yifeng and Merino, Enrique},
 year={1990},
title={Self-organizational origin of agates: \?{B}anding, fiber twisting,
  composition, and dynamic crystallization model},
journal={Geochim. Cosmochim. Acta},
 volume={54},
 number={6},
  pages={1627--1638},
 orgname={Elsevier},
doi={10.1016/0016-7037(90)90396-3},
}

@article{Nacken1948,
SerialNo={254},
author={Nacken, R.},
 year={1948},
title={\"Uber die Nachbildung von Chalcedon-Mandeln},
journal={{Natur} {Folk}},
 volume={78},
  pages={2--8},
}

@article{White1961,
SerialNo={255},
author={White, J.F. and Corwin, J.F.},
 year={1961},
title={Synthesis and origin of chalcedony},
journal={{Amer.} {Miner.}: {J.} Earth {Planet.} {Mater.}},
 volume={46},
 number={1--2},
  pages={112--119},
 orgname={Mineralogical Society of America},
}

@article{Fallick1985,
SerialNo={256},
author={Fallick, A. E. and Jocelyn, J. and Donnelly, T. and Guy, M. and Behan, C.},
 year={1985},
title={Origin of agates in volcanic rocks from {\?{Scotland}}},
journal={Nature},
 volume={313},
 number={6004},
  pages={672--674},
 orgname={Nature Publishing Group UK London},
doi={10.1038/313672a0},
}

@article{Emmanuel2007,
SerialNo={257},
author={Emmanuel, Simon and Berkowitz, Brian},
 year={2007},
title={Effects of pore-size controlled solubility on reactive transport in
  heterogeneous rock},
journal={Geophys. Res. Lett.},
 volume={34},
 number={6},
  pages={L06404},
 orgname={Wiley Online Library},
doi={10.1029/2006GL028962},
}

@article{Theeuwes1975,
SerialNo={258},
author={Theeuwes, Felix},
 year={1975},
title={Elementary osmotic pump},
journal={J. Pharm. Sci.},
 volume={64},
 number={12},
  pages={1987--1991},
 orgname={Elsevier},
doi={10.1002/jps.2600641218},
}

@article{Goodchild1899,
SerialNo={259},
author={Goodchild, J. G.},
 year={1899},
title={On the genesis of some {Scottish} minerals},
journal={{Proc.} {Roy.} {Phys.} {Soc.} {Edin.}},
 volume={14},
  pages={181--220},
 orgname={Elsevier},
}

@article{Schlossmacher1960,
SerialNo={260},
author={Schlossmacher, K.},
 year={1960},
title={{Die Entstehung der Achate, {Teil} \?{II}}},
journal={Z. {Deutsch} Ges. Edelsteinkde.},
 volume={33},
  pages={11--16},
}

@article{Hou2023,
SerialNo={261},
author={Hou, Zhaoliang and Wo{\'s}, Dawid and Tschegg, Cornelius and Rogowitz,
  Anna and Rice, A Hugh N and Nasdala, Lutz and Fusseis, Florian and Szymczak,
  Piotr and Grasemann, Bernhard},
 year={2023},
title={Three-dimensional mineral dendrites reveal a nonclassical
  crystallization pathway},
journal={Geology},
 volume={51},
  pages={626--630},
 orgname={Geological Society of America},
doi={10.1130/G51127.1},
}

@misc{Pliny77,
SerialNo={262},
author={{Pliny the Elder}, {}},
 year={77},
title={Naturalis historia},
commn={{Naturalis historia, 77}},
}

@misc{Scheuchzer1700,
mypattern={[authors,atitle,midc][booktag,bktitle,yr,endbooktag,bibpages,bibdoi][post,myehost]},
SerialNo={263},
author={Scheuchzer, J. J.},
 year={1700},
title={Dissertatio epistolica Acarnanis de Dendritis aliisque lapidibus, qui
  in superficie sua plantarum, foliorum, florum figuras exprimunt},

post={Miscellanea Curiosa Sive Ephemeridum   Medico-Physicarum Germanicarum Academiae Caesareo-Leopoldinae Naturae   Curiosorum Decuriae III. Annus Quintus Et Sextus, Anni M. DC. XCVII \&   XCVIII. Continens Celeberrimorum Virorum Tum Medicorum tum aliorum Eruditorum   in Germania \& extra eam Observationes   Medico-Physico-Anatomico-Botanico-Mathematicas Cum Appendice \& Privilegio   Sac. Caes. Majestatis},
  pages={57--80},
}

@book{Mylius1709,
title={Memorabilium Saxoni{\ae}Subterrane\ae},
SerialNo={264},
author={{Mylius}, G. F.},
 year={1709},
booktitle={Memorabilium Saxoni{\ae}Subterrane\ae},
 orgname={F. Groschuff},
}

@book{daCosta1757,
title={A Natural History of Fossils},
SerialNo={265},
author={{Mendes da Costa}, E.},
 year={1757},
booktitle={A Natural History of Fossils},
 orgname={L. Davis and C. Reymers},
}

@article{Straaten1978,
SerialNo={266},
author={{Van Straaten}, LMJU},
 year={1978},
title={Dendrites},
journal={{J.} {Geol.} {Soc.}},
 volume={135},
 number={1},
  pages={137--151},
 orgname={The Geological Society of London},
doi={10.1144/gsjgs.135.1.0137},
}

@article{Garcia1994,
SerialNo={267},
author={Garc{\'\i}a-Ruiz, Juan M and Ot{\'a}lora, Ferm{\'\i}n and
  Sanchez-Navas, Antonio and Higes-Rolando, Francisco J.},
 year={1994},
title={The formation of manganese dendrites as the mineral record of flow
  structures},
journal={Fractals {Dynamic} {Sys.} {Geosci.}},
  pages={307--318},
 orgname={Springer},
doi={10.1007/978-3-662-07304-9\_23},
}

@article{Trivedi1994,
SerialNo={268},
author={Trivedi, R. and Kurz, W.},
 year={1994},
title={Dendritic growth},
journal={Int. Mater. Rev.},
 volume={39},
 number={2},
  pages={49--74},
 orgname={Taylor \& Francis},
doi={10.1179/imr.1994.39.2.49},
}

@article{Fowler1989,
SerialNo={269},
author={Fowler, A. D. and Stanley, H. E. and Daccord, G.},
 year={1989},
title={Disequilibrium silicate mineral textures: \?{F}ractal and non-fractal
  features},
journal={Nature},
 volume={341},
  pages={134--138},
doi={10.1038/341134a0},
}

@article{Welsch2013,
SerialNo={270},
author={Welsch, B. and Faure, F. and Famin, V. and Baronnet, A. and
  Bach{\`e}lery, P.},
 year={2013},
title={Dendritic crystallization: \?{A} single process for all the textures of
  olivine in basalts?},
chsep={\chsep[atitle]{~}},
journal={J. Petrol.},
 volume={54},
  pages={539--574},
doi={10.1093/petrology/egs077},
}

@article{Barbey2019,
SerialNo={271},
author={Barbey, P. and Faure, F. and Paquette, J. L. and Pistre, K. and
  Delangle, C. and Gremilliet, J. P.},
 year={2019},
title={Skeletal quartz and dendritic biotite: \?{W}itnesses of primary
  disequilibrium growth textures in an alkali-feldspar granite},
journal={Lithos},
 volume={348--349},
doi={10.1016/j.lithos.2019.105202},
}

@article{Chopard1991,
SerialNo={272},
author={Chopard, B. and Herrmann, H.J. and Vicsek, T.},
 year={1991},
title={Structure and growth mechanism of mineral dendrites},
journal={Nature},
 volume={353},
 number={6343},
  pages={409--412},
 orgname={Nature Publishing Group UK London},
doi={10.1038/353409a0},
}

@article{Li2014,
SerialNo={273},
author={Li, Dongyan and Yang, Jun and Tang, Wenxiang and Wu, Xiaofeng and Wei,
  Lianqi and Chen, Yunfa},
 year={2014},
title={Controlled synthesis of hierarchical {\?{MnO}}2 microspheres with hollow
  interiors for the removal of benzene},
journal={RSC {Adv.}},
 volume={4},
 number={51},
  pages={26796--26803},
 orgname={Royal Society of Chemistry},
doi={10.1039/C4RA01146E},
}

@article{Huang2015,
SerialNo={274},
author={Huang, Ming and Li, Fei and Dong, Fan and Zhang, Yu Xin and Zhang, Li
  Li},
 year={2015},
title={{\?{MnO}}2-based nanostructures for high-performance supercapacitors},
journal={{J.} {Mater.} {Chem.} A},
 volume={3},
 number={43},
  pages={21380--21423},
 orgname={Royal Society of Chemistry},
doi={10.1039/C5TA05523G},
}

@book{Coelfen2008,
title={Mesocrystals and Nonclassical Crystallization},
SerialNo={275},
author={C{\"o}elfen, H. and Antonietti, M.},
 year={2008},
booktitle={Mesocrystals and Nonclassical Crystallization},
 orgname={Wiley},
}

@article{deYoreo2015,
SerialNo={276},
author={{De Yoreo}, James J and Gilbert, Pupa UPA and Sommerdijk, Nico AJM and
  Penn, R Lee and Whitelam, Stephen and Joester, Derk and Zhang, Hengzhong and
  Rimer, Jeffrey D and Navrotsky, Alexandra and Banfield, Jillian F and Adam F.
  Wallace and F. Marc Michel and Fiona C. Meldrum and Helmut C{\"o}lfen and
  Patricia M. Dove},
 year={2015},
title={Crystallization by particle attachment in synthetic, biogenic, and
  geologic environments},
journal={Science},
 volume={349},
 number={6247},
  pages={aaa6760},
 orgname={American Association for the Advancement of Science},
doi={10.1126/science.aaa6760},
}

@article{Mathiesen2006,
SerialNo={277},
author={Mathiesen, Joachim and Procaccia, Itamar and Swinney, Harry L and
  Thrasher, Matthew},
 year={2006},
title={The universality class of diffusion-limited aggregation and viscous
  fingering},
journal={Europhys. Lett.},
 volume={76},
 number={2},
  pages={257},
 orgname={IOP Publishing},
doi={10.1209/epl/i2006-10246-x},
}

@article{Potter1979,
SerialNo={278},
author={Potter, Russel M. and Rossman, George R.},
 year={1979},
title={Mineralogy of manganese dendrites and coatings},
journal={Am. Miner.},
 volume={64},
 number={11--12},
  pages={1219--1226},
 orgname={Mineralogical Society of America},
}

@article{Schoenly1993,
SerialNo={279},
author={Schoenly, Paul A. and Saunders, James A.},
 year={1993},
title={Natural gold dendrites from hydrothermal {Au-Ag} deposits:
  characteristics and computer simulations},
journal={Fractals},
 volume={1},
 number={03},
  pages={585--593},
 orgname={World Scientific},
doi={10.1142/S0218348X93000617},
}

@article{Saunders2017,
SerialNo={280},
author={Saunders, James A. and Burke, Michelle},
 year={2017},
title={Formation and aggregation of gold (electrum) nanoparticles in
  epithermal ores},
journal={Minerals},
 volume={7},
 number={9},
  pages={163},
 orgname={MDPI},
doi={10.3390/min7090163},
}

@article{Saunders1994,
SerialNo={281},
author={Saunders, James A.},
 year={1994},
title={Silica and gold textures in bonanza ores of the {Sleeper Deposit,
  Humboldt County, Nevada}; evidence for colloids and implications for
  epithermal ore-forming processes},
journal={{Econom.} {Geol.}},
 volume={89},
 number={3},
  pages={628--638},
 orgname={Society of Economic Geologists},
doi={10.2113/gsecongeo.89.3.628},
}

@article{Saunders2020,
SerialNo={282},
author={Saunders, James A. and Burke, Michelle and Brueseke, Matthew E.},
 year={2020},
title={Scanning-electron-microscope imaging of gold (electrum) nanoparticles
  in middle \?{Miocene} bonanza epithermal ores from northern {Nevada}, {{\?{USA}}}},
journal={{Miner.} Deposita},
 volume={55},
  pages={389--398},
 orgname={Springer},
doi={10.1007/s00126-019-00935-y},
}

@book{Keller1990,
title={Gemstones and Their Origin},
SerialNo={283},
author={Keller, P.C.},
 year={1990},
booktitle={Gemstones and Their Origin},
 orgname={Springer},
}

@article{Papineau2020,
SerialNo={284},
author={Papineau, D.},
 year={2020},
title={Chemically oscillating reactions in the formation of botryoidal
  malachite},
journal={Am. Miner.},
 volume={105},
 number={4},
  pages={447--454},
 orgname={Mineralogical Society of America},
doi={10.2138/am-2020-7029},
}

@article{Balitsky1987,
SerialNo={285},
author={Balitsky, V. S. and Bublikova, T. M. and Sorolzjna, S. L. and
  Balitskaya, L. V. and Shteinberg, A. S.},
 year={1987},
title={Man-made jewelry malachite},
journal={Gems {Gemol.}},
 volume={23},
  pages={152--157},
doi={10.5741/GEMS.23.3.152},
}

@article{Petrov2013,
SerialNo={286},
author={Petrov, T.G. and Protopopov, E.N. and Shuyskiy, A.V.},
 year={2013},
title={Decorative grown malachite. \?{Nature} and technology},
journal={{Russi} {J.} Earth {Sci.}},
 volume={13},
  pages={ES2001},
doi={10.2205/2013ES000529},
}

@article{craw2016gold,
SerialNo={287},
author={Craw, Dave and Lilly, Kat},
 year={2016},
title={Gold nugget morphology and geochemical environments of nugget
  formation, southern {\?{New Zealand}}},
journal={Ore Geol. Rev.},
 volume={79},
  pages={301--315},
 orgname={Elsevier},
doi={10.1016/j.oregeorev.2016.06.001},
}

@article{butt2020gold,
SerialNo={288},
author={Butt, C. R. M. and Hough, R. M. and Verrall, M.},
 year={2020},
title={Gold nuggets: the inside story},
journal={Ore Energy {Resource} {Geol.}},
 volume={4},
  artnum={100009},
 orgname={Elsevier},
doi={10.1016/j.oreoa.2020.100009},
}

@article{curtis_review_2019,
SerialNo={289},
author={Curtis, Neville J. and Gascooke, Jason R. and Johnston, Martin R. and
  Pring, Allan},
 year={2019},
title={A {review} of the {classification} of {opal} with {reference} to
  {recent} {new} {localities}},
journal={Minerals},
 volume={9},
 number={5},
  pages={299},
doi={10.3390/min9050299},
}

@article{gaillou_common_2008,
SerialNo={290},
author={Gaillou, Elo\"{\i}se and Fritsch, Emmanuel and Aguilar-Reyes, Bertha and
  Rondeau, Benjamin and Post, Jeffrey and Barreau, Alain and Ostroumov,
  Mikhail},
 year={2008},
title={Common gem opal: \?{A}n investigation of micro- to nano-structure},
journal={Am. Miner.},
 volume={93},
 number={11--12},
  pages={1865--1873},
doi={10.2138/am.2008.2518},
 issn={0003-004X},
}

@article{curtis_silicon-oxygen_2021,
SerialNo={291},
author={Curtis, Neville J. and Gascooke, Jason R. and Pring, Allan},
 year={2021},
title={Silicon-{oxygen} {region} {infrared} and {\?{Raman}} {analysis} of {opals}:
  \?{T}he {effect} of {sample} {preparation} and {measurement} {type}},
journal={Minerals},
 volume={11},
 number={2},
  pages={173},
doi={10.3390/min11020173},
}

@article{ILER1965Formation,
SerialNo={292},
author={Iler, R. K.},
 year={1965},
title={Formation of Precious Opal},
journal={Nature},
 volume={207},
 number={4996},
  pages={472--473},
doi={10.1038/207472a0},
 issn={0028-0836},
}

@article{Norris2004Opaline,
SerialNo={293},
author={Norris, D. J. and Arlinghaus, E. G. and Meng, L. and Heiny, R. and
  Scriven, L. E.},
 year={2004},
title={Opaline Photonic Crystals: \?{H}ow Does Self-Assembly Work?},
chsep={\chsep[atitle]{~}},
journal={Adv. Mater.},
 volume={16},
 number={16},
  pages={1393--1399},
doi={10.1002/adma.200400455},
 issn={0935-9648},
}

@article{Stewart2010Self,
SerialNo={294},
author={Stewart, A.M. and Chadderton, Lewis T. and Senior, Brian R.},
 year={2010},
title={Self-assembly in the growth of precious opal},
journal={J. Cryst. Growth},
 volume={312},
 number={3},
  pages={391--396},
doi={10.1016/j.jcrysgro.2009.09.042},
 issn={0022-0248},
}

@article{stober_controlled_1968,
SerialNo={295},
author={St{\"o}ber, Werner and Fink, Arthur and Bohn, Ernst},
 year={1968},
title={Controlled growth of monodisperse silica spheres in the micron size
  range},
journal={J. Colloid Interface Sci.},
 volume={26},
 number={1},
  pages={62--69},
doi={10.1016/0021-9797(68)90272-5},
 issn={0021-9797},
}

@article{bogush_preparation_1988,
SerialNo={296},
author={Bogush, G. H. and Tracy, M. A. and Zukoski, C. F.},
 year={1988},
title={Preparation of monodisperse silica particles: \?{C}ontrol of size and
  mass fraction},
journal={J. Non Cryst. Solids},
 volume={104},
 number={1},
  pages={95--106},
doi={10.1016/0022-3093(88)90187-1},
 issn={0022-3093},
}

@article{Liesegang2014Australian,
SerialNo={297},
author={Liesegang, M. and Milke, R.},
 year={2014},
title={{Australia}n sedimentary opal-\?A and its associated minerals:
  \?{I}mplications for natural silica sphere formation},
journal={Am. Miner.},
 volume={99},
 number={7},
  pages={1488--1499},
 orgname={Mineralogical Society of America},
doi={10.2138/am.2014.4791},
 issn={0003-004X},
}

@article{gao_facile_2016,
SerialNo={298},
author={Gao, Weihong and Rigout, Muriel and Owens, Huw},
 year={2016},
title={Facile control of silica nanoparticles using a novel solvent varying
  method for the fabrication of artificial opal photonic crystals},
journal={{J.} Nanoparticle {Res.}},
 volume={18},
 number={12},
  pages={387},
doi={10.1007/s11051-016-3691-8},
 issn={1572-896X},
}

@article{Migaszewski2006,
SerialNo={299},
author={Migaszewski, Zdzis{\l}aw M and Ga{\l}uszka, Agnieszka and Durakiewicz,
  Tomasz and Starnawska, Ewa},
 year={2006},
title={{Middle Oxfordian--Lower Kimmeridgian} chert nodules in the {Holy Cross
  Mountains}, south-central {\?{Poland}}},
journal={{Sedimentary} {Geol.}},
 volume={187},
 number={1--2},
  pages={11--28},
 orgname={Elsevier},
doi={10.1016/j.sedgeo.2005.12.003},
}

\end{document}